# Automated Tool Support for Category-Partition Testing—Design Decisions, UI and Examples of Use

**Yvan Labiche, IEEE Senior Member**
**Carleton University, Department of Systems and Computer Engineering**
**Ottawa, ON, Canada**
*yvan.labiche@carleton.ca*

*Abstract*—Category-Partition is a functional testing technique that is based on the idea that the input domain of the system under test can be divided into sub-domains, with the assumption that inputs that belong to the same sub-domain trigger a similar behaviour and that therefore it is sufficient to select one input from each sub-domain. Category-Partition proceeds in several steps, from the identification of so-called categories and choices, possibly constrained, which are subsequently used to form test frames, i.e., combinations of choices, and eventually test cases. This paper reports on an ongoing attempt to automate as many of those steps as possible, with graphical-user interface tool support. Specifically, the user interface allows the user to specify parameters as well as so-called environment variables, further specify categories and choices with optional constraints. Choices are provided with precise specifications with operations specific to their types (e.g., Boolean, Integer, Real, String). Then, the tool automates the construction of test frames, which are combinations of choices, according to alternative selection criteria, and the identification of input values for parameters and environment variables for these test frames, thereby producing test cases. The paper illustrates the capabilities of the tool with the use of nine different case studies.

## I. INTRODUCTION

Functional testing aims to identify discrepancies between the behaviour actually implemented in a software system and the desired behaviour of this system as provided in its functional specification. In practice, functional testing is performed on the basis of a specification that describes the expected system behaviour, and test cases are derived and executed accordingly. It is typically complemented by a confirmation that structural coverage objectives have been reached, and more functional testing should take place if this is not the case [1]. Various modeling and specification languages can be used to draft a functional specification for testing purposes, including plain language [2].

Category-partition (CP) [3] is a functional testing technique that specifically targets plain language specifications, and has shown to be useful in several contexts [3, 4, 5, 6, 7, 8, 9]. With CP [3] one first identifies functions/functionalities to be tested in isolation and for each one of them identifies input parameters. One also identifies so-called environment conditions/variables, which are aspects of the environment of execution of the tested function/functionality that can potentially impact its behaviour. For instance, if the function under test takes a file as input, a common interpretation is to consider the name of the file as a parameter and the contents of the file as an environment variable [3]. Parameters and environment variables are further specified with categories, which are characteristics of the input parameter or environment variable that are deemed important from a testing point of view. Categories provide alternative, orthogonal perspectives on what is deemed interesting from a testing point of view about a parameter (or an environment variable). Each category leads to the definition of choices, which are specifying how the domain of values implicitly specified by the category is split into sub-domains, for instance using equivalence class partitioning and boundary value analysis [10]. Choices from different categories are then combined according to a selection criterion to form test frames. For instance, the Each-Choice criterion requires that the set of test frames exercises each choice of each category at least once. Test frames are also called test case specifications since they specify conditions (specified as choices) that test inputs should satisfy [11]; solving the conditions imposed by a test frame leads to identifying test inputs (i.e., a test case). Constraints can also be used to specify that some choices from two different categories should always be used together, can never be used together, or can only be used together under certain circumstances, or that it will be sufficient to involve some choices only once in the entire set of test frames. We refer to the information comprising parameters, environment variables, categories, choices and constraints as the CP specification. Once test frames have been identified from a CP specification, such test case specifications are used to identify test inputs for input parameters and environment variables, to eventually help produce test scripts. We note that prior to producing test frames and test cases, one should verify the CP specification; among other things, one should for instance verify that no two choices of a category overlap and that all the categories of a choice split the entire domain of values defined by the category, in other words that any value in the domain defined by the category belongs to one and only one choice of that category.

As can be noticed from this description, CP involves several steps during which critical decisions are made. Producing a satisfactory CP specification is notoriously difficult, which has led some authors to recommend good practices to prevent pitfalls [8, 12, 13]. This calls for automated support to help users create CP specifications and verify them according to precise criteria. Additionally, the impact on cost, for instance in terms of number of tests eventually created, and effectiveness at revealing faults, of selection criteria one can use to create test frames is relatively unknown. Similarly, strategies to identify test inputs from test frames have also not been studied thoroughly, with a very few exceptions [14, 15, 16]. Again, without tool support it is essentially impossible to answer those questions. Last, issues with the CP specification may only become apparent once test inputs have been created [3]; the more automation, the sooner these situations can be discovered and the CP specification corrected.

This paper describes an on-going attempt to automate the use of CP. With our solution, once a CP specification is provided thanks to a user interface, the user can automatically verify the CP specification (e.g., verifying that choices in a category do not overlap), can automatically obtain test frames according to either one of four selection criteria, and then automatically obtain test cases (i.e., test input values) for those test frames, provided the input parameters and environment variables have appropriate types (i.e., Integer, Real, Boolean, String, Character, Set of integers, Enumeration). As evidenced later in the paper, it is, to the best of our knowledge, the only tool support of CP to date that achieves such a level of automation.

The rest of the paper is structured as follows. First (Section II), we provide additional background on CP. We then discuss related work (Section III), specifically focusing on automated support for CP. Section IV discusses our tool support for CP. Section V describes examples of use. We conclude in Section VI.

## II. BACKGROUND ON CATEGORY-PARTITION

We illustrate the use of CP and the concepts of parameter, category, choice and constraints on two examples and Figure 1: one defines a CP specification from the specification of the software under test (step 1 in the figure), and verifies the CP specification (step 2) before using it to produce test frames (step 3) and then test cases (step 4).

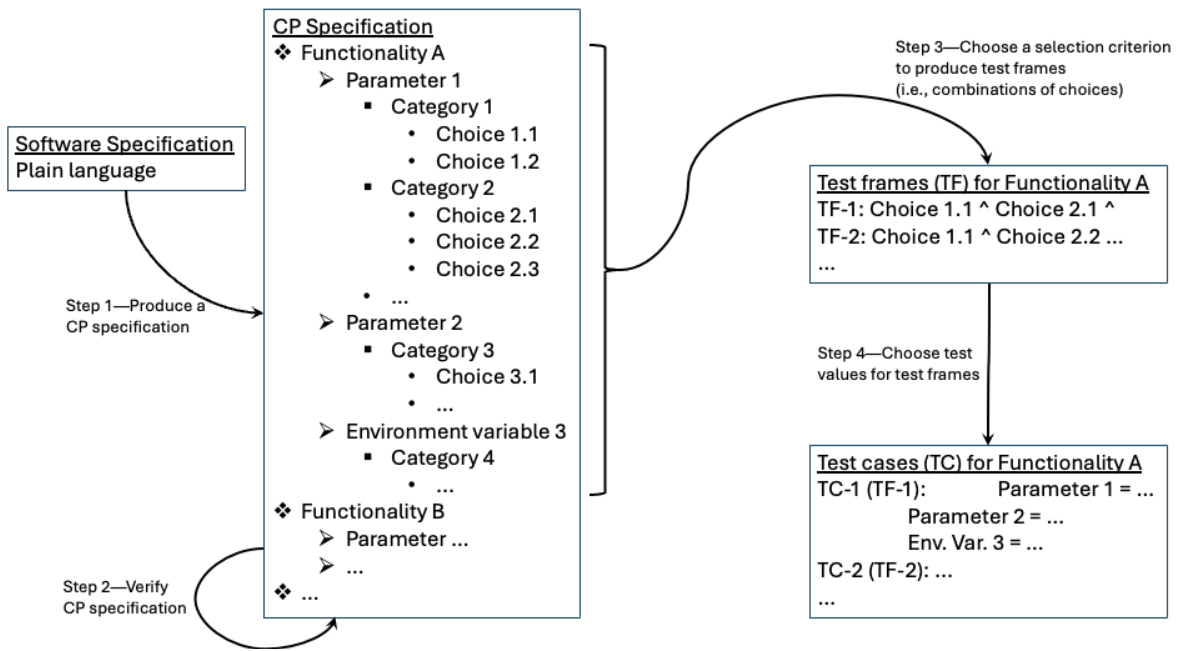

Figure 1.    Illustrating the flow of CP and the main CP concepts

## II.A.    *Example 1*

The problem of the triangle classification is typically used in textbooks to illustrate CP [11]. The software specification states that the function accepts three side lengths as inputs and determines whether the three values form a triangle and, if they do, also determines the type of triangle. Three lengths form a valid triangle if they satisfy the triangle inequalities, which state that the sum of the lengths of any two sides must be greater than the length of the remaining side. If any of these conditions is not met, the lengths cannot form a valid triangle. Assuming the lengths form a triangle, the triangle is either equilateral if all three lengths are equal, isosceles if exactly two sides are equal (but not three), or scalene if all three sides are different.

Applying CP from this specification (step 1 in Figure 1), we identify the program has only one functionality with three input parameters, SA, SB and SC. No condition in the environment of execution of the function under test can foreseeably impact its execution, so there is no environment variable. We define a first category (Cat1) for parameter SA, which is about its values; the category specifies a domain of values which we split (choices) into two: choice 1.1 (choice 1 of category 1) is for strictly positive SA values (`SA>0`) and choice 1.2 is for negative or zero values (`SA≤0`).

We constrain choice 1.2 as an "error" choice, meaning that a test script using an SA value in this domain should expect an error message. Admittedly, providing a value of zero is not an error since mathematically a length can be zero (as opposed to negative), and one could therefore consider defining choice 1.2 as all strictly negative values (`SA<0`), which would still be an "error" choice, and create a third choice 1.3 for the singleton value zero (`SA=0`) and constrain this choice as "single" (meaning that it should be sufficient to exercise this boundary value only once in a complete testing campaign); Providing a value of zero is however not very interesting for the triangle problem and we consider having two choices as defined, instead of three, is sufficient. Indeed, if any value equals zero, there is no triangle. Deciding whether one CP specification, here with two choices for category 1, is better than another, i.e., with three choices, is often a personal decision of the user. We note that thanks to automated support for CP such as the one we created, one can study the impact of decisions one makes during the construction of the CP specification, such as the impact on the number of test frames and therefore test cases that will eventually be created. Specifically, one may be generous with the specification of many choices for boundary values and values just outside boundaries [17], resulting in many test frames, realize that this number of test frames (and therefore test cases) is not practical for the system under test, and therefore decide to revisit the CP specification to reduce the number of choices. As several iterations of this process may be needed, automated support is key.

We also define a category (Cat2) to compare values of SA to the sum of the other two lengths (SB and SC), and identify two choices: choice 2.1 specifies that SA is strictly smaller than the sum of the other two values (`SA<SB+SC`); choice 2.2 specifies that SA is greater than or equal to that sum (`SA≥SB+SC`). Last, we identify a third category for SA, which is to compare the value of SA to that of SB, with two choices: choice 3.1 specifies that SA equals to SB (`SA=SB`) whereas choice 3.2 specifies they have different values (`SA≠SB`).

We proceed similarly for length SB: category 1 is about its values with two choices (strictly positive values, negative or zero values); category 2 is to compare SB to SC with two choices (either the lengths are equal or they are different); category 3 compares SB to SA+SC with two choices (the triangle inequality holds or not). We proceed similarly for parameter SC: three categories, each one having two choices. The CP specification has therefore three parameters, a total of nine categories and 18 choices.

A slightly different CP specification could be considered by for instance singling out boundary conditions. Specifically, choice 2.2 for parameter SA, which specifies that `SA≥SB+SC` could be split into two choices, one specifying that `SA=SB+SC` and the other that `SA>SB+SC`. Applying this principle to each choice which has an inequality leads us to defining six additional choices, thus resulting in a total of 24 choices for the same number of parameters and categories. The complete CP specification with 18 choices, along with test frames and test cases generated by our tool, can be found in Appendix A.

One criterion for the construction of test frames (step 3 in Figure 1), specifically the base choice criterion, asks that one and only one of the choices in a category be tagged as a base choice [11, 18]. A base choice is one that is expected to be most often selected when the subject under test is used. In the CP specification discussed above, choices ensuring that triangle inequalities hold are base choices.

Even on a simple example such as the Triangle problem, constraints between choices must be specified to prevent certain choices combinations; recall that a test frame, which is a test case specification, is a combination of choices. For example, the three choices that specify that `SA>SB+SC`, `SC>0`, and `SA=SB` should not be combined in a test frame since otherwise they together lead to a contradiction (specifically, that `SC>0` and `SC<0`) for the identification of input parameter values that satisfy the conditions of a test frame that involves them.

## II.B.    *Example 2*

The second example we wish to briefly discuss (see further details in Section V) is that of a function under test that takes two parameters as input, one being a string and the second being a character; the function identifies the position of the first occurrence of the character in the string.

Applying CP, we identify two parameters and no environment variable. The first category we identify for the input string parameter is about its length; the first choice of the category (choice 1.1) specifies an empty string whereas the second choice specifies a nonempty string (choice 1.2). A second category of the input string parameter is about the number of occurrences of the character in the string; the character can be absent from the string (choice 2.1), it can appear only once (choice 2.2), or it can appear more than once strictly (choice 2.3). We identify only one category for the character parameter which is about the position of its first occurrence in the string; the first occurrence of the character can be the first character in the string (choice 3.1), the last character in the string (choice 3.2), or somewhere else in the string, i.e., neither the first nor the last character (choice 3.3).

This second example is interesting since it demonstrates that a category really discusses a characteristic of the corresponding parameter that is deemed important from a testing point of view rather than necessarily possible values of the parameter itself. For example, the category for the character parameter does not specify values of the character, such as being an upper-case character, a digit, …; instead, it is about a characteristic of a character, specifically the position of its first occurrence in the string. Consequently, the set of values of the domain specified by a category and split into choices is not necessarily the domain of values of the corresponding parameter: here, the category specifying the position of the character specifies a domain that is not the domain of possible values for the character.

This example also allows us to further illustrate constraints between choices. For instance, it does not make sense to discuss the occurrence of the character in the string if the string is empty; it doesn't make sense to specify that the character appears more than once if the first occurrence of the character is the last character of the string. Each situation leads to creating constraints on the corresponding choices to prevent their use together in a test frame. Such constraints can be specified with predicates [3] as illustrated later in this paper (Section V).

Step 2 in Figure 1 is about the verification of the CP specification. When using equivalence class partitioning during the specification of choices, it is important to specify that (1) choices do not overlap, that is no value can belong to two different choices, and (2) choices altogether represent all the possible values of the domain specified by the category. For short CP specifications this is relatively easy to achieve by hand, although this has been recognized as a main source of errors in CP specifications [8, 12, 13]. For larger specifications and/or when choices' specifications are complex, tool support is necessary. In the case of the triangle problem, this verification is trivial: it is clear for instance that choices 1.1 (`SA>0`) and 1.2 (`SA≤0`) do not overlap (it is impossible for a value of SA to satisfy both at the same time) and represent the entire domain of values (the set of values that satisfy neither conditions is empty). In our second example, this is a bit trickier: choices "the first occurrence of the character is the first character in the string", "the first occurrence of the character is the last character of the string" and "the first occurrence of the character is neither the first nor the last character of the string" clearly do not overlap and represent the entire domain of values. Providing tool support so that software can confirm this automatically is a challenge, unless the choices are precisely specified as we will discuss in section IV.

## III. RELATED WORK

Since we describe an attempt to automatically support as many steps of CP [3] as possible, we focus our discussion of related work to similar attempts and put aside descriptions of the use of Category-Partition. To identify related work, we proceeded in two stages. First, we searched on Scopus and Web of Science for work that discuss tool support for Category-Partition; we searched in abstracts, titles and keywords with the following query, limiting results to sources written in English but not limiting the search period: "`category partition`" and "`tool*`" and "`test*`". As we already knew of two attempts to provide tool support for Category-Partition, the one by the original contributors of Category-Partition [4] and that of Chen and colleagues [19] we also conducted a forward snowballing search from these references. Altogether, we obtained a list of 162 papers, which we analyzed by considering the title and reading the abstract, and when needed also reading the entire paper.

A large majority of the papers we identified are not relevant to our purpose and sometimes come from other fields (e.g., biology). Others simply mention CP in passing or to provide some functional testing context. Yet another significant proportion of the set are applying CP without discussing any tool support. We are left with four different threads of research that discuss tool support for CP, which we analyze next.

Ostrand and colleagues defined the Test Specification Language (TSL) and tool support alongside CP [3, 4]. In their original work the user manually writes the CP specification in a text file, following the TSL syntax. The authors provide little guidance as to how choices should be combined, that is which selection criteria are to be considered for producing test frames. The only selection criterion being implicitly used is All-Choices-Combinations, which requires that all feasible (i.e., accounting for choice constraints) combinations of choices be exercised by the set of test frames. The user can optionally provide one (typical) value for each choice in the CP specification and those values are used when transforming test frames into test cases. Our tool differs a great deal from Ostrand and colleagues' solution. It provides a graphical user interface (UI) for the definition of the CP specification, which should be more convenient than editing a possibly long text file; Although our tool does not support the All-Choices-Combinations criterion, adding it is entirely feasible as discussed later in the paper. Ostrand and colleagues' decision to rely on one possible value for each choice to be used when creating test cases from test frames may turn out to pause problems since it can be difficult to a-priori know which conditions input parameters and environment variables should satisfy when accounting for the specification of all the choices involved in a test frame. If possible values are not provided, their solution delegates the task of identifying input values from test frames to the user. Instead, our tool delegates the task of identifying test inputs for test frames to a Satisfiability Modulo Theory (SMT) solver. Our solution also supports the verification of the CP specification, which Ostrand and colleagues did not consider.

An attempt to automate CP [20] relies on the SMT Z3 solver [21] and a custom-made genetic algorithm to generate test frames and test data. There is however no clear description of how constraints between choices are handled; we instead rely on well-established solutions from combinatorics to create test frames, thereby attempting to minimize the number of test frames that satisfy a selection criterion while accounting for choice constraints. Their solution does not offer the user the possibility to select among alternative selection criteria when creating test frames and does not provide any UI, especially for the creation of the CP specification. We instead use covering array technologies as they have been specifically designed for the purpose of combining alternatives (in our case, choices), while accounting for constraints, to produce a minimum number of combinations (tests). Their solution is also limited to subjects with input parameters of type integer, whereas we support other types. Last, that solution does not take care of verifying a CP specification satisfies adequate properties.

Wiederseiner and colleagues provide tool support to combine choices (although they do not use this terminology as they do not attempt to support CP) using predefined template choice specifications for select input parameter types [22]. For instance, the int8 and int32 types define boundaries for maximum and minimum integer values that can be used in a template for any variable of these types. Choices are not further specified according to the specification of the software under test: e.g., the software may have different behaviours if the int32 values is greater than, say 20 or not. Choices are combined irrespective of possible constraints (the UI does not allow the specification of constraints). Despite this reduced set of capabilities, the definition of template choice definitions, similarly to what others have suggested in the context of CP [12, 13, 23, 24], is interesting from a tool support point of view to speed up the correct construction of a CP specification.

The work of Poon and colleagues [8, 9, 13, 19, 24, 25] is the one most related to ours. They also provide a UI to help the user record categories and choices. Although they recognize that users of CP can easily produce erroneous CP specifications, they did not attempt to provide tool support for the verification of the CP specification. We note that, contrary to our solution, choices are only given a description, whereas in our solution we allow the user to provide a precise specification, which allows us to automatically verify properties of the CP specification and to automatically produce test inputs for test frames, whereas Poon and colleagues let the user deal with these difficult tasks. Their solution only supports the All-Choices-Combinations criterion which, although leading to a comprehensive list of test frames, often leads to a very large number of test frames and therefore test cases. This can become impractical for complex situations, which led the authors to discuss a user-driven prioritization solution of test frames during the generation of test cases. They provide their own implementation for this criterion without recognizing that covering array algorithms exist for this kind of problem (i.e., solutions exist to satisfy a selection criterion while minimizing the number of test frames and therefore test cases); Although this should be studied further, we believe their solution might not produce a minimum number of test frames, and therefore test cases for this criterion. Additionally, support for alternative selection criteria can give sufficient information to the user to decide how much effort (e.g., how many test cases) they are willing to spend on the testing activity and tool support to study the impact of alternative criteria is then crucial.

Last, we believe their solution may be a slight, though important, deviation from the original CP method. Indeed, in their 2004 paper [24], they write: "Let $[X_K]$ denote the category associated with the factor K;" where a factor represents either a parameter or an environment variable. This text indicates that each parameter or environment variable can only have one category. On the contrary, the CP method does not include this restriction; and indeed, our experience with CP suggests that a parameter (or environment variable) can have more than one category. This issue is somewhat confirmed in another of their papers [9] where the authors write: "A test case is formed from every valid choice combination … [i.e., test frame] by randomly selecting and combining an instance from each choice …" This selection of input values separately from each choice may work if each parameter (or environment variable) has only one category or when values in domains specified by different categories and therefore sub-domains specified by choices do not have any semantic relations that constrain the respective values. Otherwise, the statement reported above will not always lead to a feasible test case, that is input values for the parameters and environment variables that together satisfy the specification of each combined choice. For example, as reported later in this paper, when testing an implementation of the Triangle problem, we obtain a test frame that combines choice specifications "Edge1 > 0", "Edge1 == Edge2", and "Edge2 > 0" with other choices, meaning that those conditions must be satisfied at the same time by the input values of the parameters "Edge1", "Edge2" … If a value is randomly selected separately for each choice and then combined, as Poon and colleagues suggest (see the quote above), then one (or an algorithm supporting this step of CP) may select "Edge1 = 1" for the first choice, "Edge1 = Edge2 = 2" for the second choice and "Edge2 = 3" for the third choice, which clearly leads to inconsistent input values for a test case. Of course, this is a non-issue when the test input identification is performed by a user since that person can take into account all the information carried by a test frame when identifying input values; the procedure however does not necessarily work if we wish to automate the task, as we discuss in this paper. We argue that the specifications of all the choices involved in a test frame must be considered together when looking for input values and this is an important difference our solution has with Poon and colleagues'. Of course, on such a simple (Triangle) example, the user would not make the above mistake; however, with a more complex system under test and a larger CP specification, it would be easier to make the mistake.

To the best of our knowledge, the above are the only attempts to automate the use of CP. Our analysis of these attempts is summarize in Table 1, where we compare them against our solution (last column) along seven dimensions (rows): whether and to what extent a user interface is available to support CP; whether a solution supports the verification of a CP specification; which selection criteria are used to produce test frames; what technology is used to produce those test frames; how constraints on choices are handled; how test inputs for test frames are identified; which types of specification parameters are handled. From the table, it is clear that our solution provides the highest level of automation for CP.

Table 1    Summary of characteristics of CP tool supports

| | Ostrand and colleagues [3, 4] | Chimisliu and Wotawa [20] | Wiederseiner [22] | Poon and colleagues [8, 9, 13, 19, 24, 25] | Our solution |
|---|---|---|---|---|---|
| User interface | No (STL language) | No (XML file) | Partial (e.g., no specification of constraints) | Partial (e.g., no specification of constraints) | Yes (Full support of CP) |
| CP specification verification | Manual (by user) | Manual (by user) | Manual (by user) | Manual (by user) | Automated |
| Test frame construction criteria | All combinations | All combination | All combinations | All combinations | Each-Choice, Pairwise, three-way, Base Choice |
| Test frame construction tool | Not relevant (manual) | Specifically crafted genetic algorithm | Microsoft TestAPI | Specifically crafted algorithm | Combinatorics tool |
| Choice constraints handling | Manual | Unclear | No | Unclear | Automated |
| Test input values | User provided | Z3 | User provided | User provided | Z3 |
| Specification parameter types | Not relevant (manual) | Integer | int8, int32 | Not relevant (manual) | Integer, Character, Enumeration, String, Boolean, Real, Integer Set |

CP has of course been used, sometimes with necessary, though incomplete for our purpose, tool support. Worthy of mention are the following. TDE/UML [6, 7] relies on CP data embedded in a Unified Modeling Language (UML) model to drive the selection of test inputs; TDE/UML focuses on the identification of categories and choices, along with values, for variables that are used in UML diagrams. It also allows the specification of a probability value for choices, which is used during test case generation. Ammann and Offutt [18] map CP constructs to Z constructs to automate the construction of test frames for the Base-Choice criterion they define. The mapping to Z helps ensure that choice combinations are feasible. Our tool offers a more general procedure to generate feasible test frames, thanks to its use of third-party software producing covering arrays. Grochtmann and colleagues' CTL tool (now Testona[1]) supports the Classification Tree functional testing technique [26, 27]. Similarly to CP, CTL relies on user-defined typical values for choices. Z can be used to help identify and formalize categories and choices in a CTL tree [28]. Classification trees have also been extensively studied by Chen and colleagues: e.g., [25]. The method and tool has also evolved to offer additional capabilities such as promoting the use of specific choices over others or prioritizing tests [29]. From our perspective, in this paper, one main difference with CP is the graphical representation of the specification of parameters, categories and choices under the form of a tree and the precise specification of the categories/choices tests exercise so as to allow automated test input identification for various parameter types (CTL/Testona leaves this test input identification task to the user). Our solution provides a higher level of automation. Briand and colleagues rely on CP to map existing tests, i.e. test input data, to test frames and learn the input-output relationships that are actually exercised by an existing test suite [30]. This information can then be used to help the user refine the categories and choices as well as refine the test suite (e.g., trimming redundant tests, adding needed tests to exercise other combinations of choices). The solution does not support CP per se, i.e., producing test frames and corresponding test input values; it does not support the construction of adequate test cases by following the CP method and is more limited than our solution in terms of parameter types it can handle.

---

[1] https://products.expleogroup.com/testona/

Chen and colleagues have developed a series of recommendations for the identification of appropriate categories and choices, some of which could be supported by software [9, 13, 19, 24]. Their tool support does not implement these recommendations whereas our solution provides automated support for some of them. Khalsa [31] developed algorithms to ensure that the Base-Choice criterion [18] subsumes the Each-Choice criterion even in the presence of complex constraints between choices. These pieces of work are complementary to our solution since they help solve individual CP specification construction steps without providing a solution that is as complete as ours.

## IV. AUTOMATING THE USE OF CP—THE MELBA APPROACH

The UML package diagram of Figure 2 shows the main third-party packages Melba relies on to automate as much of CP as possible. First, Melba has a Java FX UI (version 17.0.10), which brings a current look and feel and will allow our solution to eventually be available on the web. Melba also has a command line option. Our solution relies on Simple[2] (version 2.7.1) for high performance XML serialization of CP specification data into (and from) files. It also uses Janino[3] (version 2.5.15), a small, fast Java compiler to compile user-provided Java expressions (section IV.B). Last, Melba relies on CASA [32, 33] or ACTS [34], which implement covering array solutions, for test frame generation, and SMT solver Z3 [21] (version 4.15.3) for test input generation from test frames: see Sections IV.D and IV.E, respectively. We also use Z3 and Janino when verifying a CP specification (Section IV.C).

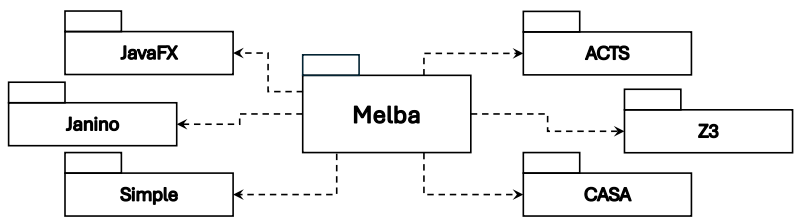

Figure 2.        Melba's Architecture (UML)

Not surprisingly, the model classes of Melba represent the CP specification concepts (Figure 3): a system under test has a CP specification which contains parameters (although not shown, Melba distinguishes between parameters and environment variables) which have a type (Melba supports various alternative types as discussed later in section IV.B) and categories; a category has choices, one of which is a base choice; a choice has a plain text description, has an expression specification, can be single or error (or none of those), has a selector (to describe which other choices it should or should not be combined with), and defines properties (to be used in other choices' selectors).

We illustrate the use of the tool and some of its design aspects through its UI: overview of the UI (Section IV.A), definition of a CP specification (Section IV.B), verification of the CP specification (Section IV.C), generation of test frames (Section IV.D) and test inputs (Section IV.E). When providing screenshots of Melba's UI we cropped images so they can easily fit on a page while making important aspects discussed in the text legible.

Although our decision to illustrate our tool through its UI may make the tool seem simple, this is primarily due to the UI. The technologies we rely on are not simple and the overall design of the tool is original. The software is written in Java and JavaScript (for the FX UI) and made of more than 5,300 LOC of Java and

---

2,500 LOC of JavaScript (not counting comments or blank lines). These numbers do not count third party libraries. When illustrating our solution, we will use the well-known Triangle problem (Section II).

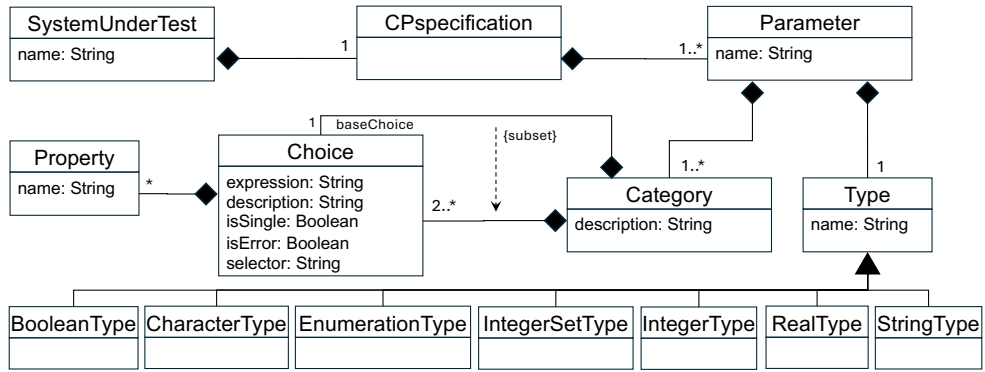

Figure 3.        Melba's model classes (UML class diagram)

## IV.A.    Overview of the UI

The menus of the user interface (Figure 4) are not surprising since they allow the user to follow the steps of CP. The *File* menu, Figure 4 (a), allows the user to define a new CP specification from scratch, save/export the one that is currently being defined to a file, or load one from a file. The *Define CP* menu, Figure 4 (b), allows the user to specify parameters and environment variables, categories, and their choices (Section IV.B). The *Actions* menu, Figure 4 (c), is used to generate test frames, according to a selection criterion (Section IV.D), and test inputs for produced test frames (Section IV.E), as well as trigger the verification of the CP specification (Section IV.C). The *View* menu, Figure 4 (d), allows the user to visualize the CP specification in a pretty-printed text format, the generated test frames, the generated test cases (input values), or the log for any error message. The *Help* menu (not shown) provides guidance to the user and credits.

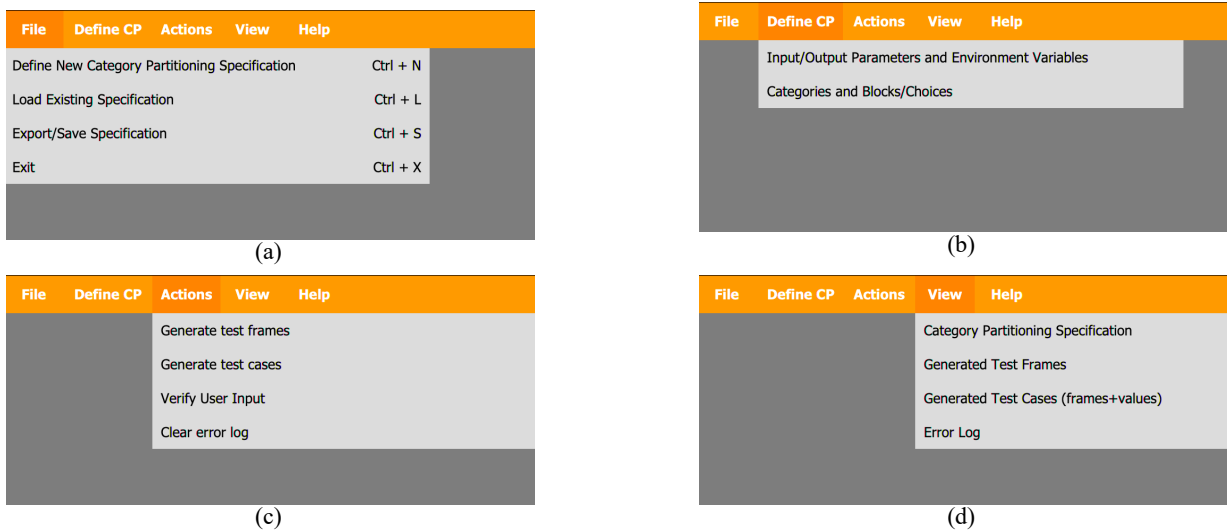

Figure 4.        Menus of Melba's User Interface

## IV.B.    Defining a CP Specification

Input parameters and environment variables have a name and a type. The parameter types Melba currently supports (drop-down list in Figure 5) are Boolean, Character, Enumeration, Integer, IntegerSet, Real and String. All types are self-explanatory except perhaps IntegerSet; a variable of this type holds a set of integer values. The user interface uses collapsible cards to help focus on specific elements: see symbols ˅ to expand a card and ⌄ to collapse a card at the top-right of a card. The **+** symbol allows the user to add a card (in Figure 5, either an input parameter or an environment variable) while the ⊟ and ⊟ symbols are used to save and delete a card (in Figure 5, a parameter or an environment variable), respectively.

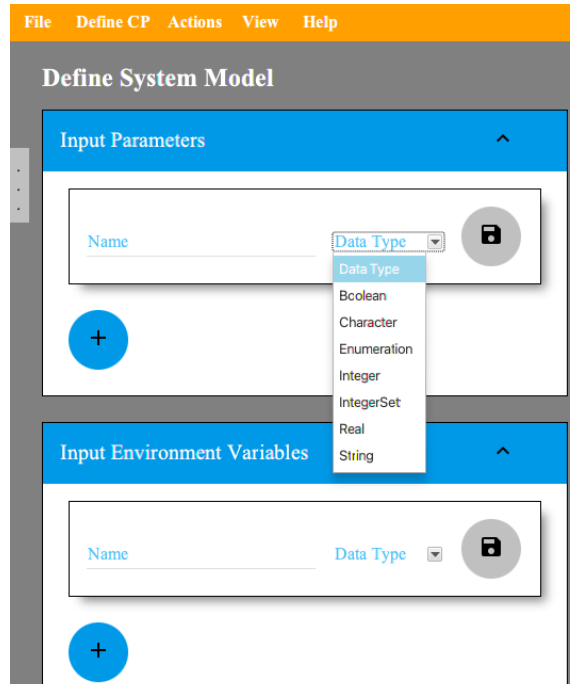

Figure 5.    Definition of a parameter or environment variable (name and type)

Each parameter (or environment variable) is specified with categories and each category has a plain language description, for easy reference, and choices (also referred to as blocks [11]): Figure 6 (a). The figure shows the specification of three categories for the input parameter SA of type Integer: category "Values of SA", with two choices called "strictly positive" and "negative or zero", category "SA compared to sum of other two", and category "SA compared to SB" (collapsed card). Each category has its own card (collapsible, with save and delete symbols), and additional categories can be created for the parameter. The user can choose, among the specified choices, which of the choices of the category is the *base* one: choice "strictly positive" is the base choice for category "Value of SA". Recall the base choice of a category is the most "important" one, according to the expert; for instance, it is to be tested more often than other choices of the category; it corresponds to the "normal mode" of execution of the software under test [18]. This information is used with the Base-Choice selection criterion when generating test frames as well as for other selection criteria (Section IV.D).

Figure 6 (b) shows how a choice/block is specified. This is where the tool gets the core of the information to eventually construct test frames and identify test inputs for test frames. Each choice has its own card (collapsible, with save and delete symbols). A choice is given a plain language description for easy reference. Since a category implicitly specifies a domain of values and choices are used to split this domain into sub-domains, the user can specify the sub-domain of a choice with an expression. Different options are available to the user but the most general form for such a specification of a choice is a Java Boolean expression: for instance, choice "Strictly positive" is specified by expression `SA>0`. Alternatively, Melba provides shortcuts for typical needs depending on the type of the parameter, using a drop-down list (see the specification of choice "Negative or zero"): here, given the parameter is an integer, a choice can specify whether the value is odd or even; when using any of the comparison operators (e.g., <, <=), the user simply provides a value; with the interval shortcuts the user is presented with a UI to provide upper bound and lower bound values of the interval. If a shortcut is used, Melba automatically produces the corresponding Java expression: e.g., an inclusive interval specified with values 10 and 18 for SA would become `SA>=10 && SA<=18`. Expressions specifying choices being Java expressions, the right-hand side of a comparison operator for an integer parameter does not have to be an integer literal; it can be the result of a more complex expression, returning an integer value, possibly involving another parameter (see later in the paper). Melba checks expressions are valid Java Boolean expression thanks to the use of Janino.

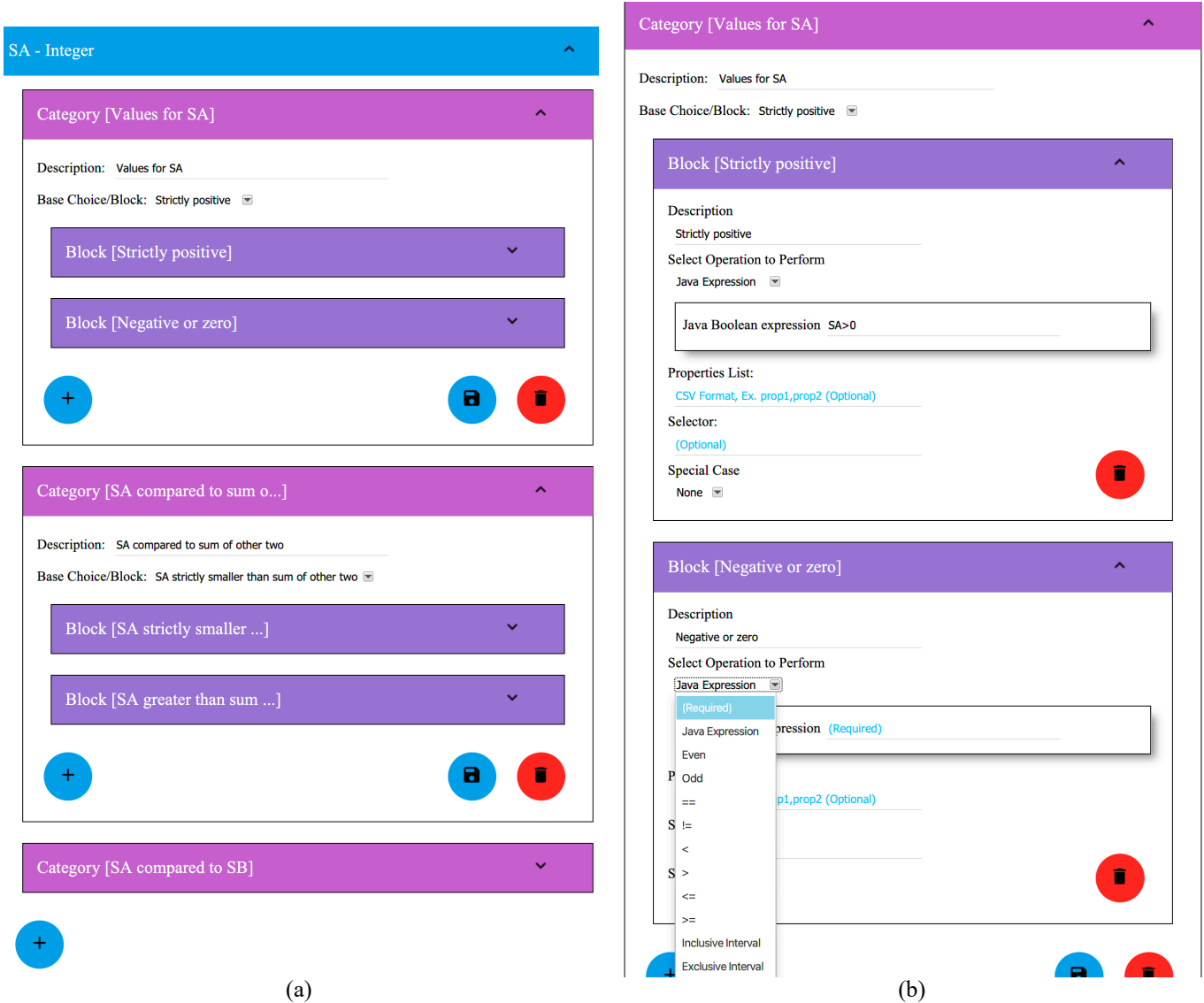

(a)                                                                 (b)

Figure 6.        Definition of categories and choices

Melba provides a list of shortcuts that is specific to parameter types. For a string parameter (class `StringType` in Figure 3), Melba supports typical string operations with Java's semantics such as `length()`, `concat()`, `indexOf()`, `includes()`, `endsWith()`, `isEmpty()`, but also more exotic operations such as `AllLowerCase()` (specifying the string contains only lower case characters), `MixedCase()` (the string contains a mix of lower case and upper case characters) or `AllHexadecimalCharacters` (the string contains only hexadecimals characters). See the complete list of string operations Melba supports in Table 2; character operations are available in Table 3. For the `IntegerSet` type (Figure 3), Melba supports standard set operations: inclusion, intersection, difference, equality, size, union. Because a choice specification specifies a condition that must hold, it has to be a Java expression that evaluates to a Boolean value, which Melba also checks thanks to Janino. A choice expression can be more complex and involve several parameters of the CP specification. Referring back to the earlier example about the first occurrence of a character in a string, a choice specifying the character appears only once in the string can be specified as (assuming the string parameter is S and the character parameter is C):

```
S.indexOf(C)>=0 && S.substring(S.indexOf(C)+1).contains(C)==false.
```

Table 2    Operations Melba supports for type String

| | |
|---|---|
| `charAt(integer i)` | returns the Character at the given index in the string |
| `concat(string str)` | returns the concatenation of the string to which this applies with str |
| `contains(string str)` | returns true if the string contains str |
| `endsWith(string str)` | returns true if the string ends with str |
| `equals(string str)` | returns true if the string equals str (case sensitive) |
| `equalsIgnoreCase(string str)` | returns true if the string equals str (case insensitive) |
| `indexOf(string str)` | returns the first index at which the string contains str |
| `indexOf(string str, integer from)` | returns the first index at which the string contains str, after the offset |
| `isEmpty()` | returns true if the string is empty |
| `lastIndexOf(string str)` | returns the last index of str in the string |
| `length()` | returns the length of the string |
| `startsWith(string str)` | returns true if the string starts with str |
| `substring(integer end)` | returns the substring from the beginning of the string (index 0) to the end index |
| `substring(integer begin, integer end)` | returns the substring from the start index to the end index |
| `Alphanumerical()` | returns true if the string contains only alphanumerical characters |
| `AllLowercaseCharacters()` | returns true if all the characters in the string are lower case letter |
| `AllUppercaseCharacters()` | returns true if all the characters in the string are upper case letters |
| `AllNumericalCharacters()` | returns true if the string contains only digits |
| `MixedCase()` | returns true if the string contains mixed case letters |
| `MixedCaseHexadecimalCharacters()` | returns true if the string contains mixed case hexadecimal characters |
| `AllLowercaseHexadecimal()` | returns true if the string only contains lower case hexadecimal characters |
| `AllUppercaseHexadecimal()` | returns true if the string only contains upper case hexadecimal characters |
| `NoHexadecimalCharacters()` | returns true if the string does not contain any hexadecimal character |
| `AllHexadecimalCharacters()` | returns true if the string contains only hexadecimal characters |
| `OddNumberofHexadecimalCharacters()` | returns true if the string contains an odd number of hexadecimal characters |
| `EvenNumberofHexadecimalCharacters()` | returns true if the string contains an even number of hexadecimal characters |
| `MixHexadecimalNonHexadecimalCharacters()` | returns true if the string contains hexadecimal and non-hexadecimal characters |
| `NoHexadecimalCharactersinFirstChars(integer i)` | same as `NoHexadecimalCharacters` but checks only the first i characters |
| `AllHexadecimalCharactersinFirstChars(integer i)` | same as `AllHexadecimalCharacters` but checks only the first i characters |
| `MixHexadecimalNonHexadecimal CharactersinFirstChars(integer i)` | same as `MixHexadecimalNonHexadecimalCharacters` but checks only the first i characters |
| `EvenNumberofHexadecimal CharactersinFirstChars(integer i)` | same as `EvenNumberofHexadecimalCharacters` but checks only the first i characters |
| `OddNumberofHexadecimal CharactersinFirstChars(integer i)` | same as `OddNumberofHexadecimalCharacters` but checks only the first i characters |

Table 3    Operations Melba supports for type Character

| `isDigit()` | returns true if the character is a digit |
|---|---|
| `isLetter()` | returns true if the character is a letter |
| `isLetterOrDigit()` | returns true if the character is either a letter or a digit |
| `isLowerCase()` | returns true if the character is lower case |
| `isLegal()` | returns true if the character is either a letter, a digit or a space |
| `isUpperCase()` | returns true if the character is upper case |
| `isSpace()` | returns true if the character is space |
| `isHexadecimal()` | returns true if the character is an hexadecimal character |
| `asString()` | returns a string representation of the character (a string of length 1) |

CP uses the notions of properties and selectors to help the user specify under which conditions choices should (or should not) be combined into test frames. Properties and selectors are user-defined and optionally specify the choices of a category. Choice "Strictly positive" in Figure 6 (b) does not have any property or selector defined. A Property can be thought of as a Boolean variable, with a user-defined name, that is true when the corresponding choice is used in a test frame. A selector specifies a condition that must be satisfied for the corresponding choice to be used in a test frame in terms of which other choices or choice combination should be used or not. For instance, Figure 7 indicates that choice "SA greater than sum of other two" can be involved in a test frame, i.e., a combination of choices, if and only if a complex condition involving four properties (of four other choices/blocks) is satisfied (a.k.a. a selector). The selector specifies that choice "SA greater than sum of other two", for which we have this selector, cannot be combined with the choice which property is `SBgreaterThanSumOthers`, cannot be combined with the choice which property is `SCgreaterThanSumOthers`, but can be combined with the choice which property is `SCpositive`, …

Figure 7.    Illustrating a complex selector

Last, the "Special Case" option allows the specification of "error" and "single" choices. Choice "Strictly positive" in Figure 6 (b) has no such constraint whereas choice "SA greater than sum of other two" in Figure 7 is specified as Single. Having a choice specified as "error" or "single" means that it will be sufficient to involve the choice once in the entire set of test frames. The difference is that with an "error" choice the user should expect an exception or an error message when executing the test case whereas with "single" the user indicates that it is their opinion triggering this condition once in the entire test suite should be sufficient.

*IV.C.   Verifying the CP Specification*

First, we note that the CP specification should also be validated; it must make sense given the plain language specification that was used to create it. We do not provide any technical solution to this problem of validating the CP specification against the specification of the system under test. However, our experience with Melba, its verification capabilities but also the generation of test frames and test inputs, indicates that Melba can help validate the CP specification. For instance, obtaining a test frame for which it is impossible to find inputs (see Section IV.E) has helped us correct validation problems with CP specifications.

Making sure the CP specification is correct is an important step before using it in an automated fashion for test frame and test case generation. Our solution automatically helps check some crucial properties of the CP specification, a capability that we have not seen in any similar tool support for CP. Beyond trivial checks such as the fact the CP specification contains at least two parameters (or environment variables), that parameters, environment variables, categories, choices have a name and other prescribed characteristics (e.g., a choice has a description), that no categories have the same name, that each category has at least two choices, that property names used in selectors are defined in the CP specification for at least a choice in another category, our solution also checks the following:

- A choice specification is a syntactically correct Java Boolean expression. We use Janino for this aim. Specifically, Melba provides each choice specification, i.e., Java expression to Janino and Janino determines whether the Java syntax is correct and whether it evaluates to a Boolean expression. Janino can indeed work on default Java library classes and operations (like operation `length()` of class `String`), but can also work on other classes and operations such as the Melba type classes (subclasses of class `Type` in Figure 3). Melba-specific types, such as `IntegerSet`, and Melba specific operations such as `size()` for class `IntegerSet`, or `AllHexadecimalCharacters()` for class `StringType` are defined in the appropriate Melba classes and are understood by Janino.

- No two choices of a category overlap, that is no value in the domain implicitly specified by the category belongs to two choices of that category. To check this, we transform the Java expression defining each choice of a category into a corresponding Z3 constraint (recall such Java expressions are Boolean expressions) and ask Z3 to find a value that satisfies each pair of these expressions together. Z3 should respond "not satisfiable" in a correct CP specification. Otherwise, if Z3 finds a solution, it has found a value that satisfies the specification of two choices of the class category, which should not happen. Recall choices are equivalence classes. Since Z3 is also used for identifying test data for test frames, Melba's use of Z3 is further discussed in section IV.E.

- All the choices of a category together represent the entire domain of values specified by the category, that is any value of the domain should belong to a choice. We ask Z3 to find a value that does not belong to any choice. Specifically, assuming a category has three choices, each provided with a Java Boolean expression `expr1`, `expr2`, `expr3`, we first create the expression `!expr1 && !expr2 && !expr3`, translate this expression into Z3 instructions (see section IV.E) and ask Z3 to solve the problem. Z3 should respond "not satisfiable" in a correct CP specification since, otherwise, there is a value in the domain implicitly specified by the category that does not fit in any of the category's choices. Note that this is not necessarily an error but we at least flag this to the user. Indeed, if, for instance, a category specifies the length of a string parameter, the type of the length values is Integer; the domain of values implicitly specified by the category is the set of integers. However, one would not include a choice specifying negative values for the string length; one would typically specify a first choice `length()>0` and a second choice `length()==0`. In this case, when providing expression `!(length()>0) && !(length()==0)` to Z3 (actually a translation of this Java expression into Z3 instructions—section IV.E), Z3 would find it is possible to find a value (a negative one) that does not fit any of the category's choices.

- All the base choices can be combined in a test frame, that is their properties and selectors are satisfiable together. Specifically, Melba creates a Java (Boolean) expression that is the conjunction of the selectors of all the base choices. The property names involved in this expression are specified with Janino as Boolean variables. If any of those names is a property defined for a base choice the corresponding variable value is set to true, and it is false otherwise. The expression, the variables with their type and value, are passed to Janino and Janino returns true if these variable values make the expression true, which is the expected outcome.
- It is possible to find a value for the test frame that is the combination of all the base choices. We use Z3 in a way similar to the two earlier cases: we produce the conjunction of the Java expression of all the base choices and ask Z3 to solve it; Z3 should find this staisfiable. This condition and the previous one should hold since base choices together describe a nominal, typical use of the function/functionality under test.

If any of the verifications fails (e.g., the combination of base choices is infeasible, choices in a category overlap), the user is informed and a detailed report is provided (View→Error log menu in Figure 4 (d)). In the case two choices of a category overlap for instance, Melba can report on a value that satisfies the two expressions at the same time.

*IV.D. Generating Test Frames*

Test frames are combinations of choices produced using a selection criterion [11]. The Each-Choice criterion requires that the set of produced test frames exercises each choice at least once. The Pairwise criterion requires that the set of test frames exercises each pair of choices (from two categories) at least once. The Three-way criterion requires that the set of test frames exercises each triplet of choices (from three categories) at least once. The All-Choices-Combinations requires that all the possible combinations of choices are exercised at least once by a set of test frames. This is essentially the n-way criterion where n equals the number of categories in the CP specification. The Base Choice criterion proceeds as follows: (i) a first test frame is created with all the base choices, i.e., one base choice per category; (ii) Other test frames are created from this initial, base test frame by holding all but one base choice constant and using each non-base choice once for the one non-constant choice.

Of course, regardless of the chosen selection criterion, test frames should satisfy constraints on choices (i.e., selectors); combining choices according to a criterion without accounting for such constraints is likely to result into unfeasible test frames, that is test frames for which it is impossible to identify test inputs [35]. For instance, without accounting for constraints in the Triangle CP specification we may produce a test frame that combines the choices with specifications `Edge1>0`, `Edge2<=0`, `Edge3<=0`, and `Edge1 < Edge2 + Edge3`.

Combining choices according to these selection criteria, except Base Choice, is akin to the use of combinatorial design in software testing. There is one fundamental difference though. Contrary to combinatorial testing where one wants to combine values of parameters [36], in our case we want to combine choices of categories; we do not combine values, we combine descriptions of ranges of values and those values are not that of parameters of the system under test, they are ranges of values implicitly specified by categories. Therefore, parameters in combinatorial design are categories in CP, values in combinatorial design are choices in CP (i.e., ranges of values).

Our solution is therefore not comparable from a testing point of view to combinatorial testing. However, because of the similarity of the problem of producing test frames and combinatorial testing, including the presence of constraints, we decided to rely on technologies that can produce covering arrays while accounting for constraints, instead of creating our own solution to this combination problem. This has the advantages of simplifying our task, relying on proven technologies, and benefitting from existing technologies that attempt to minimize the number of test frames while satisfying a criterion and constraints. (Previous works [9, 20] have suggested their own heuristics for this problem and we suspect, although this needs to be empirically confirmed, they are not as optimal in terms of number of test frames as combinatorial testing solutions and therefore test cases produced.)

Melba interfaces with either CASA [33, 32] or ACTS [34] for this purpose, since these have been specifically designed to solve combinatorial testing problems under constraints, are well-known, and support Each-Choice, Pairwise, Three-way: Figure 8 (Melba also offers Base Choice). Both CASA and ACTS support constraints and have been shown to be good at reducing the number of solutions that satisfy a criterion with good performance. While CASA uses a meta-heuristic engine to find a solution, ACTS relies on the IPOG algorithm (In-Parameter-Order-General). While highly efficient, IPOG can sometimes produce more test cases than the theoretical minimum because it prioritizes computational speed over absolute minima. In general the smallest number of rows of a covering array for an arbitrary problem is unknown. We chose to support both covering array tools so we can study their performance when integrated into Melba.

Melba automatically transforms the choices' selectors into a format that can be understood by either CASA or ACTS. Although our solution does not currently support All-Choices-Combinations, we are confident that adding this capability is straightforward. Indeed, solutions exist for this problem [9, 20], we can develop our own, and integration in Melba will be facilitated by the use of software design patterns such as Bridge and Adapter.

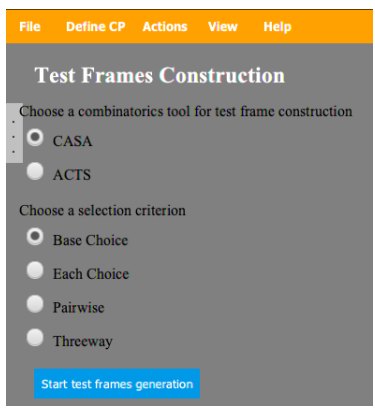

Figure 8.        Criteria and Generation Technologies for Test Frames

CASA does not support the Base Choice criterion; although ACTS does support a criterion called Base Choice, its definition is not exactly that generally accepted for Base Choice in CP [37]. Melba therefore provides its own implementation of Base Choice, aligned with that accepted in CP, instead of delegating the task to either CASA or ACTS. Should Melba be interacting with a third-party covering array software that handles Base Choice the way the criterion is defined for CP, the task could be entirely delegated to that third-party software.

CASA and ACTS do not have a notion of "error" and "single" choices as in CP. Melba therefore offers its own mechanism to handle such cases, keeping in mind that "error" and "single" choices only need to be involved in one test frame. Melba proceeds in two stages. It first produces a CP specification in which "error" and "single" choices are omitted and asks the covering array solution from third-party software selected by the user to produce combinations from this reduced set of choices. (The CP specification provided by the user is not changed; only the one provided to the third-party covering array software selected by the user is.) This ensures that the user-selected selection criterion is satisfied for all choices except "error" and "single" ones. In a second stage, for "error" choices, Melba adds one test frame for each "error" choice and this test frame only contains one choice, the "error" choice. Although an "error" choice can in theory be combined with other choices in the CP specification, we opted for a simple solution since the specification of the system under test rarely specifies whether values corresponding to choices other than one with an "error" choice would matter during execution. (For instance, if input data corresponding to an error choice is first checked in some defense-programming code, then other input values do not matter.) An alternative could be to combine an "error" choice with the base choice of all the other categories, while satisfying constraints so as to eventually obtain nominal values for everything except the "error" choice. To generate test frames for "single" choices in its second stage, Melba provides the complete CP specification, without "error" choices but including "single" choices, to the third-party covering array software selected by the user. The result is a series of test frames, some of which involve "single" choices. Melba visits this set of test frames and, for each "single" choice, selects the first test frame it finds that involves it. This is admittedly not optimal in terms of executions since the third-party covering array software is executed twice; but we opted for this solution as we then rely on the capability of that software to deal with constraints between choices (and Melba does not need to take care of this) and produce a test frame that satisfies choice constraints for each "single" choice. An alternative could be, similarly to "error" choices, to combine each "single" choice with base choices of other categories, while satisfying constraints. We note that, as far as we know, existing tool supports for CP do not support "error" and "single" choices.

Last, it may happen, as we have experienced [38], that, due to choice constraints, using a criterion does not produce test frames that involve every single choice in the CP specification. In other words, if we are not careful, a Pairwise set of test frames may not be Each-Choice adequate, which is counter intuitive. To overcome this problem, Melba creates a new test frame for each choice that is not involved by the use of a selection criterion. To create this test frame, Melba attempts to combine this initially missing choice with as many base choices as possible, while satisfying constraints.

This means that, although the specification of Base Choices is optional when using CP, Melba relies on this information to produce test frames. Of course alternatives solutions could be envisioned, should Base Choices not be identified in the CP specification.

*IV.E. Generating Test Data*

Identifying test data for a test frame means finding values for the input parameters and environment variables that satisfy the conditions imposed by the test frame' choices. Recall that a choice specifies a sub-domain of the domain of values implicitly specified by the corresponding category, and that, with our solution, this sub-domain is specified under the form of a syntactically correct Java Boolean expression (which Janino confirms during CP specification verification). Finding input values for a test frame then amounts to finding values that satisfy all together (conjunction) the Java Boolean expressions for the choices involved in that test frame.

To find those values, our tool automatically transforms those Java expressions into an expression that can be understood by a solver and asks that solver to solve it. Given that parameters and environment variables in our tool can have various types, we considered SMT solvers that support several theories. We created a grammar of Java expressions (Appendix A), inspired by the Java mathematical expression parser and evaluator "Jep Java"[4], which production rules transform a Java Boolean expression into commands of the selected solver. Melba currently only supports Z3. Melba currently does not support the entire amount of the Java syntax a user could consider when specifying choices. For instance, it does not support regular expressions on strings. We used JavaCC[5], a well-known Java Compiler Compiler, that consumes our grammar and automatically creates the Java code that performs the translation.

---

[4] https://www.singularsys.com/jep/
[5] https://javacc.github.io/javacc/

Let us succinctly illustrate the translation. Recall the CP specification of the Triangle problem (section II.A), where we created a category (Cat1) for parameter SA with two choices: choice 1.1 (`SA>0`) and choice 1.2 (`SA≤0`). Recall (section IV.C) that one CP specification verification step is to check that the choices of a category do not overlap, that is, it is not possible to find a value that satisfies the conditions of any two choices of a category at the same time. In the case of Cat1, this amounts to attempting to solve the Java expression `SA>0 && SA<=0`. Melba uses the grammar code to translate this expression into calls to the Z3 API. Without going into the details of the production rules, which can be cumbersome, these calls are essentially equivalent to the use of the Z3 API as illustrated on the left side of Figure 9. The actual model Z3 is asked to use by Melba appears on the right hand side of the figure: the set of constraints contains more labels and functions than necessary (more than for only SA); this is because we use Z3's mechanism ("assert and track") to monitor the status of assertions and report, when a result is unsatisfiable, about which assertions conflict; Melba can use this information to then tell the user what went wrong. For instance, it may happen that, despite all efforts and despite passing the CP specification verification phase, Z3 fails to identify test data for a test frame; when this happens, Melba can report on which choices in the test frame conflict, which likely means that the CP specification needs additional choice constraints, or that those choices' expression should be revisited. In the particular Triangle example, with the solver in the figure, Z3 confirms this is unsatisfiable: the two choices do not overlap.

| | |
|---|---|
| ```<br>saVar = mkIntConst("SA");<br>e1 = mkGt(saVar, mkInt(0)); // SA>0<br>e2 = mkLe(saVar, mkInt(0)); // SA<=0<br>final = mkAnd(e1,e2); // conjunction<br>solver = mkSolver();<br>solver.add(final);<br>``` | ```<br>solver=<br>  (declare-fun label102 () Bool)<br>  (declare-fun SA () Int)<br>  (declare-fun label100 () Bool)<br>  (declare-fun label101 () Bool)<br>  (assert (=> label102 true))<br>  (assert (=> label100 (> SA 0)))<br>  (assert (=> label101 (<= SA 0)))<br>``` |
| (a)　　Equivalent series of Z3 API calls (omitting types and the use of a Z3 context when calling the API functions) | (b)　　The actual assertions being used through Melba |

Figure 9.　　Illustrating the use of Z3 on example 1 (section II.A)

When Melba attempts to identify values of parameters and environment variables for a test frame, it creates the conjunction of all the Java expressions describing the choices involved in the test frame. For the example of the search of the index of the first occurrence of a character in a string (section II.B) a test frame Melba generates is made of the following choices (descriptions): the string has a small number of characters (which we specified as having 3 or 4 characters); the character appears more than once in the string; the first occurrence of the character is not the first character of the string nor its last character. Melba then creates the conjunction of the choice expressions: top part of Figure 10 (assuming the string parameter is called `strParam`, and the character parameter is called `charParam`). Melba's use of the grammar instructs Z3 to use the assertions in Figure 10 (second part): lines 6-8 asks for a solution made of hexadecimal characters (this is automatically added by Melba for String parameters in the CP specification); lines 9-19 correspond to assertion C (top of figure); line 20 corresponds to assertion A; the rest corresponds to assertion B. Z3 finds this conjunction satisfiable and identifies values "`0MM`" and '`M`' for the string and character, respectively.

```
A  strParam.length()>2 && strParam.length()<5
   &&
B  strParam.indexOf(charParam.asString())>=0 &&
      strParam.substring(strParam.indexOf(charParam.asString())+1).contains(charParam.asString())==true
   &&
C  strParam.indexOf(charParam.asString())>0 && strParam.indexOf(charParam.asString())<strParam.length()-1
```

```
1   solver=
2   (declare-fun strParam () String)      (declare-fun label106 () Bool)        (declare-fun strV68 () String)
3   (declare-fun strV16 () String)        (declare-fun charParam () Unicode)    (declare-fun label105 () Bool)
4   (declare-fun label103 () Bool)        (declare-fun strV95 () String)        (declare-fun strV36 () String)
5   (declare-fun strV54 () String)        (declare-fun label104 () Bool)
6   (assert (let
7     (( a!1 (re.* (re.union (re.union (re.range "0" "9") (re.range "a" "z")) (re.range "A" "Z"))) ))
8     (=> label106 (and true (str.in_re strParam a!1))) ) )
9   (assert
10    (let  ( (a!1 (str.indexof (str.substr strParam 0 (- (str.len strParam) 0)) strV16 0) )
11           (a!2 (str.indexof (str.substr strParam 0 (- (str.len strParam) 0)) strV68 0) )  )
12     (let  ( (a!3 (and (= (str.len strV16) 1)
13                       (= (seq.nth strV16 0) (char.from_bv (char.to_bv charParam)))
14                       (= (str.len strV68) 1)
15                       (= (seq.nth strV68 0) (char.from_bv (char.to_bv charParam)))
16                       (> (+ a!1 0) 0)
17                       (< (+ a!2 0) (+ (str.len strParam) (- 1)))  )  )  )
18     (=> label105 a!3)
19     ) ) )
20  (assert (=> label103 (and (> (str.len strParam) 2) (< (str.len strParam) 5))))
21  (assert
22    (let  ( (a!1 (str.indexof (str.substr strParam 0 (- (str.len strParam) 0)) strV54 0) )
23           (a!2 (str.indexof (str.substr strParam 0 (- (str.len strParam) 0)) strV36 0) )  )
24     (let  ( (a!3 (str.contains (str.substr strParam (+ a!2 0 1) (- (str.len strParam) (+ a!2 0 1))) strV95)))
25     (let ((a!4 (and (= (str.len strV54) 1)
26                     (= (seq.nth strV54 0) (char.from_bv (char.to_bv charParam)))
27                     (= (str.len strV36) 1)
28                     (= (seq.nth strV36 0) (char.from_bv (char.to_bv charParam)))
29                     (= (str.len strV95) 1)
30                     (= (seq.nth strV95 0) (char.from_bv (char.to_bv charParam)))
31                     (>= (+ a!1 0) 0)
32                     (= a!3 true)  )  )  )
33     (=> label104 a!4)
34     ))))
```

Figure 10.        Illustrating the use of Z3 on example 2 (section II.B)

It is very possible for complex sets of assertions that Z3 times out without finding a solution. Currently Melba informs the user. In the future, Melba will allow the user to set a different timeout value for Z3 so they can try to run the search for input values again.

Melba currently produces only one test input per test frame since this is the initial intent of the CP specification [3, 4] and we have evidence test frames can be regarded as equivalence classes [14, 16]. Considering alternative strategies [15] is entirely possible and we already have experience with this practice [14, 16].

## V.    EXAMPLES

Below we illustrate the use of Melba on nine different case studies. They help illustrate the use of various parameter types, sometimes combined in the same CP specification, specifically Boolean, Character, Enumeration, Integer, Integer Set, Real, and String. Some of these CP specifications are adapted from earlier work by graduate students [16, 39, 30, 40, 31, 38, 35, 41]. Below we only mention a sample of test frames and test cases generated by our tool. Details about each study, including the plain language specification, the CP specification, and all the test frames and test cases produced for each selection criterion (Each-Choice, Pairwise, Three-way, Base Choice) and combinatorics tool (CASA, ACTS) can be found in Appendices.

## V.A.    Triangle

The screenshots we used earlier in the paper were produced while using Melba for the Triangle problem, famously used in software testing education: a function accepts three integers, SA, SB, and SC, and returns the type of the triangle these three length values specify (scalene, isosceles, equilateral, or not a triangle). With our tool, we created a CP specification (Appendix A) made of three parameters of type Integer (no environment variable), nine categories and 18 choices. (Five choices have a Boolean expression constraint; three choices are "single" and three are "error".) When using CASA with the Each-Choice criterion, Melba produces eight test frames (Appendix A).

As an example, a test frame combines the following choices:

- Choice 1 of category 1 for parameter SA, which description is "Strictly positive", and which (Java expression) specification is `SA>0`;
- Choice 1 of category 2 for parameter SA, which description is "SA strictly smaller than sum of other two", and which specification is `SA<SB+SC`;
- Choice 2 of category 3 for parameter SA, which description is "SA and SB are different", and specification is `SA!=SB`;
- Choice 1 of category 1 for parameter SB, which description is "Strictly positive", and specification is `SB>0`;
- Choice 2 of category 2 for parameter SB, which description is "SB and SC are different", and specification is `SB!=SC`;
- Choice 1 of category 3 for parameter SB, which description is "SB strictly smaller than sum of other two", and specification is `SB<SA+SC`;
- Choice 2 of category 1 for parameter SC, which description is "SC different from SA", and specification is `SC!=SA`;
- Choice 1 of category 2 for parameter SC, which description is "SC strictly smaller than sum of other two", and specification is `SC<SA+SB`;
- Choice 2 of category 3 for parameter SC, which description is "SC strictly positive", and specification is `SC>0`.

The Boolean expression to solve is therefore the conjunction of all these choice specifications: `SA>0 && SA<SB+SC && SA!=SB && SB>0 && SB!=SC && SB<SA+SC && SC!=SA && SC<SA+SB && SC>0`, which is a Java expression Melba automatically translates into Z3 instructions. Z3 identifies that this is solvable and returns values 3, 2 and 4 for SA, SB and SC, respectively.

## V.B.    Character search

Another example that is used in textbooks and articles on CP is to test a function that returns the position of the first occurrence of a character in a string. This typically calls for categories about the length of the string (string is empty or not), the number of occurrences of the character in the string (e.g., 0, 1, or more) and the position of the first occurrence of the character in the string (start of the string, end of the string, elsewhere in the string). Using Melba, we produced a CP specification with two parameters, three categories and eleven choices: Appendix C. One parameter is of type String while the other is of type Character. Using ACTS and the Pairwise criterion generates nine test frames for which Z3 identifies values—Appendix C.

An example test frame calls for input values that satisfy the following conditions:

- The input string has a nominal values, which we specified as "`InputString.length()>=5`";
- The character appears more than once in the string, which we specified as

```
InputString.indexOf(InputCharacter.asString())>=0
&& InputString.substring(InputString.indexOf(InputCharacter.asString())+1)
.contains(InputCharacter.asString())==true;
```

- The position of the first occurrence of the character in the string is "middle", meaning it is not the first nor the last character, which we specified as
```
InputString.indexOf(InputCharacter.asString())>0
&& InputString.indexOf(InputCharacter.asString())<InputString.length()-1.
```

With these conditions, Z3 identified values "`P733z`" for InputString and '`3`' for InputCharacter.

*V.C.    PackHexChar*

The third example further illustrates the handling of string input parameters with complex operations supported by shortcuts. The PackHexChar function compacts a string of characters representing hexadecimal digits into a binary format. The function processes an input string (called stringS), ignoring non-hexadecimal characters, and outputs an array of bytes containing the compacted information alongside an integer value. The behaviour of the program varies depending on the number of hexadecimal characters in the input string. If the input string contains an even number of hexadecimal characters, all hexadecimal characters are compacted (in pairs) into the byte array, and the program returns an integer value of -1. For an odd number of hexadecimal characters, the function compacts all but the last hexadecimal character into the byte array and returns the remaining hexadecimal character as a separate output. Additionally, the program allows the user to specify a substring of the input string to be analyzed using the input parameter RLEN. The function then only analyzes the first RLEN characters of the input string. If RLEN is invalid (negative or greater than the length of the input string), the function returns -2. The program also provides a mechanism for handling leftover characters from previous calls using the ODD_DIGIT input parameter. This parameter allows the user to append a trailing hexadecimal character from a previous result to the beginning of the input string for subsequent processing. If ODD_DIGIT is set to -1, no character is appended. However, if ODD_DIGIT is invalid (less than -1 or not a hexadecimal value), the function returns -3. This feature is particularly useful for processing strings split across multiple calls, ensuring consistent handling of characters between executions.

We created a CP specification with three input parameters and a total of five categories and 16 choices (Appendix D): two Integer parameters and one String parameter. Using the Three-way criterion and CASA, we obtain twelve test frames from which Z3 can find values for the input parameters. One of the test frames specifies the following conditions for input parameter values:

- Parameter stringS contains an even number of hexadecimals in its first RLEN characters;
- ODD_DIGIT has an hexadecimal value;
- The first RLEN characters of stringS are a mix of hexadecimals and non-hexadecimals;
- RLEN's value is within the string (strictly positive and smaller than stringS' length);
- stringS is not empty.

These choices lead to the following conjunction that is transformed into a Z3 input:

```
stringS.EvenNumberofHexadecimalCharactersinFirstChars(RLEN)
&& ( (ODD_DIGIT>=48 && ODD_DIGIT<=57) || (ODD_DIGIT>=65 && ODD_DIGIT<=70) ||
(ODD_DIGIT>=97 && ODD_DIGIT<=102) )
&& stringS.MixHexadecimalNonHexadecimalCharactersinFirstChars(RLEN)
&& RLEN>0
&& RLEN<stringS.length()
&& stringS.length()>0
```

Z3 identifies the following values for this test frame: `ODD_DIGIT='48'`, `RLEN='3'`, `stringS="Gb3"`.

## V.D. Integer set difference

Another subject that helps illustrate the various data types Melba can support is a function that determines the difference between two sets of integer values—Appendix E. The function takes two sets of integers as input, Set1 and Set2, and returns the difference Set1\Set2. The CP specification has two parameters of type Integer Set, five categories and 14 choices. Using the Each-Choice criterion with ACTS produces five test frames. One of them requires inputs such that Set1 is larger than Set2, neither Set1 include Set2 nor Set2 include Set1, Set1 is large (we specified three choices for set size: empty, small, i.e., between 1 and 10 elements, and large, i.e., at least 11 elements), the two sets have an non-empty intersection, and Set2 is small, which led Z3 to identify `Set1 = {0, 1, 2, 3, 4, 5, 6, 7, 8, 9, 13, 14, 15, 16, 17, 18, 19, 20, 21, 23}`, and `Set2 = {1, 4, 7, 16}`.

## V.E. Next date

Yet another example often used in academic references is the Next Date function which, given a date as an input, computes and returns the next date: Appendix F. The function takes a date as input, i.e., a 3-tuple with the year, the month, and the day, and returns the date of the following day, while accounting for leap years. We created a CP specification with three parameters (all of type Integer), four categories and 16 choices. One CASA generated test frame for the Pairwise criterion requires that the input date be the 28th day of a 30-days months in a non-leap year, for which Z3 identified the following values—year=1601, month = 4 (April), day = 28.

## V.F. Daylight saving

The function under test takes parameters to determine the cost of a phone call, given a start time and either a duration or an end time. The function applies a simple fee model: a call that lasts less than 20mn is billed at a rate that is different than that used if a call lasts more than 20mn; a call cannot last more than two hours (a longer than two hours call is simply dropped). The function also handled daylight saving periods, in the Fall and Spring, that is the practice of advancing clocks in the Spring by one hour to make better use of the longer daylight available during Summer so that darkness falls at a later clock time, and also setting clocks back by one hour in the Fall for a similar reason. For the purpose of this study, we fixed the daylight saving dates to March 28th at 2 o-clock in the morning (at 2 o-clock on March 28th, clocks are changed to 3 o-clock) and October 30th at 3 o-clock in the morning (at 3 o-clock on October 30th, clocks are changed to 2 o-clock).

This subject allows us to illustrate that CP can be used with a different set of parameters than what the function under test expects. This can sometimes facilitate the creation of the CP specification. For this specific function, if we use the parameters expected by the function we need to reason about months, which have varying numbers of days, to trigger changes of day (i.e., a phone call that starts in a day and ends in another), and involve the computation of the duration of a call around daylight saving changes in the CP specification with values of month, day and time. Our experience is that this makes the CP specification very complex. Instead, we opted for a CP specification that counts a year in seconds: Appendix G. Given an amount of seconds, one can easily compute the time of a day in a month; for instance, if a test case uses an input value of 5,680,352 seconds, a test driver can easily determine that this refers to a time in the 65th day of the year, which is March 6 on a non-leap year, and more precisely 17h, 52mn, 32s on March 6th. (A day has 86,400 seconds, an hour has 3,600 seconds and a minute has 60 seconds.) On a non-leap year, the change dates for daylight saving are the 87th (March 28th) and the 303rd (October 30th) days of the year. The CP specification has three parameters of type Integer (a day of year, a time of day and a call duration), three categories and 20 choices.

One of the test frames CASA produces for the Each-Choice criterion requires inputs such that the month is March, the duration of the phone call is at least 20mn and the start of the call triggers the change of time (the call starts before the change of time and ends after), which led to the following input values produced by Z3: Day of year is 87 (October 30[th]), duration is 1,200 seconds (i.e., 20mn), and start of the call at 7,437,000 seconds, which is 1h50 in the morning (10mn before the change of time).

## V.G.    Taxi bill

The system under test mimics the fee calculation of cabs in Paris, France, where the cost of a cab trip depends on the starting location, either Paris' core, the suburb or beyond the suburb, the day of the week (Saturday being considered a weekday and only Sunday being different), the time of the day, and whether the day is a Holiday. We simplified the specification by considering that a trip's cost only depends on mileage; in reality, when the cab is idle that also contributes to the cost. The CP specification has four parameters (two of type Integer, one of type Boolean, and one of type Enumeration), four categories and 13 choices: Appendix H.

An ACTS test frame for the Each-Choice criterion requires a departure in the mid-day period, morning from Paris' core during a Sunday that is not a Holiday, for which Z3 found the following values: DayOfWeek is false, StartingArea is 'core', departure time is 10h (counting from 0 to 23), and Holiday is false.

## V.H.    Grocery cashier

The system under test simulates a cashier in a grocery store that computes the cost of a client's purchases depending on the type and quantity or weight of item being purchased. The item can be an ordinary grocery item, which can be purchased in various quantities (a quantity input is provided), an item that needs to be weighted (a weight input is provided), a health-care item (with a quantity). An item can also be a discount coupon. An item is uniquely identified by a code and a type; two items with different types can have the same code. When an ordinary or health-care item is purchased, or when a coupon is used, the weight input is ignored; when a weighted item is purchased, the quantity input is ignored.

The CP specification we produced (Appendix I) ignores the item code; adding that parameter is simple, with likely one category and two choices specifying whether the code for this item type exists (in the database) or not.

The CP specification has three parameters (an item type that is an enumeration, a quantity of type integer and a weight of type real), three categories, and twelve choices. The first test frame produced with the Base Choice criterion, which involves all the base choices, specifies that test inputs should be such that the item is a weighted item, the quantity parameter is ignored (which we specified as a value of zero), the weight is in an expected range for a grocery store (we specified an unexpected weight range so perhaps the system can warn the cashier), for which Z3 finds the following values: the weight is ½ and the quantity is 0.

## V.I.    TCAS

The Traffic Alert and Collision Avoidance System (TCAS) is a core safety component to modern aircraft. It helps prevent mid-air collisions between aircraft. The system monitors the surrounding airspace for intruding aircraft that could pose a threat. If a threat is detected, a Traffic Advisory is forwarded to the pilot to let them know that there is an aircraft within the vicinity of their own aircraft which could pose a threat. If the intruding aircraft comes within a certain threshold range of the TCAS equipped aircraft, the TCAS system will notify the pilot and offer a Resolution Advisory (RA) which tells the pilot the best course of action to minimize the chance of a collision. The RA can be in the form of either an Upwards RA, Downwards RA, or Unresolved. Upwards RA means the pilot should direct the aircraft in a climbing maneuver whereas Downwards RA means the pilot should begin a descending maneuver. Unresolved means that the TCAS could not find a suitable maneuver.

We created a CP specification (Appendix J) with eleven parameters, either of type Integer, Boolean or Enumeration, and one environment variable of type Integer, with a total of 12 categories and 39 choices. Z3 was able to provide test input values for all the test frames produced by either CASA or ACTS for either of the selection criteria.

*V.J.    Summary*

Table 4 summarizes data for the various examples we discussed. For each example the table reports on the types of the parameters or environment variables being used, the number of parameters and environment variables, the total numbers of categories and choices, the number and kind of constraints on choices, and the number of generated test frames (and therefore test cases) when using either one of the four selection criteria and either one of the two covering array generation tools. Overall, despite small fluctuations, CASA and ACTS appear to be equivalent in terms of number of generated test frames.

This clearly demonstrates that the tool can be used in a variety of situations. These examples could be considered of relatively small size and scalability could be questioned. For several functions provided by an industry partner we have used our tool, creating the CP specification, automatically creating test frames, but manually creating test data because the integration with Z3 was not finalized at the time. Over several functional units tested separately, we specified an overall number of 46 categories and 106 choices [38, 31]. In another partnership we used our solution in a similar manner for a commercial financial application with eight categories and 20 choices [40]. Although scalability should be studied, we posit that Category-Partition, even when applied at the system level, that is one functionality (use case) at a time, may not always require many, many more categories and choices than what we have experimented with. This needs further studies though. We conjecture that a technology such as our tool is likely to turn even more useful as the scale of the problem grows. Indeed, once the CP specification effort is over, the automation level we provide can be extremely useful to, among other things, verify the CP specification is correct (this would be an onerous and error-prone process if conducted manually as the number of categories, choices and constraints grows), simulate the impact of CP specification decisions on the number of test frames / test cases, that is on testing cost. The tool can help answer questions such as: If I specify a choice as single, by how much do I reduce the number of test cases? If I add a constraint on a choice, can I still obtain test frames (i.e., finding test frames that satisfy the selection criterion of choice is satisfiable) and, if this is the case, what is the consequence on the number of test frames? If I decide to focus on boundaries and values immediately before/below and after/above those boundaries, thereby creating more choices, what is the exact consequence on the number of test frames?

Overall, the level of automation the tool provides leads to an iterative CP testing process, which we have experienced first-hand when working on the examples. Often, an initial CP specification is not bullet-proof and includes errors and the CP specification verification step can identify them; creating test frames and thereafter test cases can lead to unsatisfiable constraints reported by Z3, which lead to refinements of the CP specification. This repeats until the user is satisfied by the outcome, i.e., the test frames and test cases.

One limitation we are aware of is not a limitation of our automated support for CP but rather a fundamental limitation of CP itself. CP is not very convenient when the number of parameters varies. Suppose for instance that a functionality under test can take at least one parameter but also up to three parameters of type integer. One can easily specify three parameters with CP and add a fourth (enumeration) parameter that specifies how many of the three parameters we wish to involve in a test case. The enumeration values are One, Two, and Three. One then specifies constraints on the choices of the second parameter so that they are only involved in test frames where enumeration value Two is involved; similarly, one specifies constraints on the choices of the third parameter so that they are only involved in test frames where enumeration value Three is involved. This is a little bit cumbersome and becomes a challenge when the number of parameters grows or the upper bound of the number of parameters is not known. As another example, suppose also that the parameters measure angles which can sometimes be measured in degrees or radians and that both units can be used in a test case. It becomes tricky to specify this with CP. A solution to this problem has recently been proposed [42], without achieving the level of automation we provide.

Table 4    Summary of results for the examples

| | Type | Parameters / Environment variables | Categories | Choices | Constraints / Single / Error | CASA | | | ACTS | | | Base |
|---|---|---|---|---|---|---|---|---|---|---|---|---|
| | | | | | | 1-way | 2-way | 3-way | 1-way | 2-way | 3-way | |
| Character search | String, Character | 2 | 3 | 11 | 6 / 2 / 0 | 7 | 10 | 14 | 7 | 9 | 13 | 9 |
| Daylight saving | Integer | 3 | 3 | 20 | 4 / 5 / 4 | 15 | 21 | 24 | 15 | 21 | 24 | 18 |
| Grocery | Boolean, Integer, Real | 3 | 3 | 12 | 4 / 3 / 2 | 8 | 9 | 9 | 8 | 8 | 8 | 10 |
| Integer set | Integer set | 2 | 5 | 14 | 6 / 2 / 0 | 5 | 8 | 12 | 6 | 8 | 12 | 10 |
| Next date | Integer | 3 | 4 | 16 | 3 / 0 / 4 | 9 | 21 | 37 | 9 | 21 | 37 | 13 |
| PackHexChar | Integer, String | 3 | 5 | 16 | 8 / 4 / 2 | 8 | 12 | 15 | 9 | 13 | 15 | 12 |
| Taxi billing | Boolean, Integer, Enumeration | 4 | 4 | 13 | 0 / 0 / 0 | 6 | 18 | 36 | 6 | 18 | 36 | 10 |
| TCAS | Boolean, Integer, Enumeration | 11 / 1 | 12 | 39 | 10 / 0 / 6 | 10 | 18 | 53 | 10 | 22 | 59 | 28 |
| Triangle | Integer | 3 | 9 | 18 | 5 / 3 / 3 | 8 | 10 | 10 | 8 | 10 | 10 | 10 |

Using the command line option to execute Melba, we also measured the execution time it takes Melba to produce Three-way adequate test suites with CASA, ACTS and Z3 for each example. We chose Three-way since this tends to be the more demanding criterion, producing the largest number of test frames (Table 4). We ran each request for test cases 100 times to produce the boxplots of Figure 11. Executions took place on a MacBook with Apple M4 chip and 32GB of RAM, running OSX 15.6; execution times were collected before and after running Melba from the command line with the 'date' command. Execution times in the figure therefore also include the firing and teardown of the Java virtual machine.

Although our purpose has not been to compare ACTS and CASA, the data show that CASA tends to take more time than ACTS, especially when the CP specification contains many choices and constraints (see results for TCAS for example). We note this is not a proper comparison because execution times also account for Z3's attempt to find test input values; however, since the number of test frames produced by ACTS and CASA is most of the time identical, differences in execution times are predominantly due to covering array construction. We also recognize that Melba interacts with ACTS directly through an API whereas CASA does not offer this capability and Melba therefore relies on files on disk to exchange information with CASA; this solely cannot however explain all the differences we notice. When using ACTS and Z3, the maximum average execution is for the Integer Set Difference example and is likely due to the type of the parameters which is demanding for Z3. Using ACTS and Z3, Melba takes on average between 0.6 seconds and 3 seconds to process a CP specification and generate test cases, which makes it an appropriate tool for practitioners. More studies and comparisons are however necessary to study scalability and confirm this conclusion.

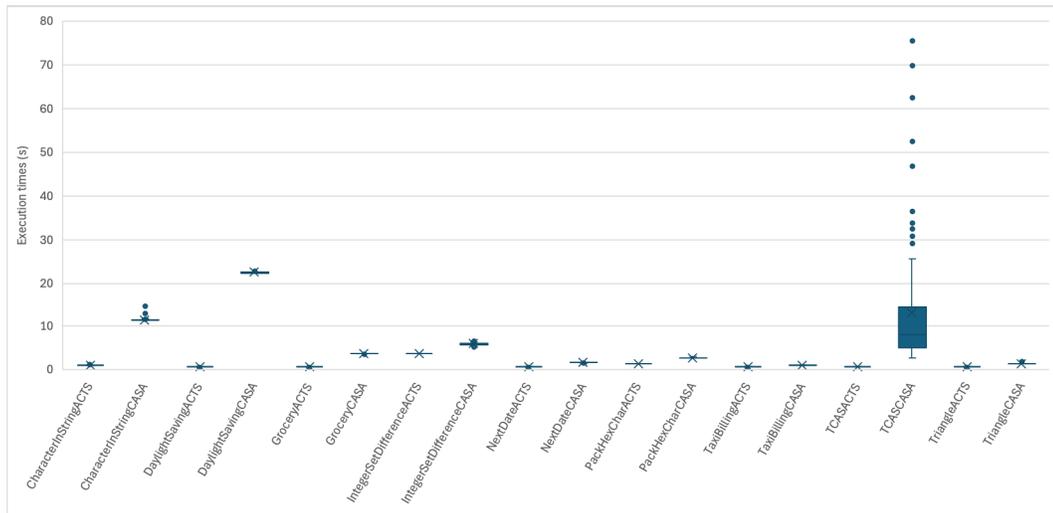

Figure 11.    Execution times (s) over 100 runs for the Three-way criterion

## VI. Conclusion

Category-Partition (CP) is a functional testing technique that has been successfully used in several contexts, both in academia and industry. It asks the user to follow a series of prescribed activities which can be cumbersome when performed solely by hand.

In this paper we describe a novel attempt to automate these activities by relying on several existing and proven technologies. To the best of our knowledge our solution provides the highest level of automation of CP among existing solutions.

The software solution we describe provides a user interface for the specification of input parameters and environment variables, their categories and choices along with constraints, which we refer to as the CP specification. Parameters are specified with a type, which currently can be either one of Boolean, Character, Enumeration, Integer, Real, String or Integer Set. Choices are precisely specified with a Boolean expression that follows the Java syntax.

The tool relies on covering array technologies to produce test frames, i.e., combinations of choices that satisfy at the same time a selection criterion prescribing that some tuples of choices be exercised and constraints between choices. As we discussed in the paper, our solution is not an alternative to combinatorial testing; we use those technologies because they are appropriate for our needs. This appears to us better than existing partly automated support of CP that rely on ad-hoc algorithms. Our tool supports the Base Choice, Each-Choice, Pairwise and Three-way selection criteria, which is a set of criteria much larger than what other tool supports offer. Other criteria can easily be supported. Our solution delegates the task of finding input values for test frames to an SMT solver, currently using Z3, which is again novel, especially considering the variety of parameter types our tool supports. To achieve this, it automatically transforms the Java choice descriptions into an input that Z3 can process. Our solution also verifies a number of properties of the CP specification which should hold for the CP specification to correctly lead to test frames and test cases. We believe our solution is the most complete tool to automatically support CP.

This paper described the JavaFX user interface of our solution, some of the design decisions we made and illustrated its use on nine examples in detail and succinctly reported on other experiences in the context of industry partnerships. Performance data indicate that the tool can make practical sense in practice. We envision that our tool will be of interest to different audiences. Instructors may find it useful when teaching CP since they can focus on CP activities which we believe are crucial to its performance, specifically the definition of the CP specification, and let our tool take care of what can be automated: CP verification, test frame construction, test input values identification. For similar reasons, students who are exposed to CP should find our solution useful when trying alternative CP specifications for a given problem and experimenting with their outcome (e.g., number of test frames).

Practitioners should find our solution useful too since it provides automated support to all the activities that can be automated. With automated support for tasks that can be automated, the user can spend more time on crucial steps of CP, such as defining the CP specification. We believe our solution should scale up to industry contexts; Its capability to record complex CP specifications has been exemplified during several Masters and Ph.D. theses (e.g., [39, 38, 40]). Of course, the more complex the CP specification the longer it will take CASA or ACTS to produce test frames and, the more complex choice specifications the longer it will take Z3 to find input values (if it does not time out). We have experimented with our tool in several industry contexts. In particular, we have shown how CP, with automated support, can help point out the extra testing cost of existing test suites [38]. Given encouraging execution time data we believe the tool can be very useful when used in an iterative process, as we have experimented when working ourselves on the examples: the user can experiment with test frame and test case construction, after CP verification, to fix possible issues with the CP specification as well as decide of an appropriate CP specification (e.g., decision on choices, on constraints) to lead to a number of test frames and then test cases that is compatible with a test budget.

Our solution should also be of interest to researchers; thanks to several instances of the use of the adapter design pattern for the automated generation of test frames for different selection criteria, and test inputs for those test frames, and the command line execution alternative, it should now be possible to experimentally study the impact of the various decisions one makes when using CP: Does using CASA or ACTS make a difference, and if there is a difference what is the scale of the difference, for instance in terms of the number of generated tests, in terms of fault detection? How do selection criteria compare in terms of cost and effectiveness at finding faults? If different CP specifications can be considered, can we characterize differences and measure their impact on cost and effectiveness to eventually guide the user towards defining effective CP specifications? These questions cannot be answered through analytical means; experiments are necessary and automation is key, while helping control experimental factors.

As we mentioned at the beginning of this paper, we report here on an on-going effort to automate CP as much as possible. Much remains to be done, including: providing additional feedback to the user so they can provide a better CP specification, in a fashion similar to what Briand and colleagues have done [30]; implementing additional recommendations for producing a good CP specification and avoiding pitfalls [9, 24]; addressing the oracle problem in the context of CP; integrating our automation ideas with a solution to handle structured input data with constraints [42].

## Acknowledgment

This research has received funding from the National Sciences and Engineering Research Council of Canada.

## Appendix A.   JavaCC grammar used by Melba

```
SKIP : { " " | "\t" | "\n" | "\r" | "\f" } /* WHITE SPACE */

TOKEN : { /* LITERALS */
      < INTEGER_LITERAL: <DECIMAL_LITERAL> >
      |
      < #DECIMAL_LITERAL: ["0"-"9"] (["0"-"9"])* >
      |
      < FLOATING_POINT_LITERAL:
            (["0"-"9"])+ "." (["0"-"9"])* (<EXPONENT>)?
          | "." (["0"-"9"])+ (<EXPONENT>)?
          | (["0"-"9"])+ <EXPONENT>
      >
      |
      < #EXPONENT: ["e","E"] (["+","-"])? (["0"-"9"])+ >
      |
      < STRING_LITERAL:
            "\""
            ( (~["\"","\\","\n","\r"])
            | ("\\" ["n","t","b","r","f","\\","'","\""]    )
            )*
            "\""
      >
      |
      < CHARACTER_LITERAL:
            "'"
            (~["'","\\"])
            "'"
      >
}

TOKEN : { /* SEPARATORS */
  < LPAREN: "(" >   |   < RPAREN: ")" >   |   < LBRACKET: "[" >   |   < RBRACKET: "]" >   |
  < COMMA: "," >    |   < DOT: "." >      |   < LBRACE: "{" >     |   < RBRACE: "}" >
}

TOKEN : { /* OPERATORS */
  < LT: "<" >   |   < GT: ">" >   |   < NOT: "!" >   |   < TILDE: "~" >   |   < EQ: "==" >
| < LE: "<=" > |   < GE: ">=" >  |   < NE: "!=" >   |   < SC_OR: "||" > |   < SC_AND: "&&" >
| < PLUS: "+" > |  < MINUS: "-" > |  < MUL: "*" >   |  < DIV: "/" >   |  < BIT_AND: "&" >
| < BIT_OR: "|" > | < POWER: "^" > |  < MOD: "%" >  |  < LSHIFT: "<<" >
}

TOKEN : { /* IDENTIFIERS */
      <IDENTIFIER: <LETTER>(<LETTER>|<DIGIT>)*>
      |
      < #LETTER: [ "A"-"Z", "_", "a"-"z" ] >
      |
      < #DIGIT: ["0"-"9"] >
}
```

## A.1    Grammar rules

```
MelbaExpression:  Expression

Expression: AndExpression ( <SC_OR> AndExpression )*

AndExpression:    EqualExpression ( <SC_AND> EqualExpression )*

EqualExpression: RelationalExpression ( <NE> RelationalExpression | <EQ>
                 RelationalExpression )*

RelationalExpression: AdditiveExpression ( <LT> AdditiveExpression | <GT>
                      AdditiveExpression | <LE> AdditiveExpression | <GE>
                      AdditiveExpression )*

AdditiveExpression: MultiplicativeExpression ( <PLUS> MultiplicativeExpression |
                    <MINUS> MultiplicativeExpression )*

MultiplicativeExpression:   UnaryExpression ( <MUL> UnaryExpression | <DOT>
                            UnaryExpression | <DIV> UnaryExpression | <MOD>
                            UnaryExpression )*

UnaryExpression: ( <PLUS> UnaryExpression | <MINUS> UnaryExpression
                | <NOT> UnaryExpression | PowerExpression )

PowerExpression:  UnaryExpressionNotPlusMinus ( <POWER> UnaryExpression )?

UnaryExpressionNotPlusMinus:
      ( AnyConstant | ( LOOKAHEAD(2) Function | Variable )
      | <LPAREN> Expression <RPAREN> | ListExpression )

ListExpression: ( <LBRACKET> Expression ( <COMMA> Expression )* <RBRACKET> )

Variable:   Identifier

Function:   Identifier <LPAREN> ArgumentList <RPAREN>

ArgumentList: ( Expression ( <COMMA> Expression )* )?

Identifier: <IDENTIFIER>

AnyConstant:  <STRING_LITERAL> | <CHARACTER_LITERAL> | RealConstant | <LBRACE>
              ArgumentList <RBRACE>

RealConstant:  <INTEGER_LITERAL> | <FLOATING_POINT_LITERAL>
```

The grammar below is a simplification of the JavaCC grammar Melba uses to transform a Java Boolean expression into Z3 instructions, using the Z3 API. The text illustrates the main elements of the grammar and omits Z3 API detailed calls but provides hints about what is being done.

```
BoolExpr MelbaExpression():
{       Expr theExpr;       }
{       theExpr = Expression() {
            // Z3 code to return a Boolean expression
        }
}

Expr Expression():
{       Expr leftExpr, rightExpr;       }
{       leftExpr = AndExpression()
        (       <SC_OR> rightExpr = AndExpression() {
                    // Z3 API calls to create an Or expression. Essentially:
                    // leftExpr = mkOr((BoolExpr)leftExpr,(BoolExpr)rightExpr)
                }
        )*
        return leftExpr;
}

Expr AndExpression():
{       Expr leftExpr, rightExpr;       }
{       leftExpr = EqualExpression()
        (       <SC_AND> rightExpr = EqualExpression() {
                    // Z3 API calls to create an And expression. Essentially:
                    // leftExpr = mkAnd((BoolExpr)leftExpr,(BoolExpr)rightExpr)
                }
        )*
        return leftExpr;
}

Expr EqualExpression():
{       Expr leftExpr, rightExpr;       }
{       leftExpr = RelationalExpression()
        (
            (
            <NE> rightExpr = RelationalExpression() {
                // Z3 API calls to produce a "not equal" expression. Essentially:
                // leftExpr = mkNot(Z3_context.mkEq(leftExpr,rightExpr))
                }
            )
            |
            (
            <EQ> rightExpr = RelationalExpression() {
                // Z3 API calls to produce an "equal" expression. Essentially:
                // leftExpr = mkEq(leftExpr,rightExpr)
                }
            )
        )*
        return leftExpr;
}
```

```
Expr RelationalExpression():
{       Expr leftExpr, rightExpr, result=null;     }
{       (
        leftExpr = AdditiveExpression()
        (
                <LT> rightExpr = AdditiveExpression() {
                        // Z3 API calls to produce a "lower than" expression:
                        // result = mkLt((ArithExpr)leftExpr, (ArithExpr)rightExpr)
                        }
                |
                <GT> rightExpr = AdditiveExpression() {
                        // Z3 API calls to produce a "greater than" expression:
                        // result = mkGt((ArithExpr)leftExpr, (ArithExpr)rightExpr)
                        }
                |
                <LE> rightExpr = AdditiveExpression() {
                        // Z3 API calls to produce a "lower or equal" expression:
                        // result = mkLe((ArithExpr)leftExpr, (ArithExpr)rightExpr)
                        }
                |
                        <GE> rightExpr = AdditiveExpression() {
                        // Z3 API calls to produce a "great or equal" expression:
                        // result = mkGe((ArithExpr)leftExpr, (ArithExpr)rightExpr)
                        }
        )*
        )
        return result;
}

Expr AdditiveExpression():
{       Expr leftExpr, rightExpr;       }
{       leftExpr = MultiplicativeExpression()
        (
                <PLUS> rightExpr = MultiplicativeExpression() {
                        // Z3 API calls to produce an addition:
                        // leftExpr = mkAdd(     (ArithExpr<ArithSort>)leftExpr,
                                                 (ArithExpr)rightExpr)
                }
                |
                <MINUS> rightExpr = MultiplicativeExpression() {
                        // Z3 API calls to create a substraction:
                        // leftExpr = Z3_context.mkAdd((ArithExpr)leftExpr,
                                        Z3_context.mkUnaryMinus((ArithExpr)rightExpr))
                }
        )*
        return leftExpr;
}
```

```
Expr MultiplicativeExpression():
{     Expr leftExpr, rightExpr;      }
{
      leftExpr = UnaryExpression()
      (
            <MUL> rightExpr = UnaryExpression() {
                  // Z3 API calls to create a multiplication:
                  // leftExpr = mkMul((ArithExpr)leftExpr, (ArithExpr)rightExpr)
                  }
            |
            {
                  // Code to handle function calls in Java expression.
                  // Pusing the expression to which the call applies to a stack.
            }
            <DOT> leftExpr = UnaryExpression()
            |
            <DIV> rightExpr = UnaryExpression() {
                  // Z3 API calls to create a division:
                  // leftExpr = mkSub((ArithExpr)leftExpr, (ArithExpr)rightExpr)
            }
            |
            <MOD> rightExpr = UnaryExpression() {
                  // Z3 API calls to create a modulo:
                  // leftExpr = mkMod((IntExpr)leftExpr, (IntExpr)rightExpr)
            }
      )*
      return leftExpr;
}

Expr UnaryExpression():
{     Expr result = null, Expr temp;      }
{
      (
            <PLUS> result = UnaryExpression()
            |
            <MINUS> temp = UnaryExpression() {
                  // Z3 API calls to create an unary minus:
                  // result = mkUnaryMinus((ArithExpr)temp)
                  }
            |
            <NOT> temp = UnaryExpression() {
                  // Z3 API calls to create a negation:
                  // result = mkNot((BoolExpr)temp)
                  }
            |
            result = PowerExpression()
      )
      return result;
}
```

```
Expr PowerExpression():
{     Expr leftExpr, rightExpr;       }
{
      leftExpr = UnaryExpressionNotPlusMinus()
      (
            <POWER> rightExpr = UnaryExpression() {
                  // Z3 API calls to create a power expression:
                  // return mkPower((ArithExpr)leftExpr, (ArithExpr)rightExpr)
            }
      )?
      return leftExpr;
}

Expr UnaryExpressionNotPlusMinus():
{     Expr result = null;       }
{
      (
      result = AnyConstant()
      |
            (
                  LOOKAHEAD(2)
                  result = Function()
                  |
                  result = Variable()
            )
      |
      <LPAREN> result = Expression() <RPAREN>
      |
      ListExpression()
      )
      return result;
}

void ListExpression():
{}
{
      ( <LBRACKET> Expression() ( <COMMA> Expression() )* <RBRACKET> )
}

Expr Variable():
{     Expr result;       }
{
      // Code to handle identifiers, i.e., variables used in the Java expression.
      // return Identifier(identifierType.REGULARID)
}

Expr Function():
{     Expr functionName;       }
{
      functionName = Identifier(identifierType.FUNCTIONID) <LPAREN>
            args = ArgumentList() <RPAREN>
      // Code to deal with function calls in the Java expression, depending
      // on the type to which the call is made: function on a string,
      // function on an integer set, function on a character
}
```

```
ArrayList<Expr> ArgumentList():
{       ArrayList<Expr> result = new ArrayList<Expr>();
        Expr anExpr;        }
{
        (
        anExpr = Expression()
        result.add(anExpr);
                (
                <COMMA>
                anExpr = Expression()
                result.add(anExpr);
                )*
        )?
        return result;
}

Expr Identifier(identifierType idType):
{       Token t;
        Expr id = null;     }
{
        t = <IDENTIFIER>
        // Code to deal with identifiers: e.g., variables, function names.
}

Expr AnyConstant():
{
        Token t;
        Expr result = null;
        ArrayList<Expr> args;
}
{
        (
        t = <STRING_LITERAL>
                {    // Z3 API calls to create a string literal
                     // result = mkString(t.image);
                }
        |
        t = <CHARACTER_LITERAL>
                {    // Z3 API calls to create a character literal from t.
                     // result = mkConst(t.image)
                }
        |
        result = RealConstant()
        |
        <LBRACE> args = ArgumentList() <RBRACE>
                {
                     result = createIntegerSetLiteral(args);
                }
        )
        return result;
}
```

```
ArithExpr RealConstant():
{       Token t;
        ArithExpr result = null;        }
{
        (
        t = <INTEGER_LITERAL>
                {       // Z3 API calls to create an integer literal
                        // result = mkInt(t.image)
                }
        |
        t = <FLOATING_POINT_LITERAL>
                {       // Z3 API calls to create a floating point literal
                        // result = mkReal(t.image)
                }
        )
        return result;
}
```

# Appendix B. TRIANGLE CASE STUDY

In the CP specification below, an asterisk ('*') beside a choice indicates the base choice for that category. Each parameter is uniquely identified by an integer identifier, which is then used in the composition of test frames; As a result, a choice for a parameter is uniquely identified by the parameter identifier, the category identifier and the choice identifier. Below we summarize the specification of the unit under test (section B.1), and provide the CP specification we used (section B.2). We then provide the test frames and corresponding test cases (inputs) we obtain when using CASA (Each-Choice in section B.3, Pairwise in section B.5, Three-way in section B.7), when using ACTS (Each-Choice in section B.4, Pairwise in section B.6, Three-way in section B.8), and when using the Base Choice criterion (section B.9). The CP specification below is adapted from earlier work [16, 38].

## B.1 Specification

The unit under test takes three input parameters, SA, SB and SC, of type integer, and identifies (output) whether the three values represent the lengths of a triangle (using triangle inequalities) and, in the event this is the case, it identifies the type of triangle they specify (either equilateral, isosceles, or scalene).

## B.2 CP Specification

```
Parameter(2638): SA of type Integer
  Category 1 - Values for SA
    *Ch 1.1  Strictly positive
             SA>0
             [properties SApositive]
     Ch 1.2  Negative or zero
             SA<=0
             [error]
  Category 2 - SA compared to sum of other two
    *Ch 2.1  SA strictly smaller than sum of other two
             SA<SB+SC
     Ch 2.2  SA greater than sum of other two
             SA>=SB+SC
             [if (!SBgreaterThanSumOthers && SCpositive) && (!SCgreaterThanSumOthers &&
             SBpositive) && (!SAequalsSB || !SCpositive) && (!SCequalsSA ||
             !SBpositive)][single][properties SAgreaterThanSumOthers]
  Category 3 - SA compared to SB
     Ch 3.1  SA equals to SB
             SA==SB
             [properties SAequalsSB]
    *Ch 3.2  SA and SB are different
             SA!=SB
             [properties SAdifferentFromSB]
Parameter(2639): SB of type Integer
  Category 1 - Values for SB
    *Ch 1.1  Strictly positive
             SB>0
             [properties SBpositive]
     Ch 1.2  Negative or zero
             SB<=0
             [error]
  Category 2 - SB compared to SC
     Ch 2.1  SB equals to SC
             SB==SC
             [properties SBequalsSC]
    *Ch 2.2  SB and SC are different
             SB!=SC
             [properties SBdifferentFromSC]
  Category 3 - SB compared to sum of other two
```

```
     *Ch 3.1   SB strictly smaller than sum of other two
               SB<SA+SC
      Ch 3.2   SB greater than sum of other two
               SB>=SA+SC
               [if (!SAgreaterThanSumOthers && SCpositive) && (!SCgreaterThanSumOthers &&
               SApositive) && (!SAequalsSB || !SCpositive) && (!SBequalsSC ||
               !SApositive)][single][properties SBgreaterThanSumOthers]
Parameter(2640): SC of type Integer
   Category 1 - SC compared to SA
      Ch 1.1   SC equals to SA
               SC==SA
               [if !(SAequalsSB && SBdifferentFromSC) && !(SAdifferentFromSB && SBdifferentFromSC)
               && !(SAdifferentFromSB && SBequalsSC)][properties SCequalsSA]
     *Ch 1.2   SC different from SA
               SC!=SA
               [if !(SAequalsSB && SBequalsSC)][properties SCdifferentFromSA]
   Category 2 - SC compared to sum of other two
     *Ch 2.1   SC strictly smaller than sum of other two
               SC<SA+SB
      Ch 2.2   SC greater than sum of other two
               SC>=SA+SB
               [if (!SBgreaterThanSumOthers && SApositive) && (!SAgreaterThanSumOthers &&
               SBpositive) && (!SCequalsSA || !SBpositive) && (!SBequalsSC ||
               !SApositive)][single][properties SCgreaterThanSumOthers]
   Category 3 - Values for SC
      Ch 3.1   SC negative or zero
               SC<=0
               [error]
     *Ch 3.2   SC strictly positive
               SC>0
               [properties SCpositive]
```

## *B.3    Each-Choice Test Frames and Corresponding Test Cases when using CASA*

**Test Case 1:**
```
choices:    2638Cat1Ch1-|-2638Cat2Ch1-|-2638Cat3Ch2-|-2639Cat1Ch1-|-2639Cat2Ch2-|-2639Cat3Ch1-
            |-2640Cat1Ch2-|-2640Cat2Ch1-|-2640Cat3Ch2
SA = Values for SA = Strictly positive|SA = SA compared to sum of other two = SA strictly
smaller than sum of other two|SA = SA compared to SB = SA and SB are different|SB = Values for
SB = Strictly positive|SB = SB compared to SC = SB and SC are different|SB = SB compared to sum
of other two = SB strictly smaller than sum of other two|SC = SC compared to SA = SC different
from SA|SC = SC compared to sum of other two = SC strictly smaller than sum of other two|SC =
Values for SC = SC strictly positive
values:     SB = '2', SA = '3', SC = '4'
```
**Test Case 2:**
```
choices:    2638Cat1Ch1-|-2638Cat2Ch1-|-2638Cat3Ch1-|-2639Cat1Ch1-|-2639Cat2Ch1-|-2639Cat3Ch1-
            |-2640Cat1Ch1-|-2640Cat2Ch1-|-2640Cat3Ch2
SA = Values for SA = Strictly positive|SA = SA compared to sum of other two = SA strictly
smaller than sum of other two|SA = SA compared to SB = SA equals to SB|SB = Values for SB =
Strictly positive|SB = SB compared to SC = SB equals to SC|SB = SB compared to sum of other two
= SB strictly smaller than sum of other two|SC = SC compared to SA = SC equals to SA|SC = SC
compared to sum of other two = SC strictly smaller than sum of other two|SC = Values for SC = SC
strictly positive
values:     SB = '1', SA = '1', SC = '1'
```
**Test Case 3:**
```
choices:    2638Cat1Ch1-|-2638Cat2Ch2-|-2638Cat3Ch2-|-2639Cat1Ch1-|-2639Cat2Ch1-|-2639Cat3Ch1-
            |-2640Cat1Ch2-|-2640Cat2Ch1-|-2640Cat3Ch2
SA = Values for SA = Strictly positive|SA = SA compared to sum of other two = SA greater than
sum of other two|SA = SA compared to SB = SA and SB are different|SB = Values for SB = Strictly
positive|SB = SB compared to SC = SB equals to SC|SB = SB compared to sum of other two = SB
strictly smaller than sum of other two|SC = SC compared to SA = SC different from SA|SC = SC
compared to sum of other two = SC strictly smaller than sum of other two|SC = Values for SC = SC
strictly positive
values:     SB = '1', SA = '2', SC = '1'
```

**Test Case 4:**
```
choices:      2638Cat1Ch1−|−2638Cat2Ch1−|−2638Cat3Ch2−|−2639Cat1Ch1−|−2639Cat2Ch2−|−2639Cat3Ch2−
              |−2640Cat1Ch2−|−2640Cat2Ch1−|−2640Cat3Ch2
```
SA = Values for SA = Strictly positive|SA = SA compared to sum of other two = SA strictly
smaller than sum of other two|SA = SA compared to SB = SA and SB are different|SB = Values for
SB = Strictly positive|SB = SB compared to SC = SB and SC are different|SB = SB compared to sum
of other two = SB greater than sum of other two|SC = SC compared to SA = SC different from SA|SC
= SC compared to sum of other two = SC strictly smaller than sum of other two|SC = Values for SC
= SC strictly positive
```
values:       SB = '3', SA = '1', SC = '2'
```
**Test Case 5:**
```
choices:      2638Cat1Ch1−|−2638Cat2Ch1−|−2638Cat3Ch1−|−2639Cat1Ch1−|−2639Cat2Ch2−|−2639Cat3Ch1−
              |−2640Cat1Ch2−|−2640Cat2Ch1−|−2640Cat3Ch2
```
SA = Values for SA = Strictly positive|SA = SA compared to sum of other two = SA strictly
smaller than sum of other two|SA = SA compared to SB = SA and SB are different|SB = Values for SB =
Strictly positive|SB = SB compared to SC = SB and SC are different|SB = SB compared to sum of
other two = SB strictly smaller than sum of other two|SC = SC compared to SA = SC different from
SA|SC = SC compared to sum of other two = SC greater than sum of other two|SC = Values for SC =
SC strictly positive
```
values:       SB = '1', SA = '1', SC = '2'
```
**Test Case 6:**
```
choices:      2638Cat1Ch2
```
SA = Values for SA = Negative or zero
```
values:       SA = '0'
```
**Test Case 7:**
```
choices:      2639Cat1Ch2
```
SB = Values for SB = Negative or zero
```
values:       SB = '0'
```
**Test Case 8:**
```
choices:      2640Cat3Ch1
```
SC = Values for SC = SC negative or zero
```
values:       SC = '0'
```

## B.4    Each-Choice Test Frames and Corresponding Test Cases when using ACTS

**Test Case 1:**
```
choices:      2638Cat3Ch1−|−2639Cat2Ch2−|−2640Cat1Ch2−|−2638Cat1Ch1−|−2638Cat2Ch1−|−2639Cat1Ch1−
              |−2639Cat3Ch1−|−2640Cat2Ch1−|−2640Cat3Ch2
```
SA = SA compared to SB = SA equals to SB|SB = SB compared to SC = SB and SC are different|SC =
SC compared to SA = SC different from SA|SA = Values for SA = Strictly positive|SA = SA compared
to sum of other two = SA strictly smaller than sum of other two|SB = Values for SB = Strictly
positive|SB = SB compared to sum of other two = SB strictly smaller than sum of other two|SC =
SC compared to sum of other two = SC strictly smaller than sum of other two|SC = Values for SC =
SC strictly positive
```
values:       SB = '2', SA = '2', SC = '3'
```
**Test Case 2:**
```
choices:      2638Cat3Ch2−|−2639Cat2Ch1−|−2640Cat1Ch2−|−2638Cat1Ch1−|−2638Cat2Ch1−|−2639Cat1Ch1−
              |−2639Cat3Ch1−|−2640Cat2Ch1−|−2640Cat3Ch2
```
SA = SA compared to SB = SA and SB are different|SB = SB compared to SC = SB equals to SC|SC =
SC compared to SA = SC different from SA|SA = Values for SA = Strictly positive|SA = SA compared
to sum of other two = SA strictly smaller than sum of other two|SB = Values for SB = Strictly
positive|SB = SB compared to sum of other two = SB strictly smaller than sum of other two|SC =
SC compared to sum of other two = SC strictly smaller than sum of other two|SC = Values for SC =
SC strictly positive
```
values:       SB = '2', SA = '1', SC = '2'
```
**Test Case 3:**
```
choices:      2638Cat3Ch1−|−2639Cat2Ch1−|−2640Cat1Ch1−|−2638Cat1Ch1−|−2638Cat2Ch1−|−2639Cat1Ch1−
              |−2639Cat3Ch1−|−2640Cat2Ch1−|−2640Cat3Ch2
```
SA = SA compared to SB = SA equals to SB|SB = SB compared to SC = SB equals to SC|SC = SC
compared to SA = SC equals to SA|SA = Values for SA = Strictly positive|SA = SA compared to sum
of other two = SA strictly smaller than sum of other two|SB = Values for SB = Strictly
positive|SB = SB compared to sum of other two = SB strictly smaller than sum of other two|SC =
SC compared to sum of other two = SC strictly smaller than sum of other two|SC = Values for SC =
SC strictly positive
```
values:       SB = '1', SA = '1', SC = '1'
```

**Test Case 4:**
```
choices:      2638Cat2Ch1−|−2638Cat3Ch2−|−2639Cat2Ch2−|−2639Cat3Ch2−|−2640Cat1Ch2−|−2640Cat2Ch1−
              |−2638Cat1Ch1−|−2639Cat1Ch1−|−2640Cat3Ch2
```
SA = SA compared to sum of other two = SA strictly smaller than sum of other two|SA = SA
compared to SB = SA and SB are different|SB = SB compared to SC = SB and SC are different|SB =
SB compared to sum of other two = SB greater than sum of other two|SC = SC compared to SA = SC
different from SA|SC = SC compared to sum of other two = SC strictly smaller than sum of other
two|SA = Values for SA = Strictly positive|SB = Values for SB = Strictly positive|SC = Values
for SC = SC strictly positive
```
values:       SB = '3', SA = '1', SC = '2'
```
**Test Case 5:**
```
choices:      2638Cat2Ch2−|−2638Cat3Ch2−|−2639Cat2Ch1−|−2639Cat3Ch1−|−2640Cat1Ch2−|−2640Cat2Ch1−
              |−2638Cat1Ch1−|−2639Cat1Ch1−|−2640Cat3Ch2
```
SA = SA compared to sum of other two = SA greater than sum of other two|SA = SA compared to SB =
SA and SB are different|SB = SB compared to SC = SB equals to SC|SB = SB compared to sum of
other two = SB strictly smaller than sum of other two|SC = SC compared to SA = SC different from
SA|SC = SC compared to sum of other two = SC strictly smaller than sum of other two|SA = Values
for SA = Strictly positive|SB = Values for SB = Strictly positive|SC = Values for SC = SC
strictly positive
```
values:       SB = '1', SA = '2', SC = '1'
```
**Test Case 6:**
```
choices:      2638Cat2Ch1−|−2638Cat3Ch1−|−2639Cat2Ch2−|−2639Cat3Ch1−|−2640Cat1Ch2−|−2640Cat2Ch2−
              |−2638Cat1Ch1−|−2639Cat1Ch1−|−2640Cat3Ch2
```
SA = SA compared to sum of other two = SA strictly smaller than sum of other two|SA = SA
compared to SB = SA equals to SB|SB = SB compared to SC = SB and SC are different|SB = SB
compared to sum of other two = SB strictly smaller than sum of other two|SC = SC compared to SA
= SC different from SA|SC = SC compared to sum of other two = SC greater than sum of other
two|SA = Values for SA = Strictly positive|SB = Values for SB = Strictly positive|SC = Values
for SC = SC strictly positive
```
values:       SB = '1', SA = '1', SC = '2'
```
**Test Case 7:**
```
choices:      2638Cat1Ch2
```
SA = Values for SA = Negative or zero
```
values:       SA = '0'
```
**Test Case 8:**
```
choices:      2639Cat1Ch2
```
SB = Values for SB = Negative or zero
```
values:       SB = '0'
```
**Test Case 9:**
```
choices:      2640Cat3Ch1
```
SC = Values for SC = SC negative or zero
```
values:       SC = '0'
```

## B.5    Pairwise Test Frames and Corresponding Test Cases when using CASA

**Test Case 1:**
```
choices:      2638Cat1Ch1−|−2638Cat2Ch1−|−2638Cat3Ch1−|−2639Cat1Ch1−|−2639Cat2Ch1−|−2639Cat3Ch1−
              |−2640Cat1Ch1−|−2640Cat2Ch1−|−2640Cat3Ch1
```
SA = Values for SA = Strictly positive|SA = SA compared to sum of other two = SA strictly
smaller than sum of other two|SA = SA compared to SB = SA equals to SB|SB = Values for SB =
Strictly positive|SB = SB compared to SC = SB equals to SC|SB = SB compared to sum of other two
= SB strictly smaller than sum of other two|SC = SC compared to SA = SC equals to SA|SC = SC
compared to sum of other two = SC strictly smaller than sum of other two|SC = Values for SC = SC
strictly positive
```
values:       SB = '1', SA = '1', SC = '1'
```
**Test Case 2:**
```
choices:      2638Cat1Ch1−|−2638Cat2Ch1−|−2638Cat3Ch2−|−2639Cat1Ch1−|−2639Cat2Ch1−|−2639Cat3Ch1−
              |−2640Cat1Ch2−|−2640Cat2Ch1−|−2640Cat3Ch2
```
SA = Values for SA = Strictly positive|SA = SA compared to sum of other two = SA strictly
smaller than sum of other two|SA = SA compared to SB = SA and SB are different|SB = Values for
SB = Strictly positive|SB = SB compared to SC = SB equals to SC|SB = SB compared to sum of other
two = SB strictly smaller than sum of other two|SC = SC compared to SA = SC different from SA|SC
= SC compared to sum of other two = SC strictly smaller than sum of other two|SC = Values for SC
= SC strictly positive

values:      SB = '2', SA = '1', SC = '2'

**Test Case 3:**
choices:      2638Cat1Ch1−|−2638Cat2Ch1−|−2638Cat3Ch2−|−2639Cat1Ch1−|−2639Cat2Ch2−|−2639Cat3Ch1−
              |−2640Cat1Ch2−|−2640Cat2Ch1−|−2640Cat3Ch2

SA = Values for SA = Strictly positive|SA = SA compared to sum of other two = SA strictly
smaller than sum of other two|SA = SA compared to SB = SA and SB are different|SB = Values for
SB = Strictly positive|SB = SB compared to SC = SB and SC are different|SB = SB compared to sum
of other two = SB strictly smaller than sum of other two|SC = SC compared to SA = SC different
from SA|SC = SC compared to sum of other two = SC strictly smaller than sum of other two|SC =
Values for SC = SC strictly positive
values:      SB = '4', SA = '2', SC = '3'

**Test Case 4:**
choices:      2638Cat1Ch1−|−2638Cat2Ch1−|−2638Cat3Ch1−|−2639Cat1Ch1−|−2639Cat2Ch2−|−2639Cat3Ch1−
              |−2640Cat1Ch2−|−2640Cat2Ch1−|−2640Cat3Ch2

SA = Values for SA = SA compared to sum of other two = SA strictly
smaller than sum of other two|SA = SA compared to SB = SA equals to SB|SB = Values for SB =
Strictly positive|SB = SB compared to SC = SB and SC are different|SB = SB compared to sum of
other two = SB strictly smaller than sum of other two|SC = SC compared to SA = SC different from
SA|SC = SC compared to sum of other two = SC strictly smaller than sum of other two|SC = Values
for SC = SC strictly positive
values:      SB = '2', SA = '2', SC = '3'

**Test Case 5:**
choices:      2638Cat1Ch1−|−2638Cat2Ch1−|−2638Cat3Ch2−|−2639Cat1Ch1−|−2639Cat2Ch2−|−2639Cat3Ch2−
              |−2640Cat1Ch2−|−2640Cat2Ch1−|−2640Cat3Ch2

SA = Values for SA = Strictly positive|SA = SA compared to sum of other two = SA strictly
smaller than sum of other two|SA = SA compared to SB = SA and SB are different|SB = Values for
SB = Strictly positive|SB = SB compared to SC = SB and SC are different|SB = SB compared to sum
of other two = SB greater than sum of other two|SC = SC compared to SA = SC different from SA|SC
= SC compared to sum of other two = SC strictly smaller than sum of other two|SC = Values for SC
= SC strictly positive
values:      SB = '3', SA = '1', SC = '2'

**Test Case 6:**
choices:      2638Cat1Ch1−|−2638Cat2Ch2−|−2638Cat3Ch2−|−2639Cat1Ch1−|−2639Cat2Ch1−|−2639Cat3Ch1−
              |−2640Cat1Ch2−|−2640Cat2Ch1−|−2640Cat3Ch2

SA = Values for SA = Strictly positive|SA = SA compared to sum of other two = SA greater than
sum of other two|SA = SA compared to SB = SA and SB are different|SB = Values for SB = Strictly
positive|SB = SB compared to SC = SB equals to SC|SB = SB compared to sum of other two = SB
strictly smaller than sum of other two|SC = SC compared to SA = SC different from SA|SC = SC
compared to sum of other two = SC strictly smaller than sum of other two|SC = Values for SC = SC
strictly positive
values:      SB = '1', SA = '2', SC = '1'

**Test Case 7:**
choices:      2638Cat1Ch1−|−2638Cat2Ch1−|−2638Cat3Ch1−|−2639Cat1Ch1−|−2639Cat2Ch2−|−2639Cat3Ch1−
              |−2640Cat1Ch2−|−2640Cat2Ch2−|−2640Cat3Ch2

SA = Values for SA = Strictly positive|SA = SA compared to sum of other two = SA strictly
smaller than sum of other two|SA = SA compared to SB = SA equals to SB|SB = Values for SB =
Strictly positive|SB = SB compared to SC = SB and SC are different|SB = SB compared to sum of
other two = SB strictly smaller than sum of other two|SC = SC compared to SA = SC different from
SA|SC = SC compared to sum of other two = SC greater than sum of other two|SC = Values for SC =
SC strictly positive
values:      SC = '2', SB = '1', SA = '1'

**Test Case 8:**
choices:      2638Cat1Ch2
SA = Values for SA = Negative or zero
values:      SA = '0'

**Test Case 9:**
choices:      2639Cat1Ch2
SB = Values for SB = Negative or zero
values:      SB = '0'

**Test Case 10:**
choices:      2640Cat3Ch1
SC = Values for SC = SC negative or zero
values:      SC = '0'

## B.6    Pairwise Test Frames and Corresponding Test Cases when using ACTS

**Test Case 1:**
choices:      2638Cat3Ch1−|−2639Cat2Ch1−|−2640Cat1Ch1−|−2638Cat1Ch1−|−2638Cat2Ch1−|−2639Cat1Ch1−
              |−2639Cat3Ch1−|−2640Cat2Ch1−|−2640Cat3Ch2
SA = SA compared to SB = SA equals to SB|SB = SB compared to SC = SB equals to SC|SC = SC
compared to SA = SC equals to SA|SA = Values for SA = Strictly positive|SA = SA compared to sum
of other two = SA strictly smaller than sum of other two|SB = Values for SB = Strictly
positive|SB = SB compared to sum of other two = SB strictly smaller than sum of other two|SC =
SC compared to sum of other two = SC strictly smaller than sum of other two|SC = Values for SC =
SC strictly positive
values:       SB = '1', SA = '1', SC = '1'

**Test Case 2:**
choices:      2638Cat3Ch1−|−2639Cat2Ch2−|−2640Cat1Ch2−|−2638Cat1Ch1−|−2638Cat2Ch1−|−2639Cat1Ch1−
              |−2639Cat3Ch1−|−2640Cat2Ch1−|−2640Cat3Ch2
SA = SA compared to SB = SA equals to SB|SB = SB compared to SC = SB and SC are different|SC =
SC compared to SA = SC different from SA|SA = Values for SA = Strictly positive|SA = SA compared
to sum of other two = SA strictly smaller than sum of other two|SB = Values for SB = Strictly
positive|SB = SB compared to sum of other two = SB strictly smaller than sum of other two|SC =
SC compared to sum of other two = SC strictly smaller than sum of other two|SC = Values for SC =
SC strictly positive
values:       SB = '2', SA = '2', SC = '3'

**Test Case 3:**
choices:      2638Cat3Ch2−|−2639Cat2Ch1−|−2640Cat1Ch2−|−2638Cat1Ch1−|−2638Cat2Ch1−|−2639Cat1Ch1−
              |−2639Cat3Ch1−|−2640Cat2Ch1−|−2640Cat3Ch2
SA = SA compared to SB = SA and SB are different|SB = SB compared to SC = SB equals to SC|SC =
SC compared to SA = SC different from SA|SA = Values for SA = Strictly positive|SA = SA compared
to sum of other two = SA strictly smaller than sum of other two|SB = Values for SB = Strictly
positive|SB = SB compared to sum of other two = SB strictly smaller than sum of other two|SC =
SC compared to sum of other two = SC strictly smaller than sum of other two|SC = Values for SC =
SC strictly positive
values:       SB = '2', SA = '1', SC = '2'

**Test Case 4:**
choices:      2638Cat3Ch2−|−2639Cat2Ch2−|−2640Cat1Ch2−|−2638Cat1Ch1−|−2638Cat2Ch1−|−2639Cat1Ch1−
              |−2639Cat3Ch1−|−2640Cat2Ch1−|−2640Cat3Ch2
SA = SA compared to SB = SA and SB are different|SB = SB compared to SC = SB and SC are
different|SC = SC compared to SA = SC different from SA|SA = Values for SA = Strictly
positive|SA = SA compared to sum of other two = SA strictly smaller than sum of other two|SB =
Values for SB = Strictly positive|SB = SB compared to sum of other two = SB strictly smaller
than sum of other two|SC = SC compared to sum of other two = SC strictly smaller than sum of
other two|SC = Values for SC = SC strictly positive
values:       SB = '2', SA = '4', SC = '3'

**Test Case 5:**
choices:      2638Cat2Ch1−|−2638Cat3Ch1−|−2639Cat3Ch1−|−2639Cat3Ch1−|−2640Cat1Ch2−|−2640Cat2Ch2−
              |−2638Cat1Ch1−|−2639Cat1Ch1−|−2640Cat3Ch2
SA = SA compared to sum of other two = SA strictly smaller than sum of other two|SA = SA
compared to SB = SA equals to SB|SB = SB compared to SC = SB and SC are different|SB = SB
compared to sum of other two = SB strictly smaller than sum of other two|SC = SC compared to SA
= SC different from SA|SC = SC compared to sum of other two = SC greater than sum of other
two|SA = Values for SA = Strictly positive|SB = Values for SB = Strictly positive|SC = Values
for SC = SC strictly positive
values:       SB = '1', SA = '1', SC = '2'

**Test Case 6:**
choices:      2638Cat2Ch2−|−2638Cat3Ch2−|−2639Cat2Ch2−|−2639Cat3Ch1−|−2640Cat1Ch2−|−2640Cat2Ch1−
              |−2638Cat1Ch1−|−2639Cat1Ch1−|−2640Cat3Ch2
SA = SA compared to sum of other two = SA greater than sum of other two|SA = SA compared to SB =
SA and SB are different|SB = SB compared to SC = SB and SC are different|SB = SB compared to sum
of other two = SB strictly smaller than sum of other two|SC = SC compared to SA = SC different
from SA|SC = SC compared to sum of other two = SC strictly smaller than sum of other two|SA =
Values for SA = Strictly positive|SB = Values for SB = Strictly positive|SC = Values for SC = SC
strictly positive
values:       SB = '1', SA = '3', SC = '2'

**Test Case 7:**
choices:      2638Cat2Ch1−|−2638Cat3Ch2−|−2639Cat2Ch2−|−2639Cat3Ch2−|−2640Cat1Ch2−|−2640Cat2Ch1−
              |−2638Cat1Ch1−|−2639Cat1Ch1−|−2640Cat3Ch2

```
SA = SA compared to sum of other two = SA strictly smaller than sum of other two|SA = SA
compared to SB = SA and SB are different|SB = SB compared to SC = SB and SC are different|SB =
SB compared to sum of other two = SB greater than sum of other two|SC = SC compared to SA = SC
different from SA|SC = SC compared to sum of other two = SC strictly smaller than sum of other
two|SA = Values for SA = Strictly positive|SB = Values for SB = Strictly positive|SC = Values
for SC = SC strictly positive
values:        SC = '2', SB = '3', SA = '1'
```
**Test Case 8:**
```
choices:       2638Cat1Ch2
SA = Values for SA = Negative or zero
values:        SA = '0'
```
**Test Case 9:**
```
choices:       2639Cat1Ch2
SB = Values for SB = Negative or zero
values:        SB = '0'
```
**Test Case 10:**
```
choices:       2640Cat3Ch1
SC = Values for SC = SC negative or zero
values:        SC = '0'
```

## B.7    Three-way Test Frames and Corresponding Test Cases when using CASA

**Test Case 1:**
```
choices:       2638Cat1Ch1−|−2638Cat2Ch1−|−2638Cat3Ch2−|−2639Cat1Ch1−|−2639Cat2Ch1−|−2639Cat3Ch1−
               |−2640Cat1Ch2−|−2640Cat2Ch1−|−2640Cat3Ch2
SA = Values for SA = Strictly positive|SA = SA compared to sum of other two = SA strictly
smaller than sum of other two|SA = SA compared to SB = SA and SB are different|SB = Values for
SB = Strictly positive|SB = SB compared to SC = SB equals to SC|SB = SB compared to sum of other
two = SB strictly smaller than sum of other two|SC = SC compared to SA = SC different from SA|SC
= SC compared to sum of other two = SC strictly smaller than sum of other two|SC = Values for SC
= SC strictly positive
values:        SB = '2', SA = '1', SC = '2'
```
**Test Case 2:**
```
choices:       2638Cat1Ch1−|−2638Cat2Ch1−|−2638Cat3Ch1−|−2639Cat1Ch1−|−2639Cat2Ch2−|−2639Cat3Ch1−
               |−2640Cat1Ch2−|−2640Cat2Ch1−|−2640Cat3Ch2
SA = Values for SA = Strictly positive|SA = SA compared to sum of other two = SA strictly
smaller than sum of other two|SA = SA compared to SB = SA equals to SB|SB = Values for SB =
Strictly positive|SB = SB compared to SC = SB and SC are different|SB = SB compared to sum of
other two = SB strictly smaller than sum of other two|SC = SC compared to SA = SC different from
SA|SC = SC compared to sum of other two = SC strictly smaller than sum of other two|SC = Values
for SC = SC strictly positive
values:        SB = '2', SA = '2', SC = '3'
```
**Test Case 3:**
```
choices:       2638Cat1Ch1−|−2638Cat2Ch1−|−2638Cat3Ch2−|−2639Cat1Ch1−|−2639Cat2Ch2−|−2639Cat3Ch1−
               |−2640Cat1Ch2−|−2640Cat2Ch1−|−2640Cat3Ch2
SA = Values for SA = Strictly positive|SA = SA compared to sum of other two = SA strictly
smaller than sum of other two|SA = SA compared to SB = SA and SB are different|SB = Values for
SB = Strictly positive|SB = SB compared to SC = SB and SC are different|SB = SB compared to sum
of other two = SB strictly smaller than sum of other two|SC = SC compared to SA = SC different
from SA|SC = SC compared to sum of other two = SC strictly smaller than sum of other two|SC =
Values for SC = SC strictly positive
values:        SB = '4', SA = '2', SC = '3'
```
**Test Case 4:**
```
choices:       2638Cat1Ch1−|−2638Cat2Ch1−|−2638Cat3Ch1−|−2639Cat1Ch1−|−2639Cat2Ch1−|−2639Cat3Ch1−
               |−2640Cat1Ch1−|−2640Cat2Ch1−|−2640Cat3Ch2
SA = Values for SA = Strictly positive|SA = SA compared to sum of other two = SA strictly
smaller than sum of other two|SA = SA compared to SB = SA equals to SB|SB = Values for SB =
Strictly positive|SB = SB compared to SC = SB equals to SC|SB = SB compared to sum of other two
= SB strictly smaller than sum of other two|SC = SC compared to SA = SC equals to SA|SC = SC
compared to sum of other two = SC strictly smaller than sum of other two|SC = Values for SC = SC
strictly positive
values:        SB = '1', SA = '1', SC = '1'
```
**Test Case 5:**
```
choices:       2638Cat1Ch1−|−2638Cat2Ch1−|−2638Cat3Ch2−|−2639Cat1Ch1−|−2639Cat2Ch2−|−2639Cat3Ch2−
               |−2640Cat1Ch2−|−2640Cat2Ch1−|−2640Cat3Ch2
```

```
SA = Values for SA = Strictly positive|SA = SA compared to sum of other two = SA strictly
smaller than sum of other two|SA = SA compared to SB = SA and SB are different|SB = Values for
SB = Strictly positive|SB = SB compared to SC = SB and SC are different|SB = SB compared to sum
of other two = SB greater than sum of other two|SC = SC compared to SA = SC different from SA|SC
= SC compared to sum of other two = SC strictly smaller than sum of other two|SC = Values for SC
= SC strictly positive
values:     SB = '3', SA = '1', SC = '2'
```

**Test Case 6:**
```
choices:    2638Cat1Ch1−|−2638Cat2Ch2−|−2638Cat3Ch2−|−2639Cat1Ch1−|−2639Cat2Ch2−|−2639Cat3Ch1−
            |−2640Cat1Ch2−|−2640Cat2Ch1−|−2640Cat3Ch2
```
```
SA = Values for SA = Strictly positive|SA = SA compared to sum of other two = SA greater than
sum of other two|SA = SA compared to SB = SA and SB are different|SB = Values for SB = Strictly
positive|SB = SB compared to SC = SB and SC are different|SB = SB compared to sum of other two =
SB strictly smaller than sum of other two|SC = SC compared to SA = SC different from SA|SC = SC
compared to sum of other two = SC strictly smaller than sum of other two|SC = Values for SC = SC
strictly positive
values:     SB = '1', SA = '3', SC = '2'
```
**Test Case 7:**
```
choices:    2638Cat1Ch1−|−2638Cat2Ch1−|−2638Cat3Ch1−|−2639Cat1Ch1−|−2639Cat2Ch2−|−2639Cat3Ch1−
            |−2640Cat1Ch2−|−2640Cat2Ch2−|−2640Cat3Ch2
```
```
SA = Values for SA = Strictly positive|SA = SA compared to sum of other two = SA strictly
smaller than sum of other two|SA = SA compared to SB = SA equals to SB|SB = Values for SB =
Strictly positive|SB = SB compared to SC = SB and SC are different|SB = SB compared to sum of
other two = SB strictly smaller than sum of other two|SC = SC compared to SA = SC different from
SA|SC = SC compared to sum of other two = SC greater than sum of other two|SC = Values for SC =
SC strictly positive
values:     SC = '2', SB = '1', SA = '1'
```
**Test Case 8:**
```
choices:    2638Cat1Ch2
SA = Values for SA = Negative or zero
values:     SA = '0'
```
**Test Case 9:**
```
choices:    2639Cat1Ch2
SB = Values for SB = Negative or zero
values:     SB = '0'
```
**Test Case 10:**
```
choices:    2640Cat3Ch1
SC = Values for SC = SC negative or zero
values:     SC = '0'
```

## B.8    Three-way Test Frames and Corresponding Test Cases when using ACTS

**Test Case 1:**
```
choices:    2638Cat3Ch1−|−2639Cat2Ch1−|−2640Cat1Ch1−|−2638Cat1Ch1−|−2638Cat2Ch1−|−2639Cat1Ch1−
            |−2639Cat3Ch1−|−2640Cat2Ch1−|−2640Cat3Ch2
```
```
SA = SA compared to SB = SA equals to SB|SB = SB compared to SC = SB equals to SC|SC = SC
compared to SA = SC equals to SA|SA = Values for SA = Strictly positive|SA = SA compared to sum
of other two = SA strictly smaller than sum of other two|SB = Values for SB = Strictly
positive|SB = SB compared to sum of other two = SB strictly smaller than sum of other two|SC =
SC compared to sum of other two = SC strictly smaller than sum of other two|SC = Values for SC =
SC strictly positive
values:     SB = '1', SA = '1', SC = '1'
```
**Test Case 2:**
```
choices:    2638Cat3Ch1−|−2639Cat2Ch2−|−2640Cat1Ch2−|−2638Cat1Ch1−|−2638Cat2Ch1−|−2639Cat1Ch1−
            |−2639Cat3Ch1−|−2640Cat2Ch1−|−2640Cat3Ch2
```
```
SA = SA compared to SB = SA equals to SB|SB = SB compared to SC = SB and SC are different|SC =
SC compared to SA = SC different from SA|SA = Values for SA = Strictly positive|SA = SA compared
to sum of other two = SA strictly smaller than sum of other two|SB = Values for SB = Strictly
positive|SB = SB compared to sum of other two = SB strictly smaller than sum of other two|SC =
SC compared to sum of other two = SC strictly smaller than sum of other two|SC = Values for SC =
SC strictly positive
values:     SB = '2', SA = '2', SC = '3'
```
**Test Case 3:**
```
choices:    2638Cat3Ch2−|−2639Cat2Ch1−|−2640Cat1Ch2−|−2638Cat1Ch1−|−2638Cat2Ch1−|−2639Cat1Ch1−
            |−2639Cat3Ch1−|−2640Cat2Ch1−|−2640Cat3Ch2
```

```
SA = SA compared to SB = SA and SB are different|SB = SB compared to SC = SB equals to SC|SC =
SC compared to SA = SC different from SA|SA = Values for SA = Strictly positive|SA = SA compared
to sum of other two = SA strictly smaller than sum of other two|SB = Values for SB = Strictly
positive|SB = SB compared to sum of other two = SB strictly smaller than sum of other two|SC =
SC compared to sum of other two = SC strictly smaller than sum of other two|SC = Values for SC =
SC strictly positive
values:      SB = '2', SA = '1', SC = '2'
```

**Test Case 4:**
```
choices:     2638Cat3Ch2—|—2639Cat2Ch2—|—2640Cat1Ch2—|—2638Cat1Ch1—|—2638Cat2Ch1—|—2639Cat1Ch1—
             |—2639Cat3Ch1—|—2640Cat2Ch1—|—2640Cat3Ch2
SA = SA compared to SB = SA and SB are different|SB = SB compared to SC = SB and SC are
different|SC = SC compared to SA = SC different from SA|SA = Values for SA = Strictly
positive|SA = SA compared to sum of other two = SA strictly smaller than sum of other two|SB =
Values for SB = Strictly positive|SB = SB compared to sum of other two = SB strictly smaller
than sum of other two|SC = SC compared to sum of other two = SC strictly smaller than sum of
other two|SC = Values for SC = SC strictly positive
values:      SB = '2', SA = '4', SC = '3'
```

**Test Case 5:**
```
choices:     2638Cat2Ch1—|—2638Cat3Ch1—|—2639Cat2Ch2—|—2639Cat3Ch1—|—2640Cat1Ch2—|—2640Cat2Ch2—
             |—2638Cat1Ch1—|—2639Cat1Ch1—|—2640Cat3Ch2
SA = SA compared to sum of other two = SA strictly smaller than sum of other two|SA = SA
compared to SB = SA equals to SB|SB = SB compared to SC = SB and SC are different|SB = SB
compared to sum of other two = SB strictly smaller than sum of other two|SC = SC compared to SA
= SC different from SA|SC = SC compared to sum of other two = SC greater than sum of other
two|SA = Values for SA = Strictly positive|SB = Values for SB = Strictly positive|SC = Values
for SC = SC strictly positive
values:      SB = '1', SA = '1', SC = '2'
```

**Test Case 6:**
```
choices:     2638Cat2Ch1—|—2638Cat3Ch2—|—2639Cat2Ch2—|—2639Cat3Ch2—|—2640Cat1Ch2—|—2640Cat2Ch1—
             |—2638Cat1Ch1—|—2639Cat1Ch1—|—2640Cat3Ch2
SA = SA compared to sum of other two = SA strictly smaller than sum of other two|SA = SA
compared to SB = SA and SB are different|SB = SB compared to SC = SB and SC are different|SB =
SB compared to sum of other two = SB greater than sum of other two|SC = SC compared to SA = SC
different from SA|SC = SC compared to sum of other two = SC strictly smaller than sum of other
two|SA = Values for SA = Strictly positive|SB = Values for SB = Strictly positive|SC = Values
for SC = SC strictly positive
values:      SB = '3', SA = '1', SC = '2'
```

**Test Case 7:**
```
choices:     2638Cat2Ch2—|—2638Cat3Ch2—|—2639Cat2Ch1—|—2639Cat3Ch1—|—2640Cat1Ch2—|—2640Cat2Ch1—
             |—2638Cat1Ch1—|—2639Cat1Ch1—|—2640Cat3Ch2
SA = SA compared to sum of other two = SA greater than sum of other two|SA = SA compared to SB =
SA and SB are different|SB = SB compared to SC = SB equals to SC|SB = SB compared to sum of
other two = SB strictly smaller than sum of other two|SC = SC compared to SA = SC different from
SA|SC = SC compared to sum of other two = SC strictly smaller than sum of other two|SA = Values
for SA = Strictly positive|SB = Values for SB = Strictly positive|SC = Values for SC = SC
strictly positive
values:      SC = '1', SB = '1', SA = '2'
```

**Test Case 8:**
```
choices:     2638Cat1Ch2
SA = Values for SA = Negative or zero
values:      SA = '0'
```

**Test Case 9:**
```
choices:     2639Cat1Ch2
SB = Values for SB = Negative or zero
values:      SB = '0'
```

**Test Case 10:**
```
choices:     2640Cat3Ch1
SC = Values for SC = SC negative or zero
values:      SC = '0'
```

## B.9     *Base Choice Test Frames and Corresponding Test Cases*

**Test Case 1:**
```
choices:     2638Cat1Ch1—|—2638Cat2Ch1—|—2638Cat3Ch2—|—2639Cat1Ch1—|—2639Cat2Ch2—|—2639Cat3Ch1—
             |—2640Cat1Ch2—|—2640Cat2Ch1—|—2640Cat3Ch2
```

SA = Values for SA = Strictly positive|SA = SA compared to sum of other two = SA strictly smaller than sum of other two|SA = SA compared to SB = SA and SB are different|SB = Values for SB = Strictly positive|SB = SB compared to SC = SB and SC are different|SB = SB compared to sum of other two = SB strictly smaller than sum of other two|SC = SC compared to SA = SC different from SA|SC = SC compared to sum of other two = SC strictly smaller than sum of other two|SC = Values for SC = SC strictly positive
values:      SB = '2', SA = '3', SC = '4'

## Test Case 2:
choices:      2638Cat1Ch2
SA = Values for SA = Negative or zero
values:      SA = '0'

## Test Case 3:
choices:      2638Cat1Ch1−|−2638Cat3Ch2−|−2639Cat1Ch1−|−2639Cat2Ch2−|−2639Cat3Ch1−|−2640Cat1Ch2−
            |−2640Cat2Ch1−|−2640Cat3Ch2−|−2638Cat2Ch2
SA = Values for SA = Strictly positive|SA = SA compared to SB = SA and SB are different|SB = Values for SB = Strictly positive|SB = SB compared to SC = SB and SC are different|SB = SB compared to sum of other two = SB strictly smaller than sum of other two|SC = SC compared to SA = SC different from SA|SC = SC compared to sum of other two = SC strictly smaller than sum of other two|SC = Values for SC = SC strictly positive|SA = SA compared to sum of other two = SA greater than sum of other two
values:      SB = '2', SA = '3', SC = '1'

## Test Case 4:
choices:      2638Cat1Ch1−|−2638Cat2Ch1−|−2639Cat1Ch1−|−2639Cat2Ch2−|−2639Cat3Ch1−|−2640Cat1Ch2−
            |−2640Cat2Ch1−|−2640Cat3Ch2−|−2638Cat3Ch1
SA = Values for SA = Strictly positive|SA = SA compared to sum of other two = SA strictly smaller than sum of other two|SB = Values for SB = Strictly positive|SB = SB compared to SC = SB and SC are different|SB = SB compared to sum of other two = SB strictly smaller than sum of other two|SC = SC compared to SA = SC different from SA|SC = SC compared to sum of other two = SC strictly smaller than sum of other two|SC = Values for SC = SC strictly positive|SA = SA compared to SB = SA equals to SB
values:      SB = '2', SA = '2', SC = '3'

## Test Case 5:
choices:      2639Cat1Ch2
SB = Values for SB = Negative or zero
values:      SB = '0'

## Test Case 6:
choices:      2638Cat1Ch1−|−2638Cat2Ch1−|−2638Cat3Ch2−|−2639Cat1Ch1−|−2639Cat3Ch1−|−2640Cat1Ch2−
            |−2640Cat2Ch1−|−2640Cat3Ch2−|−2639Cat2Ch1
SA = Values for SA = Strictly positive|SA = SA compared to sum of other two = SA strictly smaller than sum of other two|SA = SA compared to SB = SA and SB are different|SB = Values for SB = Strictly positive|SB = SB compared to sum of other two = SB strictly smaller than sum of other two|SC = SC compared to SA = SC different from SA|SC = SC compared to sum of other two = SC strictly smaller than sum of other two|SC = Values for SC = SC strictly positive|SB = SB compared to SC = SB equals to SC
values:      SB = '2', SA = '1', SC = '2'

## Test Case 7:
choices:      2638Cat1Ch1−|−2638Cat2Ch1−|−2638Cat3Ch2−|−2639Cat1Ch1−|−2639Cat2Ch2−|−2640Cat1Ch2−
            |−2640Cat2Ch1−|−2640Cat3Ch2−|−2639Cat3Ch2
SA = Values for SA = Strictly positive|SA = SA compared to sum of other two = SA strictly smaller than sum of other two|SA = SA compared to SB = SA and SB are different|SB = Values for SB = Strictly positive|SB = SB compared to SC = SB and SC are different|SC = SC compared to SA = SC different from SA|SC = SC compared to sum of other two = SC strictly smaller than sum of other two|SC = Values for SC = SC strictly positive|SB = SB compared to sum of other two = SB greater than sum of other two
values:      SB = '3', SA = '1', SC = '2'

## Test Case 8:
choices:      2638Cat3Ch1−|−2639Cat2Ch1−|−2640Cat1Ch1−|−2638Cat1Ch1−|−2638Cat2Ch1−|−2639Cat1Ch1−
            |−2639Cat3Ch1−|−2640Cat2Ch1−|−2640Cat3Ch2
SA = SA compared to SB = SA equals to SB|SB = SB compared to SC = SB equals to SC|SC = SC compared to SA = SC equals to SA|SA = Values for SA = Strictly positive|SA = SA compared to sum of other two = SA strictly smaller than sum of other two|SB = Values for SB = Strictly positive|SB = SB compared to sum of other two = SB strictly smaller than sum of other two|SC = SC compared to sum of other two = SC strictly smaller than sum of other two|SC = Values for SC = SC strictly positive
values:      SB = '1', SA = '1', SC = '1'

**Test Case 9:**

choices:        2638Cat1Ch1−|−2638Cat2Ch1−|−2638Cat3Ch2−|−2639Cat1Ch1−|−2639Cat2Ch2−|−2639Cat3Ch1−
                |−2640Cat1Ch2−|−2640Cat3Ch2−|−2640Cat2Ch2

SA = Values for SA = Strictly positive|SA = SA compared to sum of other two = SA strictly
smaller than sum of other two|SA = SA compared to SB = SA and SB are different|SB = Values for
SB = Strictly positive|SB = SB compared to SC = SB and SC are different|SB = SB compared to sum
of other two = SB strictly smaller than sum of other two|SC = SC compared to SA = SC different
from SA|SC = Values for SC = SC strictly positive|SC = SC compared to sum of other two = SC
greater than sum of other two

values:        SB = '2', SA = '1', SC = '3'

**Test Case 10:**

choices:        2640Cat3Ch1

SC = Values for SC = SC negative or zero

values:        SC = '0'

# Appendix C. CHARACTER SEARCH IN STRING CASE STUDY

In the CP specification below, an asterisk ('*') beside a choice indicates the base choice for that category. Each parameter is uniquely identified by an integer identifier, which is then used in the composition of test frames; As a result, a choice for a parameter is uniquely identified by the parameter identifier, the category identifier and the choice identifier. Below we summarize the specification of the unit under test (section C.1), and provide the CP specification we used (section C.2). We then provide the test frames and corresponding test cases (inputs) we obtain when using CASA (Each-Choice in section C.3, Pairwise in section C.5, Three-way in section C.7), when using ACTS (Each-Choice in section C.4, Pairwise in section C.6, Three-way in section C.8), and when using the Base Choice criterion (section C.9).

## C.1    Specification

The function under test takes two parameters: a string called InputString and a character called InputCharacter. The function returns the location (index) of the first occurrence of the character in the string.

## C.2    CP Specification

```
Parameter(1361890395): InputString of type String
  Category 1 - String length
    Ch 1.1: Empty string
            InputString.isEmpty()
            [properties emptyString]
    Ch 1.2: Small string
            InputString.length()>2 && InputString.length()<5
            [properties smallString]
   *Ch 1.3: Nominal length
            InputString.length()>=5
            [properties nominalStringLength]
    Ch 1.4: Length 1
            InputString.length()==1
            [single][properties stringLength1]
    Ch 1.5: Length 2
            InputString.length()==2
            [single][properties stringLength2]
  Category 2 - Number of occurences of character in string
    Ch 2.1: None
            !(InputString.contains(InputCharacter.asString()))
            [if !emptyString][properties noOccurence]
    Ch 2.2: One
            InputString.indexOf(InputCharacter.asString())>=0 &&
            InputString.substring(InputString.indexOf(InputCharacter.asString())+1).contains(InputCharacter.asString())==false
            [if !emptyString][properties oneOccurence]
   *Ch 2.3: More than one
            InputString.indexOf(InputCharacter.asString())>=0 &&
            InputString.substring(InputString.indexOf(InputCharacter.asString())+1).contains(InputCharacter.asString())==true
            [if !emptyString && !stringLength1][properties severalOccurences]
Parameter(725716321): InputCharacter of type Character
  Category 1 - Position of first occurence
    Ch 1.1: First character in string
            InputString.indexOf(InputCharacter.asString())==0
            [if !noOccurence && !emptyString]
    Ch 1.2: Last character in string
            InputString.indexOf(InputCharacter.asString())==InputString.length()-1
            [if oneOccurence && (nominalStringLength || smallString || stringLength2)]
   *Ch 1.3: In middle of string
            InputString.indexOf(InputCharacter.asString())>0 &&
            InputString.indexOf(InputCharacter.asString())<InputString.length()-1
            [if !noOccurence && (nominalStringLength || smallString)]
```

## C.3    Each-Choice Test Frames and Corresponding Test Cases when using CASA

**Test Case 1:**
```
choices:        1361890395Cat1Ch2-|−1361890395Cat2Ch3-|−725716321Cat1Ch3
InputString = String length = Small string|InputString = Number of occurences of character in
string = More than one|InputCharacter = Position of first occurence = In middle of string
values:         InputCharacter = 'M', InputString = '0MM'
```
**Test Case 2:**
```
choices:        1361890395Cat1Ch3-|−1361890395Cat2Ch2-|−725716321Cat1Ch1
InputString = String length = Nominal length|InputString = Number of occurences of character in
string = One|InputCharacter = Position of first occurence = First character in string
values:         InputString = '04kkK', InputCharacter = '0'
```
**Test Case 3:**
```
choices:        1361890395Cat1Ch3-|−1361890395Cat2Ch2-|−725716321Cat1Ch2
InputString = String length = Nominal length|InputString = Number of occurences of character in
string = One|InputCharacter = Position of first occurence = Last character in string
values:         InputString = 'PPscl', InputCharacter = 'l'
```
**Test Case 4:**
```
choices:        1361890395Cat1Ch4-|−1361890395Cat2Ch2-|−725716321Cat1Ch1
InputString = String length = Length 1|InputString = Number of occurences of character in string
= One|InputCharacter = Position of first occurence = First character in string
values:         InputString = 'T', InputCharacter = 'T'
```
**Test Case 5:**
```
choices:        1361890395Cat1Ch5-|−1361890395Cat2Ch2-|−725716321Cat1Ch2
InputString = String length = Length 2|InputString = Number of occurences of character in string
= One|InputCharacter = Position of first occurence = Last character in string
values:         InputString = '0c', InputCharacter = 'c'
```
**Test Case 6:**
```
choices:        1361890395Cat2Ch1-|−1361890395Cat1Ch3
InputString = Number of occurences of character in string = None|InputString = String length =
Nominal length
values:         InputString = 'zA2sh', InputCharacter = '3'
```
**Test Case 7:**
```
choices:        1361890395Cat1Ch1
InputString = String length = Empty string
values:         InputString = '', InputCharacter = '3'
```

## C.4    Each-Choice Test Frames and Corresponding Test Cases when using ACTS

**Test Case 1:**
```
choices:        1361890395Cat1Ch2-|−1361890395Cat2Ch2-|−725716321Cat1Ch3
InputString = String length = Small string|InputString = Number of occurences of character in
string = One|InputCharacter = Position of first occurence = In middle of string
values:         InputString = 'dl8', InputCharacter = 'l'
```
**Test Case 2:**
```
choices:        1361890395Cat1Ch3-|−1361890395Cat2Ch3-|−725716321Cat1Ch1
InputString = String length = Nominal length|InputString = Number of occurences of character in
string = More than one|InputCharacter = Position of first occurence = First character in string
values:         InputString = '333ny', InputCharacter = '3'
```
**Test Case 3:**
```
choices:        1361890395Cat1Ch3-|−1361890395Cat2Ch2-|−725716321Cat1Ch2
InputString = String length = Nominal length|InputString = Number of occurences of character in
string = One|InputCharacter = Position of first occurence = Last character in string
values:         InputString = 'RbcLP', InputCharacter = 'P'
```
**Test Case 4:**
```
choices:        1361890395Cat1Ch4-|−1361890395Cat2Ch2-|−725716321Cat1Ch1
InputString = String length = Length 1|InputString = Number of occurences of character in string
= One|InputCharacter = Position of first occurence = First character in string
values:         InputString = '4', InputCharacter = '4'
```
**Test Case 5:**
```
choices:        1361890395Cat1Ch5-|−1361890395Cat2Ch2-|−725716321Cat1Ch2
InputString = String length = Length 2|InputString = Number of occurences of character in string
= One|InputCharacter = Position of first occurence = Last character in string
values:         InputString = '0h', InputCharacter = 'h'
```

**Test Case 6:**
```
choices:      1361890395Cat2Ch1−|−1361890395Cat1Ch3
InputString = Number of occurences of character in string = None|InputString = String length =
Nominal length
values:       InputString = '889hZ', InputCharacter = '3'
```
**Test Case 7:**
```
choices:      1361890395Cat1Ch1
InputString = String length = Empty string
values:       InputString = '', InputCharacter = '3'
```

## C.5  Pairwise Test Frames and Corresponding Test Cases when using CASA

**Test Case 1:**
```
choices:      1361890395Cat1Ch3−|−1361890395Cat2Ch3−|−725716321Cat1Ch1
InputString = String length = Nominal length|InputString = Number of occurences of character in
string = More than one|InputCharacter = Position of first occurence = First character in string
values:       InputCharacter = '3', InputString = '3333B'
```
**Test Case 2:**
```
choices:      1361890395Cat1Ch2−|−1361890395Cat2Ch3−|−725716321Cat1Ch3
InputString = String length = Small string|InputString = Number of occurences of character in
string = More than one|InputCharacter = Position of first occurence = In middle of string
values:       InputString = 'LAA', InputCharacter = 'A'
```
**Test Case 3:**
```
choices:      1361890395Cat2Ch2−|−1361890395Cat2Ch2−|−725716321Cat1Ch1
InputString = String length = Small string|InputString = Number of occurences of character in
string = One|InputCharacter = Position of first occurence = First character in string
values:       InputString = 'akt', InputCharacter = 'a'
```
**Test Case 4:**
```
choices:      1361890395Cat1Ch3−|−1361890395Cat2Ch2−|−725716321Cat1Ch3
InputString = String length = Nominal length|InputString = Number of occurences of character in
string = One|InputCharacter = Position of first occurence = In middle of string
values:       InputString = '80ed9', InputCharacter = 'd'
```
**Test Case 5:**
```
choices:      1361890395Cat1Ch2−|−1361890395Cat2Ch2−|−725716321Cat1Ch2
InputString = String length = Small string|InputString = Number of occurences of character in
string = One|InputCharacter = Position of first occurence = Last character in string
values:       InputString = 'dT1', InputCharacter = '1'
```
**Test Case 6:**
```
choices:      1361890395Cat1Ch3−|−1361890395Cat2Ch2−|−725716321Cat1Ch2
InputString = String length = Nominal length|InputString = Number of occurences of character in
string = One|InputCharacter = Position of first occurence = Last character in string
values:       InputString = 'HH4s0', InputCharacter = '0'
```
**Test Case 7:**
```
choices:      1361890395Cat1Ch4−|−1361890395Cat2Ch2−|−725716321Cat1Ch1
InputString = String length = Length 1|InputString = Number of occurences of character in string
= One|InputCharacter = Position of first occurence = First character in string
values:       InputString = 'r', InputCharacter = 'r'
```
**Test Case 8:**
```
choices:      1361890395Cat1Ch5−|−1361890395Cat2Ch3−|−725716321Cat1Ch1
InputString = String length = Length 2|InputString = Number of occurences of character in string
= More than one|InputCharacter = Position of first occurence = First character in string
values:       InputString = 'dd', InputCharacter = 'd'
```
**Test Case 9:**
```
choices:      1361890395Cat2Ch1−|−1361890395Cat1Ch3
InputString = Number of occurences of character in string = None|InputString = String length =
Nominal length
values:       InputString = '8sp2p', InputCharacter = '3'
```
**Test Case 10:**
```
choices:      1361890395Cat1Ch1
InputString = String length = Empty string
values:       InputString = '', InputCharacter = '3'
```

## C.6    Pairwise Test Frames and Corresponding Test Cases when using ACTS

**Test Case 1:**
```
choices:        1361890395Cat1Ch2−|−1361890395Cat2Ch2−|−725716321Cat1Ch3
InputString = String length = Small string|InputString = Number of occurences of character in
string = One|InputCharacter = Position of first occurence = In middle of string
values:         InputCharacter = '3', InputString = 'H3z'
```
**Test Case 2:**
```
choices:        1361890395Cat1Ch2−|−1361890395Cat2Ch3−|−725716321Cat1Ch1
InputString = String length = Small string|InputString = Number of occurences of character in
string = More than one|InputCharacter = Position of first occurence = First character in string
values:         InputString = '999', InputCharacter = '9'
```
**Test Case 3:**
```
choices:        1361890395Cat1Ch3−|−1361890395Cat2Ch2−|−725716321Cat1Ch2
InputString = String length = Nominal length|InputString = Number of occurences of character in
string = One|InputCharacter = Position of first occurence = Last character in string
values:         InputString = 'DKbBh', InputCharacter = 'h'
```
**Test Case 4:**
```
choices:        1361890395Cat1Ch3−|−1361890395Cat2Ch3−|−725716321Cat1Ch3
InputString = String length = Nominal length|InputString = Number of occurences of character in
string = More than one|InputCharacter = Position of first occurence = In middle of string
values:         InputString = 'P733z', InputCharacter = '3'
```
**Test Case 5:**
```
choices:        1361890395Cat1Ch3−|−1361890395Cat2Ch2−|−725716321Cat1Ch1
InputString = String length = Nominal length|InputString = Number of occurences of character in
string = One|InputCharacter = Position of first occurence = First character in string
values:         InputString = 'xB3kN', InputCharacter = 'x'
```
**Test Case 6:**
```
choices:        1361890395Cat1Ch4−|−1361890395Cat2Ch2−|−725716321Cat1Ch1
InputString = String length = Length 1|InputString = Number of occurences of character in string
= One|InputCharacter = Position of first occurence = First character in string
values:         InputString = 'T', InputCharacter = 'T'
```
**Test Case 7:**
```
choices:        1361890395Cat1Ch5−|−1361890395Cat2Ch2−|−725716321Cat1Ch2
InputString = String length = Length 2|InputString = Number of occurences of character in string
= One|InputCharacter = Position of first occurence = Last character in string
values:         InputString = 'ls', InputCharacter = 's'
```
**Test Case 8:**
```
choices:        1361890395Cat2Ch1−|−1361890395Cat1Ch3
InputString = Number of occurences of character in string = None|InputString = String length =
Nominal length
values:         InputString = '889Pm', InputCharacter = '3'
```
**Test Case 9:**
```
choices:        1361890395Cat1Ch1
InputString = String length = Empty string
values:         InputString = '', InputCharacter = '3'
```

## C.7    Three-way Test Frames and Corresponding Test Cases when using CASA

**Test Case 1:**
```
choices:        1361890395Cat1Ch2−|−1361890395Cat2Ch3−|−725716321Cat1Ch3
InputString = String length = Small string|InputString = Number of occurences of character in
string = More than one|InputCharacter = Position of first occurence = In middle of string
values:         InputCharacter = 'i', InputString = 'pii'
```
**Test Case 2:**
```
choices:        1361890395Cat1Ch2−|−1361890395Cat2Ch2−|−725716321Cat1Ch1
InputString = String length = Small string|InputString = Number of occurences of character in
string = One|InputCharacter = Position of first occurence = First character in string
values:         InputString = '5k0', InputCharacter = '5'
```
**Test Case 3:**
```
choices:        1361890395Cat1Ch3−|−1361890395Cat2Ch2−|−725716321Cat1Ch3
InputString = String length = Nominal length|InputString = Number of occurences of character in
string = One|InputCharacter = Position of first occurence = In middle of string
values:         InputString = 'KDDdL', InputCharacter = 'd'
```

**Test Case 4:**
```
choices:       1361890395Cat1Ch2-|-1361890395Cat2Ch2-|-725716321Cat1Ch3
InputString = String length = Small string|InputString = Number of occurences of character in
string = One|InputCharacter = Position of first occurence = In middle of string
values:        InputString = 'pH0', InputCharacter = 'H'
```
**Test Case 5:**
```
choices:       1361890395Cat1Ch3-|-1361890395Cat2Ch2-|-725716321Cat1Ch1
InputString = String length = Nominal length|InputString = Number of occurences of character in
string = One|InputCharacter = Position of first occurence = First character in string
values:        InputString = 'IH3k3', InputCharacter = 'I'
```
**Test Case 6:**
```
choices:       1361890395Cat1Ch3-|-1361890395Cat2Ch3-|-725716321Cat1Ch3
InputString = String length = Nominal length|InputString = Number of occurences of character in
string = More than one|InputCharacter = Position of first occurence = In middle of string
values:        InputString = 'csZsb', InputCharacter = 's'
```
**Test Case 7:**
```
choices:       1361890395Cat1Ch2-|-1361890395Cat2Ch3-|-725716321Cat1Ch1
InputString = String length = Small string|InputString = Number of occurences of character in
string = More than one|InputCharacter = Position of first occurence = First character in string
values:        InputString = 'HHK', InputCharacter = 'H'
```
**Test Case 8:**
```
choices:       1361890395Cat1Ch3-|-1361890395Cat2Ch3-|-725716321Cat1Ch1
InputString = String length = Nominal length|InputString = Number of occurences of character in
string = More than one|InputCharacter = Position of first occurence = First character in string
values:        InputString = 'pKPph', InputCharacter = 'p'
```
**Test Case 9:**
```
choices:       1361890395Cat1Ch3-|-1361890395Cat2Ch2-|-725716321Cat1Ch2
InputString = String length = Nominal length|InputString = Number of occurences of character in
string = One|InputCharacter = Position of first occurence = Last character in string
values:        InputString = 'XqIPH', InputCharacter = 'H'
```
**Test Case 10:**
```
choices:       1361890395Cat1Ch2-|-1361890395Cat2Ch2-|-725716321Cat1Ch2
InputString = String length = Small string|InputString = Number of occurences of character in
string = One|InputCharacter = Position of first occurence = Last character in string
values:        InputString = 'HjI', InputCharacter = 'I'
```
**Test Case 11:**
```
choices:       1361890395Cat1Ch4-|-1361890395Cat2Ch2-|-725716321Cat1Ch1
InputString = String length = Length 1|InputString = Number of occurences of character in string
= One|InputCharacter = Position of first occurence = First character in string
values:        InputString = 'i', InputCharacter = 'i'
```
**Test Case 12:**
```
choices:       1361890395Cat1Ch5-|-1361890395Cat2Ch3-|-725716321Cat1Ch1
InputString = String length = Length 2|InputString = Number of occurences of character in string
= More than one|InputCharacter = Position of first occurence = First character in string
values:        InputString = '44', InputCharacter = '4'
```
**Test Case 13:**
```
choices:       1361890395Cat2Ch1-|-1361890395Cat1Ch3
InputString = Number of occurences of character in string = None|InputString = String length =
Nominal length
values:        InputString = '8Ruk0', InputCharacter = '3'
```
**Test Case 14:**
```
choices:       1361890395Cat1Ch1
InputString = String length = Empty string
values:        InputString = '', InputCharacter = '3'
```

## C.8    Three-way Test Frames and Corresponding Test Cases when using ACTS

**Test Case 1:**
```
choices:       1361890395Cat1Ch2-|-1361890395Cat2Ch2-|-725716321Cat1Ch1
InputString = String length = Small string|InputString = Number of occurences of character in
string = One|InputCharacter = Position of first occurence = First character in string
values:        InputString = '4sp', InputCharacter = '4'
```
**Test Case 2:**
```
choices:       1361890395Cat1Ch2-|-1361890395Cat2Ch2-|-725716321Cat1Ch3
```

```
InputString = String length = Small string|InputString = Number of occurences of character in
string = One|InputCharacter = Position of first occurence = In middle of string
values:      InputString = 'hzC', InputCharacter = 'z'
```
**Test Case 3:**
```
choices:     1361890395Cat1Ch2−|−1361890395Cat2Ch3−|−725716321Cat1Ch1
InputString = String length = Small string|InputString = Number of occurences of character in
string = More than one|InputCharacter = Position of first occurence = First character in string
values:      InputString = 'SSS', InputCharacter = 'S'
```
**Test Case 4:**
```
choices:     1361890395Cat1Ch2−|−1361890395Cat2Ch3−|−725716321Cat1Ch3
InputString = String length = Small string|InputString = Number of occurences of character in
string = More than one|InputCharacter = Position of first occurence = In middle of string
values:      InputString = 'H33', InputCharacter = '3'
```
**Test Case 5:**
```
choices:     1361890395Cat1Ch3−|−1361890395Cat2Ch2−|−725716321Cat1Ch1
InputString = String length = Nominal length|InputString = Number of occurences of character in
string = One|InputCharacter = Position of first occurence = First character in string
values:      InputString = '4f5dp', InputCharacter = '4'
```
**Test Case 6:**
```
choices:     1361890395Cat1Ch3−|−1361890395Cat2Ch2−|−725716321Cat1Ch2
InputString = String length = Nominal length|InputString = Number of occurences of character in
string = One|InputCharacter = Position of first occurence = Last character in string
values:      InputString = 'h9s20', InputCharacter = '0'
```
**Test Case 7:**
```
choices:     1361890395Cat1Ch3−|−1361890395Cat2Ch2−|−725716321Cat1Ch3
InputString = String length = Nominal length|InputString = Number of occurences of character in
string = One|InputCharacter = Position of first occurence = In middle of string
values:      InputString = 'Khkpz', InputCharacter = 'h'
```
**Test Case 8:**
```
choices:     1361890395Cat1Ch3−|−1361890395Cat2Ch3−|−725716321Cat1Ch1
InputString = String length = Nominal length|InputString = Number of occurences of character in
string = More than one|InputCharacter = Position of first occurence = First character in string
values:      InputString = 'S28Sl', InputCharacter = 'S'
```
**Test Case 9:**
```
choices:     1361890395Cat1Ch3−|−1361890395Cat2Ch3−|−725716321Cat1Ch3
InputString = String length = Nominal length|InputString = Number of occurences of character in
string = More than one|InputCharacter = Position of first occurence = In middle of string
values:      InputString = 'knZZk', InputCharacter = 'Z'
```
**Test Case 10:**
```
choices:     1361890395Cat1Ch4−|−1361890395Cat2Ch2−|−725716321Cat1Ch1
InputString = String length = Length 1|InputString = Number of occurences of character in string
= One|InputCharacter = Position of first occurence = First character in string
values:      InputString = 'd', InputCharacter = 'd'
```
**Test Case 11:**
```
choices:     1361890395Cat1Ch5−|−1361890395Cat2Ch2−|−725716321Cat1Ch1
InputString = String length = Length 2|InputString = Number of occurences of character in string
= One|InputCharacter = Position of first occurence = First character in string
values:      InputString = 'Ak', InputCharacter = 'A'
```
**Test Case 12:**
```
choices:     1361890395Cat2Ch1−|−1361890395Cat1Ch3
InputString = Number of occurences of character in string = None|InputString = String length =
Nominal length
values:      InputString = '88r4X', InputCharacter = '3'
```
**Test Case 13:**
```
choices:     1361890395Cat1Ch1
InputString = String length = Empty string
values:      InputString = '', InputCharacter = '3'
```

*C.9    Base Choice Test Frames and Corresponding Test Cases*

**Test Case 1:**
```
choices:     1361890395Cat1Ch3−|−1361890395Cat2Ch3−|−725716321Cat1Ch3
InputString = String length = Nominal length|InputString = Number of occurences of character in
string = More than one|InputCharacter = Position of first occurence = In middle of string
values:      InputCharacter = 'z', InputString = 'kzPz4'
```

**Test Case 2:**
```
choices:        1361890395Cat1Ch1
InputString = String length = Empty string
values:         InputCharacter = '3', InputString = ''
```
**Test Case 3:**
```
choices:        1361890395Cat2Ch3—|—725716321Cat1Ch3—|—1361890395Cat1Ch2
InputString = Number of occurences of character in string = More than one|InputCharacter =
Position of first occurence = In middle of string|InputString = String length = Small string
values:         InputString = '833', InputCharacter = '3'
```
**Test Case 4:**
```
choices:        1361890395Cat1Ch4—|—1361890395Cat2Ch2—|—725716321Cat1Ch1
InputString = String length = Length 1|InputString = Number of occurences of character in string
= One|InputCharacter = Position of first occurence = First character in string
values:         InputString = '0', InputCharacter = '0'
```
**Test Case 5:**
```
choices:        1361890395Cat1Ch5—|—1361890395Cat2Ch2—|—725716321Cat1Ch2
InputString = String length = Length 2|InputString = Number of occurences of character in string
= One|InputCharacter = Position of first occurence = Last character in string
values:         InputString = 'ln', InputCharacter = 'n'
```
**Test Case 6:**
```
choices:        1361890395Cat2Ch1—|—1361890395Cat1Ch3
InputString = Number of occurences of character in string = None|InputString = String length =
Nominal length
values:         InputString = '80MmK', InputCharacter = '3'
```
**Test Case 7:**
```
choices:        1361890395Cat1Ch3—|—1361890395Cat2Ch2
InputString = String length = Nominal length|InputCharacter = Position of first occurence = In
middle of string|InputString = Number of occurences of character in string = One
values:         InputString = 'dDfFH', InputCharacter = 'F'
```
**Test Case 8:**
```
choices:        1361890395Cat1Ch3—|—1361890395Cat2Ch3—|—725716321Cat1Ch1
InputString = String length = Nominal length|InputString = Number of occurences of character in
string = More than one|InputCharacter = Position of first occurence = First character in string
values:         InputString = 'x3dx3', InputCharacter = 'x'
```
**Test Case 9:**
```
choices:        1361890395Cat1Ch3—|—1361890395Cat2Ch2—|—725716321Cat1Ch2
InputString = String length = Nominal length|InputString = Number of occurences of character in
string = One|InputCharacter = Position of first occurence = Last character in string
values:         InputCharacter = '0', InputString = 'PDzk0'
```

# Appendix D. PACKHEXCHAR CASE STUDY

In the CP specification below, an asterisk ('*') beside a choice indicates the base choice for that category. Each parameter is uniquely identified by an integer identifier, which is then used in the composition of test frames; As a result, a choice for a parameter is uniquely identified by the parameter identifier, the category identifier and the choice identifier. Below we summarize the specification of the unit under test (section D.1), and provide the CP specification we used (section D.2). We then provide the test frames and corresponding test cases (inputs) we obtain when using CASA (Each-Choice in section D.3, Pairwise in section D.5, Three-way in section D.7), when using ACTS (Each-Choice in section D.4, Pairwise in section D.6, Three-way in section D.8), and when using the Base Choice criterion (section D.9). The CP specification below is adapted from earlier work [16, 38, 39].

## D.1 Specification

The PackHexChar problem focuses on compacting a string of characters representing hexadecimal digits into a binary format. The function processes an input string, ignoring non-hexadecimal characters, and outputs an array of bytes containing the compacted information alongside an integer value. The behaviour of the program varies depending on the number of valid hexadecimal characters in the input string.

If the input string contains an even number of hexadecimal characters, all hexadecimal characters are compacted (in pairs) into the byte array, and the program returns an integer value of -1. For an odd number of hexadecimal characters, the function compacts all but the last hexadecimal character into the byte array and returns the remaining hexadecimal character as a separate output. Additionally, the program allows the user to specify a substring of the input string to be analyzed using the input parameter RLEN. If RLEN is invalid (negative or greater than the length of the input string), the function returns -2.

The program also provides a mechanism for handling leftover characters from previous calls using the ODD_DIGIT input parameter. This parameter allows the user to append a trailing hexadecimal character from a previous result to the beginning of the input string for subsequent processing. If ODD_DIGIT is set to -1, no character is appended. If ODD_DIGIT is invalid (less than -1 or not a hexadecimal value), the function returns -3. This feature is particularly useful for processing strings split across multiple calls, ensuring consistent handling of characters between executions.

## D.2 CP Specification

```
Parameter(2518115): RLEN of type Integer
  Category 1 - RLEN values
    Ch 1.1: zero
            RLEN == 0
            [single][properties RLENisZero]
   *Ch 1.2: within string strictly
            RLEN>0 && RLEN<stringS.length()
            [if notEmpty][properties rlenOK]
    Ch 1.3: strictly negative
            RLEN < 0
            [error]
    Ch 1.4: over string length
            RLEN>stringS.length()
            [if notEmpty][error]
    Ch 1.5: equals string length
            RLEN==stringS.length()
            [if notEmpty][properties rlenMax]
Parameter(202558941): ODD_DIGIT of type Integer
  Category 1 - Values
    Ch 1.1: strictly below -1
            ODD_DIGIT < -1
            [single]
```

```
          Ch 1.2: minus 1
                  ODD_DIGIT == -1
        *Ch 1.3: hexadecimal value (ASCII)
                  (ODD_DIGIT>=48 && ODD_DIGIT<=57) || (ODD_DIGIT>=65 && ODD_DIGIT<=70) ||
                  (ODD_DIGIT>=97 && ODD_DIGIT<=102)
          Ch 1.4: Not an hexadecimal (ASCII)
                  (ODD_DIGIT>=0 && ODD_DIGIT<=47) || (ODD_DIGIT>=58 && ODD_DIGIT<=64) ||
                  (ODD_DIGIT>=71 && ODD_DIGIT<=96) || ODD_DIGIT>=103
                  [single]
Parameter(1881759134): stringS of type String
  Category 1 - empty or not
      Ch 1.1: empty
                  stringS.isEmpty()
                  [single]
        *Ch 1.2: not empty
                  stringS.length()>0
                  [properties notEmpty]
  Category 2 - types of characters in first part of string
       *Ch 2.1: all hexadecimals
                  stringS.AllHexadecimalCharactersinFirstChars(RLEN)
                  [if notEmpty && (rlenOK || rlenMax)]
        Ch 2.2: all non hexadecimals
                  stringS.NoHexadecimalCharactersinFirstChars(RLEN)
                  [if notEmpty && (rlenOK || rlenMax)][single][properties NoHexaChar]
        Ch 2.3: mix hexa and non hexa
                  stringS.MixHexadecimalNonHexadecimalCharactersinFirstChars(RLEN)
                  [if notEmpty && (rlenOK || rlenMax)][properties mixhexaNoHexa]
  Category 3 - number of hexadecimals in first part
       *Ch 3.1: Odd
                  stringS.OddNumberofHexadecimalCharactersinFirstChars(RLEN)
                  [if notEmpty && (rlenOK || rlenMax) && !NoHexaChar]
        Ch 3.2: Even
                  stringS.EvenNumberofHexadecimalCharactersinFirstChars(RLEN)
                  [if notEmpty && (rlenOK || rlenMax)]
```

## D.3    Each-Choice Test Frames and Corresponding Test Cases when using CASA

**Test Case 1:**
```
choices:     2518115Cat1Ch2-|-202558941Cat1Ch3-|-1881759134Cat1Ch2-|-1881759134Cat2Ch3-|-
             1881759134Cat3Ch2
RLEN = RLEN values = within string strictly|ODD_DIGIT = Values = hexadecimal value
(ASCII)|stringS = empty or not = not empty|stringS = types of characters in first part of string
= mix hexa and non hexa|stringS = number of hexadecimals in first part = Even
values:      RLEN = '3', ODD_DIGIT = '97', stringS = '9ga0'
```
**Test Case 2:**
```
choices:     2518115Cat1Ch5-|-202558941Cat1Ch2-|-1881759134Cat1Ch2-|-1881759134Cat2Ch1-|-
             1881759134Cat3Ch1
RLEN = RLEN values = equals string length|ODD_DIGIT = Values = minus 1|stringS = empty or not =
not empty|stringS = types of characters in first part of string = all hexadecimals|stringS =
number of hexadecimals in first part = Odd
values:      RLEN = '1', ODD_DIGIT = '-1', stringS = '0'
```
**Test Case 3:**
```
choices:     2518115Cat1Ch2-|-202558941Cat1Ch1-|-1881759134Cat1Ch2-|-1881759134Cat2Ch2-|-
             1881759134Cat3Ch2
RLEN = RLEN values = within string strictly|ODD_DIGIT = Values = strictly below -1|stringS =
empty or not = not empty|stringS = types of characters in first part of string = all non
hexadecimals|stringS = number of hexadecimals in first part = Even
values:      ODD_DIGIT = '-2', RLEN = '1', stringS = 'Q3'
```
**Test Case 4:**
```
choices:     2518115Cat1Ch2-|-202558941Cat1Ch4-|-1881759134Cat1Ch2-|-1881759134Cat2Ch3-|-
             1881759134Cat3Ch1
RLEN = RLEN values = within string strictly|ODD_DIGIT = Values = Not an hexadecimal
(ASCII)|stringS = empty or not = not empty|stringS = types of characters in first part of string
= mix hexa and non hexa|stringS = number of hexadecimals in first part = Odd
values:      ODD_DIGIT = '103', RLEN = '3', stringS = 'CgJ5'
```

**Test Case 5:**
```
choices:      2518115Cat1Ch3
RLEN = RLEN values = strictly negative
values:       RLEN = '-1'
```
**Test Case 6:**
```
choices:      2518115Cat1Ch4
RLEN = RLEN values = over string length
values:       RLEN = '1', stringS = ''
```
**Test Case 7:**
```
choices:      2518115Cat1Ch1-|-1881759134Cat1Ch2-|-202558941Cat1Ch3
RLEN = RLEN values = zero|stringS = empty or not = not empty|ODD_DIGIT = Values = hexadecimal
value (ASCII)
values:       ODD_DIGIT = '97', RLEN = '0', stringS = '4'
```
**Test Case 8:**
```
choices:      1881759134Cat1Ch1-|-202558941Cat1Ch3
stringS = empty or not = empty|ODD_DIGIT = Values = hexadecimal value (ASCII)
values:       ODD_DIGIT = '97', stringS = ''
```

## D.4    Each-Choice Test Frames and Corresponding Test Cases when using ACTS

**Test Case 1:**
```
choices:      2518115Cat1Ch2-|-202558941Cat1Ch3-|-1881759134Cat2Ch3-|-1881759134Cat3Ch2-|-
              1881759134Cat1Ch2
RLEN = RLEN values = within string strictly|ODD_DIGIT = Values = hexadecimal value
(ASCII)|stringS = types of characters in first part of string = mix hexa and non hexa|stringS =
number of hexadecimals in first part = Even|stringS = empty or not = not empty
values:       RLEN = '3', ODD_DIGIT = '97', stringS = '4Q5l'
```
**Test Case 2:**
```
choices:      2518115Cat1Ch5-|-202558941Cat1Ch2-|-1881759134Cat2Ch1-|-1881759134Cat3Ch1-|-
              1881759134Cat1Ch2
RLEN = RLEN values = equals string length|ODD_DIGIT = Values = minus 1|stringS = types of
characters in first part of string = all hexadecimals|stringS = number of hexadecimals in first
part = Odd|stringS = empty or not = not empty
values:       RLEN = '1', ODD_DIGIT = '-1', stringS = '2'
```
**Test Case 3:**
```
choices:      202558941Cat1Ch4-|-2518115Cat1Ch5-|-1881759134Cat2Ch3-|-1881759134Cat1Ch2-|-
              1881759134Cat3Ch2
ODD_DIGIT = Values = Not an hexadecimal (ASCII)|RLEN = RLEN values = equals string
length|stringS = types of characters in first part of string = mix hexa and non hexa|stringS =
empty or not = not empty|stringS = number of hexadecimals in first part = Even
values:       RLEN = '3', ODD_DIGIT = '71', stringS = 'bla'
```
**Test Case 4:**
```
choices:      2518115Cat1Ch3
RLEN = RLEN values = strictly negative
values:       RLEN = '-1'
```
**Test Case 5:**
```
choices:      2518115Cat1Ch4
RLEN = RLEN values = over string length
values:       RLEN = '1', stringS = ''
```
**Test Case 6:**
```
choices:      1881759134Cat2Ch2
stringS = types of characters in first part of string = all non hexadecimals
values:       RLEN = '2', stringS = 'h'
```
**Test Case 7:**
```
choices:      202558941Cat1Ch1-|-1881759134Cat1Ch2
ODD_DIGIT = Values = strictly below -1|stringS = empty or not = not empty
values:       ODD_DIGIT = '-2', stringS = '4'
```
**Test Case 8:**
```
choices:      2518115Cat1Ch1-|-1881759134Cat1Ch2-|-202558941Cat1Ch3
RLEN = RLEN values = zero|stringS = empty or not = not empty|ODD_DIGIT = Values = hexadecimal
value (ASCII)
values:       ODD_DIGIT = '97', RLEN = '0', stringS = '4'
```
**Test Case 9:**
```
choices:      1881759134Cat1Ch1-|-202558941Cat1Ch3
stringS = empty or not = empty|ODD_DIGIT = Values = hexadecimal value (ASCII)
```

```
values:       ODD_DIGIT = '97', stringS = ''
```

## D.5    Pairwise Test Frames and Corresponding Test Cases when using CASA

**Test Case 1:**
```
choices:    2518115Cat1Ch5-|-202558941Cat1Ch3-|-1881759134Cat1Ch2-|-1881759134Cat2Ch3-|-
            1881759134Cat3Ch1

RLEN = RLEN values = equals string length|ODD_DIGIT = Values = hexadecimal value (ASCII)|stringS
= empty or not = not empty|stringS = types of characters in first part of string = mix hexa and
non hexa|stringS = number of hexadecimals in first part = Odd
values:       RLEN = '3', ODD_DIGIT = '97', stringS = 'AHl'
```
**Test Case 2:**
```
choices:    2518115Cat1Ch5-|-202558941Cat1Ch3-|-1881759134Cat1Ch2-|-1881759134Cat2Ch1-|-
            1881759134Cat3Ch2

RLEN = RLEN values = equals string length|ODD_DIGIT = Values = hexadecimal value (ASCII)|stringS
= empty or not = not empty|stringS = types of characters in first part of string = all
hexadecimals|stringS = number of hexadecimals in first part = Even
values:       RLEN = '2', ODD_DIGIT = '65', stringS = 'dd'
```
**Test Case 3:**
```
choices:    2518115Cat1Ch2-|-202558941Cat1Ch2-|-1881759134Cat1Ch2-|-1881759134Cat2Ch3-|-
            1881759134Cat3Ch2

RLEN = RLEN values = within string strictly|ODD_DIGIT = Values = minus 1|stringS = empty or not
= not empty|stringS = types of characters in first part of string = mix hexa and non
hexa|stringS = number of hexadecimals in first part = Even
values:       ODD_DIGIT = '-1', RLEN = '3', stringS = 'fr8D'
```
**Test Case 4:**
```
choices:    2518115Cat1Ch5-|-202558941Cat1Ch2-|-1881759134Cat1Ch2-|-1881759134Cat2Ch1-|-
            1881759134Cat3Ch1

RLEN = RLEN values = equals string length|ODD_DIGIT = Values = minus 1|stringS = empty or not =
not empty|stringS = types of characters in first part of string = all hexadecimals|stringS =
number of hexadecimals in first part = Odd
values:       ODD_DIGIT = '-1', RLEN = '1', stringS = '0'
```
**Test Case 5:**
```
choices:    2518115Cat1Ch2-|-202558941Cat1Ch3-|-1881759134Cat1Ch2-|-1881759134Cat2Ch1-|-
            1881759134Cat3Ch1

RLEN = RLEN values = within string strictly|ODD_DIGIT = Values = hexadecimal value
(ASCII)|stringS = empty or not = not empty|stringS = types of characters in first part of string
= all hexadecimals|stringS = number of hexadecimals in first part = Odd
values:       ODD_DIGIT = '48', RLEN = '1', stringS = 'ce'
```
**Test Case 6:**
```
choices:    2518115Cat1Ch5-|-202558941Cat1Ch3-|-1881759134Cat1Ch2-|-1881759134Cat2Ch2-|-
            1881759134Cat3Ch2

RLEN = RLEN values = equals string length|ODD_DIGIT = Values = hexadecimal value (ASCII)|stringS
= empty or not = not empty|stringS = types of characters in first part of string = all non
hexadecimals|stringS = number of hexadecimals in first part = Even
values:       RLEN = '1', ODD_DIGIT = '48', stringS = 'i'
```
**Test Case 7:**
```
choices:    2518115Cat1Ch2-|-202558941Cat1Ch4-|-1881759134Cat1Ch2-|-1881759134Cat2Ch1-|-
            1881759134Cat3Ch1

RLEN = RLEN values = within string strictly|ODD_DIGIT = Values = Not an hexadecimal
(ASCII)|stringS = empty or not = not empty|stringS = types of characters in first part of string
= all hexadecimals|stringS = number of hexadecimals in first part = Odd
values:       RLEN = '1', ODD_DIGIT = '71', stringS = 'et'
```
**Test Case 8:**
```
choices:    2518115Cat1Ch2-|-202558941Cat1Ch1-|-1881759134Cat1Ch2-|-1881759134Cat2Ch3-|-
            1881759134Cat3Ch1

RLEN = RLEN values = within string strictly|ODD_DIGIT = Values = strictly below -1|stringS =
empty or not = not empty|stringS = types of characters in first part of string = mix hexa and
non hexa|stringS = number of hexadecimals in first part = Odd
values:       ODD_DIGIT = '-2', RLEN = '3', stringS = 'AIG0'
```
**Test Case 9:**
```
choices:    2518115Cat1Ch3
RLEN = RLEN values = strictly negative
values:       RLEN = '-1'
```

**Test Case 10:**
```
choices:      2518115Cat1Ch4
RLEN = RLEN values = over string length
values:       RLEN = '1', stringS = ''
```
**Test Case 11:**
```
choices:      2518115Cat1Ch1−|−1881759134Cat1Ch2−|−202558941Cat1Ch3
RLEN = RLEN values = zero|stringS = empty or not = not empty|ODD_DIGIT = Values = hexadecimal
value (ASCII)
values:       ODD_DIGIT = '97', RLEN = '0', stringS = '4'
```
**Test Case 12:**
```
choices:      1881759134Cat1Ch1−|−202558941Cat1Ch3
stringS = empty or not = empty|ODD_DIGIT = Values = hexadecimal value (ASCII)
values:       stringS = '', ODD_DIGIT = '97'
```

## D.6    *Pairwise Test Frames and Corresponding Test Cases when using ACTS*

**Test Case 1:**
```
choices:      2518115Cat1Ch2−|−202558941Cat1Ch2−|−1881759134Cat2Ch3−|−1881759134Cat3Ch2−|−
1881759134Cat1Ch2
RLEN = RLEN values = within string strictly|ODD_DIGIT = Values = minus 1|stringS = types of
characters in first part of string = mix hexa and non hexa|stringS = number of hexadecimals in
first part = Even|stringS = empty or not = not empty
values:       RLEN = '3', ODD_DIGIT = '-1', stringS = '0rad'
```
**Test Case 2:**
```
choices:      2518115Cat1Ch2−|−202558941Cat1Ch3−|−1881759134Cat2Ch1−|−1881759134Cat3Ch1−|−
1881759134Cat1Ch2
RLEN = RLEN values = within string strictly|ODD_DIGIT = Values = hexadecimal value
(ASCII)|stringS = types of characters in first part of string = all hexadecimals|stringS =
number of hexadecimals in first part = Odd|stringS = empty or not = not empty
values:       RLEN = '1', ODD_DIGIT = '65', stringS = 'c2'
```
**Test Case 3:**
```
choices:      2518115Cat1Ch5−|−202558941Cat1Ch2−|−1881759134Cat2Ch1−|−1881759134Cat3Ch2−|−
1881759134Cat1Ch2
RLEN = RLEN values = equals string length|ODD_DIGIT = Values = minus 1|stringS = types of
characters in first part of string = all hexadecimals|stringS = number of hexadecimals in first
part = Even|stringS = empty or not = not empty
values:       ODD_DIGIT = '-1', RLEN = '2', stringS = '0a'
```
**Test Case 4:**
```
choices:      2518115Cat1Ch5−|−202558941Cat1Ch3−|−1881759134Cat2Ch3−|−1881759134Cat3Ch1−|−
1881759134Cat1Ch2
RLEN = RLEN values = equals string length|ODD_DIGIT = Values = hexadecimal value (ASCII)|stringS
= types of characters in first part of string = mix hexa and non hexa|stringS = number of
hexadecimals in first part = Odd|stringS = empty or not = not empty
values:       ODD_DIGIT = '97', RLEN = '3', stringS = '3gX'
```
**Test Case 5:**
```
choices:      2518115Cat1Ch5−|−202558941Cat1Ch2−|−1881759134Cat2Ch1−|−1881759134Cat3Ch1−|−
1881759134Cat1Ch2
RLEN = RLEN values = equals string length|ODD_DIGIT = Values = minus 1|stringS = types of
characters in first part of string = all hexadecimals|stringS = number of hexadecimals in first
part = Odd|stringS = empty or not = not empty
values:       ODD_DIGIT = '-1', RLEN = '1', stringS = 'f'
```
**Test Case 6:**
```
choices:      2518115Cat1Ch5−|−202558941Cat1Ch3−|−1881759134Cat2Ch1−|−1881759134Cat3Ch2−|−
1881759134Cat1Ch2
RLEN = RLEN values = equals string length|ODD_DIGIT = Values = hexadecimal value (ASCII)|stringS
= types of characters in first part of string = all hexadecimals|stringS = number of
hexadecimals in first part = Even|stringS = empty or not = not empty
values:       RLEN = '2', ODD_DIGIT = '48', stringS = 'bD'
```
**Test Case 7:**
```
choices:      202558941Cat1Ch1−|−2518115Cat1Ch5−|−1881759134Cat2Ch3−|−1881759134Cat2Ch2−|−
1881759134Cat3Ch1
ODD_DIGIT = Values = strictly below -1|RLEN = RLEN values = equals string length|stringS = types
of characters in first part of string = mix hexa and non hexa|stringS = empty or not = not
empty|stringS = number of hexadecimals in first part = Odd
values:       RLEN = '3', ODD_DIGIT = '-2', stringS = 'DIG'
```

**Test Case 8:**
```
choices:      202558941Cat1Ch2−|−2518115Cat1Ch5−|−1881759134Cat2Ch2−|−1881759134Cat1Ch2−|−
1881759134Cat3Ch2
ODD_DIGIT = Values = minus 1|RLEN = RLEN values = equals string length|stringS = types of
characters in first part of string = all non hexadecimals|stringS = empty or not = not
empty|stringS = number of hexadecimals in first part = Even
values:       ODD_DIGIT = '-1', RLEN = '1', stringS = 'j'
```
**Test Case 9:**
```
choices:      202558941Cat1Ch4−|−2518115Cat1Ch5−|−1881759134Cat2Ch3−|−1881759134Cat1Ch2−|−
1881759134Cat3Ch1
ODD_DIGIT = Values = Not an hexadecimal (ASCII)|RLEN = RLEN values = equals string
length|stringS = types of characters in first part of string = mix hexa and non hexa|stringS =
empty or not = not empty|stringS = number of hexadecimals in first part = Odd
values:       ODD_DIGIT = '58', RLEN = '3', stringS = 'saJ'
```
**Test Case 10:**
```
choices:      2518115Cat1Ch3
RLEN = RLEN values = strictly negative
values:       RLEN = '-1'
```
**Test Case 11:**
```
choices:      2518115Cat1Ch4
RLEN = RLEN values = over string length
values:       stringS = '', RLEN = '1'
```
**Test Case 12:**
```
choices:      2518115Cat1Ch1−|−1881759134Cat1Ch2−|−202558941Cat1Ch3
RLEN = RLEN values = zero|stringS = empty or not = not empty|ODD_DIGIT = Values = hexadecimal
value (ASCII)
values:       ODD_DIGIT = '97', RLEN = '0', stringS = '4'
```
**Test Case 13:**
```
choices:      1881759134Cat1Ch1−|−202558941Cat1Ch3
stringS = empty or not = empty|ODD_DIGIT = Values = hexadecimal value (ASCII)
values:       stringS = '', ODD_DIGIT = '97'
```

## D.7    Three-way Test Frames and Corresponding Test Cases when using CASA

**Test Case 1:**
```
choices:      2518115Cat1Ch2−|−202558941Cat1Ch3−|−1881759134Cat1Ch2−|−1881759134Cat2Ch3−|−
1881759134Cat3Ch1
RLEN = RLEN values = within string strictly|ODD_DIGIT = Values = hexadecimal value
(ASCII)|stringS = empty or not = not empty|stringS = types of characters in first part of string
= mix hexa and non hexa|stringS = number of hexadecimals in first part = Odd
values:       RLEN = '3', ODD_DIGIT = '65', stringS = 'SCPk'
```
**Test Case 2:**
```
choices:      2518115Cat1Ch5−|−202558941Cat1Ch3−|−1881759134Cat1Ch2−|−1881759134Cat2Ch1−|−
1881759134Cat3Ch1
RLEN = RLEN values = equals string length|ODD_DIGIT = Values = hexadecimal value (ASCII)|stringS
= empty or not = not empty|stringS = types of characters in first part of string = all
hexadecimals|stringS = number of hexadecimals in first part = Odd
values:       RLEN = '1', ODD_DIGIT = '97', stringS = '0'
```
**Test Case 3:**
```
choices:      2518115Cat1Ch2−|−202558941Cat1Ch2−|−1881759134Cat1Ch2−|−1881759134Cat2Ch1−|−
1881759134Cat3Ch1
RLEN = RLEN values = within string strictly|ODD_DIGIT = Values = minus 1|stringS = empty or not
= not empty|stringS = types of characters in first part of string = all hexadecimals|stringS =
number of hexadecimals in first part = Odd
values:       ODD_DIGIT = '-1', RLEN = '1', stringS = 'ap'
```
**Test Case 4:**
```
choices:      2518115Cat1Ch2−|−202558941Cat1Ch3−|−1881759134Cat1Ch2−|−1881759134Cat2Ch1−|−
1881759134Cat3Ch2
RLEN = RLEN values = within string strictly|ODD_DIGIT = Values = hexadecimal value
(ASCII)|stringS = empty or not = not empty|stringS = types of characters in first part of string
= all hexadecimals|stringS = number of hexadecimals in first part = Even
values:       ODD_DIGIT = '48', RLEN = '2', stringS = 'BcH'
```
**Test Case 5:**
```
choices:      2518115Cat1Ch5−|−202558941Cat1Ch3−|−1881759134Cat1Ch2−|−1881759134Cat2Ch3−|−
1881759134Cat3Ch2
```

```
RLEN = RLEN values = equals string length|ODD_DIGIT = Values = hexadecimal value (ASCII)|stringS
= empty or not = not empty|stringS = types of characters in first part of string = mix hexa and
non hexa|stringS = number of hexadecimals in first part = Even
values:      ODD_DIGIT = '48', RLEN = '3', stringS = 'Gb3'
```

```
choices:     2518115Cat1Ch5-|−202558941Cat1Ch2-|−1881759134Cat1Ch2-|−1881759134Cat2Ch3-|−
1881759134Cat3Ch1
RLEN = RLEN values = equals string length|ODD_DIGIT = Values = minus 1|stringS = empty or not =
not empty|stringS = types of characters in first part of string = mix hexa and non hexa|stringS
= number of hexadecimals in first part = Odd
values:      RLEN = '3', ODD_DIGIT = '-1', stringS = 'awV'
```

```
choices:     2518115Cat1Ch5-|−202558941Cat1Ch2-|−1881759134Cat1Ch2-|−1881759134Cat2Ch1-|−
1881759134Cat3Ch2
RLEN = RLEN values = equals string length|ODD_DIGIT = Values = minus 1|stringS = empty or not =
not empty|stringS = types of characters in first part of string = all hexadecimals|stringS =
number of hexadecimals in first part = Even
values:      RLEN = '2', ODD_DIGIT = '-1', stringS = 'C5'
```

```
choices:     2518115Cat1Ch2-|−202558941Cat1Ch2-|−1881759134Cat1Ch2-|−1881759134Cat2Ch3-|−
1881759134Cat3Ch2
RLEN = RLEN values = within string strictly|ODD_DIGIT = Values = minus 1|stringS = empty or not
= not empty|stringS = types of characters in first part of string = mix hexa and non
hexa|stringS = number of hexadecimals in first part = Even
values:      ODD_DIGIT = '-1', RLEN = '3', stringS = '9pDp'
```

```
choices:     2518115Cat1Ch5-|−202558941Cat1Ch4-|−1881759134Cat1Ch2-|−1881759134Cat2Ch1-|−
1881759134Cat3Ch2
RLEN = RLEN values = equals string length|ODD_DIGIT = Values = Not an hexadecimal
(ASCII)|stringS = empty or not = not empty|stringS = types of characters in first part of string
= all hexadecimals|stringS = number of hexadecimals in first part = Even
values:      ODD_DIGIT = '58', RLEN = '2', stringS = 'Aa'
```

```
choices:     2518115Cat1Ch5-|−202558941Cat1Ch2-|−1881759134Cat1Ch2-|−1881759134Cat2Ch2-|−
1881759134Cat3Ch2
RLEN = RLEN values = equals string length|ODD_DIGIT = Values = minus 1|stringS = empty or not =
not empty|stringS = types of characters in first part of string = all non hexadecimals|stringS =
number of hexadecimals in first part = Even
values:      ODD_DIGIT = '-1', RLEN = '1', stringS = 's'
```

```
choices:     2518115Cat1Ch2-|−202558941Cat1Ch1-|−1881759134Cat1Ch2-|−1881759134Cat2Ch3-|−
1881759134Cat3Ch2
RLEN = RLEN values = within string strictly|ODD_DIGIT = Values = strictly below -1|stringS =
empty or not = not empty|stringS = types of characters in first part of string = mix hexa and
non hexa|stringS = number of hexadecimals in first part = Even
values:      RLEN = '3', ODD_DIGIT = '-2', stringS = 'at0b'
```

```
choices:     2518115Cat1Ch3
RLEN = RLEN values = strictly negative
values:      RLEN = '-1'
```

```
choices:     2518115Cat1Ch4
RLEN = RLEN values = over string length
values:      stringS = '', RLEN = '1'
```

```
choices:     2518115Cat1Ch1-|−1881759134Cat1Ch2-|−202558941Cat1Ch3
RLEN = RLEN values = zero|stringS = empty or not = not empty|ODD_DIGIT = Values = hexadecimal
value (ASCII)
values:      ODD_DIGIT = '97', RLEN = '0', stringS = '4'
```

```
choices:     1881759134Cat1Ch1-|−202558941Cat1Ch3
stringS = empty or not = empty|ODD_DIGIT = Values = hexadecimal value (ASCII)
values:      ODD_DIGIT = '97', stringS = ''
```

## D.8  Three-way Test Frames and Corresponding Test Cases when using ACTS

**Test Case 1:**
choices:      2518115Cat1Ch2−|−202558941Cat1Ch2−|−1881759134Cat2Ch1−|−1881759134Cat3Ch1−|−1881759134Cat1Ch2
RLEN = RLEN values = within string strictly|ODD_DIGIT = Values = minus 1|stringS = types of characters in first part of string = all hexadecimals|stringS = number of hexadecimals in first part = Odd|stringS = empty or not = not empty
values:       RLEN = '1', ODD_DIGIT = '-1', stringS = 'cQ'

**Test Case 2:**
choices:      2518115Cat1Ch2−|−202558941Cat1Ch2−|−1881759134Cat2Ch3−|−1881759134Cat3Ch2−|−1881759134Cat1Ch2
RLEN = RLEN values = within string strictly|ODD_DIGIT = Values = minus 1|stringS = types of characters in first part of string = mix hexa and non hexa|stringS = number of hexadecimals in first part = Even|stringS = empty or not = not empty
values:       RLEN = '3', ODD_DIGIT = '-1', stringS = 'AGCh'

**Test Case 3:**
choices:      2518115Cat1Ch2−|−202558941Cat1Ch3−|−1881759134Cat2Ch1−|−1881759134Cat3Ch2−|−1881759134Cat1Ch2
RLEN = RLEN values = within string strictly|ODD_DIGIT = Values = hexadecimal value (ASCII)|stringS = types of characters in first part of string = all hexadecimals|stringS = number of hexadecimals in first part = Even|stringS = empty or not = not empty
values:       ODD_DIGIT = '65', RLEN = '2', stringS = 'DFz'

**Test Case 4:**
choices:      2518115Cat1Ch2−|−202558941Cat1Ch3−|−1881759134Cat2Ch3−|−1881759134Cat3Ch1−|−1881759134Cat1Ch2
RLEN = RLEN values = within string strictly|ODD_DIGIT = Values = hexadecimal value (ASCII)|stringS = types of characters in first part of string = mix hexa and non hexa|stringS = number of hexadecimals in first part = Odd|stringS = empty or not = not empty
values:       ODD_DIGIT = '65', RLEN = '3', stringS = 'WbxK'

**Test Case 5:**
choices:      2518115Cat1Ch5−|−202558941Cat1Ch2−|−1881759134Cat2Ch1−|−1881759134Cat3Ch2−|−1881759134Cat1Ch2
RLEN = RLEN values = equals string length|ODD_DIGIT = Values = minus 1|stringS = types of characters in first part of string = all hexadecimals|stringS = number of hexadecimals in first part = Even|stringS = empty or not = not empty
values:       ODD_DIGIT = '-1', RLEN = '2', stringS = '08'

**Test Case 6:**
choices:      2518115Cat1Ch5−|−202558941Cat1Ch2−|−1881759134Cat2Ch3−|−1881759134Cat3Ch1−|−1881759134Cat1Ch2
RLEN = RLEN values = equals string length|ODD_DIGIT = Values = minus 1|stringS = types of characters in first part of string = mix hexa and non hexa|stringS = number of hexadecimals in first part = Odd|stringS = empty or not = not empty
values:       RLEN = '3', ODD_DIGIT = '-1', stringS = 'awV'

**Test Case 7:**
choices:      2518115Cat1Ch5−|−202558941Cat1Ch3−|−1881759134Cat2Ch1−|−1881759134Cat3Ch1−|−1881759134Cat1Ch2
RLEN = RLEN values = equals string length|ODD_DIGIT = Values = hexadecimal value (ASCII)|stringS = types of characters in first part of string = all hexadecimals|stringS = number of hexadecimals in first part = Odd|stringS = empty or not = not empty
values:       RLEN = '1', ODD_DIGIT = '97', stringS = '7'

**Test Case 8:**
choices:      2518115Cat1Ch5−|−202558941Cat1Ch3−|−1881759134Cat2Ch3−|−1881759134Cat3Ch2−|−1881759134Cat1Ch2
RLEN = RLEN values = equals string length|ODD_DIGIT = Values = hexadecimal value (ASCII)|stringS = types of characters in first part of string = mix hexa and non hexa|stringS = number of hexadecimals in first part = Even|stringS = empty or not = not empty
values:       ODD_DIGIT = '48', RLEN = '3', stringS = 'Ue7'

**Test Case 9:**
choices:      202558941Cat1Ch1−|−2518115Cat1Ch2−|−1881759134Cat2Ch1−|−1881759134Cat1Ch2−|−1881759134Cat3Ch1
ODD_DIGIT = Values = strictly below -1|RLEN = RLEN values = within string strictly|stringS = types of characters in first part of string = all hexadecimals|stringS = empty or not = not empty|stringS = number of hexadecimals in first part = Odd
values:       ODD_DIGIT = '-2', RLEN = '1', stringS = 'el'

**Test Case 10:**
```
choices:      202558941Cat1Ch2−|−2518115Cat1Ch2−|−1881759134Cat2Ch2−|−1881759134Cat1Ch2−|−
1881759134Cat3Ch2
ODD_DIGIT = Values = minus 1|RLEN = RLEN values = within string strictly|stringS = types of
characters in first part of string = all non hexadecimals|stringS = empty or not = not
empty|stringS = number of hexadecimals in first part = Even
values:       ODD_DIGIT = '-1', RLEN = '1', stringS = 'Y2'
```
**Test Case 11:**
```
choices:      202558941Cat1Ch4−|−2518115Cat1Ch2−|−1881759134Cat2Ch1−|−1881759134Cat1Ch2−|−
1881759134Cat3Ch1
ODD_DIGIT = Values = Not an hexadecimal (ASCII)|RLEN = RLEN values = within string
strictly|stringS = types of characters in first part of string = all hexadecimals|stringS =
empty or not = not empty|stringS = number of hexadecimals in first part = Odd
values:       RLEN = '1', ODD_DIGIT = '71', stringS = 'cp'
```
**Test Case 12:**
```
choices:      2518115Cat1Ch3
RLEN = RLEN values = strictly negative
values:       RLEN = '-1'
```
**Test Case 13:**
```
choices:      2518115Cat1Ch4
RLEN = RLEN values = over string length
values:       stringS = '', RLEN = '1'
```
**Test Case 14:**
```
choices:      2518115Cat1Ch1−|−1881759134Cat1Ch2−|−202558941Cat1Ch3
RLEN = RLEN values = zero|stringS = empty or not = not empty|ODD_DIGIT = Values = hexadecimal
value (ASCII)
values:       ODD_DIGIT = '97', RLEN = '0', stringS = '4'
```
**Test Case 15:**
```
choices:      1881759134Cat1Ch1−|−202558941Cat1Ch3
stringS = empty or not = empty|ODD_DIGIT = Values = hexadecimal value (ASCII)
values:       ODD_DIGIT = '97', stringS = ''
```

## *D.9    Base Choice Test Frames and Corresponding Test Cases*

**Test Case 1:**
```
choices:      2518115Cat1Ch2−|−202558941Cat1Ch3−|−1881759134Cat1Ch2−|−1881759134Cat2Ch1−|−
1881759134Cat3Ch1
RLEN = RLEN values = within string strictly|ODD_DIGIT = Values = hexadecimal value
(ASCII)|stringS = empty or not = not empty|stringS = types of characters in first part of string
= all hexadecimals|stringS = number of hexadecimals in first part = Odd
values:       RLEN = '1', ODD_DIGIT = '65', stringS = 'c1'
```
**Test Case 2:**
```
choices:      2518115Cat1Ch1−|−1881759134Cat1Ch2−|−202558941Cat1Ch3
RLEN = RLEN values = zero|stringS = empty or not = not empty|ODD_DIGIT = Values = hexadecimal
value (ASCII)
values:       RLEN = '0', ODD_DIGIT = '97', stringS = '4'
```
**Test Case 3:**
```
choices:      2518115Cat1Ch3
RLEN = RLEN values = strictly negative
values:       RLEN = '-1'
```
**Test Case 4:**
```
choices:      2518115Cat1Ch4
RLEN = RLEN values = over string length
values:       stringS = '', RLEN = '1'
```
**Test Case 5:**
```
choices:      202558941Cat1Ch3−|−1881759134Cat1Ch2−|−1881759134Cat2Ch1−|−1881759134Cat3Ch1−|−
2518115Cat1Ch5
ODD_DIGIT = Values = hexadecimal value (ASCII)|stringS = empty or not = not empty|stringS =
types of characters in first part = all hexadecimals|stringS = number of hexadecimals
in first part = Odd|RLEN = RLEN values = equals string length
values:       ODD_DIGIT = '97', RLEN = '1', stringS = '5'
```
**Test Case 6:**
```
choices:      2518115Cat1Ch2−|−1881759134Cat1Ch2−|−1881759134Cat2Ch1−|−1881759134Cat3Ch1−|−
202558941Cat1Ch1
```

```
RLEN = RLEN values = within string strictly|stringS = empty or not = not empty|stringS = types
of characters in first part of string = all hexadecimals|stringS = number of hexadecimals in
first part = Odd|ODD_DIGIT = Values = strictly below -1
values:       ODD_DIGIT = '-2', RLEN = '1', stringS = 'ap'
```

**Test Case 7:**
```
choices:     2518115Cat1Ch2−|−1881759134Cat1Ch2−|−1881759134Cat2Ch1−|−1881759134Cat3Ch1−|−
202558941Cat1Ch2
RLEN = RLEN values = within string strictly|stringS = empty or not = not empty|stringS = types
of characters in first part of string = all hexadecimals|stringS = number of hexadecimals in
first part = Odd|ODD_DIGIT = Values = minus 1
values:       ODD_DIGIT = '-1', RLEN = '1', stringS = 'ab'
```

**Test Case 8:**
```
choices:     2518115Cat1Ch2−|−1881759134Cat1Ch2−|−1881759134Cat2Ch1−|−1881759134Cat3Ch1−|−
202558941Cat1Ch4
RLEN = RLEN values = within string strictly|stringS = empty or not = not empty|stringS = types
of characters in first part of string = all hexadecimals|stringS = number of hexadecimals in
first part = Odd|ODD_DIGIT = Values = Not an hexadecimal (ASCII)
values:       RLEN = '1', ODD_DIGIT = '71', stringS = 'bT'
```

**Test Case 9:**
```
choices:     1881759134Cat1Ch1−|−202558941Cat1Ch3
stringS = empty or not = empty|ODD_DIGIT = Values = hexadecimal value (ASCII)
values:       ODD_DIGIT = '97', stringS = ''
```

**Test Case 10:**
```
choices:     1881759134Cat2Ch2
stringS = types of characters in first part of string = all non hexadecimals
values:       RLEN = '2', stringS = 's'
```

**Test Case 11:**
```
choices:     2518115Cat1Ch2−|−202558941Cat1Ch3−|−1881759134Cat1Ch2−|−1881759134Cat3Ch1−|−
1881759134Cat2Ch3
RLEN = RLEN values = within string strictly|ODD_DIGIT = Values = hexadecimal value
(ASCII)|stringS = empty or not = not empty|stringS = types of characters in first part =
Odd|stringS = types of characters in first part of string = mix hexa and non hexa
values:       ODD_DIGIT = '97', RLEN = '3', stringS = '2gkt'
```

**Test Case 12:**
```
choices:     2518115Cat1Ch2−|−202558941Cat1Ch3−|−1881759134Cat1Ch2−|−1881759134Cat2Ch1−|−
1881759134Cat3Ch2
RLEN = RLEN values = within string strictly|ODD_DIGIT = Values = hexadecimal value
(ASCII)|stringS = empty or not = not empty|stringS = types of characters in first part of string
= all hexadecimals|stringS = number of hexadecimals in first part = Even
values:       ODD_DIGIT = '48', RLEN = '2', stringS = 'e9K'
```

## Appendix E. INTEGER SET DIFFERENCE

In the CP specification below, an asterisk ('*') beside a choice indicates the base choice for that category. Each parameter is uniquely identified by an integer identifier, which is then used in the composition of test frames; As a result, a choice for a parameter is uniquely identified by the parameter identifier, the category identifier and the choice identifier. Below we summarize the specification of the unit under test (section E.1), and provide the CP specification we used (section E.2). We then provide the test frames and corresponding test cases (inputs) we obtain when using CASA (Each-Choice in section E.3, Pairwise in section E.5, Three-way in section E.7), when using ACTS (Each-Choice in section E.4, Pairwise in section E.6, Three-way in section E.8), and when using the Base Choice criterion (section E.9).

### E.1 Specification

The function under test takes as input two sets of integers, Set1 and Set2, and returns their intersection.

### E.2 CP Specification

```
Parameter(2573359): Set1 of type IntegerSet
  Category 1 - Size of Set1
    Ch 1.1: Empty set 1
            Set1.isEmpty()
            [single][properties Set1Empty]
    Ch 1.2: Small set 1
            Set1.size()>0 && Set1.size()<=10
            [properties smallSet1]
   *Ch 1.3: Large set 1
            Set1.size()>10
            [properties largeSet1]
  Category 2 - Comparing the two set sizes
   *Ch 2.1: Set1 larger than Set2
            Set1.size()>Set2.size()
            [if !Set1Empty && largeSet1 && !largeSet2][properties Set1larger]
    Ch 2.2: Set1 smaller than Set2
            Set1.size()<Set2.size()
            [if !Set2Empty && largeSet2 && !largeSet1][properties Set2larger]
    Ch 2.3: Sets of equal size
            Set1.size()==Set2.size()
            [if (smallSet1 && smallSet2) || (largeSet1 && largeSet2)]
  Category 3 - Intersection of the two sets
    Ch 3.1: Empty intersection
            Set1.intersection(Set2).isEmpty()
            [properties EmptyIntersection]
   *Ch 3.2: Non empty intersection
            Set1.intersection(Set2).size()>0
            [if !Set1Empty && !Set2Empty]
  Category 4 - Inclusion of the two sets
    Ch 4.1: Set1 includes Set2
            Set1.includes(Set2)
            [if !Set1Empty && Set1larger && !EmptyIntersection]
    Ch 4.2: Set2 includes Set1
            Set2.includes(Set1)
            [if !Set2Empty && Set2larger && !EmptyIntersection]
   *Ch 4.3: No set inclusion
            !(Set1.includes(Set2)) && !(Set2.includes(Set1))
Parameter(2573360): Set2 of type IntegerSet
  Category 1 - Size of Set2
    Ch 1.1: empty set 2
            Set2.isEmpty()
            [single][properties Set2Empty]
   *Ch 1.2: Small set 2
            Set2.size()>0 && Set2.size()<=10
            [properties smallSet2]
```

```
Ch 1.3: Large set 2
        Set2.size()>10
        [properties largeSet2]
```

## E.3     Each-Choice Test Frames and Corresponding Test Cases when using CASA

**Test Case 1:**
choices:    2573359Cat1Ch2−|−2573359Cat2Ch2−|−2573359Cat3Ch2−|−2573359Cat4Ch2−|−2573360Cat1Ch3
Set1 = Size of Set1 = Small set 1|Set1 = Comparing the two set sizes = Set1 smaller than
Set2|Set1 = Intersection of the two sets = Non empty intersection|Set1 = Inclusion of the two
sets = Set2 includes Set1|Set2 = Size of Set2 = Large set 2
values:     Set1 = '8', Set2 = '0, 4, 8, 9, 10, 11, 12, 14, 16, 19, 20, 24'

**Test Case 2:**
choices:    2573359Cat1Ch3−|−2573359Cat2Ch3−|−2573359Cat3Ch1−|−2573359Cat4Ch3−|−2573360Cat1Ch3
Set1 = Size of Set1 = Large set 1|Set1 = Comparing the two set sizes = Sets of equal size|Set1 =
Intersection of the two sets = Empty intersection|Set1 = Inclusion of the two sets = No set
inclusion|Set2 = Size of Set2 = Large set 2
values:     Set1 = '1, 2, 4, 5, 9, 11, 12, 14, 19, 23, 24', Set2 = '0, 6, 7, 8, 10, 13, 15, 18,
            20, 21, 22'

**Test Case 3:**
choices:    2573359Cat1Ch3−|−2573359Cat2Ch1−|−2573359Cat3Ch2−|−2573359Cat4Ch1−|−2573360Cat1Ch2
Set1 = Size of Set1 = Large set 1|Set1 = Comparing the two set sizes = Set1 larger than Set2
|Set1 = Intersection of the two sets = Non empty intersection|Set1 = Inclusion of the two sets =
Set1 includes Set2|Set2 = Size of Set2 = Small set 2
values:     Set1 = '0, 1, 3, 5, 8, 9, 10, 12, 16, 18, 19, 20, 22, 23, 24', Set2 = '0'

**Test Case 4:**
choices:    2573359Cat1Ch3−|−2573359Cat2Ch1−|−2573359Cat3Ch1−|−2573359Cat4Ch3−|−2573360Cat1Ch1
Set1 = Size of Set1 = Large set 1|Set1 = Comparing the two set sizes = Set1 larger than Set2
|Set1 = Intersection of the two sets = Empty intersection|Set1 = Inclusion of the two sets = No
set inclusion|Set2 = Size of Set2 = empty set 2
values:     Set1 = '1, 2, 5, 6, 7, 9, 13, 14, 15, 17, 18, 20, 21, 22, 23, 24', Set2 = ''

**Test Case 5:**
choices:    2573359Cat1Ch1−|−2573359Cat2Ch2−|−2573359Cat3Ch1−|−2573359Cat4Ch3−|−2573360Cat1Ch3
Set1 = Size of Set1 = Empty set 1|Set1 = Comparing the two set sizes = Set1 smaller than
Set2|Set1 = Intersection of the two sets = Empty intersection|Set1 = Inclusion of the two sets =
No set inclusion|Set2 = Size of Set2 = Large set 2
values:     Set1 = '', Set2 = '0, 1, 3, 4, 5, 7, 13, 14, 15, 16, 18, 19, 20, 23'

## E.4     Each-Choice Test Frames and Corresponding Test Cases when using ACTS

**Test Case 1:**
choices:    2573359Cat2Ch1−|−2573359Cat4Ch3−|−2573359Cat1Ch3−|−2573359Cat3Ch2−|−2573360Cat1Ch2
Set1 = Comparing the two set sizes = Set1 larger than Set2 |Set1 = Inclusion of the two sets =
No set inclusion|Set1 = Size of Set1 = Large set 1|Set1 = Intersection of the two sets = Non
empty intersection|Set2 = Size of Set2 = Small set 2
values:     Set1 = '0, 1, 2, 3, 4, 5, 6, 7, 8, 9, 13, 14, 15, 16, 17, 18, 19, 20, 21, 23', Set2
            = '1, 4, 7, 16'

**Test Case 2:**
choices:    2573359Cat2Ch2−|−2573359Cat4Ch2−|−2573359Cat1Ch2−|−2573359Cat3Ch2−|−2573360Cat1Ch3
Set1 = Comparing the two set sizes = Set1 smaller than Set2|Set1 = Inclusion of the two sets =
Set2 includes Set1|Set1 = Size of Set1 = Small set 1|Set1 = Intersection of the two sets = Non
empty intersection|Set2 = Size of Set2 = Large set 2
values:     Set1 = '1, 13', Set2 = '1, 4, 6, 7, 9, 11, 12, 13, 15, 17, 20, 21, 22'

**Test Case 3:**
choices:    2573359Cat2Ch3−|−2573359Cat4Ch3−|−2573359Cat1Ch3−|−2573359Cat3Ch1−|−2573360Cat1Ch3
Set1 = Comparing the two set sizes = Sets of equal size|Set1 = Inclusion of the two sets = No
set inclusion|Set1 = Size of Set1 = Large set 1|Set1 = Intersection of the two sets = Empty
intersection|Set2 = Size of Set2 = Large set 2
values:     Set1 = '3, 4, 5, 10, 11, 12, 15, 16, 21, 23, 24', Set2 = '0, 1, 2, 6, 7, 9, 14, 17,
            18, 19, 22'

**Test Case 4:**
choices:    2573359Cat2Ch1−|−2573359Cat4Ch1−|−2573359Cat1Ch3−|−2573359Cat3Ch2−|−2573360Cat1Ch2
Set1 = Comparing the two set sizes = Set1 larger than Set2 |Set1 = Inclusion of the two sets =
Set1 includes Set2|Set1 = Size of Set1 = Large set 1|Set1 = Intersection of the two sets = Non
empty intersection|Set2 = Size of Set2 = Small set 2
```



```
values:        Set1 = '0, 1, 6, 8, 9, 14, 17, 20, 21, 23, 24', Set2 = '14'
```

**Test Case 5:**
```
choices:       2573359Cat1Ch1−|−2573359Cat2Ch2−|−2573359Cat4Ch3−|−2573360Cat1Ch3−|−2573359Cat3Ch1
Set1 = Size of Set1 = Empty set 1|Set1 = Comparing the two set sizes = Set1 smaller than
Set2|Set1 = Inclusion of the two sets = No set inclusion|Set2 = Size of Set2 = Large set 2|Set1
= Intersection of the two sets = Empty intersection
values:        Set1 = '', Set2 = '2, 4, 7, 8, 9, 10, 12, 13, 15, 16, 17, 19, 21, 22'
```

**Test Case 6:**
```
choices:       2573359Cat1Ch3−|−2573359Cat2Ch1−|−2573359Cat4Ch3−|−2573360Cat1Ch1−|−2573359Cat3Ch1
Set1 = Size of Set1 = Large set 1|Set1 = Comparing the two set sizes = Set1 larger than Set2
|Set1 = Inclusion of the two sets = No set inclusion|Set2 = Size of Set2 = empty set 2|Set1 =
Intersection of the two sets = Empty intersection
values:        Set1 = '1, 2, 4, 5, 6, 7, 8, 9, 10, 11, 12, 13, 15, 17, 19, 20, 21, 22', Set2 = ''
```

## E.5    *Pairwise Test Frames and Corresponding Test Cases when using CASA*

**Test Case 1:**
```
choices:       2573359Cat1Ch3−|−2573359Cat2Ch1−|−2573359Cat3Ch2−|−2573359Cat4Ch1−|−2573360Cat1Ch2
Set1 = Size of Set1 = Large set 1|Set1 = Comparing the two set sizes = Set1 larger than Set2
|Set1 = Intersection of the two sets = Non empty intersection|Set1 = Inclusion of the two sets =
Set1 includes Set2|Set2 = Size of Set2 = Small set 2
values:        Set1 = '0, 1, 2, 4, 6, 7, 8, 9, 11, 17, 18, 19, 23, 24', Set2 = '2, 4, 7, 17, 18,
               19'
```

**Test Case 2:**
```
choices:       2573359Cat1Ch2−|−2573359Cat2Ch3−|−2573359Cat3Ch1−|−2573359Cat4Ch3−|−2573360Cat1Ch2
Set1 = Size of Set1 = Small set 1|Set1 = Comparing the two set sizes = Sets of equal size|Set1 =
Intersection of the two sets = Empty intersection|Set1 = Inclusion of the two sets = No set
inclusion|Set2 = Size of Set2 = Small set 2
values:        Set1 = '10', Set2 = '0'
```

**Test Case 3:**
```
choices:       2573359Cat1Ch3−|−2573359Cat2Ch3−|−2573359Cat3Ch2−|−2573359Cat4Ch3−|−2573360Cat1Ch3
Set1 = Size of Set1 = Large set 1|Set1 = Comparing the two set sizes = Sets of equal size|Set1 =
Intersection of the two sets = Non empty intersection|Set1 = Inclusion of the two sets = No set
inclusion|Set2 = Size of Set2 = Large set 2
values:        Set1 = '2, 3, 4, 7, 10, 11, 14, 16, 21, 22, 24', Set2 = '0, 1, 2, 3, 6, 7, 8, 17,
               21, 22, 24'
```

**Test Case 4:**
```
choices:       2573359Cat1Ch2−|−2573359Cat2Ch2−|−2573359Cat3Ch2−|−2573359Cat4Ch2−|−2573360Cat1Ch3
Set1 = Size of Set1 = Small set 1|Set1 = Comparing the two set sizes = Set1 smaller than
Set2|Set1 = Intersection of the two sets = Non empty intersection|Set1 = Inclusion of the two
sets = Set2 includes Set1|Set2 = Size of Set2 = Large set 2
values:        Set1 = '9', Set2 = '0, 1, 2, 3, 5, 6, 8, 9, 11, 12, 13, 14, 15, 18, 19, 21, 22, 23'
```

**Test Case 5:**
```
choices:       2573359Cat1Ch3−|−2573359Cat2Ch1−|−2573359Cat3Ch1−|−2573359Cat4Ch3−|−2573360Cat1Ch2
Set1 = Size of Set1 = Large set 1|Set1 = Comparing the two set sizes = Set1 larger than Set2
|Set1 = Intersection of the two sets = Empty intersection|Set1 = Inclusion of the two sets = No
set inclusion|Set2 = Size of Set2 = Small set 2
values:        Set1 = '0, 2, 3, 6, 8, 11, 15, 16, 17, 18, 24', Set2 = '9, 10, 13, 19, 20, 21'
```

**Test Case 6:**
```
choices:       2573359Cat1Ch2−|−2573359Cat2Ch2−|−2573359Cat3Ch1−|−2573359Cat4Ch3−|−2573360Cat1Ch3
Set1 = Size of Set1 = Small set 1|Set1 = Comparing the two set sizes = Set1 smaller than
Set2|Set1 = Intersection of the two sets = Empty intersection|Set1 = Inclusion of the two sets =
No set inclusion|Set2 = Size of Set2 = Large set 2
values:        Set1 = '15', Set2 = '1, 2, 3, 4, 5, 6, 7, 8, 9, 10, 11, 12, 14, 16, 18, 19, 20, 22,
               23, 24'
```

**Test Case 7:**
```
choices:       2573359Cat1Ch1−|−2573359Cat2Ch2−|−2573359Cat3Ch1−|−2573359Cat4Ch3−|−2573360Cat1Ch3
Set1 = Size of Set1 = Empty set 1|Set1 = Comparing the two set sizes = Set1 smaller than
Set2|Set1 = Intersection of the two sets = Empty intersection|Set1 = Inclusion of the two sets =
No set inclusion|Set2 = Size of Set2 = Large set 2
values:        Set1 = '', Set2 = '0, 1, 2, 6, 7, 8, 9, 11, 12, 14, 15, 16, 17, 18, 20, 21, 22, 23,
               24'
```

**Test Case 8:**
```
choices:       2573359Cat1Ch3−|−2573359Cat2Ch1−|−2573359Cat3Ch1−|−2573359Cat4Ch3−|−2573360Cat1Ch1
```



```
Set1 = Size of Set1 = Large set 1|Set1 = Comparing the two set sizes = Set1 larger than Set2
|Set1 = Intersection of the two sets = Empty intersection|Set1 = Inclusion of the two sets = No
set inclusion|Set2 = Size of Set2 = empty set 2
values:      Set1 = '1, 2, 4, 5, 7, 8, 9, 10, 11, 14, 15, 16, 17, 19, 20, 22, 23, 24', Set2 = ''
```

## E.6    Pairwise Test Frames and Corresponding Test Cases when using ACTS

**Test Case 1:**
```
choices:     2573359Cat2Ch1−|−2573359Cat4Ch1−|−2573359Cat1Ch3−|−2573359Cat3Ch2−|−2573360Cat1Ch2
Set1 = Comparing the two set sizes = Set1 larger than Set2 |Set1 = Inclusion of the two sets =
Set1 includes Set2|Set1 = Size of Set1 = Large set 1|Set1 = Intersection of the two sets = Non
empty intersection|Set2 = Size of Set2 = Small set 2
values:      Set1 = '0, 1, 2, 3, 4, 5, 7, 10, 14, 15, 16, 17, 18, 19, 20, 24', Set2 = '1, 4'
```
**Test Case 2:**
```
choices:     2573359Cat2Ch1−|−2573359Cat4Ch3−|−2573359Cat1Ch3−|−2573359Cat3Ch1−|−2573360Cat1Ch2
Set1 = Comparing the two set sizes = Set1 larger than Set2 |Set1 = Inclusion of the two sets =
No set inclusion|Set1 = Size of Set1 = Large set 1|Set1 = Intersection of the two sets = Empty
intersection|Set2 = Size of Set2 = Small set 2
values:      Set1 = '0, 1, 2, 4, 5, 6, 8, 9, 14, 15, 16, 17, 18, 19, 20, 22, 23', Set2 = '11,
             13, 21'
```
**Test Case 3:**
```
choices:     2573359Cat2Ch2−|−2573359Cat4Ch2−|−2573359Cat1Ch2−|−2573359Cat3Ch2−|−2573360Cat1Ch3
Set1 = Comparing the two set sizes = Set1 smaller than Set2|Set1 = Inclusion of the two sets =
Set2 includes Set1|Set1 = Size of Set1 = Small set 1|Set1 = Intersection of the two sets = Non
empty intersection|Set2 = Size of Set2 = Large set 2
values:      Set1 = '1, 5, 8, 13', Set2 = '1, 5, 6, 7, 8, 9, 13, 15, 17, 19, 23'
```
**Test Case 4:**
```
choices:     2573359Cat2Ch2−|−2573359Cat4Ch3−|−2573359Cat1Ch2−|−2573359Cat3Ch1−|−2573360Cat1Ch3
Set1 = Comparing the two set sizes = Set1 smaller than Set2|Set1 = Inclusion of the two sets =
No set inclusion|Set1 = Size of Set1 = Small set 1|Set1 = Intersection of the two sets = Empty
intersection|Set2 = Size of Set2 = Large set 2
values:      Set1 = '3', Set2 = '0, 2, 5, 6, 7, 9, 10, 13, 14, 16, 17, 19, 21, 22'
```
**Test Case 5:**
```
choices:     2573359Cat2Ch3−|−2573359Cat4Ch3−|−2573359Cat1Ch3−|−2573359Cat3Ch2−|−2573360Cat1Ch3
Set1 = Comparing the two set sizes = Sets of equal size|Set1 = Inclusion of the two sets = No
set inclusion|Set1 = Size of Set1 = Large set 1|Set1 = Intersection of the two sets = Non empty
intersection|Set2 = Size of Set2 = Large set 2
values:      Set1 = '2, 4, 8, 9, 10, 13, 17, 18, 20, 21, 24', Set2 = '3, 6, 7, 10, 11, 12, 13,
             17, 18, 21, 22'
```
**Test Case 6:**
```
choices:     2573359Cat2Ch3−|−2573359Cat4Ch3−|−2573359Cat1Ch2−|−2573359Cat3Ch1−|−2573360Cat1Ch2
Set1 = Comparing the two set sizes = Sets of equal size|Set1 = Inclusion of the two sets = No
set inclusion|Set1 = Size of Set1 = Small set 1|Set1 = Intersection of the two sets = Empty
intersection|Set2 = Size of Set2 = Small set 2
values:      Set1 = '24', Set2 = '22'
```
**Test Case 7:**
```
choices:     2573359Cat1Ch1−|−2573359Cat2Ch2−|−2573359Cat4Ch3−|−2573360Cat1Ch3−|−2573359Cat3Ch1
Set1 = Size of Set1 = Empty set 1|Set1 = Comparing the two set sizes = Set1 smaller than
Set2|Set1 = Inclusion of the two sets = No set inclusion|Set2 = Size of Set2 = Large set 2|Set1
= Intersection of the two sets = Empty intersection
values:      Set1 = '', Set2 = '0, 1, 2, 3, 4, 6, 8, 9, 10, 11, 12, 13, 14, 15, 17, 18, 19, 20,
             21, 23, 24'
```
**Test Case 8:**
```
choices:     2573359Cat1Ch3−|−2573359Cat2Ch1−|−2573359Cat4Ch3−|−2573360Cat1Ch1−|−2573359Cat3Ch1
Set1 = Size of Set1 = Large set 1|Set1 = Comparing the two set sizes = Set1 larger than Set2
|Set1 = Inclusion of the two sets = No set inclusion|Set2 = Size of Set2 = empty set 2|Set1 =
Intersection of the two sets = Empty intersection
values:      Set1 = '1, 3, 5, 7, 8, 10, 11, 12, 14, 16, 17, 18, 20, 22, 23, 24', Set2 = ''
```

## E.7    Three-way Test Frames and Corresponding Test Cases when using CASA

**Test Case 1:**
```
choices:     2573359Cat1Ch2−|−2573359Cat2Ch2−|−2573359Cat3Ch2−|−2573359Cat4Ch2−|−2573360Cat1Ch3
```



Set1 = Size of Set1 = Small set 1|Set1 = Comparing the two set sizes = Set1 smaller than
Set2|Set1 = Intersection of the two sets = Non empty intersection|Set1 = Inclusion of the two
sets = Set2 includes Set1|Set2 = Size of Set2 = Large set 2
values:      Set1 = '8', Set2 = '0, 4, 8, 9, 10, 11, 12, 14, 16, 19, 20, 24'

**Test Case 2:**
choices:     2573359Cat1Ch2−|−2573359Cat2Ch2−|−2573359Cat3Ch1−|−2573359Cat4Ch3−|−2573360Cat1Ch3
Set1 = Size of Set1 = Small set 1|Set1 = Comparing the two set sizes = Set1 smaller than
Set2|Set1 = Intersection of the two sets = Empty intersection|Set1 = Inclusion of the two sets =
No set inclusion|Set2 = Size of Set2 = Large set 2
values:      Set1 = '13', Set2 = '1, 2, 3, 4, 5, 6, 7, 8, 9, 10, 11, 12, 15, 16, 17, 20, 23, 24'

**Test Case 3:**
choices:     2573359Cat1Ch3−|−2573359Cat2Ch1−|−2573359Cat3Ch1−|−2573359Cat4Ch3−|−2573360Cat1Ch2
Set1 = Size of Set1 = Large set 1|Set1 = Comparing the two set sizes = Set1 larger than Set2
|Set1 = Intersection of the two sets = Empty intersection|Set1 = Inclusion of the two sets = No
set inclusion|Set2 = Size of Set2 = Small set 2
values:      Set1 = '1, 3, 4, 8, 10, 11, 13, 14, 17, 19, 21, 23, 24', Set2 = '0, 20'

**Test Case 4:**
choices:     2573359Cat1Ch3−|−2573359Cat2Ch3−|−2573359Cat3Ch2−|−2573359Cat4Ch3−|−2573360Cat1Ch3
Set1 = Size of Set1 = Large set 1|Set1 = Comparing the two set sizes = Sets of equal size|Set1 =
Intersection of the two sets = Non empty intersection|Set1 = Inclusion of the two sets = No set
inclusion|Set2 = Size of Set2 = Large set 2
values:      Set1 = '1, 2, 4, 5, 6, 7, 9, 10, 12, 13, 24', Set2 = '0, 3, 8, 12, 13, 14, 15, 17,
             19, 20, 23'

**Test Case 5:**
choices:     2573359Cat1Ch2−|−2573359Cat2Ch3−|−2573359Cat3Ch2−|−2573359Cat4Ch3−|−2573360Cat1Ch2
Set1 = Size of Set1 = Small set 1|Set1 = Comparing the two set sizes = Sets of equal size|Set1 =
Intersection of the two sets = Non empty intersection|Set1 = Inclusion of the two sets = No set
inclusion|Set2 = Size of Set2 = Small set 2
values:      Set1 = '10, 11, 16', Set2 = '10, 11, 22'

**Test Case 6:**
choices:     2573359Cat1Ch2−|−2573359Cat2Ch3−|−2573359Cat3Ch1−|−2573359Cat4Ch3−|−2573360Cat1Ch2
Set1 = Size of Set1 = Small set 1|Set1 = Comparing the two set sizes = Sets of equal size|Set1 =
Intersection of the two sets = Empty intersection|Set1 = Inclusion of the two sets = No set
inclusion|Set2 = Size of Set2 = Small set 2
values:      Set1 = '6, 12, 20, 24', Set2 = '2, 11, 17, 23'

**Test Case 7:**
choices:     2573359Cat1Ch3−|−2573359Cat2Ch1−|−2573359Cat3Ch2−|−2573359Cat4Ch3−|−2573360Cat1Ch2
Set1 = Size of Set1 = Large set 1|Set1 = Comparing the two set sizes = Set1 larger than Set2
|Set1 = Intersection of the two sets = Non empty intersection|Set1 = Inclusion of the two sets =
No set inclusion|Set2 = Size of Set2 = Small set 2
values:      Set1 = '3, 6, 7, 8, 10, 11, 13, 17, 18, 19, 23', Set2 = '11, 18'

**Test Case 8:**
choices:     2573359Cat1Ch2−|−2573359Cat2Ch2−|−2573359Cat3Ch2−|−2573359Cat4Ch3−|−2573360Cat1Ch3
Set1 = Size of Set1 = Small set 1|Set1 = Comparing the two set sizes = Set1 smaller than
Set2|Set1 = Intersection of the two sets = Non empty intersection|Set1 = Inclusion of the two
sets = No set inclusion|Set2 = Size of Set2 = Large set 2
values:      Set1 = '0, 19', Set2 = '0, 1, 3, 5, 6, 8, 9, 15, 18, 20, 21, 22, 23, 24'

**Test Case 9:**
choices:     2573359Cat1Ch3−|−2573359Cat2Ch3−|−2573359Cat3Ch1−|−2573359Cat4Ch3−|−2573360Cat1Ch3
Set1 = Size of Set1 = Large set 1|Set1 = Comparing the two set sizes = Sets of equal size|Set1 =
Intersection of the two sets = Empty intersection|Set1 = Inclusion of the two sets = No set
inclusion|Set2 = Size of Set2 = Large set 2
values:      Set1 = '5, 8, 11, 12, 13, 17, 19, 20, 21, 22, 23, 24', Set2 = '0, 1, 2, 3, 4, 6, 7,
             9, 10, 14, 15, 16'

**Test Case 10:**
choices:     2573359Cat1Ch3−|−2573359Cat2Ch1−|−2573359Cat3Ch2−|−2573359Cat4Ch1−|−2573360Cat1Ch2
Set1 = Size of Set1 = Large set 1|Set1 = Comparing the two set sizes = Set1 larger than Set2
|Set1 = Intersection of the two sets = Non empty intersection|Set1 = Inclusion of the two sets =
Set1 includes Set2|Set2 = Size of Set2 = Small set 2
values:      Set1 = '0, 3, 4, 5, 7, 10, 11, 13, 14, 15, 17, 19, 20, 21, 22, 23', Set2 = '5, 15'

**Test Case 11:**
choices:     2573359Cat1Ch1−|−2573359Cat2Ch2−|−2573359Cat3Ch1−|−2573359Cat4Ch3−|−2573360Cat1Ch3
Set1 = Size of Set1 = Empty set 1|Set1 = Comparing the two set sizes = Set1 smaller than
Set2|Set1 = Intersection of the two sets = Empty intersection|Set1 = Inclusion of the two sets =
No set inclusion|Set2 = Size of Set2 = Large set 2



```
values:          Set1 = '', Set2 = '0, 2, 3, 4, 8, 12, 16, 17, 18, 22, 24'
```

**Test Case 12:**
```
choices:          2573359Cat1Ch3−|−2573359Cat2Ch1−|−2573359Cat3Ch1−|−2573359Cat4Ch3−|−2573360Cat1Ch1
Set1 = Size of Set1 = Large set 1|Set1 = Comparing the two set sizes = Set1 larger than Set2
|Set1 = Intersection of the two sets = Empty intersection|Set1 = Inclusion of the two sets = No
set inclusion|Set2 = Size of Set2 = empty set 2
values:          Set1 = '1, 2, 3, 4, 5, 6, 7, 8, 9, 10, 11, 14, 17, 18, 19, 21, 22, 23', Set2 = ''
```

## E.8    Three-way Test Frames and Corresponding Test Cases when using ACTS

**Test Case 1:**
```
choices:          2573359Cat2Ch1−|−2573359Cat4Ch1−|−2573359Cat1Ch3−|−2573359Cat3Ch2−|−2573360Cat1Ch2
Set1 = Comparing the two set sizes = Set1 larger than Set2 |Set1 = Inclusion of the two sets =
Set1 includes Set2|Set1 = Size of Set1 = Large set 1|Set1 = Intersection of the two sets = Non
empty intersection|Set2 = Size of Set2 = Small set 2
values:          Set1 = '0, 1, 2, 3, 4, 5, 7, 10, 14, 15, 16, 17, 18, 19, 20, 24', Set2 = '1, 4'
```

**Test Case 2:**
```
choices:          2573359Cat2Ch1−|−2573359Cat4Ch3−|−2573359Cat1Ch3−|−2573359Cat3Ch1−|−2573360Cat1Ch2
Set1 = Comparing the two set sizes = Set1 larger than Set2 |Set1 = Inclusion of the two sets =
No set inclusion|Set1 = Size of Set1 = Large set 1|Set1 = Intersection of the two sets = Empty
intersection|Set2 = Size of Set2 = Small set 2
values:          Set1 = '0, 1, 2, 4, 5, 6, 8, 9, 14, 15, 16, 17, 18, 19, 20, 22, 23', Set2 = '11,
                 13, 21'
```

**Test Case 3:**
```
choices:          2573359Cat2Ch2−|−2573359Cat4Ch2−|−2573359Cat1Ch2−|−2573359Cat3Ch2−|−2573360Cat1Ch3
Set1 = Comparing the two set sizes = Set1 smaller than Set2|Set1 = Inclusion of the two sets =
Set2 includes Set1|Set1 = Size of Set1 = Small set 1|Set1 = Intersection of the two sets = Non
empty intersection|Set2 = Size of Set2 = Large set 2
values:          Set1 = '1, 5, 8, 13', Set2 = '1, 5, 6, 7, 8, 9, 13, 15, 17, 19, 23'
```

**Test Case 4:**
```
choices:          2573359Cat2Ch2−|−2573359Cat4Ch3−|−2573359Cat1Ch2−|−2573359Cat3Ch1−|−2573360Cat1Ch3
Set1 = Comparing the two set sizes = Set1 smaller than Set2|Set1 = Inclusion of the two sets =
No set inclusion|Set1 = Size of Set1 = Small set 1|Set1 = Intersection of the two sets = Empty
intersection|Set2 = Size of Set2 = Large set 2
values:          Set1 = '3', Set2 = '0, 2, 5, 6, 7, 9, 10, 13, 14, 16, 17, 19, 21, 22'
```

**Test Case 5:**
```
choices:          2573359Cat2Ch3−|−2573359Cat4Ch3−|−2573359Cat1Ch2−|−2573359Cat3Ch2−|−2573360Cat1Ch2
Set1 = Comparing the two set sizes = Sets of equal size|Set1 = Inclusion of the two sets = No
set inclusion|Set1 = Size of Set1 = Small set 1|Set1 = Intersection of the two sets = Non empty
intersection|Set2 = Size of Set2 = Small set 2
values:          Set1 = '2', Set2 = '2'
```

**Test Case 6:**
```
choices:          2573359Cat2Ch3−|−2573359Cat4Ch3−|−2573359Cat1Ch3−|−2573359Cat3Ch1−|−2573360Cat1Ch3
Set1 = Comparing the two set sizes = Sets of equal size|Set1 = Inclusion of the two sets = No
set inclusion|Set1 = Size of Set1 = Large set 1|Set1 = Intersection of the two sets = Empty
intersection|Set2 = Size of Set2 = Large set 2
values:          Set1 = '0, 1, 9, 11, 13, 14, 17, 20, 21, 22, 23', Set2 = '2, 3, 4, 6, 7, 8, 10, 15,
                 16, 18, 19'
```

**Test Case 7:**
```
choices:          2573359Cat2Ch3−|−2573359Cat4Ch3−|−2573359Cat1Ch3−|−2573359Cat3Ch2−|−2573360Cat1Ch3
Set1 = Comparing the two set sizes = Sets of equal size|Set1 = Inclusion of the two sets = No
set inclusion|Set1 = Size of Set1 = Large set 1|Set1 = Intersection of the two sets = Non empty
intersection|Set2 = Size of Set2 = Large set 2
values:          Set1 = '5, 6, 7, 8, 10, 14, 17, 18, 20, 23, 24', Set2 = '0, 4, 6, 9, 11, 13, 18,
                 19, 22, 23, 24'
```

**Test Case 8:**
```
choices:          2573359Cat2Ch3−|−2573359Cat4Ch3−|−2573359Cat1Ch2−|−2573359Cat3Ch1−|−2573360Cat1Ch2
Set1 = Comparing the two set sizes = Sets of equal size|Set1 = Inclusion of the two sets = No
set inclusion|Set1 = Size of Set1 = Large set 1|Set1 = Intersection of the two sets = Empty
intersection|Set2 = Size of Set2 = Small set 2
values:          Set1 = '4, 5, 6, 7, 8, 13, 16, 19, 20, 24', Set2 = '1, 3, 9, 11, 12, 14, 15, 17,
                 21, 22'
```

**Test Case 9:**
```
choices:          2573359Cat2Ch1−|−2573359Cat4Ch3−|−2573359Cat1Ch3−|−2573359Cat3Ch2−|−2573360Cat1Ch2
```



```
Set1 = Comparing the two set sizes = Set1 larger than Set2 |Set1 = Inclusion of the two sets =
No set inclusion|Set1 = Size of Set1 = Large set 1|Set1 = Intersection of the two sets = Non
empty intersection|Set2 = Size of Set2 = Small set 2
values:      Set1 = '0, 1, 2, 3, 4, 6, 7, 8, 10, 11, 12, 15, 16, 17, 19, 22, 23, 24', Set2 = '7,
             8, 11, 12, 17, 24'
```

<u>**Test Case 10:**</u>
```
choices:     2573359Cat2Ch2-|-2573359Cat4Ch3-|-2573359Cat1Ch2-|-2573359Cat3Ch2-|-2573360Cat1Ch3
Set1 = Comparing the two set sizes = Set1 smaller than Set2|Set1 = Inclusion of the two sets =
No set inclusion|Set1 = Size of Set1 = Small set 1|Set1 = Intersection of the two sets = Non
empty intersection|Set2 = Size of Set2 = Large set 2
values:      Set1 = '7, 17, 18', Set2 = '0, 1, 4, 5, 8, 9, 10, 11, 13, 17, 18, 24'
```

<u>**Test Case 11:**</u>
```
choices:     2573359Cat1Ch1-|-2573359Cat2Ch2-|-2573359Cat4Ch3-|-2573360Cat1Ch3-|-2573359Cat3Ch1
Set1 = Size of Set1 = Empty set 1|Set1 = Comparing the two set sizes = Set1 smaller than
Set2|Set1 = Inclusion of the two sets = No set inclusion|Set2 = Size of Set2 = Large set 2|Set1
= Intersection of the two sets = Empty intersection
values:      Set1 = '', Set2 = '1, 2, 4, 5, 6, 8, 11, 12, 13, 14, 15, 20, 22'
```

<u>**Test Case 12:**</u>
```
choices:     2573359Cat1Ch3-|-2573359Cat2Ch1-|-2573359Cat4Ch3-|-2573360Cat1Ch1-|-2573359Cat3Ch1
Set1 = Size of Set1 = Large set 1|Set1 = Comparing the two set sizes = Set1 larger than Set2
|Set1 = Inclusion of the two sets = No set inclusion|Set2 = Size of Set2 = empty set 2|Set1 =
Intersection of the two sets = Empty intersection
values:      Set1 = '1, 2, 3, 4, 6, 7, 11, 12, 17, 18, 20, 22, 23, 24', Set2 = ''
```

## *E.9    Base Choice Test Frames and Corresponding Test Cases*

<u>**Test Case 1:**</u>
```
choices:     2573359Cat1Ch3-|-2573359Cat2Ch1-|-2573359Cat3Ch2-|-2573359Cat4Ch3-|-2573360Cat1Ch2
Set1 = Size of Set1 = Large set 1|Set1 = Comparing the two set sizes = Set1 larger than Set2
|Set1 = Intersection of the two sets = Non empty intersection|Set1 = Inclusion of the two sets =
No set inclusion|Set2 = Size of Set2 = Small set 2
values:      Set1 = '0, 1, 2, 4, 5, 8, 11, 16, 17, 18, 19, 20, 22', Set2 = '5, 8, 13, 16, 20,
             22'
```

<u>**Test Case 2:**</u>
```
choices:     2573359Cat1Ch1-|-2573359Cat2Ch2-|-2573359Cat4Ch3-|-2573360Cat1Ch3-|-2573359Cat3Ch1
Set1 = Size of Set1 = Empty set 1|Set1 = Comparing the two set sizes = Set1 smaller than
Set2|Set1 = Inclusion of the two sets = No set inclusion|Set2 = Size of Set2 = Large set 2|Set1
= Intersection of the two sets = Empty intersection
values:      Set1 = '', Set2 = '0, 1, 2, 5, 6, 7, 10, 13, 14, 18, 19, 20, 21, 24'
```

<u>**Test Case 3:**</u>
```
choices:     2573359Cat2Ch2-|-2573359Cat4Ch2-|-2573359Cat1Ch2-|-2573359Cat3Ch2-|-2573360Cat1Ch3
Set1 = Comparing the two set sizes = Set1 smaller than Set2|Set1 = Inclusion of the two sets =
Set2 includes Set1|Set1 = Size of Set1 = Small set 1|Set1 = Intersection of the two sets = Non
empty intersection|Set2 = Size of Set2 = Large set 2
values:      Set1 = '2, 3, 4, 6, 9, 15', Set2 = '0, 1, 2, 3, 4, 5, 6, 7, 9, 11, 14, 15, 16, 17,
             19, 20, 21, 23'
```

<u>**Test Case 4:**</u>
```
choices:     2573359Cat2Ch2-|-2573359Cat4Ch2-|-2573359Cat1Ch2-|-2573359Cat3Ch2-|-2573360Cat1Ch3
Set1 = Comparing the two set sizes = Set1 smaller than Set2|Set1 = Inclusion of the two sets =
Set2 includes Set1|Set1 = Size of Set1 = Small set 1|Set1 = Intersection of the two sets = Non
empty intersection|Set2 = Size of Set2 = Large set 2
values:      Set1 = '2, 3, 4, 6, 9, 15', Set2 = '0, 1, 2, 3, 4, 5, 6, 7, 9, 11, 14, 15, 16, 17,
             19, 20, 21, 23'
```

<u>**Test Case 5:**</u>
```
choices:     2573359Cat2Ch3-|-2573359Cat4Ch3-|-2573359Cat1Ch3-|-2573359Cat3Ch1-|-2573360Cat1Ch3
Set1 = Comparing the two set sizes = Sets of equal size|Set1 = Inclusion of the two sets = No
set inclusion|Set1 = Size of Set1 = Large set 1|Set1 = Intersection of the two sets = Empty
intersection|Set2 = Size of Set2 = Large set 2
values:      Set1 = '0, 3, 5, 7, 11, 16, 17, 19, 20, 23, 24', Set2 = '1, 2, 4, 8, 9, 10, 12, 13,
             14, 18, 22'
```

<u>**Test Case 6:**</u>
```
choices:     2573359Cat1Ch3-|-2573359Cat2Ch1-|-2573359Cat4Ch3-|-2573360Cat1Ch2-|-2573359Cat3Ch1
Set1 = Size of Set1 = Large set 1|Set1 = Comparing the two set sizes = Set1 larger than Set2
|Set1 = Inclusion of the two sets = No set inclusion|Set2 = Size of Set2 = Small set 2|Set1 =
Intersection of the two sets = Empty intersection
```



```
values:       Set1 = '0, 1, 2, 3, 5, 7, 10, 13, 14, 16, 19, 20, 21, 22, 24', Set2 = '9, 11, 12,
              17, 18'
```

**Test Case 7:**
```
choices:      2573359Cat1Ch3—|—2573359Cat2Ch1—|—2573359Cat3Ch2—|—2573360Cat1Ch2—|—2573359Cat4Ch1
Set1 = Size of Set1 = Large set 1|Set1 = Comparing the two set sizes = Set1 larger than Set2
|Set1 = Intersection of the two sets = Non empty intersection|Set2 = Size of Set2 = Small set
2|Set1 = Inclusion of the two sets = Set1 includes Set2
values:       Set1 = '0, 3, 5, 7, 9, 10, 11, 14, 20, 22, 23, 24', Set2 = '0, 9, 24'
```

**Test Case 8:**
```
choices:      2573359Cat2Ch2—|—2573359Cat4Ch2—|—2573359Cat1Ch2—|—2573359Cat3Ch2—|—2573360Cat1Ch3
Set1 = Comparing the two set sizes = Set1 smaller than Set2|Set1 = Inclusion of the two sets =
Set2 includes Set1|Set1 = Size of Set1 = Small set 1|Set1 = Intersection of the two sets = Non
empty intersection|Set2 = Size of Set2 = Large set 2
values:       Set1 = '2, 3, 4, 6, 9, 15', Set2 = '0, 1, 2, 3, 4, 5, 6, 7, 9, 11, 14, 15, 16, 17,
              19, 20, 21, 23'
```

**Test Case 9:**
```
choices:      2573359Cat1Ch3—|—2573359Cat2Ch1—|—2573359Cat4Ch3—|—2573360Cat1Ch1—|—2573359Cat3Ch1
Set1 = Size of Set1 = Large set 1|Set1 = Comparing the two set sizes = Set1 larger than Set2
|Set1 = Inclusion of the two sets = No set inclusion|Set2 = Size of Set2 = empty set 2|Set1 =
Intersection of the two sets = Empty intersection
values:       Set1 = '0, 2, 5, 6, 7, 9, 10, 11, 13, 14, 15, 16, 17, 18, 21, 22, 23, 24', Set2 =
              ''
```

**Test Case 10:**
```
choices:      2573359Cat2Ch2—|—2573359Cat4Ch2—|—2573359Cat1Ch2—|—2573359Cat3Ch2—|—2573360Cat1Ch3
Set1 = Comparing the two set sizes = Set1 smaller than Set2|Set1 = Inclusion of the two sets =
Set2 includes Set1|Set1 = Size of Set1 = Small set 1|Set1 = Intersection of the two sets = Non
empty intersection|Set2 = Size of Set2 = Large set 2
values:       Set1 = '2, 3, 4, 6, 9, 15', Set2 = '0, 1, 2, 3, 4, 5, 6, 7, 9, 11, 14, 15, 16, 17,
              19, 20, 21, 23'
```



## Appendix F. NEXT DATE

In the CP specification below, an asterisk ('*') beside a choice indicates the base choice for that category. Each parameter is uniquely identified by an integer identifier, which is then used in the composition of test frames; As a result, a choice for a parameter is uniquely identified by the parameter identifier, the category identifier and the choice identifier. Below we summarize the specification of the unit under test (section F.1), and provide the CP specification we used (section F.2). We then provide the test frames and corresponding test cases (inputs) we obtain when using CASA (Each-Choice in section F.3, Pairwise in section F.5, Three-way in section F.7), when using ACTS (Each-Choice in section F.4, Pairwise in section F.6, Three-way in section F.8), and when using the Base Choice criterion (section F.9). The CP specification below is adapted from earlier work [38].

### F.1  Specification

The Next Date function returns the specifics (year, month, day of the month) of the day that immediately follows the day provided as input under the form of a triplet of integer values (year, month, day). The function accounts for leap years in the Gregorian calendar, i.e., starting in year 1582. (Only for the purpose of limiting possible values for the year, we consider a year beyond 2100 an invalid input.)

### F.2  CP Specification

```
Parameter(2751581): Year of type Integer
  Category 1 - Year range
     Ch 1.1: Too early
             Year < 1582
             [error]
     Ch 1.2: Too late
             Year > 2100
             [error]
    *Ch 1.3: Normal
             Year>=1582 && Year<=2100
  Category 2 - Leap year or not
     Ch 2.1: Leap year
             (Year % 400 == 0) || ((Year % 4 == 0) && (Year % 100 != 0))
             [properties isLeap]
    *Ch 2.2: Not a leap year
             (Year % 400 != 0) && ((Year % 4 != 0) || (Year % 100 == 0))
Parameter(74527328): Month of type Integer
  Category 1 - Month values
     Ch 1.1: Illegal value
             Month<1 || Month>12
             [error]
    *Ch 1.2: 30 days months
             Month==4 ||  Month==6 || Month==9 || Month==11
             [properties is30Days]
     Ch 1.3: 31 days months except december
             Month==1 || Month==3 || Month==5 || Month==7 || Month==8 || Month==10
             [properties is31Days]
     Ch 1.4: February
             Month==2
             [properties isFebruary]
     Ch 1.5: December
             Month==12
             [properties isDecember]
Parameter(68476): Day of type Integer
  Category 1 - Day values
     Ch 1.1: Illegal value
             Day<=0 || Day>31
             [error]
```



```
    *Ch 1.2: Value for every month
            Day>=1 && Day<28
     Ch 1.3: 28th day
            Day == 28
     Ch 1.4: 29th day
            Day == 29
            [if !isFebruary || isLeap]
     Ch 1.5: 30th day
            Day == 30
            [if !isFebruary]
     Ch 1.6: 31st day
            Day == 31
            [if !isFebruary && !is30Days]
```

## *F.3    Each-Choice Test Frames and Corresponding Test Cases when using CASA*

**Test Case 1:**
choices:      2751581Cat1Ch3−|−2751581Cat2Ch2−|−74527328Cat1Ch4−|−68476Cat1Ch3
Year = Year range = Normal|Year = Leap year or not = Not a leap year|Month = Month values =
February|Day = Day values = 28th day
values:      Day = '28', Year = '1599', Month = '2'

**Test Case 2:**
choices:      2751581Cat1Ch3−|−2751581Cat2Ch2−|−74527328Cat1Ch2−|−68476Cat1Ch2
Year = Year range = Normal|Year = Leap year or not = Not a leap year|Month = Month values = 30
days months|Day = Day values = Value for every month
values:      Day = '1', Year = '1582', Month = '4'

**Test Case 3:**
choices:      2751581Cat1Ch3−|−2751581Cat2Ch2−|−74527328Cat1Ch3−|−68476Cat1Ch6
Year = Year range = Normal|Year = Leap year or not = Not a leap year|Month = Month values = 31
days months except december|Day = Day values = 31st day
values:      Day = '31', Year = '1582', Month = '1'

**Test Case 4:**
choices:      2751581Cat1Ch3−|−2751581Cat2Ch2−|−74527328Cat1Ch5−|−68476Cat1Ch5
Year = Year range = Normal|Year = Leap year or not = Not a leap year|Month = Month values =
December|Day = Day values = 30th day
values:      Day = '30', Year = '1599', Month = '12'

**Test Case 5:**
choices:      2751581Cat1Ch3−|−2751581Cat2Ch1−|−74527328Cat1Ch3−|−68476Cat1Ch4
Year = Year range = Normal|Year = Leap year or not = Leap year|Month = Month values = 31 days
months except december|Day = Day values = 29th day
values:      Day = '29', Year = '1584', Month = '1'

**Test Case 6:**
choices:      2751581Cat1Ch1
Year = Year range = Too early
values:      Year = '0'

**Test Case 7:**
choices:      2751581Cat1Ch2
Year = Year range = Too late
values:      Year = '2101'

**Test Case 8:**
choices:      74527328Cat1Ch1
Month = Month values = Illegal value
values:      Month = '13'

**Test Case 9:**
choices:      68476Cat1Ch1
Day = Day values = Illegal value
values:      Day = '32'

## *F.4    Each-Choice Test Frames and Corresponding Test Cases when using ACTS*

**Test Case 1:**
choices:      68476Cat1Ch2−|−74527328Cat1Ch3−|−2751581Cat2Ch2−|−2751581Cat1Ch3
Day = Day values = Value for every month|Month = Month values = 31 days months except
december|Year = Leap year or not = Not a leap year|Year = Year range = Normal
values:      Day = '1', Year = '1599', Month = '1'
```

**Test Case 2:**
```
choices:       68476Cat1Ch3—|—74527328Cat1Ch4—|—2751581Cat2Ch1—|—2751581Cat1Ch3
Day = Day values = 28th day|Month = Month values = February|Year = Leap year or not = Leap
year|Year = Year range = Normal
values:        Day = '28', Year = '1584', Month = '2'
```
**Test Case 3:**
```
choices:       68476Cat1Ch4—|—74527328Cat1Ch5—|—2751581Cat2Ch2—|—2751581Cat1Ch3
Day = Day values = 29th day|Month = Month values = December|Year = Leap year or not = Not a leap
year|Year = Year range = Normal
values:        Day = '29', Year = '1599', Month = '12'
```
**Test Case 4:**
```
choices:       68476Cat1Ch5—|—74527328Cat1Ch2—|—2751581Cat2Ch2—|—2751581Cat1Ch3
Day = Day values = 30th day|Month = Month values = 30 days months|Year = Leap year or not = Not
a leap year|Year = Year range = Normal
values:        Day = '30', Year = '1582', Month = '4'
```
**Test Case 5:**
```
choices:       68476Cat1Ch6—|—74527328Cat1Ch3—|—2751581Cat2Ch2—|—2751581Cat1Ch3
Day = Day values = 31st day|Month = Month values = 31 days months except december|Year = Leap
year or not = Not a leap year|Year = Year range = Normal
values:        Day = '31', Year = '1599', Month = '1'
```
**Test Case 6:**
```
choices:       2751581Cat1Ch1
Year = Year range = Too early
values:        Year = '0'
```
**Test Case 7:**
```
choices:       2751581Cat1Ch2
Year = Year range = Too late
values:        Year = '2101'
```
**Test Case 8:**
```
choices:       74527328Cat1Ch1
Month = Month values = Illegal value
values:        Month = '13'
```
**Test Case 9:**
```
choices:       68476Cat1Ch1
Day = Day values = Illegal value
values:        Day = '32'
```

## F.5    *Pairwise Test Frames and Corresponding Test Cases when using CASA*

**Test Case 1:**
```
choices:       2751581Cat1Ch3—|—2751581Cat2Ch2—|—74527328Cat1Ch2—|—68476Cat1Ch3
Year = Year range = Normal|Year = Leap year or not = Not a leap year|Month = Month values = 30
days months|Day = Day values = 28th day
values:        Day = '28', Year = '1601', Month = '4'
```
**Test Case 2:**
```
choices:       2751581Cat1Ch3—|—2751581Cat2Ch2—|—74527328Cat1Ch3—|—68476Cat1Ch4
Year = Year range = Normal|Year = Leap year or not = Not a leap year|Month = Month values = 31
days months except december|Day = Day values = 29th day
values:        Day = '29', Year = '1599', Month = '1'
```
**Test Case 3:**
```
choices:       2751581Cat1Ch3—|—2751581Cat2Ch2—|—74527328Cat1Ch5—|—68476Cat1Ch5
Year = Year range = Normal|Year = Leap year or not = Not a leap year|Month = Month values =
December|Day = Day values = 30th day
values:        Day = '30', Year = '1582', Month = '12'
```
**Test Case 4:**
```
choices:       2751581Cat1Ch3—|—2751581Cat2Ch2—|—74527328Cat1Ch4—|—68476Cat1Ch2
Year = Year range = Normal|Year = Leap year or not = Not a leap year|Month = Month values =
February|Day = Day values = Value for every month
values:        Day = '1', Year = '1599', Month = '2'
```
**Test Case 5:**
```
choices:       2751581Cat1Ch3—|—2751581Cat2Ch1—|—74527328Cat1Ch2—|—68476Cat1Ch5
Year = Year range = Normal|Year = Leap year or not = Leap year|Month = Month values = 30 days
months|Day = Day values = 30th day
values:        Day = '30', Year = '1584', Month = '4'
```

**Test Case 6:**
```
choices:      2751581Cat1Ch3-|-2751581Cat2Ch1-|-74527328Cat1Ch2-|-68476Cat1Ch2
Year = Year range = Normal|Year = Leap year or not = Leap year|Month = Month values = 30 days
months|Day = Day values = Value for every month
values:       Day = '1', Year = '1584', Month = '4'
```
**Test Case 7:**
```
choices:      2751581Cat1Ch3-|-2751581Cat2Ch1-|-74527328Cat1Ch5-|-68476Cat1Ch2
Year = Year range = Normal|Year = Leap year or not = Leap year|Month = Month values =
December|Day = Day values = Value for every month
values:       Day = '1', Year = '1584', Month = '12'
```
**Test Case 8:**
```
choices:      2751581Cat1Ch3-|-2751581Cat2Ch1-|-74527328Cat1Ch5-|-68476Cat1Ch6
Year = Year range = Normal|Year = Leap year or not = Leap year|Month = Month values =
December|Day = Day values = 31st day
values:       Day = '31', Year = '1584', Month = '12'
```
**Test Case 9:**
```
choices:      2751581Cat1Ch3-|-2751581Cat2Ch2-|-74527328Cat1Ch3-|-68476Cat1Ch6
Year = Year range = Normal|Year = Leap year or not = Not a leap year|Month = Month values = 31
days months except december|Day = Day values = 31st day
values:       Day = '31', Month = '1', Year = '1599'
```
**Test Case 10:**
```
choices:      2751581Cat1Ch3-|-2751581Cat2Ch1-|-74527328Cat1Ch4-|-68476Cat1Ch4
Year = Year range = Normal|Year = Leap year or not = Leap year|Month = Month values =
February|Day = Day values = 29th day
values:       Day = '29', Year = '1584', Month = '2'
```
**Test Case 11:**
```
choices:      2751581Cat1Ch3-|-2751581Cat2Ch2-|-74527328Cat1Ch3-|-68476Cat1Ch3
Year = Year range = Normal|Year = Leap year or not = Not a leap year|Month = Month values = 31
days months except december|Day = Day values = 28th day
values:       Day = '28', Year = '1599', Month = '1'
```
**Test Case 12:**
```
choices:      2751581Cat1Ch3-|-2751581Cat2Ch2-|-74527328Cat1Ch3-|-68476Cat1Ch2
Year = Year range = Normal|Year = Leap year or not = Not a leap year|Month = Month values = 31
days months except december|Day = Day values = Value for every month
values:       Day = '1', Year = '1599', Month = '1'
```
**Test Case 13:**
```
choices:      2751581Cat1Ch3-|-2751581Cat2Ch2-|-74527328Cat1Ch5-|-68476Cat1Ch3
Year = Year range = Normal|Year = Leap year or not = Not a leap year|Month = Month values =
December|Day = Day values = 28th day
values:       Day = '28', Year = '1582', Month = '12'
```
**Test Case 14:**
```
choices:      2751581Cat1Ch3-|-2751581Cat2Ch1-|-74527328Cat1Ch4-|-68476Cat1Ch3
Year = Year range = Normal|Year = Leap year or not = Leap year|Month = Month values =
February|Day = Day values = 28th day
values:       Day = '28', Year = '2000', Month = '2'
```
**Test Case 15:**
```
choices:      2751581Cat1Ch3-|-2751581Cat2Ch1-|-74527328Cat1Ch5-|-68476Cat1Ch4
Year = Year range = Normal|Year = Leap year or not = Leap year|Month = Month values =
December|Day = Day values = 29th day
values:       Day = '29', Year = '2000', Month = '12'
```
**Test Case 16:**
```
choices:      2751581Cat1Ch3-|-2751581Cat2Ch1-|-74527328Cat1Ch2-|-68476Cat1Ch4
Year = Year range = Normal|Year = Leap year or not = Leap year|Month = Month values = 30 days
months|Day = Day values = 29th day
values:       Day = '29', Year = '1584', Month = '4'
```
**Test Case 17:**
```
choices:      2751581Cat1Ch3-|-2751581Cat2Ch1-|-74527328Cat1Ch3-|-68476Cat1Ch5
Year = Year range = Normal|Year = Leap year or not = Leap year|Month = Month values = 31 days
months except december|Day = Day values = 30th day
values:       Day = '30', Year = '1584', Month = '1'
```
**Test Case 18:**
```
choices:      2751581Cat1Ch1
Year = Year range = Too early
values:       Year = '0'
```

```
Test Case 19:
choices:      2751581Cat1Ch2
Year = Year range = Too late
values:       Year = '2101'
Test Case 20:
choices:      74527328Cat1Ch1
Month = Month values = Illegal value
values:       Month = '13'
Test Case 21:
choices:      68476Cat1Ch1
Day = Day values = Illegal value
values:       Day = '32'
```

## F.6    Pairwise Test Frames and Corresponding Test Cases when using ACTS

```
Test Case 1:
choices:      68476Cat1Ch2−|−74527328Cat1Ch2−|−2751581Cat2Ch2−|−2751581Cat1Ch3
Day = Day values = Value for every month|Month = Month values = 30 days months|Year = Leap year
or not = Not a leap year|Year = Year range = Normal
values:       Day = '1', Year = '1582', Month = '4'
Test Case 2:
choices:      68476Cat1Ch2−|−74527328Cat1Ch3−|−2751581Cat2Ch1−|−2751581Cat1Ch3
Day = Day values = Value for every month|Month = Month values = 31 days months except
december|Year = Leap year or not = Leap year|Year = Year range = Normal
values:       Day = '1', Year = '1584', Month = '1'
Test Case 3:
choices:      68476Cat1Ch2−|−74527328Cat1Ch4−|−2751581Cat2Ch2−|−2751581Cat1Ch3
Day = Day values = Value for every month|Month = Month values = February|Year = Leap year or not
= Not a leap year|Year = Year range = Normal
values:       Day = '1', Year = '1599', Month = '2'
Test Case 4:
choices:      68476Cat1Ch2−|−74527328Cat1Ch5−|−2751581Cat2Ch1−|−2751581Cat1Ch3
Day = Day values = Value for every month|Month = Month values = December|Year = Leap year or not
= Leap year|Year = Year range = Normal
values:       Day = '1', Year = '1584', Month = '12'
Test Case 5:
choices:      68476Cat1Ch3−|−74527328Cat1Ch2−|−2751581Cat2Ch1−|−2751581Cat1Ch3
Day = Day values = 28th day|Month = Month values = 30 days months|Year = Leap year or not = Leap
year|Year = Year range = Normal
values:       Day = '28', Year = '1584', Month = '4'
Test Case 6:
choices:      68476Cat1Ch3−|−74527328Cat1Ch3−|−2751581Cat2Ch2−|−2751581Cat1Ch3
Day = Day values = 28th day|Month = Month values = 31 days months except december|Year = Leap
year or not = Not a leap year|Year = Year range = Normal
values:       Day = '28', Year = '1599', Month = '1'
Test Case 7:
choices:      68476Cat1Ch3−|−74527328Cat1Ch4−|−2751581Cat2Ch1−|−2751581Cat1Ch3
Day = Day values = 28th day|Month = Month values = February|Year = Leap year or not = Leap
year|Year = Year range = Normal
values:       Day = '28', Year = '1584', Month = '2'
Test Case 8:
choices:      68476Cat1Ch3−|−74527328Cat1Ch5−|−2751581Cat2Ch2−|−2751581Cat1Ch3
Day = Day values = 28th day|Month = Month values = December|Year = Leap year or not = Not a leap
year|Year = Year range = Normal
values:       Day = '28', Year = '1582', Month = '12'
Test Case 9:
choices:      68476Cat1Ch4−|−74527328Cat1Ch2−|−2751581Cat2Ch1−|−2751581Cat1Ch3
Day = Day values = 29th day|Month = Month values = 30 days months|Year = Leap year or not = Leap
year|Year = Year range = Normal
values:       Day = '29', Year = '1584', Month = '4'
Test Case 10:
choices:      68476Cat1Ch4−|−74527328Cat1Ch3−|−2751581Cat2Ch2−|−2751581Cat1Ch3
Day = Day values = 29th day|Month = Month values = 31 days months except december|Year = Leap
year or not = Not a leap year|Year = Year range = Normal
values:       Day = '29', Year = '1582', Month = '1'
```

**Test Case 11:**
```
choices:      68476Cat1Ch4—|—74527328Cat1Ch4—|—2751581Cat2Ch1—|—2751581Cat1Ch3
Day = Day values = 29th day|Month = Month values = February|Year = Leap year or not = Leap
year|Year = Year range = Normal
values:       Day = '29', Year = '2000', Month = '2'
```
**Test Case 12:**
```
choices:      68476Cat1Ch4—|—74527328Cat1Ch5—|—2751581Cat2Ch1—|—2751581Cat1Ch3
Day = Day values = 29th day|Month = Month values = December|Year = Leap year or not = Leap
year|Year = Year range = Normal
values:       Day = '29', Year = '2000', Month = '12'
```
**Test Case 13:**
```
choices:      68476Cat1Ch5—|—74527328Cat1Ch2—|—2751581Cat2Ch1—|—2751581Cat1Ch3
Day = Day values = 30th day|Month = Month values = 30 days months|Year = Leap year or not = Leap
year|Year = Year range = Normal
values:       Day = '30', Year = '1584', Month = '4'
```
**Test Case 14:**
```
choices:      68476Cat1Ch5—|—74527328Cat1Ch3—|—2751581Cat2Ch2—|—2751581Cat1Ch3
Day = Day values = 30th day|Month = Month values = 31 days months except december|Year = Leap
year or not = Not a leap year|Year = Year range = Normal
values:       Day = '30', Year = '1599', Month = '1'
```
**Test Case 15:**
```
choices:      68476Cat1Ch5—|—74527328Cat1Ch5—|—2751581Cat2Ch1—|—2751581Cat1Ch3
Day = Day values = 30th day|Month = Month values = December|Year = Leap year or not = Leap
year|Year = Year range = Normal
values:       Day = '30', Year = '1584', Month = '12'
```
**Test Case 16:**
```
choices:      68476Cat1Ch6—|—74527328Cat1Ch3—|—2751581Cat2Ch1—|—2751581Cat1Ch3
Day = Day values = 31st day|Month = Month values = 31 days months except december|Year = Leap
year or not = Leap year|Year = Year range = Normal
values:       Day = '31', Year = '1584', Month = '1'
```
**Test Case 17:**
```
choices:      68476Cat1Ch6—|—74527328Cat1Ch5—|—2751581Cat2Ch2—|—2751581Cat1Ch3
Day = Day values = 31st day|Month = Month values = December|Year = Leap year or not = Not a leap
year|Year = Year range = Normal
values:       Day = '31', Year = '1599', Month = '12'
```
**Test Case 18:**
```
choices:      2751581Cat1Ch1
Year = Year range = Too early
values:       Year = '0'
```
**Test Case 19:**
```
choices:      2751581Cat1Ch2
Year = Year range = Too late
values:       Year = '2101'
```
**Test Case 20:**
```
choices:      74527328Cat1Ch1
Month = Month values = Illegal value
values:       Month = '13'
```
**Test Case 21:**
```
choices:      68476Cat1Ch1
Day = Day values = Illegal value
values:       Day = '32'
```

## *F.7    Three-way Test Frames and Corresponding Test Cases when using CASA*

**Test Case 1:**
```
choices:      2751581Cat1Ch3—|—2751581Cat2Ch1—|—74527328Cat1Ch4—|—68476Cat1Ch4
Year = Year range = Normal|Year = Leap year or not = Leap year|Month = Month values =
February|Day = Day values = 29th day
values:       Day = '29', Year = '1584', Month = '2'
```
**Test Case 2:**
```
choices:      2751581Cat1Ch3—|—2751581Cat2Ch2—|—74527328Cat1Ch2—|—68476Cat1Ch4
Year = Year range = Normal|Year = Leap year or not = Not a leap year|Month = Month values = 30
days months|Day = Day values = 29th day
values:       Day = '29', Year = '1582', Month = '4'
```

**Test Case 3:**
choices:      2751581Cat1Ch3−|−2751581Cat2Ch1−|−74527328Cat1Ch2−|−68476Cat1Ch5
Year = Year range = Normal|Year = Leap year or not = Leap year|Month = Month values = 30 days months|Day = Day values = 30th day
values:       Day = '30', Year = '1584', Month = '4'

**Test Case 4:**
choices:      2751581Cat1Ch3−|−2751581Cat2Ch1−|−74527328Cat1Ch4−|−68476Cat1Ch3
Year = Year range = Normal|Year = Leap year or not = Leap year|Month = Month values = February|Day = Day values = 28th day
values:       Day = '28', Year = '1584', Month = '2'

**Test Case 5:**
choices:      2751581Cat1Ch3−|−2751581Cat2Ch2−|−74527328Cat1Ch5−|−68476Cat1Ch2
Year = Year range = Normal|Year = Leap year or not = Not a leap year|Month = Month values = December|Day = Day values = Value for every month
values:       Day = '1', Year = '1599', Month = '12'

**Test Case 6:**
choices:      2751581Cat1Ch3−|−2751581Cat2Ch2−|−74527328Cat1Ch3−|−68476Cat1Ch3
Year = Year range = Normal|Year = Leap year or not = Not a leap year|Month = Month values = 31 days months except december|Day = Day values = 28th day
values:       Day = '28', Year = '1599', Month = '1'

**Test Case 7:**
choices:      2751581Cat1Ch3−|−2751581Cat2Ch2−|−74527328Cat1Ch3−|−68476Cat1Ch4
Year = Year range = Normal|Year = Leap year or not = Not a leap year|Month = Month values = 31 days months except december|Day = Day values = 29th day
values:       Day = '29', Year = '1582', Month = '1'

**Test Case 8:**
choices:      2751581Cat1Ch3−|−2751581Cat2Ch2−|−74527328Cat1Ch4−|−68476Cat1Ch2
Year = Year range = Normal|Year = Leap year or not = Not a leap year|Month = Month values = February|Day = Day values = Value for every month
values:       Day = '1', Year = '1599', Month = '2'

**Test Case 9:**
choices:      2751581Cat1Ch3−|−2751581Cat2Ch1−|−74527328Cat1Ch5−|−68476Cat1Ch4
Year = Year range = Normal|Year = Leap year or not = Leap year|Month = Month values = December|Day = Day values 29th day
values:       Day = '29', Month = '12', Year = '2000'

**Test Case 10:**
choices:      2751581Cat1Ch3−|−2751581Cat2Ch2−|−74527328Cat1Ch2−|−68476Cat1Ch3
Year = Year range = Normal|Year = Leap year or not = Not a leap year|Month = Month values = 30 days months|Day = Day values = 28th day
values:       Day = '28', Year = '1599', Month = '4'

**Test Case 11:**
choices:      2751581Cat1Ch3−|−2751581Cat2Ch1−|−74527328Cat1Ch2−|−68476Cat1Ch2
Year = Year range = Normal|Year = Leap year or not = Leap year|Month = Month values = 30 days months|Day = Day values = Value for every month
values:       Day = '1', Year = '1584', Month = '4'

**Test Case 12:**
choices:      2751581Cat1Ch3−|−2751581Cat2Ch1−|−74527328Cat1Ch2−|−68476Cat1Ch4
Year = Year range = Normal|Year = Leap year or not = Leap year|Month = Month values = 30 days months|Day = Day values = 29th day
values:       Day = '29', Year = '1584', Month = '4'

**Test Case 13:**
choices:      2751581Cat1Ch3−|−2751581Cat2Ch2−|−74527328Cat1Ch3−|−68476Cat1Ch2
Year = Year range = Normal|Year = Leap year or not = Not a leap year|Month = Month values = 31 days months except december|Day = Day values = Value for every month
values:       Day = '1', Year = '1599', Month = '1'

**Test Case 14:**
choices:      2751581Cat1Ch3−|−2751581Cat2Ch2−|−74527328Cat1Ch3−|−68476Cat1Ch6
Year = Year range = Normal|Year = Leap year or not = Not a leap year|Month = Month values = 31 days months except december|Day = Day values = 31st day
values:       Day = '31', Year = '1582', Month = '1'

**Test Case 15:**
choices:      2751581Cat1Ch3−|−2751581Cat2Ch1−|−74527328Cat1Ch3−|−68476Cat1Ch6
Year = Year range = Normal|Year = Leap year or not = Leap year|Month = Month values = 31 days months except december|Day = Day values = 31st day
values:       Day = '31', Year = '1584', Month = '1'

**Test Case 16:**
choices:     2751581Cat1Ch3−|−2751581Cat2Ch2−|−74527328Cat1Ch5−|−68476Cat1Ch4
Year = Year range = Normal|Year = Leap year or not = Not a leap year|Month = Month values =
December|Day = Day values = 29th day
values:     Day = '29', Year = '1599', Month = '12'
**Test Case 17:**
choices:     2751581Cat1Ch3−|−2751581Cat2Ch2−|−74527328Cat1Ch5−|−68476Cat1Ch5
Year = Year range = Normal|Year = Leap year or not = Not a leap year|Month = Month values =
December|Day = Day values = 30th day
values:     Day = '30', Year = '1599', Month = '12'
**Test Case 18:**
choices:     2751581Cat1Ch3−|−2751581Cat2Ch1−|−74527328Cat1Ch5−|−68476Cat1Ch3
Year = Year range = Normal|Year = Leap year or not = Leap year|Month = Month values =
December|Day = Day values = 28th day
values:     Day = '28', Year = '1584', Month = '12'
**Test Case 19:**
choices:     2751581Cat1Ch3−|−2751581Cat2Ch1−|−74527328Cat1Ch3−|−68476Cat1Ch5
Year = Year range = Normal|Year = Leap year or not = Leap year|Month = Month values = 31 days
months except december|Day = Day values = 30th day
values:     Day = '30', Year = '1584', Month = '1'
**Test Case 20:**
choices:     2751581Cat1Ch3−|−2751581Cat2Ch1−|−74527328Cat1Ch2−|−68476Cat1Ch3
Year = Year range = Normal|Year = Leap year or not = Leap year|Month = Month values = 30 days
months|Day = Day values = 28th day
values:     Day = '28', Year = '1584', Month = '4'
**Test Case 21:**
choices:     2751581Cat1Ch3−|−2751581Cat2Ch2−|−74527328Cat1Ch2−|−68476Cat1Ch5
Year = Year range = Normal|Year = Leap year or not = Not a leap year|Month = Month values = 30
days months|Day = Day values = 30th day
values:     Day = '30', Year = '1601', Month = '4'
**Test Case 22:**
choices:     2751581Cat1Ch3−|−2751581Cat2Ch1−|−74527328Cat1Ch5−|−68476Cat1Ch6
Year = Year range = Normal|Year = Leap year or not = Leap year|Month = Month values =
December|Day = Day values = 31st day
values:     Day = '31', Year = '2000', Month = '12'
**Test Case 23:**
choices:     2751581Cat1Ch3−|−2751581Cat2Ch1−|−74527328Cat1Ch3−|−68476Cat1Ch4
Year = Year range = Normal|Year = Leap year or not = Leap year|Month = Month values = 31 days
months except december|Day = Day values = 29th day
values:     Day = '29', Year = '1584', Month = '1'
**Test Case 24:**
choices:     2751581Cat1Ch3−|−2751581Cat2Ch2−|−74527328Cat1Ch4−|−68476Cat1Ch3
Year = Year range = Normal|Year = Leap year or not = Not a leap year|Month = Month values =
February|Day = Day values = 28th day
values:     Day = '28', Year = '1582', Month = '2'
**Test Case 25:**
choices:     2751581Cat1Ch3−|−2751581Cat2Ch2−|−74527328Cat1Ch5−|−68476Cat1Ch6
Year = Year range = Normal|Year = Leap year or not = Not a leap year|Month = Month values =
December|Day = Day values = 31st day
values:     Day = '31', Year = '1582', Month = '12'
**Test Case 26:**
choices:     2751581Cat1Ch3−|−2751581Cat2Ch2−|−74527328Cat1Ch3−|−68476Cat1Ch5
Year = Year range = Normal|Year = Leap year or not = Not a leap year|Month = Month values = 31
days months except december|Day = Day values = 30th day
values:     Day = '30', Year = '1599', Month = '1'
**Test Case 27:**
choices:     2751581Cat1Ch3−|−2751581Cat2Ch1−|−74527328Cat1Ch4−|−68476Cat1Ch2
Year = Year range = Normal|Year = Leap year or not = Leap year|Month = Month values =
February|Day = Day values = Value for every month
values:     Day = '1', Year = '1584', Month = '2'
**Test Case 28:**
choices:     2751581Cat1Ch3−|−2751581Cat2Ch1−|−74527328Cat1Ch5−|−68476Cat1Ch2
Year = Year range = Normal|Year = Leap year or not = Leap year|Month = Month values =
December|Day = Day values = Value for every month
values:     Day = '1', Year = '1584', Month = '12'

**Test Case 29:**
choices:     2751581Cat1Ch3—|—2751581Cat2Ch2—|—74527328Cat1Ch2—|—68476Cat1Ch2
Year = Year range = Normal|Year = Leap year or not = Not a leap year|Month = Month values = 30
days months|Day = Day values = Value for every month
values:     Day = '1', Year = '1599', Month = '4'

**Test Case 30:**
choices:     2751581Cat1Ch3—|—2751581Cat2Ch1—|—74527328Cat1Ch5—|—68476Cat1Ch5
Year = Year range = Normal|Year = Leap year or not = Leap year|Month = Month values =
December|Day = Day values = 30th day
values:     Day = '30', Year = '1584', Month = '12'

**Test Case 31:**
choices:     2751581Cat1Ch3—|—2751581Cat2Ch1—|—74527328Cat1Ch3—|—68476Cat1Ch2
Year = Year range = Normal|Year = Leap year or not = Leap year|Month = Month values = 31 days
months except december|Day = Day values = Value for every month
values:     Day = '1', Month = '1', Year = '1584'

**Test Case 32:**
choices:     2751581Cat1Ch3—|—2751581Cat2Ch1—|—74527328Cat1Ch3—|—68476Cat1Ch3
Year = Year range = Normal|Year = Leap year or not = Leap year|Month = Month values = 31 days
months except december|Day = Day values = 28th day
values:     Day = '28', Year = '1584', Month = '1'

**Test Case 33:**
choices:     2751581Cat1Ch3—|—2751581Cat2Ch2—|—74527328Cat1Ch5—|—68476Cat1Ch3
Year = Year range = Normal|Year = Leap year or not = Not a leap year|Month = Month values =
December|Day = Day values = 28th day
values:     Day = '28', Year = '1599', Month = '12'

**Test Case 34:**
choices:     2751581Cat1Ch1
Year = Year range = Too early
values:     Year = '0'

**Test Case 35:**
choices:     2751581Cat1Ch2
Year = Year range = Too late
values:     Year = '2101'

**Test Case 36:**
choices:     74527328Cat1Ch1
Month = Month values = Illegal value
values:     Month = '13'

**Test Case 37:**
choices:     68476Cat1Ch1
Day = Day values = Illegal value
values:     Day = '32'

## *F.8    Three-way Test Frames and Corresponding Test Cases when using ACTS*

**Test Case 1:**
choices:     68476Cat1Ch2—|—74527328Cat1Ch2—|—2751581Cat2Ch1—|—2751581Cat1Ch3
Day = Day values = Value for every month|Month = Month values = 30 days months|Year = Leap year
or not = Leap year|Year = Year range = Normal
values:     Day = '1', Year = '1584', Month = '4'

**Test Case 2:**
choices:     68476Cat1Ch2—|—74527328Cat1Ch2—|—2751581Cat2Ch2—|—2751581Cat1Ch3
Day = Day values = Value for every month|Month = Month values = 30 days months|Year = Leap year
or not = Not a leap year|Year = Year range = Normal
values:     Day = '1', Year = '1582', Month = '4'

**Test Case 3:**
choices:     68476Cat1Ch2—|—74527328Cat1Ch3—|—2751581Cat2Ch1—|—2751581Cat1Ch3
Day = Day values = Value for every month|Month = Month values = 31 days months except
december|Year = Leap year or not = Leap year|Year = Year range = Normal
values:     Day = '1', Year = '1584', Month = '1'

**Test Case 4:**
choices:     68476Cat1Ch2—|—74527328Cat1Ch3—|—2751581Cat2Ch2—|—2751581Cat1Ch3
Day = Day values = Value for every month|Month = Month values = 31 days months except
december|Year = Leap year or not = Not a leap year|Year = Year range = Normal
values:     Day = '1', Year = '1599', Month = '1'

**Test Case 5:**
choices:     68476Cat1Ch2—|—74527328Cat1Ch4—|—2751581Cat2Ch1—|—2751581Cat1Ch3
Day = Day values = Value for every month|Month = Month values = February|Year = Leap year or not
= Leap year|Year = Year range = Normal
values:     Day = '1', Year = '1584', Month = '2'

**Test Case 6:**
choices:     68476Cat1Ch2—|—74527328Cat1Ch4—|—2751581Cat2Ch2—|—2751581Cat1Ch3
Day = Day values = Value for every month|Month = Month values = February|Year = Leap year or not
= Not a leap year|Year = Year range = Normal
values:     Day = '1', Year = '1582', Month = '2'

**Test Case 7:**
choices:     68476Cat1Ch2—|—74527328Cat1Ch5—|—2751581Cat2Ch1—|—2751581Cat1Ch3
Day = Day values = Value for every month|Month = Month values = December|Year = Leap year or not
= Leap year|Year = Year range = Normal
values:     Day = '1', Year = '1584', Month = '12'

**Test Case 8:**
choices:     68476Cat1Ch2—|—74527328Cat1Ch5—|—2751581Cat2Ch2—|—2751581Cat1Ch3
Day = Day values = Value for every month|Month = Month values = December|Year = Leap year or not
= Not a leap year|Year = Year range = Normal
values:     Day = '1', Year = '1582', Month = '12'

**Test Case 9:**
choices:     68476Cat1Ch3—|—74527328Cat1Ch2—|—2751581Cat2Ch1—|—2751581Cat1Ch3
Day = Day values = 28th day|Month = Month values = 30 days months|Year = Leap year or not = Leap
year|Year = Year range = Normal
values:     Day = '28', Year = '1584', Month = '4'

**Test Case 10:**
choices:     68476Cat1Ch3—|—74527328Cat1Ch2—|—2751581Cat2Ch2—|—2751581Cat1Ch3
Day = Day values = 28th day|Month = Month values = 30 days months|Year = Leap year or not = Not
a leap year|Year = Year range = Normal
values:     Day = '28', Year = '1582', Month = '4'

**Test Case 11:**
choices:     68476Cat1Ch3—|—74527328Cat1Ch3—|—2751581Cat2Ch1—|—2751581Cat1Ch3
Day = Day values = 28th day|Month = Month values = 31 days months except december|Year = Leap
year or not = Leap year|Year = Year range = Normal
values:     Day = '28', Year = '1584', Month = '1'

**Test Case 12:**
choices:     68476Cat1Ch3—|—74527328Cat1Ch3—|—2751581Cat2Ch2—|—2751581Cat1Ch3
Day = Day values = 28th day|Month = Month values = 31 days months except december|Year = Leap
year or not = Not a leap year|Year = Year range = Normal
values:     Day = '28', Year = '1599', Month = '1'

**Test Case 13:**
choices:     68476Cat1Ch3—|—74527328Cat1Ch4—|—2751581Cat2Ch1—|—2751581Cat1Ch3
Day = Day values = 28th day|Month = Month values = February|Year = Leap year or not = Leap
year|Year = Year range = Normal
values:     Day = '28', Year = '1584', Month = '2'

**Test Case 14:**
choices:     68476Cat1Ch3—|—74527328Cat1Ch4—|—2751581Cat2Ch2—|—2751581Cat1Ch3
Day = Day values = 28th day|Month = Month values = February|Year = Leap year or not = Not a leap
year|Year = Year range = Normal
values:     Day = '28', Year = '1599', Month = '2'

**Test Case 15:**
choices:     68476Cat1Ch3—|—74527328Cat1Ch5—|—2751581Cat2Ch1—|—2751581Cat1Ch3
Day = Day values = 28th day|Month = Month values = December|Year = Leap year or not = Leap
year|Year = Year range = Normal
values:     Day = '28', Year = '1584', Month = '12'

**Test Case 16:**
choices:     68476Cat1Ch3—|—74527328Cat1Ch5—|—2751581Cat2Ch2—|—2751581Cat1Ch3
Day = Day values = 28th day|Month = Month values = December|Year = Leap year or not = Not a leap
year|Year = Year range = Normal
values:     Day = '28', Year = '1582', Month = '12'

**Test Case 17:**
choices:     68476Cat1Ch4—|—74527328Cat1Ch2—|—2751581Cat2Ch1—|—2751581Cat1Ch3
Day = Day values = 29th day|Month = Month values = 30 days months|Year = Leap year or not = Leap
year|Year = Year range = Normal
values:     Day = '29', Year = '1584', Month = '4'

**Test Case 18:**
choices:        68476Cat1Ch4-|-74527328Cat1Ch2-|-2751581Cat2Ch2-|-2751581Cat1Ch3
Day = Day values = 29th day|Month = Month values = 30 days months|Year = Leap year or not = Not a leap year|Year = Year range = Normal
values:         Day = '29', Year = '1582', Month = '4'

**Test Case 19:**
choices:        68476Cat1Ch4-|-74527328Cat1Ch3-|-2751581Cat2Ch1-|-2751581Cat1Ch3
Day = Day values = 29th day|Month = Month values = 31 days months except december|Year = Leap year or not = Leap year|Year = Year range = Normal
values:         Day = '29', Year = '1584', Month = '1'

**Test Case 20:**
choices:        68476Cat1Ch4-|-74527328Cat1Ch3-|-2751581Cat2Ch2-|-2751581Cat1Ch3
Day = Day values = 29th day|Month = Month values = 31 days months except december|Year = Leap year or not = Not a leap year|Year = Year range = Normal
values:         Day = '29', Year = '1599', Month = '1'

**Test Case 21:**
choices:        68476Cat1Ch4-|-74527328Cat1Ch4-|-2751581Cat2Ch1-|-2751581Cat1Ch3
Day = Day values = 29th day|Month = Month values = February|Year = Leap year or not = Leap year|Year = Year range = Normal
values:         Day = '29', Year = '1584', Month = '2'

**Test Case 22:**
choices:        68476Cat1Ch4-|-74527328Cat1Ch5-|-2751581Cat2Ch1-|-2751581Cat1Ch3
Day = Day values = 29th day|Month = Month values = December|Year = Leap year or not = Leap year|Year = Year range = Normal
values:         Day = '29', Year = '1584', Month = '12'

**Test Case 23:**
choices:        68476Cat1Ch4-|-74527328Cat1Ch5-|-2751581Cat2Ch2-|-2751581Cat1Ch3
Day = Day values = 29th day|Month = Month values = December|Year = Leap year or not = Not a leap year|Year = Year range = Normal
values:         Day = '29', Year = '1599', Month = '12'

**Test Case 24:**
choices:        68476Cat1Ch5-|-74527328Cat1Ch2-|-2751581Cat2Ch1-|-2751581Cat1Ch3
Day = Day values = 30th day|Month = Month values = 30 days months|Year = Leap year or not = Leap year|Year = Year range = Normal
values:         Day = '30', Year = '1584', Month = '4'

**Test Case 25:**
choices:        68476Cat1Ch5-|-74527328Cat1Ch2-|-2751581Cat2Ch2-|-2751581Cat1Ch3
Day = Day values = 30th day|Month = Month values = 30 days months|Year = Leap year or not = Not a leap year|Year = Year range = Normal
values:         Day = '30', Year = '1599', Month = '4'

**Test Case 26:**
choices:        68476Cat1Ch5-|-74527328Cat1Ch3-|-2751581Cat2Ch1-|-2751581Cat1Ch3
Day = Day values = 30th day|Month = Month values = 31 days months except december|Year = Leap year or not = Leap year|Year = Year range = Normal
values:         Day = '30', Year = '1584', Month = '1'

**Test Case 27:**
choices:        68476Cat1Ch5-|-74527328Cat1Ch3-|-2751581Cat2Ch2-|-2751581Cat1Ch3
Day = Day values = 30th day|Month = Month values = 31 days months except december|Year = Leap year or not = Not a leap year|Year = Year range = Normal
values:         Day = '30', Year = '1582', Month = '1'

**Test Case 28:**
choices:        68476Cat1Ch5-|-74527328Cat1Ch5-|-2751581Cat2Ch1-|-2751581Cat1Ch3
Day = Day values = 30th day|Month = Month values = December|Year = Leap year or not = Leap year|Year = Year range = Normal
values:         Day = '30', Year = '2000', Month = '12'

**Test Case 29:**
choices:        68476Cat1Ch5-|-74527328Cat1Ch5-|-2751581Cat2Ch2-|-2751581Cat1Ch3
Day = Day values = 30th day|Month = Month values = December|Year = Leap year or not = Not a leap year|Year = Year range = Normal
values:         Day = '30', Year = '1582', Month = '12'

**Test Case 30:**
choices:        68476Cat1Ch6-|-74527328Cat1Ch3-|-2751581Cat2Ch1-|-2751581Cat1Ch3
Day = Day values = 31st day|Month = Month values = 31 days months except december|Year = Leap year or not = Leap year|Year = Year range = Normal
values:         Day = '31', Year = '1584', Month = '1'

**Test Case 31:**
```
choices:      68476Cat1Ch6−|−74527328Cat1Ch3−|−2751581Cat2Ch2−|−2751581Cat1Ch3
Day = Day values = 31st day|Month = Month values = 31 days months except december|Year = Leap
year or not = Not a leap year|Year = Year range = Normal
values:      Day = '31', Year = '1599', Month = '1'
```
**Test Case 32:**
```
choices:      68476Cat1Ch6−|−74527328Cat1Ch5−|−2751581Cat2Ch1−|−2751581Cat1Ch3
Day = Day values = 31st day|Month = Month values = December|Year = Leap year or not = Leap
year|Year = Year range = Normal
values:      Day = '31', Year = '1584', Month = '12'
```
**Test Case 33:**
```
choices:      68476Cat1Ch6−|−74527328Cat1Ch5−|−2751581Cat2Ch2−|−2751581Cat1Ch3
Day = Day values = 31st day|Month = Month values = December|Year = Leap year or not = Not a leap
year|Year = Year range = Normal
values:      Day = '31', Year = '1601', Month = '12'
```
**Test Case 34:**
```
choices:      2751581Cat1Ch1
Year = Year range = Too early
values:      Year = '0'
```
**Test Case 35:**
```
choices:      2751581Cat1Ch2
Year = Year range = Too late
values:      Year = '2101'
```
**Test Case 36:**
```
choices:      74527328Cat1Ch1
Month = Month values = Illegal value
values:      Month = '13'
```
**Test Case 37:**
```
choices:      68476Cat1Ch1
Day = Day values = Illegal value
values:      Day = '32'
```

## *F.9*    *Base Choice Test Frames and Corresponding Test Cases*

**Test Case 1:**
```
choices:      2751581Cat1Ch3−|−2751581Cat2Ch2−|−74527328Cat1Ch2−|−68476Cat1Ch2
Year = Year range = Normal|Year = Leap year or not = Not a leap year|Month = Month values = 30
days months|Day = Day values = Value for every month
values:      Day = '1', Year = '1601', Month = '4'
```
**Test Case 2:**
```
choices:      2751581Cat1Ch1
Year = Year range = Too early
values:      Year = '0'
```
**Test Case 3:**
```
choices:      2751581Cat1Ch2
Year = Year range = Too late
values:      Year = '2101'
```
**Test Case 4:**
```
choices:      2751581Cat1Ch3−|−74527328Cat1Ch2−|−68476Cat1Ch2−|−2751581Cat2Ch1
Year = Year range = Normal|Month = Month values = 30 days months|Day = Day values = Value for
every month|Year = Leap year or not = Leap year
values:      Day = '1', Year = '1584', Month = '4'
```
**Test Case 5:**
```
choices:      74527328Cat1Ch1
Month = Month values = Illegal value
values:      Month = '13'
```
**Test Case 6:**
```
choices:      2751581Cat1Ch3−|−2751581Cat2Ch2−|−68476Cat1Ch2−|−74527328Cat1Ch3
Year = Year range = Normal|Year = Leap year or not = Not a leap year|Day = Day values = Value
for every month|Month = Month values = 31 days months except december
values:      Day = '1', Year = '1599', Month = '1'
```
**Test Case 7:**
```
choices:      2751581Cat1Ch3−|−2751581Cat2Ch2−|−68476Cat1Ch2−|−74527328Cat1Ch4
Year = Year range = Normal|Year = Leap year or not = Not a leap year|Day = Day values = Value
for every month|Month = Month values = February
```

```
values:     Day = '1', Year = '1599', Month = '2'
```

**Test Case 8:**
```
choices:    2751581Cat1Ch3—|—2751581Cat2Ch2—|—68476Cat1Ch2—|—74527328Cat1Ch5
Year = Year range = Normal|Year = Leap year or not = Not a leap year|Day = Day values = Value
for every month|Month = Month values = December
values:     Day = '1', Year = '1582', Month = '12'
```

**Test Case 9:**
```
choices:    68476Cat1Ch1
Day = Day values = Illegal value
values:     Day = '32'
```

**Test Case 10:**
```
choices:    2751581Cat1Ch3—|—2751581Cat2Ch2—|—74527328Cat1Ch2—|—68476Cat1Ch3
Year = Year range = Normal|Year = Leap year or not = Not a leap year|Month = Month values = 30
days months|Day = Day values = 28th day
values:     Day = '28', Year = '1601', Month = '4'
```

**Test Case 11:**
```
choices:    2751581Cat1Ch3—|—2751581Cat2Ch2—|—74527328Cat1Ch2—|—68476Cat1Ch4
Year = Year range = Normal|Year = Leap year or not = Not a leap year|Month = Month values = 30
days months|Day = Day values = 29th day
values:     Day = '29', Year = '1601', Month = '4'
```

**Test Case 12:**
```
choices:    2751581Cat1Ch3—|—2751581Cat2Ch2—|—74527328Cat1Ch2—|—68476Cat1Ch5
Year = Year range = Normal|Year = Leap year or not = Not a leap year|Month = Month values = 30
days months|Day = Day values = 30th day
values:     Day = '30', Year = '1601', Month = '4'
```

**Test Case 13:**
```
choices:    68476Cat1Ch6—|—74527328Cat1Ch3—|—2751581Cat2Ch2—|—2751581Cat1Ch3
Day = Day values = 31st day|Month = Month values = 31 days months except december|Year = Leap
year or not = Not a leap year|Year = Year range = Normal
values:     Day = '31', Year = '1599', Month = '1'
```

# Appendix G. DAYLIGHT SAVING

In the CP specification below, an asterisk ('*') beside a choice indicates the base choice for that category. Each parameter is uniquely identified by an integer identifier, which is then used in the composition of test frames; As a result, a choice for a parameter is uniquely identified by the parameter identifier, the category identifier and the choice identifier. Below we summarize the specification of the unit under test (section G.1), and provide the CP specification we used (section G.2). We then provide the test frames and corresponding test cases (inputs) we obtain when using CASA (Each-Choice in section B.3, Pairwise in section G.5, Three-way in section G.7), when using ACTS (Each-Choice in section G.4, Pairwise in section G.6, Three-way in section G.8), and when using the Base Choice criterion (section G.9). The CP specification below is adapted from earlier work [16].

## G.1 Specification

The function under test takes parameters to determine the cost of a phone call, given a start time and either a duration or an end time. The function applies a simple fee model: a call that lasts less than 20mn is billed at a rate that is different than that used if a call lasts more than 20mn; a call cannot last more than two hours (a longer than two hours call ius simply dropped). The function also handled daylight saving periods, in the Fall and Spring, that is the practice of advancing clocks in the Spring by one hour to make better use of the longer daylight available during summer so that darkness falls at a later clock time, and also setting clocks back by one hour in the Fall for a similar reason. For the purpose of this study, we fixed the daylight saving dates to March $28^{th}$ at 2 o-clock in the morning (at 2 o-clock on March $28^{th}$, clocks are changed to 3 o-clock) and October $30^{th}$ at 3 o-clock in the morning (at 3 o-clock on October $30^{th}$, clocks are changed to 2 o-clock).

This subject allows us to illustrate that CP can be used with a different set of parameters than what the function under test expects. This can sometime facilitate the creation of the CP specification. For this specific function, if we use the parameters expected by the function we need to reason about months, which have varying numbers of days, to trigger changes of day (i.e., a phone call that starts in a day and ends in another), an involve the computation of the duration of a call around daylight saving changes in the CP specification with values of month, day and time. Instead, we opted for a CP specification that counts a year in seconds. Given an amount of seconds, one can easily compute the time of a day in a month; for instance, if a test case is expected to use an input value of 5,680,352 seconds, a test driver can easily determine that this refers to a time in the $65^{th}$ day of the year, which is March 6 on a non-leap year, and more precisely 17h, 52mn, 32s on March $6^{th}$. (A day has 86,400 seconds, an hour has 3,600 seconds and a minute has 60 seconds.) On a non-leap year, the change dates for daylight saving are the $87^{th}$ (March $28^{th}$) and the $303^{rd}$ (October $30^{th}$) days of the year.

## G.2 CP Specification

```
Parameter(1273475408): DayOfYear of type Integer
  Category 1 - The day number in the year
    Ch 1.1: Illegal value for day
            DayOfYear<=0 || DayOfYear>365
            [error][properties IllegalDay]
   *Ch 1.2: Normal day
            DayOfYear>=1 && DayOfYear<=365 && DayOfYear!=87 && DayOfYear!=303
            [properties NormalDay]
    Ch 1.3: October change day
            DayOfYear == 303
            [properties OctoberChangeDay]
    Ch 1.4: March change day
            DayOfYear == 87
            [properties MarchChangeDay]
Parameter(1927368268): Duration of type Integer
  Category 1 - Duration of a call
```

```
        Ch 1.1: Illegal value
                Duration <= 2
                [error][properties NegativeDuration]
        Ch 1.2: Call stopped
                Duration >= 7200
                [error][properties DurationTooLong]
       *Ch 1.3: Less than 20mn call
                Duration>=3 && Duration<=1199
        Ch 1.4: At least 20mn call
                Duration>=1200 && Duration<=7199
Parameter(946816132): CallStart of type Integer
  Category 1 - Start time of a call
       *Ch 1.1: Call within a normal day
                CallStart>=(DayOfYear-1)*86400 && CallStart<=DayOfYear*86400-1-Duration
                [if NormalDay && !NegativeDuration && !DurationTooLong]
                [properties WithinDay]
        Ch 1.2: Change to next normal day during call
                CallStart>=DayOfYear*86400-Duration && CallStart<=DayOfYear*86400-1 &&
                2*CallStart+Duration==2*DayOfYear*86400
                [if NormalDay && !NegativeDuration && !DurationTooLong]
        Ch 1.3: Illegal value
                CallStart <(DayOfYear-1)*86400  || CallStart>DayOfYear*86400-1
        Ch 1.4: Call entirely before March change
                CallStart>=(DayOfYear-1)*86400 && CallStart<=(DayOfYear-1)*86400+7200-Duration-1
                [if MarchChangeDay && !NegativeDuration && !DurationTooLong][single]
        Ch 1.5: Call over March change
                CallStart>=(DayOfYear-1)*86400+7200-Duration && CallStart<=(DayOfYear-
                1)*86400+7200-1 && 2*CallStart+Duration==2*(DayOfYear-1)*86400+2*7200
                [if MarchChangeDay && !NegativeDuration && !DurationTooLong]
        Ch 1.6: Illegal missed hour in March
                CallStart>(DayOfYear-1)*86400+7199 && CallStart<(DayOfYear-1)*86400+10800
                [if MarchChangeDay && !NegativeDuration && !DurationTooLong][single]
        Ch 1.7: Rest of day in March change
                CallStart>=(DayOfYear-1)*86400+10800 && CallStart<=DayOfYear*86400-1-Duration
                [if MarchChangeDay && !NegativeDuration && !DurationTooLong][error]
        Ch 1.8: Over to next day for March change
                CallStart>=DayOfYear*86400-Duration && CallStart<=DayOfYear*86400-1 &&
                2*CallStart+Duration==2*DayOfYear*86400
                [if MarchChangeDay && !NegativeDuration && !DurationTooLong][single]
        Ch 1.9: Call entirely before October change
                CallStart>=(DayOfYear-1)*86400 && CallStart<=(DayOfYear-1)*86400+10800-Duration-1
                [if OctoberChangeDay && !NegativeDuration && !DurationTooLong][single]
        Ch 1.10:Call over October change
                CallStart>=(DayOfYear-1)*86400+10800-Duration && CallStart<=(DayOfYear-
                1)*86400+10800-1 && 2*CallStart+Duration==2*(DayOfYear-1)*86400+2*10800
                [if OctoberChangeDay && !NegativeDuration && !DurationTooLong]
        Ch 1.11:Rest of day in October change
                CallStart>=(DayOfYear-1)*86400+10800 && CallStart<=DayOfYear*86400-1-Duration
                [if OctoberChangeDay && !NegativeDuration && !DurationTooLong][single]
        Ch 1.12:Over to next day for October change
                CallStart>=DayOfYear*86400-Duration && CallStart<=DayOfYear*86400-1 &&
                2*CallStart+Duration==2*DayOfYear*86400
                [if OctoberChangeDay && !NegativeDuration && !DurationTooLong][single]
```

## G.3    Each-Choice Test Frames and Corresponding Test Cases when using CASA

**Test Case 1:**
```
choices:  1273475408Cat1Ch2—|—1927368268Cat1Ch3—|—946816132Cat1Ch1
DayOfYear = The day number in the year = Normal day|Duration = Duration of a call = Less than
20mn call|CallStart = Start time of a call = Call within a normal day
values:   Duration = '3',CallStart = '0',DayOfYear = '1'
```
**Test Case 2:**
```
choices:  1273475408Cat1Ch2—|—1927368268Cat1Ch3—|—946816132Cat1Ch2
DayOfYear = The day number in the year = Normal day|Duration = Duration of a call = Less than
20mn call|CallStart = Start time of a call = Change to next normal day during call
```

```
values:    Duration = '4',CallStart = '86398',DayOfYear = '1'
```

**Test Case 3:**
```
choices:   1273475408Cat1Ch4—|—1927368268Cat1Ch4—|—946816132Cat1Ch5
DayOfYear = The day number in the year = March change day|Duration = Duration of a call = At
least 20mn call|CallStart = Start time of a call = Call over March change
values:    Duration = '1200',CallStart = '7437000',DayOfYear = '87'
```

**Test Case 4:**
```
choices:   1273475408Cat1Ch3—|—1927368268Cat1Ch4—|—946816132Cat1Ch10
DayOfYear = The day number in the year = October change day|Duration = Duration of a call = At
least 20mn call|CallStart = Start time of a call = Call over October change
values:    Duration = '1200',CallStart = '26103000',DayOfYear = '303'
```

**Test Case 5:**
```
choices:   1273475408Cat1Ch3—|—1927368268Cat1Ch4—|—946816132Cat1Ch3
DayOfYear = The day number in the year = October change day|Duration = Duration of a call = At
least 20mn call|CallStart = Start time of a call = Illegal value
values:    Duration = '1200',CallStart = '26179200',DayOfYear = '303'
```

**Test Case 6:**
```
choices:   1273475408Cat1Ch3—|—1927368268Cat1Ch3—|—946816132Cat1Ch9
DayOfYear = The day number in the year = October change day|Duration = Duration of a call = Less
than 20mn call|CallStart = Start time of a call = Call entirely before October change
values:    Duration = '3',CallStart = '26092800',DayOfYear = '303'
```

**Test Case 7:**
```
choices:   1273475408Cat1Ch3—|—1927368268Cat1Ch3—|—946816132Cat1Ch11
DayOfYear = The day number in the year = October change day|Duration = Duration of a call = Less
than 20mn call|CallStart = Start time of a call = Rest of day in October change
values:    Duration = '3',CallStart = '26103600',DayOfYear = '303'
```

**Test Case 8:**
```
choices:   1273475408Cat1Ch4—|—1927368268Cat1Ch4—|—946816132Cat1Ch4
DayOfYear = The day number in the year = March change day|Duration = Duration of a call = At
least 20mn call|CallStart = Start time of a call = Call entirely before March change
values:    Duration = '1200',CallStart = '7430400',DayOfYear = '87'
```

**Test Case 9:**
```
choices:   1273475408Cat1Ch4—|—1927368268Cat1Ch4—|—946816132Cat1Ch6
DayOfYear = The day number in the year = March change day|Duration = Duration of a call = At
least 20mn call|CallStart = Start time of a call = Illegal missed hour in March
values:    Duration = '1200',CallStart = '7441199',DayOfYear = '87'
```

**Test Case 10:**
```
choices:   1273475408Cat1Ch4—|—1927368268Cat1Ch3—|—946816132Cat1Ch8
DayOfYear = The day number in the year = March change day|Duration = Duration of a call = Less
than 20mn call|CallStart = Start time of a call = Over to next day for March change
values:    Duration = '4',CallStart = '7516798',DayOfYear = '87'
```

**Test Case 11:**
```
choices:   1273475408Cat1Ch3—|—1927368268Cat1Ch4—|—946816132Cat1Ch12
DayOfYear = The day number in the year = October change day|Duration = Duration of a call = At
least 20mn call|CallStart = Start time of a call = Over to next day for October change
values:    Duration = '1200',CallStart = '26178600',DayOfYear = '303'
```

**Test Case 12:**
```
choices:   1273475408Cat1Ch1
DayOfYear = The day number in the year = Illegal value for day
values:    DayOfYear = '366'
```

**Test Case 13:**
```
choices:   1927368268Cat1Ch1
Duration = Duration of a call = Illegal value
values:    Duration = '0'
```

**Test Case 14:**
```
choices:   1927368268Cat1Ch2
Duration = Duration of a call = Call stopped
values:    Duration = '7200'
```

**Test Case 15:**
```
choices:   946816132Cat1Ch7
CallStart = Start time of a call = Rest of day in March change
values:        Duration = '-1',DayOfYear = '0',CallStart = '0'
```

## G.4    Each-Choice Test Frames and Corresponding Test Cases when using ACTS

**Test Case 1:**
choices:   946816132Cat1Ch1—|—1273475408Cat1Ch2—|—1927368268Cat1Ch4
CallStart = Start time of a call = Call within a normal day|DayOfYear = The day number in the
year = Normal day|Duration = Duration of a call = At least 20mn call
values:    Duration = '1200',CallStart = '0',DayOfYear = '1'

**Test Case 2:**
choices:   946816132Cat1Ch2—|—1273475408Cat1Ch2—|—1927368268Cat1Ch3
CallStart = Start time of a call = Change to next normal day during call|DayOfYear = The day
number in the year = Normal day|Duration = Duration of a call = Less than 20mn call
values:    Duration = '4',CallStart = '26265598',DayOfYear = '304'

**Test Case 3:**
choices:   946816132Cat1Ch3—|—1273475408Cat1Ch3—|—1927368268Cat1Ch4
CallStart = Start time of a call = Illegal value|DayOfYear = The day number in the year =
October change day|Duration = Duration of a call = At least 20mn call
values:    Duration = '1200',CallStart = '26179200',DayOfYear = '303'

**Test Case 4:**
choices:   946816132Cat1Ch5—|—1273475408Cat1Ch4—|—1927368268Cat1Ch3
CallStart = Start time of a call = Call over March change|DayOfYear = The day number in the year
= March change day|Duration = Duration of a call = Less than 20mn call
values:    Duration = '4',CallStart = '7437598',DayOfYear = '87'

**Test Case 5:**
choices:   946816132Cat1Ch10—|—1273475408Cat1Ch3—|—1927368268Cat1Ch4
CallStart = Start time of a call = Call over October change|DayOfYear = The day number in the
year = October change day|Duration = Duration of a call = At least 20mn call
values:    Duration = '1200',CallStart = '26103000',DayOfYear = '303'

**Test Case 6:**
choices:   946816132Cat1Ch4—|—1273475408Cat1Ch4—|—1927368268Cat1Ch3
CallStart = Start time of a call = Call entirely before March change|DayOfYear = The day number
in the year = March change day|Duration = Duration of a call = Less than 20mn call
values:    Duration = '3',CallStart = '7437596',DayOfYear = '87'

**Test Case 7:**
choices:   946816132Cat1Ch6—|—1273475408Cat1Ch4—|—1927368268Cat1Ch4
CallStart = Start time of a call = Illegal missed hour in March|DayOfYear = The day number in
the year = March change day|Duration = Duration of a call = At least 20mn call
values:    Duration = '1200',CallStart = '7437600',DayOfYear = '87'

**Test Case 8:**
choices:   946816132Cat1Ch8—|—1273475408Cat1Ch4—|—1927368268Cat1Ch3
CallStart = Start time of a call = Over to next day for March change|DayOfYear = The day number
in the year = March change day|Duration = Duration of a call = Less than 20mn call
values:    Duration = '4',CallStart = '7516798',DayOfYear = '87'

**Test Case 9:**
choices:   946816132Cat1Ch9—|—1273475408Cat1Ch3—|—1927368268Cat1Ch4
CallStart = Start time of a call = Call entirely before October change|DayOfYear = The day
number in the year = October change day|Duration = Duration of a call = At least 20mn call
values:    Duration = '1200',CallStart = '26092800',DayOfYear = '303'

**Test Case 10:**
choices:   946816132Cat1Ch11—|—1273475408Cat1Ch3—|—1927368268Cat1Ch3
CallStart = Start time of a call = Rest of day in October change|DayOfYear = The day number in
the year = October change day|Duration = Duration of a call = Less than 20mn call
values:    Duration = '3',CallStart = '26103600',DayOfYear = '303'

**Test Case 11:**
choices:   946816132Cat1Ch12—|—1273475408Cat1Ch3—|—1927368268Cat1Ch4
CallStart = Start time of a call = Over to next day for October change|DayOfYear = The day
number in the year = October change day|Duration = Duration of a call = At least 20mn call
values:    Duration = '1200',CallStart = '26178600',DayOfYear = '303'

**Test Case 12:**
choices:   1273475408Cat1Ch1
DayOfYear = The day number in the year = Illegal value for day
values:    DayOfYear = '366'

**Test Case 13:**
choices:   1927368268Cat1Ch1
Duration = Duration of a call = Illegal value
values:    Duration = '0'

```
Test Case 14:
choices:  1927368268Cat1Ch2
Duration = Duration of a call = Call stopped
values:   Duration = '7200'
Test Case 15:
choices:  946816132Cat1Ch7
CallStart = Start time of a call = Rest of day in March change
values:       Duration = '-1',DayOfYear = '0',CallStart = '0'
```

## G.5    *Pairwise Test Frames and Corresponding Test Cases when using CASA*

```
Test Case 1:
choices:  1273475408Cat1Ch3−|−1927368268Cat1Ch4−|−946816132Cat1Ch10
DayOfYear = The day number in the year = October change day|Duration = Duration of a call = At
least 20mn call|CallStart = Start time of a call = Call over October change
values:   Duration = '1200',CallStart = '26103000',DayOfYear = '303'
Test Case 2:
choices:  1273475408Cat1Ch4−|−1927368268Cat1Ch4−|−946816132Cat1Ch3
DayOfYear = The day number in the year = March change day|Duration = Duration of a call = At
least 20mn call|CallStart = Start time of a call = Illegal value
values:   Duration = '1200',CallStart = '7516800',DayOfYear = '87'
Test Case 3:
choices:  1273475408Cat1Ch4−|−1927368268Cat1Ch3−|−946816132Cat1Ch5
DayOfYear = The day number in the year = March change day|Duration = Duration of a call = Less
than 20mn call|CallStart = Start time of a call = Call over March change
values:   Duration = '4',CallStart = '7437598',DayOfYear = '87'
Test Case 4:
choices:  1273475408Cat1Ch3−|−1927368268Cat1Ch3−|−946816132Cat1Ch10
DayOfYear = The day number in the year = October change day|Duration = Duration of a call = Less
than 20mn call|CallStart = Start time of a call = Call over October change
values:   Duration = '4',CallStart = '26103598',DayOfYear = '303'
Test Case 5:
choices:  1273475408Cat1Ch4−|−1927368268Cat1Ch4−|−946816132Cat1Ch5
DayOfYear = The day number in the year = March change day|Duration = Duration of a call = At
least 20mn call|CallStart = Start time of a call = Call over March change
values:   Duration = '1200',CallStart = '7437000',DayOfYear = '87'
Test Case 6:
choices:  1273475408Cat1Ch2−|−1927368268Cat1Ch3−|−946816132Cat1Ch2
DayOfYear = The day number in the year = Normal day|Duration = Duration of a call = Less than
20mn call|CallStart = Start time of a call = Change to next normal day during call
values:   Duration = '4',CallStart = '86398',DayOfYear = '1'
Test Case 7:
choices:  1273475408Cat1Ch2−|−1927368268Cat1Ch4−|−946816132Cat1Ch1
DayOfYear = The day number in the year = Normal day|Duration = Duration of a call = At least
20mn call|CallStart = Start time of a call = Call within a normal day
values:   Duration = '1200',CallStart = '0',DayOfYear = '1'
Test Case 8:
choices:  1273475408Cat1Ch3−|−1927368268Cat1Ch3−|−946816132Cat1Ch3
DayOfYear = The day number in the year = October change day|Duration = Duration of a call = Less
than 20mn call|CallStart = Start time of a call = Illegal value
values:   Duration = '3',CallStart = '26179200',DayOfYear = '303'
Test Case 9:
choices:  1273475408Cat1Ch2−|−1927368268Cat1Ch3−|−946816132Cat1Ch3
DayOfYear = The day number in the year = Normal day|Duration = Duration of a call = Less than
20mn call|CallStart = Start time of a call = Illegal value
values:   Duration = '3',CallStart = '26265600',DayOfYear = '304'
Test Case 10:
choices:  1273475408Cat1Ch2−|−1927368268Cat1Ch3−|−946816132Cat1Ch1
DayOfYear = The day number in the year = Normal day|Duration = Duration of a call = Less than
20mn call|CallStart = Start time of a call = Call within a normal day
values:   Duration = '3',CallStart = '0',DayOfYear = '1'
Test Case 11:
choices:  1273475408Cat1Ch2−|−1927368268Cat1Ch4−|−946816132Cat1Ch2
DayOfYear = The day number in the year = Normal day|Duration = Duration of a call = At least
20mn call|CallStart = Start time of a call = Change to next normal day during call
```

```
values:    Duration = '1200',CallStart = '85800',DayOfYear = '1'
```
**Test Case 12:**
```
choices:   1273475408Cat1Ch3-|-1927368268Cat1Ch4-|-946816132Cat1Ch9
DayOfYear = The day number in the year = October change day|Duration = Duration of a call = At
least 20mn call|CallStart = Start time of a call = Call entirely before October change
values:    Duration = '1200',CallStart = '26092800',DayOfYear = '303'
```
**Test Case 13:**
```
choices:   1273475408Cat1Ch4-|-1927368268Cat1Ch4-|-946816132Cat1Ch8
DayOfYear = The day number in the year = March change day|Duration = Duration of a call = At
least 20mn call|CallStart = Start time of a call = Over to next day for March change
values:    Duration = '1200',CallStart = '7516200',DayOfYear = '87'
```
**Test Case 14:**
```
choices:   1273475408Cat1Ch3-|-1927368268Cat1Ch4-|-946816132Cat1Ch11
DayOfYear = The day number in the year = October change day|Duration = Duration of a call = At
least 20mn call|CallStart = Start time of a call = Rest of day in October change
values:    Duration = '1200',CallStart = '26103600',DayOfYear = '303'
```
**Test Case 15:**
```
choices:   1273475408Cat1Ch4-|-1927368268Cat1Ch3-|-946816132Cat1Ch6
DayOfYear = The day number in the year = March change day|Duration = Duration of a call = Less
than 20mn call|CallStart = Start time of a call = Illegal missed hour in March
values:    Duration = '3',CallStart = '7441199',DayOfYear = '87'
```
**Test Case 16:**
```
choices:   1273475408Cat1Ch3-|-1927368268Cat1Ch4-|-946816132Cat1Ch12
DayOfYear = The day number in the year = October change day|Duration = Duration of a call = At
least 20mn call|CallStart = Start time of a call = Over to next day for October change
values:    Duration = '1200',CallStart = '26178600',DayOfYear = '303'
```
**Test Case 17:**
```
choices:   1273475408Cat1Ch4-|-1927368268Cat1Ch3-|-946816132Cat1Ch4
DayOfYear = The day number in the year = March change day|Duration = Duration of a call = Less
than 20mn call|CallStart = Start time of a call = Call entirely before March change
values:    Duration = '3',CallStart = '7437596',DayOfYear = '87'
```
**Test Case 18:**
```
choices:   1273475408Cat1Ch1
DayOfYear = The day number in the year = Illegal value for day
values:    DayOfYear = '366'
```
**Test Case 19:**
```
choices:   1927368268Cat1Ch1
Duration = Duration of a call = Illegal value
values:    Duration = '0'
```
**Test Case 20:**
```
choices:   1927368268Cat1Ch2
Duration = Duration of a call = Call stopped
values:    Duration = '7200'
```
**Test Case 21:**
```
choices:   946816132Cat1Ch7
CallStart = Start time of a call = Rest of day in March change
values:       CallStart = '0',DayOfYear = '0',Duration = '-1'
```

## G.6    Pairwise Test Frames and Corresponding Test Cases when using ACTS

**Test Case 1:**
```
choices:   946816132Cat1Ch1-|-1273475408Cat1Ch2-|-1927368268Cat1Ch4
CallStart = Start time of a call = Call within a normal day|DayOfYear = The day number in the
year = Normal day|Duration = Duration of a call = At least 20mn call
values:    Duration = '1200',CallStart = '0',DayOfYear = '1'
```
**Test Case 2:**
```
choices:   946816132Cat1Ch2-|-1273475408Cat1Ch2-|-1927368268Cat1Ch3
CallStart = Start time of a call = Change to next normal day during call|DayOfYear = The day
number in the year = Normal day|Duration = Duration of a call = Less than 20mn call
values:    Duration = '4',CallStart = '26265598',DayOfYear = '304'
```
**Test Case 3:**
```
choices:   946816132Cat1Ch3-|-1273475408Cat1Ch2-|-1927368268Cat1Ch4
CallStart = Start time of a call = Illegal value|DayOfYear = The day number in the year = Normal
day|Duration = Duration of a call = At least 20mn call
values:    Duration = '1200',CallStart = '86400',DayOfYear = '1'
```

**Test Case 4:**
choices:   946816132Cat1Ch3-|—1273475408Cat1Ch3-|—1927368268Cat1Ch3
CallStart = Start time of a call = Illegal value|DayOfYear = The day number in the year =
October change day|Duration = Duration of a call = Less than 20mn call
values:   Duration = '3',CallStart = '26179200',DayOfYear = '303'
**Test Case 5:**
choices:   946816132Cat1Ch3-|—1273475408Cat1Ch4-|—1927368268Cat1Ch4
CallStart = Start time of a call = Illegal value|DayOfYear = The day number in the year = March
change day|Duration = Duration of a call = At least 20mn call
values:   Duration = '1200',CallStart = '7516800',DayOfYear = '87'
**Test Case 6:**
choices:   946816132Cat1Ch5-|—1273475408Cat1Ch4-|—1927368268Cat1Ch3
CallStart = Start time of a call = Call over March change|DayOfYear = The day number in the year
= March change day|Duration = Duration of a call = Less than 20mn call
values:   Duration = '4',CallStart = '7437598',DayOfYear = '87'
**Test Case 7:**
choices:   946816132Cat1Ch10-|—1273475408Cat1Ch3-|—1927368268Cat1Ch4
CallStart = Start time of a call = Call over October change|DayOfYear = The day number in the
year = October change day|Duration = Duration of a call = At least 20mn call
values:   Duration = '7198',CallStart = '26100001',DayOfYear = '303'
**Test Case 8:**
choices:   946816132Cat1Ch1-|—1273475408Cat1Ch2-|—1927368268Cat1Ch3
CallStart = Start time of a call = Call within a normal day|DayOfYear = The day number in the
year = Normal day|Duration = Duration of a call = Less than 20mn call
values:   Duration = '3',CallStart = '0',DayOfYear = '1'
**Test Case 9:**
choices:   946816132Cat1Ch10-|—1273475408Cat1Ch3-|—1927368268Cat1Ch3
CallStart = Start time of a call = Call over October change|DayOfYear = The day number in the
year = October change day|Duration = Duration of a call = Less than 20mn call
values:   Duration = '4',CallStart = '26103598',DayOfYear = '303'
**Test Case 10:**
choices:   946816132Cat1Ch2-|—1273475408Cat1Ch2-|—1927368268Cat1Ch4
CallStart = Start time of a call = Change to next normal day during call|DayOfYear = The day
number in the year = Normal day|Duration = Duration of a call = At least 20mn call
values:   Duration = '1200',CallStart = '85800',DayOfYear = '1'
**Test Case 11:**
choices:   946816132Cat1Ch5-|—1273475408Cat1Ch4-|—1927368268Cat1Ch4
CallStart = Start time of a call = Call over March change|DayOfYear = The day number in the year
= March change day|Duration = Duration of a call = At least 20mn call
values:   Duration = '1200',CallStart = '7437000',DayOfYear = '87'
**Test Case 12:**
choices:   946816132Cat1Ch4-|—1273475408Cat1Ch4-|—1927368268Cat1Ch3
CallStart = Start time of a call = Call entirely before March change|DayOfYear = The day number
in the year = March change day|Duration = Duration of a call = Less than 20mn call
values:   Duration = '3',CallStart = '7437596',DayOfYear = '87'
**Test Case 13:**
choices:   946816132Cat1Ch6-|—1273475408Cat1Ch3-|—1927368268Cat1Ch3
CallStart = Start time of a call = Illegal missed hour in March|DayOfYear = The day number in
the year = March change day|Duration = Duration of a call = Less than 20mn call
values:   Duration = '3',CallStart = '7437600',DayOfYear = '87'
**Test Case 14:**
choices:   946816132Cat1Ch8-|—1273475408Cat1Ch4-|—1927368268Cat1Ch4
CallStart = Start time of a call = Over to next day for March change|DayOfYear = The day number
in the year = March change day|Duration = Duration of a call = At least 20mn call
values:   Duration = '1200',CallStart = '7516200',DayOfYear = '87'
**Test Case 15:**
choices:   946816132Cat1Ch9-|—1273475408Cat1Ch3-|—1927368268Cat1Ch4
CallStart = Start time of a call = Call entirely before October change|DayOfYear = The day
number in the year = October change day|Duration = Duration of a call = At least 20mn call
values:   Duration = '1200',CallStart = '26092800',DayOfYear = '303'
**Test Case 16:**
choices:   946816132Cat1Ch11-|—1273475408Cat1Ch3-|—1927368268Cat1Ch3
CallStart = Start time of a call = Rest of day in October change|DayOfYear = The day number in
the year = October change day|Duration = Duration of a call = Less than 20mn call
values:   Duration = '3',CallStart = '26103600',DayOfYear = '303'

**Test Case 17:**
```
choices:   946816132Cat1Ch12−|−1273475408Cat1Ch3−|−1927368268Cat1Ch4
CallStart = Start time of a call = Over to next day for October change|DayOfYear = The day
number in the year = October change day|Duration = Duration of a call = At least 20mn call
values:    Duration = '1200',CallStart = '26178600',DayOfYear = '303'
```
**Test Case 18:**
```
choices:   1273475408Cat1Ch1
DayOfYear = The day number in the year = Illegal value for day
values:    DayOfYear = '366'
```
**Test Case 19:**
```
choices:   1927368268Cat1Ch1
Duration = Duration of a call = Illegal value
values:    Duration = '0'
```
**Test Case 20:**
```
choices:   1927368268Cat1Ch2
Duration = Duration of a call = Call stopped
values:    Duration = '7200'
```
**Test Case 21:**
```
choices:   946816132Cat1Ch7
CallStart = Start time of a call = Rest of day in March change
values:     CallStart = '0',DayOfYear = '0',Duration = '-1'
```

## G.7    Three-way Test Frames and Corresponding Test Cases when using CASA

**Test Case 1:**
```
choices:   1273475408Cat1Ch2−|−1927368268Cat1Ch3−|−946816132Cat1Ch2
DayOfYear = The day number in the year = Normal day|Duration = Duration of a call = Less than
20mn call|CallStart = Start time of a call = Change to next normal day during call
values:    Duration = '4',CallStart = '86398',DayOfYear = '1'
```
**Test Case 2:**
```
choices:   1273475408Cat1Ch3−|−1927368268Cat1Ch3−|−946816132Cat1Ch10
DayOfYear = The day number in the year = October change day|Duration = Duration of a call = Less
than 20mn call|CallStart = Start time of a call = Call over October change
values:    Duration = '4',CallStart = '26103598',DayOfYear = '303'
```
**Test Case 3:**
```
choices:   1273475408Cat1Ch3−|−1927368268Cat1Ch4−|−946816132Cat1Ch3
DayOfYear = The day number in the year = October change day|Duration = Duration of a call = At
least 20mn call|CallStart = Start time of a call = Illegal value
values:    Duration = '1200',CallStart = '26179200',DayOfYear = '303'
```
**Test Case 4:**
```
choices:   1273475408Cat1Ch2−|−1927368268Cat1Ch3−|−946816132Cat1Ch1
DayOfYear = The day number in the year = Normal day|Duration = Duration of a call = Less than
20mn call|CallStart = Start time of a call = Call within a normal day
values:    Duration = '3',CallStart = '26179200',DayOfYear = '304'
```
**Test Case 5:**
```
choices:   1273475408Cat1Ch4−|−1927368268Cat1Ch4−|−946816132Cat1Ch3
DayOfYear = The day number in the year = March change day|Duration = Duration of a call = At
least 20mn call|CallStart = Start time of a call = Illegal value
values:    Duration = '1200',CallStart = '7516800',DayOfYear = '87'
```
**Test Case 6:**
```
choices:   1273475408Cat1Ch4−|−1927368268Cat1Ch4−|−946816132Cat1Ch5
DayOfYear = The day number in the year = March change day|Duration = Duration of a call = At
least 20mn call|CallStart = Start time of a call = Call over March change
values:    Duration = '1200',CallStart = '7437000',DayOfYear = '87'
```
**Test Case 7:**
```
choices:   1273475408Cat1Ch2−|−1927368268Cat1Ch4−|−946816132Cat1Ch1
DayOfYear = The day number in the year = Normal day|Duration = Duration of a call = At least
20mn call|CallStart = Start time of a call = Call within a normal day
values:    Duration = '1200',CallStart = '0',DayOfYear = '1'
```
**Test Case 8:**
```
choices:   1273475408Cat1Ch4−|−1927368268Cat1Ch3−|−946816132Cat1Ch5
DayOfYear = The day number in the year = March change day|Duration = Duration of a call = Less
than 20mn call|CallStart = Start time of a call = Call over March change
values:    Duration = '4',CallStart = '7437598',DayOfYear = '87'
```
**Test Case 9:**

```
choices:   1273475408Cat1Ch3—|—1927368268Cat1Ch4—|—946816132Cat1Ch10
DayOfYear = The day number in the year = October change day|Duration = Duration of a call = At
least 20mn call|CallStart = Start time of a call = Call over October change
values:   Duration = '1200',CallStart = '26103000',DayOfYear = '303'
```
**Test Case 10:**
```
choices:   1273475408Cat1Ch2—|—1927368268Cat1Ch4—|—946816132Cat1Ch3
DayOfYear = The day number in the year = Normal day|Duration = Duration of a call = At least
20mn call|CallStart = Start time of a call = Illegal value
values:   Duration = '1200',CallStart = '86400',DayOfYear = '1'
```
**Test Case 11:**
```
choices:   1273475408Cat1Ch2—|—1927368268Cat1Ch4—|—946816132Cat1Ch2
DayOfYear = The day number in the year = Normal day|Duration = Duration of a call = At least
20mn call|CallStart = Start time of a call = Change to next normal day during call
values:   Duration = '1200',CallStart = '85800',DayOfYear = '1'
```
**Test Case 12:**
```
choices:   1273475408Cat1Ch2—|—1927368268Cat1Ch3—|—946816132Cat1Ch3
DayOfYear = The day number in the year = Normal day|Duration = Duration of a call = Less than
20mn call|CallStart = Start time of a call = Illegal value
values:   Duration = '3',CallStart = '86400',DayOfYear = '1'
```
**Test Case 13:**
```
choices:   1273475408Cat1Ch3—|—1927368268Cat1Ch3—|—946816132Cat1Ch3
DayOfYear = The day number in the year = October change day|Duration = Duration of a call = Less
than 20mn call|CallStart = Start time of a call = Illegal value
values:   Duration = '3',CallStart = '26179200',DayOfYear = '303'
```
**Test Case 14:**
```
choices:   1273475408Cat1Ch4—|—1927368268Cat1Ch4—|—946816132Cat1Ch3
DayOfYear = The day number in the year = March change day|Duration = Duration of a call = Less
than 20mn call|CallStart = Start time of a call = Illegal value
values:   Duration = '3',CallStart = '7516800',DayOfYear = '87'
```
**Test Case 15:**
```
choices:   1273475408Cat1Ch3—|—1927368268Cat1Ch3—|—946816132Cat1Ch11
DayOfYear = The day number in the year = October change day|Duration = Duration of a call = Less
than 20mn call|CallStart = Start time of a call = Rest of day in October change
values:   Duration = '3',CallStart = '26179196',DayOfYear = '303'
```
**Test Case 16:**
```
choices:   1273475408Cat1Ch4—|—1927368268Cat1Ch4—|—946816132Cat1Ch6
DayOfYear = The day number in the year = March change day|Duration = Duration of a call = At
least 20mn call|CallStart = Start time of a call = Illegal missed hour in March
values:   Duration = '1200',CallStart = '7437600',DayOfYear = '87'
```
**Test Case 17:**
```
choices:   1273475408Cat1Ch3—|—1927368268Cat1Ch3—|—946816132Cat1Ch12
DayOfYear = The day number in the year = October change day|Duration = Duration of a call = Less
than 20mn call|CallStart = Start time of a call = Over to next day for October change
values:   Duration = '4',CallStart = '26179198',DayOfYear = '303'
```
**Test Case 18:**
```
choices:   1273475408Cat1Ch3—|—1927368268Cat1Ch4—|—946816132Cat1Ch9
DayOfYear = The day number in the year = October change day|Duration = Duration of a call = At
least 20mn call|CallStart = Start time of a call = Call entirely before October change
values:   Duration = '1200',CallStart = '26102399',DayOfYear = '303'
```
**Test Case 19:**
```
choices:   1273475408Cat1Ch4—|—1927368268Cat1Ch3—|—946816132Cat1Ch4
DayOfYear = The day number in the year = March change day|Duration = Duration of a call = Less
than 20mn call|CallStart = Start time of a call = Call entirely before March change
values:   Duration = '3',CallStart = '7430400',DayOfYear = '87'
```
**Test Case 20:**
```
choices:   1273475408Cat1Ch4—|—1927368268Cat1Ch3—|—946816132Cat1Ch8
DayOfYear = The day number in the year = March change day|Duration = Duration of a call = Less
than 20mn call|CallStart = Start time of a call = Over to next day for March change
values:   Duration = '4',CallStart = '7516798',DayOfYear = '87'
```
**Test Case 21:**
```
choices:   1273475408Cat1Ch1
DayOfYear = The day number in the year = Illegal value for day
values:   DayOfYear = '366'
```
**Test Case 22:**
```
choices:   1927368268Cat1Ch1
```

```
Duration = Duration of a call = Illegal value
values:    Duration = '0'
```
**Test Case 23:**
```
choices:   1927368268Cat1Ch2
Duration = Duration of a call = Call stopped
values:    Duration = '7200'
```
**Test Case 24:**
```
choices:   946816132Cat1Ch7
CallStart = Start time of a call = Rest of day in March change
values:       DayOfYear = '0',CallStart = '0',Duration = '-1'
```

## G.8     Three-way Test Frames and Corresponding Test Cases when using ACTS

**Test Case 1:**
```
choices:   946816132Cat1Ch1-|-1273475408Cat1Ch2-|-1927368268Cat1Ch3
CallStart = Start time of a call = Call within a normal day|DayOfYear = The day number in the
year = Normal day|Duration = Duration of a call = Less than 20mn call
values:    Duration = '3',CallStart = '0',DayOfYear = '1'
```
**Test Case 2:**
```
choices:   946816132Cat1Ch1-|-1273475408Cat1Ch2-|-1927368268Cat1Ch4
CallStart = Start time of a call = Call within a normal day|DayOfYear = The day number in the
year = Normal day|Duration = Duration of a call = At least 20mn call
values:    Duration = '1200',CallStart = '26179200',DayOfYear = '304'
```
**Test Case 3:**
```
choices:   946816132Cat1Ch2-|-1273475408Cat1Ch2-|-1927368268Cat1Ch3
CallStart = Start time of a call = Change to next normal day during call|DayOfYear = The day
number in the year = Normal day|Duration = Duration of a call = Less than 20mn call
values:    Duration = '4',CallStart = '86398',DayOfYear = '1'
```
**Test Case 4:**
```
choices:   946816132Cat1Ch2-|-1273475408Cat1Ch2-|-1927368268Cat1Ch4
CallStart = Start time of a call = Change to next normal day during call|DayOfYear = The day
number in the year = Normal day|Duration = Duration of a call = At least 20mn call
values:    Duration = '1200',CallStart = '26265000',DayOfYear = '304'
```
**Test Case 5:**
```
choices:   946816132Cat1Ch3-|-1273475408Cat1Ch2-|-1927368268Cat1Ch3
CallStart = Start time of a call = Illegal value|DayOfYear = The day number in the year = Normal
day|Duration = Duration of a call = Less than 20mn call
values:    Duration = '3',CallStart = '86400',DayOfYear = '1'
```
**Test Case 6:**
```
choices:   946816132Cat1Ch3-|-1273475408Cat1Ch2-|-1927368268Cat1Ch4
CallStart = Start time of a call = Illegal value|DayOfYear = The day number in the year = Normal
day|Duration = Duration of a call = At least 20mn call
values:    Duration = '1200',CallStart = '26265600',DayOfYear = '304'
```
**Test Case 7:**
```
choices:   946816132Cat1Ch3-|-1273475408Cat1Ch3-|-1927368268Cat1Ch3
CallStart = Start time of a call = Illegal value|DayOfYear = The day number in the year =
October change day|Duration = Duration of a call = Less than 20mn call
values:    Duration = '3',CallStart = '26179200',DayOfYear = '303'
```
**Test Case 8:**
```
choices:   946816132Cat1Ch3-|-1273475408Cat1Ch3-|-1927368268Cat1Ch4
CallStart = Start time of a call = Illegal value|DayOfYear = The day number in the year =
October change day|Duration = Duration of a call = At least 20mn call
values:    Duration = '1200',CallStart = '26179200',DayOfYear = '303'
```
**Test Case 9:**
```
choices:   946816132Cat1Ch3-|-1273475408Cat1Ch4-|-1927368268Cat1Ch3
CallStart = Start time of a call = Illegal value|DayOfYear = The day number in the year = March
change day|Duration = Duration of a call = Less than 20mn call
values:    Duration = '3',CallStart = '7516800',DayOfYear = '87'
```
**Test Case 10:**
```
choices:   946816132Cat1Ch3-|-1273475408Cat1Ch4-|-1927368268Cat1Ch4
CallStart = Start time of a call = Illegal value|DayOfYear = The day number in the year = March
change day|Duration = Duration of a call = At least 20mn call
values:    Duration = '1200',CallStart = '7516800',DayOfYear = '87'
```
**Test Case 11:**
```
choices:   946816132Cat1Ch5-|-1273475408Cat1Ch4-|-1927368268Cat1Ch3
```

```
CallStart = Start time of a call = Call over March change|DayOfYear = The day number in the year
= March change day|Duration = Duration of a call = Less than 20mn call
values:    Duration = '4',CallStart = '7437598',DayOfYear = '87'
```
**Test Case 12:**
```
choices:  946816132Cat1Ch5−|−1273475408Cat1Ch4−|−1927368268Cat1Ch4
CallStart = Start time of a call = Call over March change|DayOfYear = The day number in the year
= March change day|Duration = Duration of a call = At least 20mn call
values:    Duration = '1200',CallStart = '7437000',DayOfYear = '87'
```
**Test Case 13:**
```
choices:  946816132Cat1Ch10−|−1273475408Cat1Ch3−|−1927368268Cat1Ch3
CallStart = Start time of a call = Call over October change|DayOfYear = The day number in the
year = October change day|Duration = Duration of a call = Less than 20mn call
values:    Duration = '4',CallStart = '26103598',DayOfYear = '303'
```
**Test Case 14:**
```
choices:  946816132Cat1Ch10−|−1273475408Cat1Ch3−|−1927368268Cat1Ch4
CallStart = Start time of a call = Call over October change|DayOfYear = The day number in the
year = October change day|Duration = Duration of a call = At least 20mn call
values:    Duration = '1200',CallStart = '26103000',DayOfYear = '303'
```
**Test Case 15:**
```
choices:  946816132Cat1Ch4−|−1273475408Cat1Ch4−|−1927368268Cat1Ch3
CallStart = Start time of a call = Call entirely before March change|DayOfYear = The day number
in the year = March change day|Duration = Duration of a call = Less than 20mn call
values:    Duration = '3',CallStart = '7430400',DayOfYear = '87'
```
**Test Case 16:**
```
choices:  946816132Cat1Ch6−|−1273475408Cat1Ch4−|−1927368268Cat1Ch3
CallStart = Start time of a call = Illegal missed hour in March|DayOfYear = The day number in
the year = March change day|Duration = Duration of a call = Less than 20mn call
values:    Duration = '3',CallStart = '7437600',DayOfYear = '87'
```
**Test Case 17:**
```
choices:  946816132Cat1Ch8−|−1273475408Cat1Ch4−|−1927368268Cat1Ch3
CallStart = Start time of a call = Over to next day for March change|DayOfYear = The day number
in the year = March change day|Duration = Duration of a call = Less than 20mn call
values:    Duration = '4',CallStart = '7516798',DayOfYear = '87'
```
**Test Case 18:**
```
choices:  946816132Cat1Ch9−|−1273475408Cat1Ch3−|−1927368268Cat1Ch3
CallStart = Start time of a call = Call entirely before October change|DayOfYear = The day
number in the year = October change day|Duration = Duration of a call = Less than 20mn call
values:    Duration = '3',CallStart = '26092800',DayOfYear = '303'
```
**Test Case 19:**
```
choices:  946816132Cat1Ch11−|−1273475408Cat1Ch3−|−1927368268Cat1Ch3
CallStart = Start time of a call = Rest of day in October change|DayOfYear = The day number in
the year = October change day|Duration = Duration of a call = Less than 20mn call
values:    Duration = '3',CallStart = '26103600',DayOfYear = '303'
```
**Test Case 20:**
```
choices:  946816132Cat1Ch12−|−1273475408Cat1Ch3−|−1927368268Cat1Ch3
CallStart = Start time of a call = Over to next day for October change|DayOfYear = The day
number in the year = October change day|Duration = Duration of a call = Less than 20mn call
values:    Duration = '4',CallStart = '26179198',DayOfYear = '303'
```
**Test Case 21:**
```
choices:  1273475408Cat1Ch1
DayOfYear = The day number in the year = Illegal value for day
values:    DayOfYear = '366'
```
**Test Case 22:**
```
choices:  1927368268Cat1Ch1
Duration = Duration of a call = Illegal value
values:    Duration = '0'
```
**Test Case 23:**
```
choices:  1927368268Cat1Ch2
Duration = Duration of a call = Call stopped
values:    Duration = '7200'
```
**Test Case 24:**
```
choices:  946816132Cat1Ch7
CallStart = Start time of a call = Rest of day in March change
values:       DayOfYear = '0',CallStart = '0',Duration = '-1'
```

## G.9    Base Choice Test Frames and Corresponding Test Cases

**Test Case 1:**
choices:   1273475408Cat1Ch2−|−1927368268Cat1Ch3−|−946816132Cat1Ch1
DayOfYear = The day number in the year = Normal day|Duration = Duration of a call = Less than
20mn call|CallStart = Start time of a call = Call within a normal day
values:    Duration = '3',CallStart = '0',DayOfYear = '1'

**Test Case 2:**
choices:   1273475408Cat1Ch1
DayOfYear = The day number in the year = Illegal value for day
values:    DayOfYear = '366'

**Test Case 3:**
choices:   946816132Cat1Ch3−|−1273475408Cat1Ch3−|−1927368268Cat1Ch4
CallStart = Start time of a call = Illegal value|DayOfYear = The day number in the year =
October change day|Duration = Duration of a call = At least 20mn call
values:    Duration = '1200',CallStart = '26179200',DayOfYear = '303'

**Test Case 4:**
choices:   946816132Cat1Ch5−|−1273475408Cat1Ch4−|−1927368268Cat1Ch3
CallStart = Start time of a call = Call over March change|DayOfYear = The day number in the year
= March change day|Duration = Duration of a call = Less than 20mn call
values:    Duration = '4',CallStart = '7437598',DayOfYear = '87'

**Test Case 5:**
choices:   1927368268Cat1Ch1
Duration = Duration of a call = Illegal value
values:    Duration = '0'

**Test Case 6:**
choices:   1927368268Cat1Ch2
Duration = Duration of a call = Call stopped
values:    Duration = '7200'

**Test Case 7:**
choices:   1273475408Cat1Ch2−|−946816132Cat1Ch1−|−1927368268Cat1Ch4
DayOfYear = The day number in the year = Normal day|CallStart = Start time of a call = Call
within a normal day|Duration = Duration of a call = At least 20mn call
values:    Duration = '1200',CallStart = '26179200',DayOfYear = '304'

**Test Case 8:**
choices:   1273475408Cat1Ch2−|−1927368268Cat1Ch3−|−946816132Cat1Ch2
DayOfYear = The day number in the year = Normal day|Duration = Duration of a call = Less than
20mn call|CallStart = Start time of a call = Change to next normal day during call
values:    Duration = '4',CallStart = '86398',DayOfYear = '1'

**Test Case 9:**
choices:   1273475408Cat1Ch2−|−1927368268Cat1Ch3−|−946816132Cat1Ch3
DayOfYear = The day number in the year = Normal day|Duration = Duration of a call = Less than
20mn call|CallStart = Start time of a call = Illegal value
values:    Duration = '3',CallStart = '26265600',DayOfYear = '304'

**Test Case 10:**
choices:   946816132Cat1Ch4−|−1273475408Cat1Ch4−|−1927368268Cat1Ch3
CallStart = Start time of a call = Call entirely before March change|DayOfYear = The day number
in the year = March change day|Duration = Duration of a call = Less than 20mn call
values:    Duration = '3',CallStart = '7430400',DayOfYear = '87'

**Test Case 11:**
choices:   946816132Cat1Ch5−|−1273475408Cat1Ch4−|−1927368268Cat1Ch3
CallStart = Start time of a call = Call over March change|DayOfYear = The day number in the year
= March change day|Duration = Duration of a call = Less than 20mn call
values:    Duration = '4',CallStart = '7437598',DayOfYear = '87'

**Test Case 12:**
choices:   946816132Cat1Ch6−|−1273475408Cat1Ch4−|−1927368268Cat1Ch4
CallStart = Start time of a call = Illegal missed hour in March|DayOfYear = The day number in
the year = March change day|Duration = Duration of a call = At least 20mn call
values:    Duration = '1200',CallStart = '7437600',DayOfYear = '87'

**Test Case 13:**
choices:   946816132Cat1Ch7
CallStart = Start time of a call = Rest of day in March change
values:    Duration = '-1',CallStart = '0',DayOfYear = '0'

**Test Case 14:**
choices:   946816132Cat1Ch8−|−1273475408Cat1Ch4−|−1927368268Cat1Ch3

```
CallStart = Start time of a call = Over to next day for March change|DayOfYear = The day number
in the year = March change day|Duration = Duration of a call = Less than 20mn call
values:    Duration = '4',CallStart = '7516798',DayOfYear = '87'
```
**Test Case 15:**
```
choices:  946816132Cat1Ch9—|—1273475408Cat1Ch3—|—1927368268Cat1Ch4
CallStart = Start time of a call = Call entirely before October change|DayOfYear = The day
number in the year = October change day|Duration = Duration of a call = At least 20mn call
values:    Duration = '1200',CallStart = '26092800',DayOfYear = '303'
```
**Test Case 16:**
```
choices:  946816132Cat1Ch10—|—1273475408Cat1Ch3—|—1927368268Cat1Ch4
CallStart = Start time of a call = Call over October change|DayOfYear = The day number in the
year = October change day|Duration = Duration of a call = At least 20mn call
values:    Duration = '1200',CallStart = '26103000',DayOfYear = '303'
```
**Test Case 17:**
```
choices:  946816132Cat1Ch11—|—1273475408Cat1Ch3—|—1927368268Cat1Ch3
CallStart = Start time of a call = Rest of day in October change|DayOfYear = The day number in
the year = October change day|Duration = Duration of a call = Less than 20mn call
values:    Duration = '3',CallStart = '26103600',DayOfYear = '303'
```
**Test Case 18:**
```
choices:  946816132Cat1Ch12—|—1273475408Cat1Ch3—|—1927368268Cat1Ch4
CallStart = Start time of a call = Over to next day for October change|DayOfYear = The day
number in the year = October change day|Duration = Duration of a call = At least 20mn call
values:        Duration = '1200',CallStart = '26178600',DayOfYear = '303'
```

# Appendix H.  Taxi billing Case Study

In the CP specification below, an asterisk ('*') beside a choice indicates the base choice for that category. Each parameter is uniquely identified by an integer identifier, which is then used in the composition of test frames; As a result, a choice for a parameter is uniquely identified by the parameter identifier, the category identifier and the choice identifier. Below we summarize the specification of the unit under test (section H.1), and provide the CP specification we used (section H.2). We then provide the test frames and corresponding test cases (inputs) we obtain when using CASA (Each-Choice in section H.3, Pairwise in section H.5, Three-way in section H.7), when using ACTS (Each-Choice in section H.4, Pairwise in section H.6, Three-way in section H.8), and when using the Base Choice criterion (section H.9). The CP specification below is adapted from earlier work [41].

## H.1  Specification

The system under test mimics the fee calculation of cabs in Paris, France: https://www.g7.fr/tarifs-taxis-paris. The cost of a cab trip depends on the starting location, either Paris' core, the suburb or beyond the suburb, the day of the week (Saturday being considered a weekday and only Sunday being different), the time of the day, and whether the day is a Holiday. We simplified the specification by considering that a trip's cost only depends on mileage; in reality, if an idle cab also contributes to the cost.

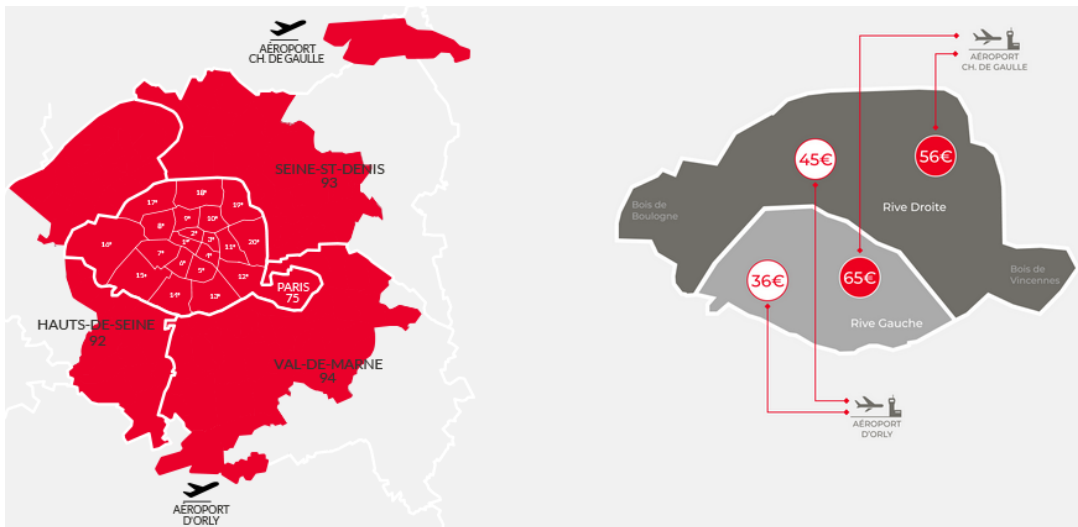

| DATES | HORAIRES | ZONE URBAINE PARIS | ZONE SUBURBAINE | AU-DELÀ DE ZONE SUBURBAINE |
|---|---|---|---|---|
| Lundi au samedi | De 0h à 7h | B | C | |
| | De 7h à 10h | | | |
| | De 10h à 17h | A | B | C |
| | De 17h à 19h | | | |
| | De 19h à 24h | B | C | |
| Dimanche | De 0h à 7h | C | C | C |
| | De 7h à 00h | B | | |
| Jours fériés | De 00h à 00h | B | C | C |

## H.2    CP Specification

```
Parameter(1538408392): Holiday of type Boolean
  Category 1 - Is the day a holiday
     Ch 1.1: Holiday
             Holiday == true
    *Ch 1.2: Not a Holiday
             Holiday == false
Parameter(1273415943): DayOfWeek of type Boolean
  Category 1 - Day of week value
    *Ch 1.1: Week day (incl.Saturday)
             DayOfWeek == true
     Ch 1.2: Sunday
             DayOfWeek == false
Parameter(633365953): DepartureTime of type Integer
  Category 1 - Departure time value
     Ch 1.1: Early morning
             DepartureTime>=0 && DepartureTime<7
     Ch 1.2: Morning
             DepartureTime>=7 && DepartureTime<10
    *Ch 1.3: Midday
             DepartureTime>=10 && DepartureTime<17
     Ch 1.4: Early evening
             DepartureTime>=17 && DepartureTime<19
     Ch 1.5: Night
             DepartureTime>=19 && DepartureTime<24
     Ch 1.6: Illegal
             DepartureTime<0 || DepartureTime>=24
Parameter(310204685): StartingArea of type Enumeration
  Category 1 - Area to start the trip from
    *Ch 1.1: Paris core
             StartingArea == Core
             [properties Core]
     Ch 1.2: Paris suburb
             StartingArea == Suburb
             [properties Suburb]
     Ch 1.3: Outside Paris suburb
             StartingArea == Outside
             [properties Outside]
```

## H.3    Each-Choice Test Frames and Corresponding Test Cases when using CASA

**Test Case 1:**
```
choices:   1538408392Cat1Ch1-|-1273415943Cat1Ch2-|-633365953Cat1Ch6-|-310204685Cat1Ch3
Holiday = Is the day a holiday = Holiday|DayOfWeek = Day of week value = Sunday|DepartureTime =
Departure time value = Illegal|StartingArea = Area to start the trip from = Outside paris suburb
values:      DayOfWeek = 'false', DepartureTime = '-1', StartingArea = 'Outside', Holiday =
             'true'
```

**Test Case 2:**
```
choices:   1538408392Cat1Ch2-|-1273415943Cat1Ch1-|-633365953Cat1Ch2-|-310204685Cat1Ch1
Holiday = Is the day a holiday = Not a Holiday|DayOfWeek = Day of week value = Week day
(incl.Saturday)|DepartureTime = Departure time value = Morning|StartingArea = Area to start the
trip from = Paris core
values:      DayOfWeek = 'true', StartingArea = 'Core', DepartureTime = '7', Holiday = 'false'
```

**Test Case 3:**
```
choices:   1538408392Cat1Ch2-|-1273415943Cat1Ch2-|-633365953Cat1Ch4-|-310204685Cat1Ch1
Holiday = Is the day a holiday = Not a Holiday|DayOfWeek = Day of week value =
Sunday|DepartureTime = Departure time value = Early evening|StartingArea = Area to start the
trip from = Paris core
values:      DayOfWeek = 'false', StartingArea = 'Core', DepartureTime = '17', Holiday = 'false'
```

**Test Case 4:**
```
choices:   1538408392Cat1Ch1-|-1273415943Cat1Ch2-|-633365953Cat1Ch3-|-310204685Cat1Ch1
Holiday = Is the day a holiday = Holiday|DayOfWeek = Day of week value = Sunday|DepartureTime =
Departure time value = Midday|StartingArea = Area to start the trip from = Paris core
values:      DayOfWeek = 'false', StartingArea = 'Core', DepartureTime = '10', Holiday = 'true'
```

**Test Case 5:**
choices:      1538408392Cat1Ch1−|−1273415943Cat1Ch2−|−633365953Cat1Ch1−|−310204685Cat1Ch2
Holiday = Is the day a holiday = Holiday|DayOfWeek = Day of week value = Sunday|DepartureTime = Departure time value = Early morning|StartingArea = Area to start the trip from = Paris suburb
values:      DayOfWeek = 'false', StartingArea = 'Suburb', DepartureTime = '0', Holiday = 'true'

**Test Case 6:**
choices:      1538408392Cat1Ch2−|−1273415943Cat1Ch2−|−633365953Cat1Ch5−|−310204685Cat1Ch1
Holiday = Is the day a holiday = Not a Holiday|DayOfWeek = Day of week value = Sunday|DepartureTime = Departure time value = Night|StartingArea = Area to start the trip from = Paris core
values:      DayOfWeek = 'false', StartingArea = 'Core', DepartureTime = '19', Holiday = 'false'

## H.4    Each-Choice Test Frames and Corresponding Test Cases when using ACTS

**Test Case 1:**
choices:      633365953Cat1Ch1−|−310204685Cat1Ch2−|−1538408392Cat1Ch2−|−1273415943Cat1Ch2
DepartureTime = Departure time value = Early morning|StartingArea = Area to start the trip from = Paris suburb|Holiday = Is the day a holiday = Not a Holiday|DayOfWeek = Day of week value = Sunday
values:      DayOfWeek = 'false', DepartureTime = '0', StartingArea = 'Suburb', Holiday = 'false'

**Test Case 2:**
choices:      633365953Cat1Ch2−|−310204685Cat1Ch3−|−1538408392Cat1Ch1−|−1273415943Cat1Ch1
DepartureTime = Departure time value = Morning|StartingArea = Area to start the trip from = Outside paris suburb|Holiday = Is the day a holiday = Holiday|DayOfWeek = Day of week value = Week day (incl.Saturday)
values:      DayOfWeek = 'true', StartingArea = 'Outside', DepartureTime = '7', Holiday = 'true'

**Test Case 3:**
choices:      633365953Cat1Ch3−|−310204685Cat1Ch1−|−1538408392Cat1Ch2−|−1273415943Cat1Ch2
DepartureTime = Departure time value = Midday|StartingArea = Area to start the trip from = Paris core|Holiday = Is the day a holiday = Not a Holiday|DayOfWeek = Day of week value = Sunday
values:      DayOfWeek = 'false', StartingArea = 'Core', DepartureTime = '10', Holiday = 'false'

**Test Case 4:**
choices:      633365953Cat1Ch4−|−310204685Cat1Ch2−|−1538408392Cat1Ch2−|−1273415943Cat1Ch2
DepartureTime = Departure time value = Early evening|StartingArea = Area to start the trip from = Paris suburb|Holiday = Is the day a holiday = Not a Holiday|DayOfWeek = Day of week value = Sunday
values:      DayOfWeek = 'false', StartingArea = 'Suburb', DepartureTime = '17', Holiday = 'false'

**Test Case 5:**
choices:      633365953Cat1Ch5−|−310204685Cat1Ch3−|−1538408392Cat1Ch2−|−1273415943Cat1Ch1
DepartureTime = Departure time value = Night|StartingArea = Area to start the trip from = Outside paris suburb|Holiday = Is the day a holiday = Not a Holiday|DayOfWeek = Day of week value = Week day (incl.Saturday)
values:      DayOfWeek = 'true', StartingArea = 'Outside', DepartureTime = '19', Holiday = 'false'

**Test Case 6:**
choices:      633365953Cat1Ch6−|−310204685Cat1Ch1−|−1538408392Cat1Ch2−|−1273415943Cat1Ch1
DepartureTime = Departure time value = Illegal|StartingArea = Area to start the trip from = Paris core|Holiday = Is the day a holiday = Not a Holiday|DayOfWeek = Day of week value = Week day (incl.Saturday)
values:      DayOfWeek = 'true', StartingArea = 'Core', DepartureTime = '-1', Holiday = 'false'

## H.5    Pairwise Test Frames and Corresponding Test Cases when using CASA

**Test Case 1:**
choices:      1538408392Cat1Ch1−|−1273415943Cat1Ch1−|−633365953Cat1Ch3−|−310204685Cat1Ch2
Holiday = Is the day a holiday = Holiday|DayOfWeek = Day of week value = Week day (incl.Saturday)|DepartureTime = Departure time value = Midday|StartingArea = Area to start the trip from = Paris suburb
values:      DayOfWeek = 'true', DepartureTime = '10', StartingArea = 'Suburb', Holiday = 'true'

**Test Case 2:**
choices:      1538408392Cat1Ch2−|−1273415943Cat1Ch2−|−633365953Cat1Ch5−|−310204685Cat1Ch1

```
Holiday = Is the day a holiday = Not a Holiday|DayOfWeek = Day of week value =
Sunday|DepartureTime = Departure time value = Night|StartingArea = Area to start the trip from =
Paris core
values:          DayOfWeek = 'false', StartingArea = 'Core', DepartureTime = '19', Holiday = 'false'
```

**Test Case 3:**
```
choices:     1538408392Cat1Ch2−|−1273415943Cat1Ch2−|−633365953Cat1Ch1−|−310204685Cat1Ch3
Holiday = Is the day a holiday = Not a Holiday|DayOfWeek = Day of week value =
Sunday|DepartureTime = Departure time value = Early morning|StartingArea = Area to start the
trip from = Outside paris suburb
values:          DayOfWeek = 'false', StartingArea = 'Outside', DepartureTime = '0', Holiday =
                 'false'
```

**Test Case 4:**
```
choices:     1538408392Cat1Ch2−|−1273415943Cat1Ch1−|−633365953Cat1Ch6−|−310204685Cat1Ch3
Holiday = Is the day a holiday = Not a Holiday|DayOfWeek = Day of week value = Week day
(incl.Saturday)|DepartureTime = Departure time value = Illegal|StartingArea = Area to start the
trip from = Outside paris suburb
values:          DayOfWeek = 'true', StartingArea = 'Outside', DepartureTime = '-1', Holiday =
                 'false'
```

**Test Case 5:**
```
choices:     1538408392Cat1Ch1−|−1273415943Cat1Ch2−|−633365953Cat1Ch6−|−310204685Cat1Ch2
Holiday = Is the day a holiday = Holiday|DayOfWeek = Day of week value = Sunday|DepartureTime =
Departure time value = Illegal|StartingArea = Area to start the trip from = Paris suburb
values:          DayOfWeek = 'false', StartingArea = 'Suburb', DepartureTime = '-1', Holiday =
                 'true'
```

**Test Case 6:**
```
choices:     1538408392Cat1Ch1−|−1273415943Cat1Ch1−|−633365953Cat1Ch1−|−310204685Cat1Ch1
Holiday = Is the day a holiday = Holiday|DayOfWeek = Day of week value = Week day
(incl.Saturday)|DepartureTime = Departure time value = Early morning|StartingArea = Area to
start the trip from = Paris core
values:          DayOfWeek = 'true', StartingArea = 'Core', DepartureTime = '0', Holiday = 'true'
```

**Test Case 7:**
```
choices:     1538408392Cat1Ch1−|−1273415943Cat1Ch2−|−633365953Cat1Ch2−|−310204685Cat1Ch2
Holiday = Is the day a holiday = Holiday|DayOfWeek = Day of week value = Sunday|DepartureTime =
Departure time value = Morning|StartingArea = Area to start the trip from = Paris suburb
values:          DayOfWeek = 'false', StartingArea = 'Suburb', DepartureTime = '7', Holiday = 'true'
```

**Test Case 8:**
```
choices:     1538408392Cat1Ch2−|−1273415943Cat1Ch1−|−633365953Cat1Ch2−|−310204685Cat1Ch1
Holiday = Is the day a holiday = Not a Holiday|DayOfWeek = Day of week value = Week day
(incl.Saturday)|DepartureTime = Departure time value = Morning|StartingArea = Area to start the
trip from = Paris core
values:          DayOfWeek = 'true', StartingArea = 'Core', DepartureTime = '7', Holiday = 'false'
```

**Test Case 9:**
```
choices:     1538408392Cat1Ch2−|−1273415943Cat1Ch2−|−633365953Cat1Ch3−|−310204685Cat1Ch3
Holiday = Is the day a holiday = Not a Holiday|DayOfWeek = Day of week value =
Sunday|DepartureTime = Departure time value = Midday|StartingArea = Area to start the trip from
= Outside paris suburb
values:          DayOfWeek = 'false', StartingArea = 'Outside', DepartureTime = '10', Holiday =
                 'false'
```

**Test Case 10:**
```
choices:     1538408392Cat1Ch2−|−1273415943Cat1Ch1−|−633365953Cat1Ch4−|−310204685Cat1Ch2
Holiday = Is the day a holiday = Not a Holiday|DayOfWeek = Day of week value = Week day
(incl.Saturday)|DepartureTime = Departure time value = Early evening|StartingArea = Area to
start the trip from = Paris suburb
values:          DayOfWeek = 'true', StartingArea = 'Suburb', DepartureTime = '17', Holiday =
                 'false'
```

**Test Case 11:**
```
choices:     1538408392Cat1Ch2−|−1273415943Cat1Ch2−|−633365953Cat1Ch3−|−310204685Cat1Ch1
Holiday = Is the day a holiday = Not a Holiday|DayOfWeek = Day of week value =
Sunday|DepartureTime = Departure time value = Midday|StartingArea = Area to start the trip from
= Paris core
values:          DayOfWeek = 'false', StartingArea = 'Core', DepartureTime = '10', Holiday = 'false'
```

**Test Case 12:**
```
choices:     1538408392Cat1Ch1−|−1273415943Cat1Ch1−|−633365953Cat1Ch5−|−310204685Cat1Ch3
```

```
Holiday = Is the day a holiday = Holiday|DayOfWeek = Day of week value = Week day
(incl.Saturday)|DepartureTime = Departure time value = Night|StartingArea = Area to start the
trip from = Outside paris suburb
values:        DayOfWeek = 'true', StartingArea = 'Outside', DepartureTime = '19', Holiday =
               'true'
```

<u>**Test Case 13:**</u>
```
choices:       1538408392Cat1Ch1-|-1273415943Cat1Ch1-|-633365953Cat1Ch2-|-310204685Cat1Ch3
Holiday = Is the day a holiday = Holiday|DayOfWeek = Day of week value = Week day
(incl.Saturday)|DepartureTime = Departure time value = Morning|StartingArea = Area to start the
trip from = Outside paris suburb
values:        DayOfWeek = 'true', StartingArea = 'Outside', DepartureTime = '7', Holiday = 'true'
```

<u>**Test Case 14:**</u>
```
choices:       1538408392Cat1Ch2-|-1273415943Cat1Ch2-|-633365953Cat1Ch4-|-310204685Cat1Ch3
Holiday = Is the day a holiday = Not a Holiday|DayOfWeek = Day of week value =
Sunday|DepartureTime = Departure time value = Early evening|StartingArea = Area to start the
trip from = Outside paris suburb
values:        DayOfWeek = 'false', StartingArea = 'Outside', DepartureTime = '17', Holiday =
               'false'
```

<u>**Test Case 15:**</u>
```
choices:       1538408392Cat1Ch1-|-1273415943Cat1Ch2-|-633365953Cat1Ch4-|-310204685Cat1Ch1
Holiday = Is the day a holiday = Holiday|DayOfWeek = Day of week value = Sunday|DepartureTime =
Departure time value = Early evening|StartingArea = Area to start the trip from = Paris core
values:        DayOfWeek = 'false', StartingArea = 'Core', DepartureTime = '17', Holiday = 'true'
```

<u>**Test Case 16:**</u>
```
choices:       1538408392Cat1Ch1-|-1273415943Cat1Ch1-|-633365953Cat1Ch6-|-310204685Cat1Ch1
Holiday = Is the day a holiday = Holiday|DayOfWeek = Day of week value = Week day
(incl.Saturday)|DepartureTime = Departure time value = Illegal|StartingArea = Area to start the
trip from = Paris core
values:        DayOfWeek = 'true', StartingArea = 'Core', DepartureTime = '-1', Holiday = 'true'
```

<u>**Test Case 17:**</u>
```
choices:       1538408392Cat1Ch2-|-1273415943Cat1Ch1-|-633365953Cat1Ch1-|-310204685Cat1Ch2
Holiday = Is the day a holiday = Not a Holiday|DayOfWeek = Day of week value = Week day
(incl.Saturday)|DepartureTime = Departure time value = Early morning|StartingArea = Area to
start the trip from = Paris suburb
values:        DayOfWeek = 'true', StartingArea = 'Suburb', DepartureTime = '0', Holiday = 'false'
```

<u>**Test Case 18:**</u>
```
choices:       1538408392Cat1Ch1-|-1273415943Cat1Ch1-|-633365953Cat1Ch5-|-310204685Cat1Ch2
Holiday = Is the day a holiday = Holiday|DayOfWeek = Day of week value = Week day
(incl.Saturday)|DepartureTime = Departure time value = Night|StartingArea = Area to start the
trip from = Paris suburb
values:        DayOfWeek = 'true', StartingArea = 'Suburb', DepartureTime = '19', Holiday = 'true'
```

## *H.6    Pairwise Test Frames and Corresponding Test Cases when using ACTS*

<u>**Test Case 1:**</u>
```
choices:       633365953Cat1Ch1-|-310204685Cat1Ch1-|-1538408392Cat1Ch2-|-1273415943Cat1Ch2
DepartureTime = Departure time value = Early morning|StartingArea = Area to start the trip from
= Paris core|Holiday = Is the day a holiday = Not a Holiday|DayOfWeek = Day of week value =
Sunday
values:        DayOfWeek = 'false', DepartureTime = '0', StartingArea = 'Core', Holiday = 'false'
```

<u>**Test Case 2:**</u>
```
choices:       633365953Cat1Ch1-|-310204685Cat1Ch2-|-1538408392Cat1Ch1-|-1273415943Cat1Ch1
DepartureTime = Departure time value = Early morning|StartingArea = Area to start the trip from
= Paris suburb|Holiday = Is the day a holiday = Holiday|DayOfWeek = Day of week value = Week day
(incl.Saturday)
values:        DayOfWeek = 'true', StartingArea = 'Suburb', DepartureTime = '0', Holiday = 'true'
```

<u>**Test Case 3:**</u>
```
choices:       633365953Cat1Ch1-|-310204685Cat1Ch3-|-1538408392Cat1Ch2-|-1273415943Cat1Ch1
DepartureTime = Departure time value = Early morning|StartingArea = Area to start the trip from
= Outside paris suburb|Holiday = Is the day a holiday = Not a Holiday|DayOfWeek = Day of week
value = Week day (incl.Saturday)
values:        DayOfWeek = 'true', StartingArea = 'Outside', DepartureTime = '0', Holiday =
               'false'
```

<u>**Test Case 4:**</u>
```
choices:       633365953Cat1Ch2-|-310204685Cat1Ch1-|-1538408392Cat1Ch1-|-1273415943Cat1Ch2
```

DepartureTime = Departure time value = Morning|StartingArea = Area to start the trip from =
Paris core|Holiday = Is the day a holiday = Holiday|DayOfWeek = Day of week value = Sunday
values:        DayOfWeek = 'false', StartingArea = 'Core', DepartureTime = '7', Holiday = 'true'
**Test Case 5:**
choices:       633365953Cat1Ch2—|—310204685Cat1Ch2—|—1538408392Cat1Ch2—|—1273415943Cat1Ch1
DepartureTime = Departure time value = Morning|StartingArea = Area to start the trip from =
Paris suburb|Holiday = Is the day a holiday = Not a Holiday|DayOfWeek = Day of week value = Week
day (incl.Saturday)
values:        DayOfWeek = 'true', StartingArea = 'Suburb', DepartureTime = '7', Holiday = 'false'
**Test Case 6:**
choices:       633365953Cat1Ch2—|—310204685Cat1Ch3—|—1538408392Cat1Ch1—|—1273415943Cat1Ch2
DepartureTime = Departure time value = Morning|StartingArea = Area to start the trip from =
Outside paris suburb|Holiday = Is the day a holiday = Holiday|DayOfWeek = Day of week value =
Sunday
values:        DayOfWeek = 'false', StartingArea = 'Outside', DepartureTime = '7', Holiday =
               'true'
**Test Case 7:**
choices:       633365953Cat1Ch3—|—310204685Cat1Ch1—|—1538408392Cat1Ch2—|—1273415943Cat1Ch1
DepartureTime = Departure time value = Midday|StartingArea = Area to start the trip from = Paris
core|Holiday = Is the day a holiday = Not a Holiday|DayOfWeek = Day of week value = Week day
(incl.Saturday)
values:        DayOfWeek = 'true', StartingArea = 'Core', DepartureTime = '10', Holiday = 'false'
**Test Case 8:**
choices:       633365953Cat1Ch3—|—310204685Cat1Ch2—|—1538408392Cat1Ch1—|—1273415943Cat1Ch2
DepartureTime = Departure time value = Midday|StartingArea = Area to start the trip from = Paris
suburb|Holiday = Is the day a holiday = Holiday|DayOfWeek = Day of week value = Sunday
values:        DayOfWeek = 'false', Holiday = 'true', StartingArea = 'Suburb', DepartureTime =
               '10'
**Test Case 9:**
choices:       633365953Cat1Ch3—|—310204685Cat1Ch3—|—1538408392Cat1Ch2—|—1273415943Cat1Ch2
DepartureTime = Departure time value = Midday|StartingArea = Area to start the trip from =
Outside paris suburb|Holiday = Is the day a holiday = Not a Holiday|DayOfWeek = Day of week
value = Sunday
values:        DayOfWeek = 'false', StartingArea = 'Outside', DepartureTime = '10', Holiday =
               'false'
**Test Case 10:**
choices:       633365953Cat1Ch4—|—310204685Cat1Ch1—|—1538408392Cat1Ch2—|—1273415943Cat1Ch1
DepartureTime = Departure time value = Early evening|StartingArea = Area to start the trip from
= Paris core|Holiday = Is the day a holiday = Not a Holiday|DayOfWeek = Day of week value = Week
day (incl.Saturday)
values:        DayOfWeek = 'true', StartingArea = 'Core', DepartureTime = '17', Holiday = 'false'
**Test Case 11:**
choices:       633365953Cat1Ch4—|—310204685Cat1Ch2—|—1538408392Cat1Ch1—|—1273415943Cat1Ch2
DepartureTime = Departure time value = Early evening|StartingArea = Area to start the trip from
= Paris suburb|Holiday = Is the day a holiday = Holiday|DayOfWeek = Day of week value = Sunday
values:        DayOfWeek = 'false', StartingArea = 'Suburb', DepartureTime = '17', Holiday =
               'true'
**Test Case 12:**
choices:       633365953Cat1Ch4—|—310204685Cat1Ch3—|—1538408392Cat1Ch1—|—1273415943Cat1Ch2
DepartureTime = Departure time value = Early evening|StartingArea = Area to start the trip from
= Outside paris suburb|Holiday = Is the day a holiday = Holiday|DayOfWeek = Day of week value =
Sunday
values:        DayOfWeek = 'false', StartingArea = 'Outside', DepartureTime = '17', Holiday =
               'true'
**Test Case 13:**
choices:       633365953Cat1Ch5—|—310204685Cat1Ch1—|—1538408392Cat1Ch2—|—1273415943Cat1Ch1
DepartureTime = Departure time value = Night|StartingArea = Area to start the trip from = Paris
core|Holiday = Is the day a holiday = Not a Holiday|DayOfWeek = Day of week value = Week day
(incl.Saturday)
values:        DayOfWeek = 'true', StartingArea = 'Core', DepartureTime = '19', Holiday = 'false'
**Test Case 14:**
choices:       633365953Cat1Ch5—|—310204685Cat1Ch2—|—1538408392Cat1Ch1—|—1273415943Cat1Ch2
DepartureTime = Departure time value = Night|StartingArea = Area to start the trip from = Paris
suburb|Holiday = Is the day a holiday = Holiday|DayOfWeek = Day of week value = Sunday

```
values:      DayOfWeek = 'false', StartingArea = 'Suburb', DepartureTime = '19', Holiday =
             'true'
```

**Test Case 15:**
```
choices:    633365953Cat1Ch5-|—310204685Cat1Ch3-|—1538408392Cat1Ch2-|—1273415943Cat1Ch1
DepartureTime = Departure time value = Night|StartingArea = Area to start the trip from =
Outside paris suburb|Holiday = Is the day a holiday = Not a Holiday|DayOfWeek = Day of week
value = Week day (incl.Saturday)
values:      DayOfWeek = 'true', StartingArea = 'Outside', DepartureTime = '19', Holiday =
             'false'
```

**Test Case 16:**
```
choices:    633365953Cat1Ch6-|—310204685Cat1Ch1-|—1538408392Cat1Ch2-|—1273415943Cat1Ch1
DepartureTime = Departure time value = Illegal|StartingArea = Area to start the trip from =
Paris core|Holiday = Is the day a holiday = Not a Holiday|DayOfWeek = Day of week value = Week
day (incl.Saturday)
values:      DayOfWeek = 'true', StartingArea = 'Core', DepartureTime = '-1', Holiday = 'false'
```

**Test Case 17:**
```
choices:    633365953Cat1Ch6-|—310204685Cat1Ch2-|—1538408392Cat1Ch2-|—1273415943Cat1Ch2
DepartureTime = Departure time value = Illegal|StartingArea = Area to start the trip from =
Paris suburb|Holiday = Is the day a holiday = Holiday|DayOfWeek = Day of week value = Sunday
values:      DayOfWeek = 'false', StartingArea = 'Suburb', DepartureTime = '-1', Holiday =
             'true'
```

**Test Case 18:**
```
choices:    633365953Cat1Ch6-|—310204685Cat1Ch3-|—1538408392Cat1Ch2-|—1273415943Cat1Ch1
DepartureTime = Departure time value = Illegal|StartingArea = Area to start the trip from =
Outside paris suburb|Holiday = Is the day a holiday = Not a Holiday|DayOfWeek = Day of week
value = Week day (incl.Saturday)
values:      DayOfWeek = 'true', StartingArea = 'Outside', DepartureTime = '-1', Holiday = 'false'
```

## *H.7*    *Three-way Test Frames and Corresponding Test Cases when using CASA*

**Test Case 1:**
```
choices:    1538408392Cat1Ch1-|—1273415943Cat1Ch1-|—633365953Cat1Ch3-|—310204685Cat1Ch2
Holiday = Is the day a holiday = Holiday|DayOfWeek = Day of week value = Week day
(incl.Saturday)|DepartureTime = Departure time value = Midday|StartingArea = Area to start the
trip from = Paris suburb
values:      DayOfWeek = 'true', DepartureTime = '10', StartingArea = 'Suburb', Holiday = 'true'
```

**Test Case 2:**
```
choices:    1538408392Cat1Ch1-|—1273415943Cat1Ch2-|—633365953Cat1Ch1-|—310204685Cat1Ch2
Holiday = Is the day a holiday = Holiday|DayOfWeek = Day of week value = Sunday|DepartureTime =
Departure time value = Early morning|StartingArea = Area to start the trip from = Paris suburb
values:      DayOfWeek = 'false', StartingArea = 'Suburb', DepartureTime = '0', Holiday = 'true'
```

**Test Case 3:**
```
choices:    1538408392Cat1Ch2-|—1273415943Cat1Ch1-|—633365953Cat1Ch1-|—310204685Cat1Ch2
Holiday = Is the day a holiday = Not a Holiday|DayOfWeek = Day of week value = Week day
(incl.Saturday)|DepartureTime = Departure time value = Early morning|StartingArea = Area to
start the trip from = Paris suburb
values:      DayOfWeek = 'true', StartingArea = 'Suburb', DepartureTime = '0', Holiday = 'false'
```

**Test Case 4:**
```
choices:    1538408392Cat1Ch1-|—1273415943Cat1Ch1-|—633365953Cat1Ch6-|—310204685Cat1Ch3
Holiday = Is the day a holiday = Holiday|DayOfWeek = Day of week value = Week day
(incl.Saturday)|DepartureTime = Departure time value = Illegal|StartingArea = Area to start the
trip from = Outside paris suburb
values:      DayOfWeek = 'true', StartingArea = 'Outside', DepartureTime = '-1', Holiday =
             'true'
```

**Test Case 5:**
```
choices:    1538408392Cat1Ch1-|—1273415943Cat1Ch2-|—633365953Cat1Ch2-|—310204685Cat1Ch2
Holiday = Is the day a holiday = Holiday|DayOfWeek = Day of week value = Sunday|DepartureTime =
Departure time value = Morning|StartingArea = Area to start the trip from = Paris suburb
values:      DayOfWeek = 'false', StartingArea = 'Suburb', DepartureTime = '7', Holiday = 'true'
```

**Test Case 6:**
```
choices:    1538408392Cat1Ch2-|—1273415943Cat1Ch2-|—633365953Cat1Ch5-|—310204685Cat1Ch3
Holiday = Is the day a holiday = Not a Holiday|DayOfWeek = Day of week value =
Sunday|DepartureTime = Departure time value = Night|StartingArea = Area to start the trip from =
Outside paris suburb
```

```
values:       DayOfWeek = 'false', StartingArea = 'Outside', DepartureTime = '19', Holiday =
              'false'
```

**Test Case 7:**
```
choices:      1538408392Cat1Ch2−|−1273415943Cat1Ch2−|−633365953Cat1Ch6−|−310204685Cat1Ch3
Holiday = Is the day a holiday = Not a Holiday|DayOfWeek = Day of week value =
Sunday|DepartureTime = Departure time value = Illegal|StartingArea = Area to start the trip from
= Outside paris suburb
values:       DayOfWeek = 'false', StartingArea = 'Outside', DepartureTime = '-1', Holiday =
              'false'
```

**Test Case 8:**
```
choices:      1538408392Cat1Ch2−|−1273415943Cat1Ch1−|−633365953Cat1Ch2−|−310204685Cat1Ch2
Holiday = Is the day a holiday = Not a Holiday|DayOfWeek = Day of week value = Week day
(incl.Saturday)|DepartureTime = Departure time value = Morning|StartingArea = Area to start the
trip from = Paris suburb
values:       DayOfWeek = 'true', StartingArea = 'Suburb', DepartureTime = '7', Holiday = 'false'
```

**Test Case 9:**
```
choices:      1538408392Cat1Ch1−|−1273415943Cat1Ch1−|−633365953Cat1Ch5−|−310204685Cat1Ch3
Holiday = Is the day a holiday = Holiday|DayOfWeek = Day of week value = Week day
(incl.Saturday)|DepartureTime = Departure time value = Night|StartingArea = Area to start the
trip from = Outside paris suburb
values:       DayOfWeek = 'true', StartingArea = 'Outside', DepartureTime = '19', Holiday =
              'true'
```

**Test Case 10:**
```
choices:      1538408392Cat1Ch2−|−1273415943Cat1Ch2−|−633365953Cat1Ch3−|−310204685Cat1Ch2
Holiday = Is the day a holiday = Not a Holiday|DayOfWeek = Day of week value =
Sunday|DepartureTime = Departure time value = Midday|StartingArea = Area to start the trip from
= Paris suburb
values:       DayOfWeek = 'false', StartingArea = 'Suburb', DepartureTime = '10', Holiday =
              'false'
```

**Test Case 11:**
```
choices:      1538408392Cat1Ch1−|−1273415943Cat1Ch1−|−633365953Cat1Ch2−|−310204685Cat1Ch3
Holiday = Is the day a holiday = Holiday|DayOfWeek = Day of week value = Week day
(incl.Saturday)|DepartureTime = Departure time value = Morning|StartingArea = Area to start the
trip from = Outside paris suburb
values:       DayOfWeek = 'true', StartingArea = 'Outside', DepartureTime = '7', Holiday = 'true'
```

**Test Case 12:**
```
choices:      1538408392Cat1Ch1−|−1273415943Cat1Ch2−|−633365953Cat1Ch3−|−310204685Cat1Ch3
Holiday = Is the day a holiday = Holiday|DayOfWeek = Day of week value = Sunday|DepartureTime =
Departure time value = Midday|StartingArea = Area to start the trip from = Outside paris suburb
values:       DayOfWeek = 'false', StartingArea = 'Outside', DepartureTime = '10', Holiday =
              'true'
```

**Test Case 13:**
```
choices:      1538408392Cat1Ch1−|−1273415943Cat1Ch2−|−633365953Cat1Ch5−|−310204685Cat1Ch2
Holiday = Is the day a holiday = Holiday|DayOfWeek = Day of week value = Sunday|DepartureTime =
Departure time value = Night|StartingArea = Area to start the trip from = Paris suburb
values:       DayOfWeek = 'false', StartingArea = 'Suburb', DepartureTime = '19', Holiday =
              'true'
```

**Test Case 14:**
```
choices:      1538408392Cat1Ch1−|−1273415943Cat1Ch1−|−633365953Cat1Ch1−|−310204685Cat1Ch1
Holiday = Is the day a holiday = Holiday|DayOfWeek = Day of week value = Week day
(incl.Saturday)|DepartureTime = Departure time value = Early morning|StartingArea = Area to
start the trip from = Paris core
values:       DayOfWeek = 'true', StartingArea = 'Core', DepartureTime = '0', Holiday = 'true'
```

**Test Case 15:**
```
choices:      1538408392Cat1Ch2−|−1273415943Cat1Ch2−|−633365953Cat1Ch2−|−310204685Cat1Ch1
Holiday = Is the day a holiday = Not a Holiday|DayOfWeek = Day of week value =
Sunday|DepartureTime = Departure time value = Morning|StartingArea = Area to start the trip from
= Paris core
values:       DayOfWeek = 'false', StartingArea = 'Core', DepartureTime = '7', Holiday = 'false'
```

**Test Case 16:**
```
choices:      1538408392Cat1Ch2−|−1273415943Cat1Ch2−|−633365953Cat1Ch1−|−310204685Cat1Ch1
Holiday = Is the day a holiday = Not a Holiday|DayOfWeek = Day of week value =
Sunday|DepartureTime = Departure time value = Early morning|StartingArea = Area to start the
trip from = Paris core
values:       DayOfWeek = 'false', StartingArea = 'Core', DepartureTime = '0', Holiday = 'false'
```

**Test Case 17:**
choices:      1538408392Cat1Ch2−|−1273415943Cat1Ch1−|−633365953Cat1Ch6−|−310204685Cat1Ch2
Holiday = Is the day a holiday = Not a Holiday|DayOfWeek = Day of week value = Week day
(incl.Saturday)|DepartureTime = Departure time value = Illegal|StartingArea = Area to start the
trip from = Paris suburb
values:       DayOfWeek = 'true', StartingArea = 'Suburb', DepartureTime = '-1', Holiday =
              'false'

**Test Case 18:**
choices:      1538408392Cat1Ch1−|−1273415943Cat1Ch2−|−633365953Cat1Ch6−|−310204685Cat1Ch2
Holiday = Is the day a holiday = Holiday|DayOfWeek = Day of week value = Sunday|DepartureTime =
Departure time value = Illegal|StartingArea = Area to start the trip from = Paris suburb
values:       DayOfWeek = 'false', StartingArea = 'Suburb', DepartureTime = '-1', Holiday =
              'true'

**Test Case 19:**
choices:      1538408392Cat1Ch2−|−1273415943Cat1Ch2−|−633365953Cat1Ch2−|−310204685Cat1Ch3
Holiday = Is the day a holiday = Not a Holiday|DayOfWeek = Day of week value =
Sunday|DepartureTime = Departure time value = Morning|StartingArea = Area to start the trip from
= Outside paris suburb
values:       Holiday = 'false', DayOfWeek = 'false', StartingArea = 'Outside', DepartureTime =
              '7'

**Test Case 20:**
choices:      1538408392Cat1Ch1−|−1273415943Cat1Ch2−|−633365953Cat1Ch5−|−310204685Cat1Ch1
Holiday = Is the day a holiday = Holiday|DayOfWeek = Day of week value = Sunday|DepartureTime =
Departure time value = Night|StartingArea = Area to start the trip from = Paris core
values:       DayOfWeek = 'false', StartingArea = 'Core', DepartureTime = '19', Holiday = 'true'

**Test Case 21:**
choices:      1538408392Cat1Ch1−|−1273415943Cat1Ch1−|−633365953Cat1Ch1−|−310204685Cat1Ch3
Holiday = Is the day a holiday = Holiday|DayOfWeek = Day of week value = Week day
(incl.Saturday)|DepartureTime = Departure time value = Early morning|StartingArea = Area to
start the trip from = Outside paris suburb
values:       DayOfWeek = 'true', StartingArea = 'Outside', DepartureTime = '0', Holiday = 'true'

**Test Case 22:**
choices:      1538408392Cat1Ch1−|−1273415943Cat1Ch1−|−633365953Cat1Ch2−|−310204685Cat1Ch1
Holiday = Is the day a holiday = Holiday|DayOfWeek = Day of week value = Week day
(incl.Saturday)|DepartureTime = Departure time value = Morning|StartingArea = Area to start the
trip from = Paris core
values:       DayOfWeek = 'true', StartingArea = 'Core', DepartureTime = '7', Holiday = 'true'

**Test Case 23:**
choices:      1538408392Cat1Ch2−|−1273415943Cat1Ch1−|−633365953Cat1Ch5−|−310204685Cat1Ch1
Holiday = Is the day a holiday = Not a Holiday|DayOfWeek = Day of week value = Week day
(incl.Saturday)|DepartureTime = Departure time value = Night|StartingArea = Area to start the
trip from = Paris core
values:       DayOfWeek = 'true', StartingArea = 'Core', DepartureTime = '19', Holiday = 'false'

**Test Case 24:**
choices:      1538408392Cat1Ch2−|−1273415943Cat1Ch1−|−633365953Cat1Ch5−|−310204685Cat1Ch2
Holiday = Is the day a holiday = Not a Holiday|DayOfWeek = Day of week value = Week day
(incl.Saturday)|DepartureTime = Departure time value = Night|StartingArea = Area to start the
trip from = Paris suburb
values:       DayOfWeek = 'true', StartingArea = 'Suburb', DepartureTime = '19', Holiday =
              'false'

**Test Case 25:**
choices:      1538408392Cat1Ch1−|−1273415943Cat1Ch2−|−633365953Cat1Ch3−|−310204685Cat1Ch1
Holiday = Is the day a holiday = Holiday|DayOfWeek = Day of week value = Sunday|DepartureTime =
Departure time value = Midday|StartingArea = Area to start the trip from = Paris core
values:       DayOfWeek = 'false', StartingArea = 'Core', DepartureTime = '10', Holiday = 'true'

**Test Case 26:**
choices:      1538408392Cat1Ch2−|−1273415943Cat1Ch1−|−633365953Cat1Ch3−|−310204685Cat1Ch1
Holiday = Is the day a holiday = Not a Holiday|DayOfWeek = Day of week value = Week day
(incl.Saturday)|DepartureTime = Departure time value = Midday|StartingArea = Area to start the
trip from = Paris core
values:       DayOfWeek = 'true', Holiday = 'false', StartingArea = 'Core', DepartureTime = '10'

**Test Case 27:**
choices:      1538408392Cat1Ch2−|−1273415943Cat1Ch1−|−633365953Cat1Ch6−|−310204685Cat1Ch1

```
Holiday = Is the day a holiday = Not a Holiday|DayOfWeek = Day of week value = Week day
(incl.Saturday)|DepartureTime = Departure time value = Illegal|StartingArea = Area to start the
trip from = Paris core
values:        DayOfWeek = 'true', StartingArea = 'Core', DepartureTime = '-1', Holiday = 'false'
```
**Test Case 28:**
```
choices:        1538408392Cat1Ch1-|−1273415943Cat1Ch2-|−633365953Cat1Ch6-|−310204685Cat1Ch1
Holiday = Is the day a holiday = Holiday|DayOfWeek = Day of week value = Sunday|DepartureTime =
Departure time value = Illegal|StartingArea = Area to start the trip from = Paris core
values:        DayOfWeek = 'false', StartingArea = 'Core', DepartureTime = '-1', Holiday = 'true'
```
**Test Case 29:**
```
choices:        1538408392Cat1Ch2-|−1273415943Cat1Ch1-|−633365953Cat1Ch3-|−310204685Cat1Ch3
Holiday = Is the day a holiday = Not a Holiday|DayOfWeek = Day of week value = Week day
(incl.Saturday)|DepartureTime = Departure time value = Midday|StartingArea = Area to start the
trip from = Outside paris suburb
values:        DayOfWeek = 'true', StartingArea = 'Outside', DepartureTime = '10', Holiday =
               'false'
```
**Test Case 30:**
```
choices:        1538408392Cat1Ch1-|−1273415943Cat1Ch2-|−633365953Cat1Ch4-|−310204685Cat1Ch1
Holiday = Is the day a holiday = Holiday|DayOfWeek = Day of week value = Sunday|DepartureTime =
Departure time value = Early evening|StartingArea = Area to start the trip from = Paris core
values:        Holiday = 'true', DayOfWeek = 'false', StartingArea = 'Core', DepartureTime = '17'
```
**Test Case 31:**
```
choices:        1538408392Cat1Ch2-|−1273415943Cat1Ch1-|−633365953Cat1Ch4-|−310204685Cat1Ch1
Holiday = Is the day a holiday = Not a Holiday|DayOfWeek = Day of week value = Week day
(incl.Saturday)|DepartureTime = Departure time value = Early evening|StartingArea = Area to
start the trip from = Paris core
values:        DayOfWeek = 'true', StartingArea = 'Core', DepartureTime = '17', Holiday = 'false'
```
**Test Case 32:**
```
choices:        1538408392Cat1Ch2-|−1273415943Cat1Ch2-|−633365953Cat1Ch1-|−310204685Cat1Ch3
Holiday = Is the day a holiday = Not a Holiday|DayOfWeek = Day of week value =
Sunday|DepartureTime = Departure time value = Early morning|StartingArea = Area to start the
trip from = Outside paris suburb
values:        DayOfWeek = 'false', StartingArea = 'Outside', DepartureTime = '0', Holiday =
               'false'
```
**Test Case 33:**
```
choices:        1538408392Cat1Ch2-|−1273415943Cat1Ch2-|−633365953Cat1Ch4-|−310204685Cat1Ch2
Holiday = Is the day a holiday = Not a Holiday|DayOfWeek = Day of week value =
Sunday|DepartureTime = Departure time value = Early evening|StartingArea = Area to start the
trip from = Paris suburb
values:        DayOfWeek = 'false', StartingArea = 'Suburb', DepartureTime = '17', Holiday =
               'false'
```
**Test Case 34:**
```
choices:        1538408392Cat1Ch1-|−1273415943Cat1Ch1-|−633365953Cat1Ch4-|−310204685Cat1Ch2
Holiday = Is the day a holiday = Holiday|DayOfWeek = Day of week value = Week day
(incl.Saturday)|DepartureTime = Departure time value = Early evening|StartingArea = Area to
start the trip from = Paris suburb
values:        DayOfWeek = 'true', StartingArea = 'Suburb', DepartureTime = '17', Holiday = 'true'
```
**Test Case 35:**
```
choices:        1538408392Cat1Ch2-|−1273415943Cat1Ch1-|−633365953Cat1Ch4-|−310204685Cat1Ch3
Holiday = Is the day a holiday = Not a Holiday|DayOfWeek = Day of week value = Week day
(incl.Saturday)|DepartureTime = Departure time value = Early evening|StartingArea = Area to
start the trip from = Outside paris suburb
values:        DayOfWeek = 'true', StartingArea = 'Outside', DepartureTime = '17', Holiday =
               'false'
```
**Test Case 36:**
```
choices:        1538408392Cat1Ch1-|−1273415943Cat1Ch2-|−633365953Cat1Ch4-|−310204685Cat1Ch3
Holiday = Is the day a holiday = Holiday|DayOfWeek = Day of week value = Sunday|DepartureTime =
Departure time value = Early evening|StartingArea = Area to start the trip from = Outside paris
suburb
values:        DayOfWeek = 'false', StartingArea = 'Outside', DepartureTime = '17', Holiday = 'true'
```

## H.8    Three-way Test Frames and Corresponding Test Cases when using ACTS

**Test Case 1:**
```
choices:        633365953Cat1Ch1-|−310204685Cat1Ch1-|−1538408392Cat1Ch1-|−1273415943Cat1Ch1
```

DepartureTime = Departure time value = Early morning|StartingArea = Area to start the trip from = Paris core|Holiday = Is the day a holiday = Holiday|DayOfWeek = Day of week value = Week day (incl.Saturday)
values:      DayOfWeek = 'true', DepartureTime = '0', StartingArea = 'Core', Holiday = 'true'

**Test Case 2:**
choices:     633365953Cat1Ch1−|−310204685Cat1Ch1−|−1538408392Cat1Ch2−|−1273415943Cat1Ch2
DepartureTime = Departure time value = Early morning|StartingArea = Area to start the trip from = Paris core|Holiday = Is the day a holiday = Not a Holiday|DayOfWeek = Day of week value = Sunday
values:      DayOfWeek = 'false', StartingArea = 'Core', DepartureTime = '0', Holiday = 'false'

**Test Case 3:**
choices:     633365953Cat1Ch1−|−310204685Cat1Ch2−|−1538408392Cat1Ch1−|−1273415943Cat1Ch2
DepartureTime = Departure time value = Early morning|StartingArea = Area to start the trip from = Paris suburb|Holiday = Is the day a holiday = Holiday|DayOfWeek = Day of week value = Sunday
values:      DayOfWeek = 'false', StartingArea = 'Suburb', DepartureTime = '0', Holiday = 'true'

**Test Case 4:**
choices:     633365953Cat1Ch1−|−310204685Cat1Ch2−|−1538408392Cat1Ch2−|−1273415943Cat1Ch1
DepartureTime = Departure time value = Early morning|StartingArea = Area to start the trip from = Paris suburb|Holiday = Is the day a holiday = Not a Holiday|DayOfWeek = Day of week value = Week day (incl.Saturday)
values:      DayOfWeek = 'true', StartingArea = 'Suburb', DepartureTime = '0', Holiday = 'false'

**Test Case 5:**
choices:     633365953Cat1Ch1−|−310204685Cat1Ch3−|−1538408392Cat1Ch1−|−1273415943Cat1Ch1
DepartureTime = Departure time value = Early morning|StartingArea = Area to start the trip from = Outside paris suburb|Holiday = Is the day a holiday = Holiday|DayOfWeek = Day of week value = Week day (incl.Saturday)
values:      DayOfWeek = 'true', StartingArea = 'Outside', DepartureTime = '0', Holiday = 'true'

**Test Case 6:**
choices:     633365953Cat1Ch1−|−310204685Cat1Ch3−|−1538408392Cat1Ch2−|−1273415943Cat1Ch2
DepartureTime = Departure time value = Early morning|StartingArea = Area to start the trip from = Outside paris suburb|Holiday = Is the day a holiday = Not a Holiday|DayOfWeek = Day of week value = Sunday
values:      DayOfWeek = 'false', StartingArea = 'Outside', DepartureTime = '0', Holiday = 'false'

**Test Case 7:**
choices:     633365953Cat1Ch2−|−310204685Cat1Ch1−|−1538408392Cat1Ch1−|−1273415943Cat1Ch2
DepartureTime = Departure time value = Morning|StartingArea = Area to start the trip from = Paris core|Holiday = Is the day a holiday = Holiday|DayOfWeek = Day of week value = Sunday
values:      DayOfWeek = 'false', StartingArea = 'Core', DepartureTime = '7', Holiday = 'true'

**Test Case 8:**
choices:     633365953Cat1Ch2−|−310204685Cat1Ch1−|−1538408392Cat1Ch1−|−1273415943Cat1Ch1
DepartureTime = Departure time value = Morning|StartingArea = Area to start the trip from = Paris core|Holiday = Is the day a holiday = Not a Holiday|DayOfWeek = Day of week value = Week day (incl.Saturday)
values:      DayOfWeek = 'true', Holiday = 'false', StartingArea = 'Core', DepartureTime = '7'

**Test Case 9:**
choices:     633365953Cat1Ch2−|−310204685Cat1Ch2−|−1538408392Cat1Ch1−|−1273415943Cat1Ch1
DepartureTime = Departure time value = Morning|StartingArea = Area to start the trip from = Paris suburb|Holiday = Is the day a holiday = Holiday|DayOfWeek = Day of week value = Week day (incl.Saturday)
values:      DayOfWeek = 'true', StartingArea = 'Suburb', DepartureTime = '7', Holiday = 'true'

**Test Case 10:**
choices:     633365953Cat1Ch2−|−310204685Cat1Ch2−|−1538408392Cat1Ch2−|−1273415943Cat1Ch2
DepartureTime = Departure time value = Morning|StartingArea = Area to start the trip from = Paris suburb|Holiday = Is the day a holiday = Not a Holiday|DayOfWeek = Day of week value = Sunday
values:      DayOfWeek = 'false', StartingArea = 'Suburb', DepartureTime = '7', Holiday = 'false'

**Test Case 11:**
choices:     633365953Cat1Ch2−|−310204685Cat1Ch3−|−1538408392Cat1Ch1−|−1273415943Cat1Ch2
DepartureTime = Departure time value = Morning|StartingArea = Area to start the trip from = Outside paris suburb|Holiday = Is the day a holiday = Holiday|DayOfWeek = Day of week value = Sunday
values:      DayOfWeek = 'false', StartingArea = 'Outside', DepartureTime = '7', Holiday = 'true'

**Test Case 12:**
```
choices:        633365953Cat1Ch2-|-310204685Cat1Ch3-|-1538408392Cat1Ch2-|-1273415943Cat1Ch1
DepartureTime = Departure time value = Morning|StartingArea = Area to start the trip from =
Outside paris suburb|Holiday = Is the day a holiday = Not a Holiday|DayOfWeek = Day of week
value = Week day (incl.Saturday)
values:         DayOfWeek = 'true', StartingArea = 'Outside', DepartureTime = '7', Holiday =
                'false'
```
**Test Case 13:**
```
choices:        633365953Cat1Ch3-|-310204685Cat1Ch1-|-1538408392Cat1Ch1-|-1273415943Cat1Ch1
DepartureTime = Departure time value = Midday|StartingArea = Area to start the trip from = Paris
core|Holiday = Is the day a holiday = Holiday|DayOfWeek = Day of week value = Week day
(incl.Saturday)
values:         DayOfWeek = 'true', StartingArea = 'Core', DepartureTime = '10', Holiday = 'true'
```
**Test Case 14:**
```
choices:        633365953Cat1Ch3-|-310204685Cat1Ch1-|-1538408392Cat1Ch2-|-1273415943Cat1Ch2
DepartureTime = Departure time value = Midday|StartingArea = Area to start the trip from = Paris
core|Holiday = Is the day a holiday = Not a Holiday|DayOfWeek = Day of week value = Sunday
values:         DayOfWeek = 'false', StartingArea = 'Core', DepartureTime = '10', Holiday = 'false'
```
**Test Case 15:**
```
choices:        633365953Cat1Ch3-|-310204685Cat1Ch1-|-1538408392Cat1Ch1-|-1273415943Cat1Ch2
DepartureTime = Departure time value = Midday|StartingArea = Area to start the trip from = Paris
suburb|Holiday = Is the day a holiday = Holiday|DayOfWeek = Day of week value = Sunday
values:         DayOfWeek = 'false', StartingArea = 'Suburb', DepartureTime = '10', Holiday =
                'true'
```
**Test Case 16:**
```
choices:        633365953Cat1Ch3-|-310204685Cat1Ch2-|-1538408392Cat1Ch2-|-1273415943Cat1Ch1
DepartureTime = Departure time value = Midday|StartingArea = Area to start the trip from = Paris
suburb|Holiday = Is the day a holiday = Not a Holiday|DayOfWeek = Day of week value = Week day
(incl.Saturday)
values:         DayOfWeek = 'true', StartingArea = 'Suburb', DepartureTime = '10', Holiday =
                'false'
```
**Test Case 17:**
```
choices:        633365953Cat1Ch3-|-310204685Cat1Ch3-|-1538408392Cat1Ch1-|-1273415943Cat1Ch1
DepartureTime = Departure time value = Midday|StartingArea = Area to start the trip from =
Outside paris suburb|Holiday = Is the day a holiday = Holiday|DayOfWeek = Day of week value =
Week day (incl.Saturday)
values:         DayOfWeek = 'true', StartingArea = 'Outside', DepartureTime = '10', Holiday =
                'true'
```
**Test Case 18:**
```
choices:        633365953Cat1Ch3-|-310204685Cat1Ch3-|-1538408392Cat1Ch2-|-1273415943Cat1Ch2
DepartureTime = Departure time value = Midday|StartingArea = Area to start the trip from =
Outside paris suburb|Holiday = Is the day a holiday = Not a Holiday|DayOfWeek = Day of week
value = Sunday
values:         DayOfWeek = 'false', StartingArea = 'Outside', DepartureTime = '10', Holiday =
                'false'
```
**Test Case 19:**
```
choices:        633365953Cat1Ch4-|-310204685Cat1Ch1-|-1538408392Cat1Ch1-|-1273415943Cat1Ch1
DepartureTime = Departure time value = Early evening|StartingArea = Area to start the trip from
= Paris core|Holiday = Is the day a holiday = Holiday|DayOfWeek = Day of week value = Week day
(incl.Saturday)
values:         DayOfWeek = 'true', StartingArea = 'Core', DepartureTime = '17', Holiday = 'true'
```
**Test Case 20:**
```
choices:        633365953Cat1Ch4-|-310204685Cat1Ch1-|-1538408392Cat1Ch2-|-1273415943Cat1Ch2
DepartureTime = Departure time value = Early evening|StartingArea = Area to start the trip from
= Paris core|Holiday = Is the day a holiday = Not a Holiday|DayOfWeek = Day of week value =
Sunday
values:         DayOfWeek = 'false', StartingArea = 'Core', DepartureTime = '17', Holiday = 'false'
```
**Test Case 21:**
```
choices:        633365953Cat1Ch4-|-310204685Cat1Ch2-|-1538408392Cat1Ch1-|-1273415943Cat1Ch2
DepartureTime = Departure time value = Early evening|StartingArea = Area to start the trip from
= Paris suburb|Holiday = Is the day a holiday = Holiday|DayOfWeek = Day of week value = Sunday
values:         DayOfWeek = 'false', StartingArea = 'Suburb', DepartureTime = '17', Holiday =
                'true'
```
**Test Case 22:**
```
choices:        633365953Cat1Ch4-|-310204685Cat1Ch2-|-1538408392Cat1Ch2-|-1273415943Cat1Ch1
```

```
DepartureTime = Departure time value = Early evening|StartingArea = Area to start the trip from
= Paris suburb|Holiday = Is the day a holiday = Not a Holiday|DayOfWeek = Day of week value =
Week day (incl.Saturday)
values:        DayOfWeek = 'true', StartingArea = 'Suburb', DepartureTime = '17', Holiday =
               'false'
```

**Test Case 23:**
```
choices:       633365953Cat1Ch4—|—310204685Cat1Ch3—|—1538408392Cat1Ch1—|—1273415943Cat1Ch1
DepartureTime = Departure time value = Early evening|StartingArea = Area to start the trip from
= Outside paris suburb|Holiday = Is the day a holiday = Holiday|DayOfWeek = Day of week value =
Week day (incl.Saturday)
values:        DayOfWeek = 'true', StartingArea = 'Outside', DepartureTime = '17', Holiday =
               'true'
```

**Test Case 24:**
```
choices:       633365953Cat1Ch4—|—310204685Cat1Ch3—|—1538408392Cat1Ch2—|—1273415943Cat1Ch2
DepartureTime = Departure time value = Early evening|StartingArea = Area to start the trip from
= Outside paris suburb|Holiday = Is the day a holiday = Not a Holiday|DayOfWeek = Day of week
value = Sunday
values:        DayOfWeek = 'false', StartingArea = 'Outside', DepartureTime = '17', Holiday =
               'false'
```

**Test Case 25:**
```
choices:       633365953Cat1Ch5—|—310204685Cat1Ch1—|—1538408392Cat1Ch1—|—1273415943Cat1Ch1
DepartureTime = Departure time value = Night|StartingArea = Area to start the trip from = Paris
core|Holiday = Is the day a holiday = Holiday|DayOfWeek = Day of week value = Week day
(incl.Saturday)
values:        DayOfWeek = 'true', StartingArea = 'Core', DepartureTime = '19', Holiday = 'true'
```

**Test Case 26:**
```
choices:       633365953Cat1Ch5—|—310204685Cat1Ch1—|—1538408392Cat1Ch2—|—1273415943Cat1Ch2
DepartureTime = Departure time value = Night|StartingArea = Area to start the trip from = Paris
core|Holiday = Is the day a holiday = Not a Holiday|DayOfWeek = Day of week value = Sunday
values:        DayOfWeek = 'false', Holiday = 'false', StartingArea = 'Core', DepartureTime = '19'
```

**Test Case 27:**
```
choices:       633365953Cat1Ch5—|—310204685Cat1Ch2—|—1538408392Cat1Ch1—|—1273415943Cat1Ch2
DepartureTime = Departure time value = Night|StartingArea = Area to start the trip from = Paris
suburb|Holiday = Is the day a holiday = Holiday|DayOfWeek = Day of week value = Sunday
values:        DayOfWeek = 'false', StartingArea = 'Suburb', DepartureTime = '19', Holiday =
               'true'
```

**Test Case 28:**
```
choices:       633365953Cat1Ch5—|—310204685Cat1Ch2—|—1538408392Cat1Ch2—|—1273415943Cat1Ch1
DepartureTime = Departure time value = Night|StartingArea = Area to start the trip from = Paris
suburb|Holiday = Is the day a holiday = Not a Holiday|DayOfWeek = Day of week value = Week day
(incl.Saturday)
values:        DayOfWeek = 'true', StartingArea = 'Suburb', DepartureTime = '19', Holiday =
               'false'
```

**Test Case 29:**
```
choices:       633365953Cat1Ch5—|—310204685Cat1Ch3—|—1538408392Cat1Ch1—|—1273415943Cat1Ch1
DepartureTime = Departure time value = Night|StartingArea = Area to start the trip from =
Outside paris suburb|Holiday = Is the day a holiday = Holiday|DayOfWeek = Day of week value =
Week day (incl.Saturday)
values:        DayOfWeek = 'true', StartingArea = 'Outside', DepartureTime = '19', Holiday =
               'true'
```

**Test Case 30:**
```
choices:       633365953Cat1Ch5—|—310204685Cat1Ch3—|—1538408392Cat1Ch2—|—1273415943Cat1Ch2
DepartureTime = Departure time value = Night|StartingArea = Area to start the trip from =
Outside paris suburb|Holiday = Is the day a holiday = Not a Holiday|DayOfWeek = Day of week
value = Sunday
values:        Holiday = 'false', DayOfWeek = 'false', StartingArea = 'Outside', DepartureTime =
               '19'
```

**Test Case 31:**
```
choices:       633365953Cat1Ch6—|—310204685Cat1Ch1—|—1538408392Cat1Ch1—|—1273415943Cat1Ch1
DepartureTime = Departure time value = Illegal|StartingArea = Area to start the trip from =
Paris core|Holiday = Is the day a holiday = Holiday|DayOfWeek = Day of week value = Week day
(incl.Saturday)
values:        DayOfWeek = 'true', StartingArea = 'Core', DepartureTime = '-1', Holiday = 'true'
```

**Test Case 32:**
```
choices:       633365953Cat1Ch6—|—310204685Cat1Ch1—|—1538408392Cat1Ch2—|—1273415943Cat1Ch2
```

```
DepartureTime = Departure time value = Illegal|StartingArea = Area to start the trip from =
Paris core|Holiday = Is the day a holiday = Not a Holiday|DayOfWeek = Day of week value = Sunday
values:       DayOfWeek = 'false', StartingArea = 'Core', DepartureTime = '-1', Holiday = 'false'
```

**Test Case 33:**
```
choices:      633365953Cat1Ch6-|-310204685Cat1Ch2-|-1538408392Cat1Ch1-|-1273415943Cat1Ch2
DepartureTime = Departure time value = Illegal|StartingArea = Area to start the trip from =
Paris suburb|Holiday = Is the day a holiday = Holiday|DayOfWeek = Day of week value = Sunday
values:       DayOfWeek = 'false', StartingArea = 'Suburb', DepartureTime = '-1', Holiday =
              'true'
```

**Test Case 34:**
```
choices:      633365953Cat1Ch6-|-310204685Cat1Ch2-|-1538408392Cat1Ch2-|-1273415943Cat1Ch1
DepartureTime = Departure time value = Illegal|StartingArea = Area to start the trip from =
Paris suburb|Holiday = Is the day a holiday = Not a Holiday|DayOfWeek = Day of week value = Week
day (incl.Saturday)
values:       DayOfWeek = 'true', StartingArea = 'Suburb', DepartureTime = '-1', Holiday =
              'false'
```

**Test Case 35:**
```
choices:      633365953Cat1Ch6-|-310204685Cat1Ch3-|-1538408392Cat1Ch1-|-1273415943Cat1Ch1
DepartureTime = Departure time value = Illegal|StartingArea = Area to start the trip from =
Outside paris suburb|Holiday = Is the day a holiday = Holiday|DayOfWeek = Day of week value =
Week day (incl.Saturday)
values:       DayOfWeek = 'true', StartingArea = 'Outside', DepartureTime = '-1', Holiday =
              'true'
```

**Test Case 36:**
```
choices:      633365953Cat1Ch6-|-310204685Cat1Ch3-|-1538408392Cat1Ch2-|-1273415943Cat1Ch2
DepartureTime = Departure time value = Illegal|StartingArea = Area to start the trip from =
Outside paris suburb|Holiday = Is the day a holiday = Not a Holiday|DayOfWeek = Day of week
value = Sunday
values:       DayOfWeek = 'false', StartingArea = 'Outside', DepartureTime = '-1', Holiday = 'false'
```

## H.9    Base Choice Test Frames and Corresponding Test Cases

**Test Case 1:**
```
choices:      1538408392Cat1Ch2-|-1273415943Cat1Ch1-|-633365953Cat1Ch3-|-310204685Cat1Ch1
Holiday = Is the day a holiday = Not a Holiday|DayOfWeek = Day of week value = Week day
(incl.Saturday)|DepartureTime = Departure time value = Midday|StartingArea = Area to start the
trip from = Paris core
values:       DayOfWeek = 'true', DepartureTime = '10', StartingArea = 'Core', Holiday = 'false'
```

**Test Case 2:**
```
choices:      1273415943Cat1Ch1-|-633365953Cat1Ch3-|-310204685Cat1Ch1-|-1538408392Cat1Ch1
DayOfWeek = Day of week value = Week day (incl.Saturday)|DepartureTime = Departure time value =
Midday|StartingArea = Area to start the trip from = Paris core|Holiday = Is the day a holiday =
Holiday
values:       DayOfWeek = 'true', StartingArea = 'Core', DepartureTime = '10', Holiday = 'true'
```

**Test Case 3:**
```
choices:      1538408392Cat1Ch2-|-633365953Cat1Ch3-|-310204685Cat1Ch1-|-1273415943Cat1Ch2
Holiday = Is the day a holiday = Not a Holiday|DepartureTime = Departure time value =
Midday|StartingArea = Area to start the trip from = Paris core|DayOfWeek = Day of week value =
Sunday
values:       DayOfWeek = 'false', StartingArea = 'Core', DepartureTime = '10', Holiday = 'false'
```

**Test Case 4:**
```
choices:      1538408392Cat1Ch2-|-1273415943Cat1Ch1-|-310204685Cat1Ch1-|-633365953Cat1Ch1
Holiday = Is the day a holiday = Not a Holiday|DayOfWeek = Day of week value = Week day
(incl.Saturday)|StartingArea = Area to start the trip from = Paris core|DepartureTime =
Departure time value = Early morning
values:       DayOfWeek = 'true', StartingArea = 'Core', DepartureTime = '0', Holiday = 'false'
```

**Test Case 5:**
```
choices:      1538408392Cat1Ch2-|-1273415943Cat1Ch1-|-310204685Cat1Ch1-|-633365953Cat1Ch2
Holiday = Is the day a holiday = Not a Holiday|DayOfWeek = Day of week value = Week day
(incl.Saturday)|StartingArea = Area to start the trip from = Paris core|DepartureTime =
Departure time value = Morning
values:       DayOfWeek = 'true', StartingArea = 'Core', DepartureTime = '7', Holiday = 'false'
```

**Test Case 6:**
```
choices:      1538408392Cat1Ch2-|-1273415943Cat1Ch1-|-310204685Cat1Ch1-|-633365953Cat1Ch4
```

```
Holiday = Is the day a holiday = Not a Holiday|DayOfWeek = Day of week value = Week day
(incl.Saturday)|StartingArea = Area to start the trip from = Paris core|DepartureTime =
Departure time value = Early evening
values:      DayOfWeek = 'true', StartingArea = 'Core', DepartureTime = '17', Holiday = 'false'
```

**Test Case 7:**
```
choices:     1538408392Cat1Ch2−|−1273415943Cat1Ch1−|−310204685Cat1Ch1−|−633365953Cat1Ch5
Holiday = Is the day a holiday = Not a Holiday|DayOfWeek = Day of week value = Week day
(incl.Saturday)|StartingArea = Area to start the trip from = Paris core|DepartureTime =
Departure time value = Night
values:      DayOfWeek = 'true', StartingArea = 'Core', DepartureTime = '19', Holiday = 'false'
```

**Test Case 8:**
```
choices:     1538408392Cat1Ch2−|−1273415943Cat1Ch1−|−310204685Cat1Ch1−|−633365953Cat1Ch6
Holiday = Is the day a holiday = Not a Holiday|DayOfWeek = Day of week value = Week day
(incl.Saturday)|StartingArea = Area to start the trip from = Paris core|DepartureTime =
Departure time value = Illegal
values:      DayOfWeek = 'true', StartingArea = 'Core', DepartureTime = '-1', Holiday = 'false'
```

**Test Case 9:**
```
choices:     1538408392Cat1Ch2−|−1273415943Cat1Ch1−|−633365953Cat1Ch3−|−310204685Cat1Ch2
Holiday = Is the day a holiday = Not a Holiday|DayOfWeek = Day of week value = Week day
(incl.Saturday)|DepartureTime = Departure time value = Midday|StartingArea = Area to start the
trip from = Paris suburb
values:      DayOfWeek = 'true', StartingArea = 'Suburb', DepartureTime = '10', Holiday =
             'false'
```

**Test Case 10:**
```
choices:     1538408392Cat1Ch2−|−1273415943Cat1Ch1−|−633365953Cat1Ch3−|−310204685Cat1Ch3
Holiday = Is the day a holiday = Not a Holiday|DayOfWeek = Day of week value = Week day
(incl.Saturday)|DepartureTime = Departure time value = Midday|StartingArea = Area to start the
trip from = Outside paris suburb
values:      DayOfWeek = 'true', StartingArea = 'Outside', DepartureTime = '10', Holiday = 'false'
```

# Appendix I. GROCERY

In the CP specification below, an asterisk ('*') beside a choice indicates the base choice for that category. Each parameter is uniquely identified by an integer identifier, which is then used in the composition of test frames; As a result, a choice for a parameter is uniquely identified by the parameter identifier, the category identifier and the choice identifier. Below we summarize the specification of the unit under test (section I.1), and provide the CP specification we used (section I.2). We then provide the test frames and corresponding test cases (inputs) we obtain when using CASA (Each-Choice in section I.3, Pairwise in section I.5, Three-way in section I.7), when using ACTS (Each-Choice in section I.4, Pairwise in section I.6, Three-way in section I.8), and when using the Base Choice criterion (section I.9).

## I.1    Specification

The system under test simulates a cashier in a grocery store that computes the cost of a client's purchases depending on the type and quantity or weight of item being purchases. The item can be an ordinary grocery item, which can be purchased in various quantities (a quantity input is provided), an item that needs to be weighted (a weight input is provided), a health-care item (with a quantity). An item can also be a discount coupon. An item is uniquely identified by a code and a type; two items with different types can have the same code. When an ordinary or health-care item is purchased, or when a coupon is used, the weight input is ignored; when a weighted item is purchased, the quantity input is ignored.

The CP specification we produced ignores the item code; adding that parameter is simple, with likely one category and two choices specifying whether the code for this item type exists (in the database) or not.

## I.2    CP Specification

```
Parameter(1242177805): ItemType of type Enumeration
  Category 1 - Item type values
     Ch 1.1: Ordinary item
             ItemType == OrdinaryItem
             [properties O]
    *Ch 1.2: Weight item
             ItemType == WeightItem
             [properties W]
     Ch 1.3: Health item
             ItemType == HealthItem
             [properties H]
     Ch 1.4: Coupon
             ItemType == Coupon
             [single][properties C]
Parameter(1220360021): Quantity of type Integer
  Category 1 - Quantity values
     Ch 1.1: Illegal quantity
             Quantity < 0
             [if O || H][error]
     Ch 1.2: Small quantity
             Quantity>0 && Quantity<6
             [if O || H]
     Ch 1.3: Large
             Quantity >= 6
             [if O || H][single]
    *Ch 1.4: Ignored quantity
             Quantity == 0
             [if W]
Parameter(1707725160): Weight of type Real
  Category 1 - Weight values
     Ch 1.1: Illegal weight
             Weight < 0
             [if W][error]
```

```
   *Ch 1.2: Expected weight range
           Weight>0 && Weight<=10
           [if W]
    Ch 1.3: suspicious weight value
           Weight > 10
           [if W][single]
    Ch 1.4: Ignored weight
           Weight == 0
           [if O||H||C]
```

## I.3    Each-Choice Test Frames and Corresponding Test Cases when using CASA

**Test Case 1:**
```
choices:    1242177805Cat1Ch3-|-1220360021Cat1Ch2-|-1707725160Cat1Ch4
ItemType = Item type values = Health item|Quantity = Quantity values = Small quantity|Weight =
Weight values = Ignored weight
values:      Weight = '0', Quantity = '1', ItemType = 'HealthItem'
```
**Test Case 2:**
```
choices:    1242177805Cat1Ch1-|-1220360021Cat1Ch2-|-1707725160Cat1Ch4
ItemType = Item type values = Ordinary item|Quantity = Quantity values = Small quantity|Weight =
Weight values = Ignored weight
values:      Weight = '0', Quantity = '1', ItemType = 'OrdinaryItem'
```
**Test Case 3:**
```
choices:    1242177805Cat1Ch2-|-1220360021Cat1Ch4-|-1707725160Cat1Ch2
ItemType = Item type values = Weight item|Quantity = Quantity values = Ignored quantity|Weight =
Weight values = Expected weight range
values:      Weight = '1/2', Quantity = '0', ItemType = 'WeightItem'
```
**Test Case 4:**
```
choices:    1242177805Cat1Ch1-|-1220360021Cat1Ch3-|-1707725160Cat1Ch4
ItemType = Item type values = Ordinary item|Quantity = Quantity values = Large|Weight = Weight
values = Ignored weight
values:      Weight = '0', Quantity = '6', ItemType = 'OrdinaryItem'
```
**Test Case 5:**
```
choices:    1242177805Cat1Ch2-|-1220360021Cat1Ch4-|-1707725160Cat1Ch3
ItemType = Item type values = Weight item|Quantity = Quantity values = Ignored quantity|Weight =
Weight values = suspicious weight value
values:      Weight = '11', Quantity = '0', ItemType = 'WeightItem'
```
**Test Case 6:**
```
choices:    1220360021Cat1Ch1
Quantity = Quantity values = Illegal quantity
values:      Quantity = '-1'
```
**Test Case 7:**
```
choices:    1707725160Cat1Ch1
Weight = Weight values = Illegal weight
values:      Weight = '-1'
```
**Test Case 8:**
```
choices:    1242177805Cat1Ch4
ItemType = Item type values = Coupon
values:      ItemType = 'Coupon'
```

## I.4    Each-Choice Test Frames and Corresponding Test Cases when using ACTS

**Test Case 1:**
```
choices:    1242177805Cat1Ch1-|-1220360021Cat1Ch2-|-1707725160Cat1Ch4
ItemType = Item type values = Ordinary item|Quantity = Quantity values = Small quantity|Weight =
Weight values = Ignored weight
values:      Weight = '0', Quantity = '1', ItemType = 'OrdinaryItem'
```
**Test Case 2:**
```
choices:    1242177805Cat1Ch2-|-1220360021Cat1Ch4-|-1707725160Cat1Ch2
ItemType = Item type values = Weight item|Quantity = Quantity values = Ignored quantity|Weight =
Weight values = Expected weight range
values:      Weight = '1/2', Quantity = '0', ItemType = 'WeightItem'
```
**Test Case 3:**
```
choices:    1242177805Cat1Ch1-|-1220360021Cat1Ch3-|-1707725160Cat1Ch4
```

```
ItemType = Item type values = Ordinary item|Quantity = Quantity values = Large|Weight = Weight
values = Ignored weight
values:      Weight = '0', Quantity = '6', ItemType = 'OrdinaryItem'
```

**Test Case 4:**
```
choices:     1242177805Cat1Ch2-|-1220360021Cat1Ch4-|-1707725160Cat1Ch3
ItemType = Item type values = Weight item|Quantity = Quantity values = Ignored quantity|Weight =
Weight values = suspicious weight value
values:      Weight = '11', Quantity = '0', ItemType = 'WeightItem'
```

**Test Case 5:**
```
choices:     1220360021Cat1Ch1
Quantity = Quantity values = Illegal quantity
values:      Quantity = '-1'
```

**Test Case 6:**
```
choices:     1707725160Cat1Ch1
Weight = Weight values = Illegal weight
values:      Weight = '-1'
```

**Test Case 7:**
```
choices:     1242177805Cat1Ch4
ItemType = Item type values = Coupon
values:      ItemType = 'Coupon'
```

**Test Case 8:**
```
choices:     1242177805Cat1Ch3
ItemType = Item type values = Health item
values:      ItemType = 'HealthItem'
```

## I.5    Pairwise Test Frames and Corresponding Test Cases when using CASA

**Test Case 1:**
```
choices:     1242177805Cat1Ch2-|-1220360021Cat1Ch4-|-1707725160Cat1Ch4
ItemType = Item type values = Weight item|Quantity = Quantity values = Ignored quantity|Weight =
Weight values = Ignored weight
values:      Weight = '0', Quantity = '0', ItemType = 'WeightItem'
```

**Test Case 2:**
```
choices:     1242177805Cat1Ch1-|-1220360021Cat1Ch2-|-1707725160Cat1Ch4
ItemType = Item type values = Ordinary item|Quantity = Quantity values = Small quantity|Weight =
Weight values = Ignored weight
values:      Weight = '0', Quantity = '1', ItemType = 'OrdinaryItem'
```

**Test Case 3:**
```
choices:     1242177805Cat1Ch3-|-1220360021Cat1Ch2-|-1707725160Cat1Ch4
ItemType = Item type values = Health item|Quantity = Quantity values = Small quantity|Weight =
Weight values = Ignored weight
values:      Weight = '0', Quantity = '1', ItemType = 'HealthItem'
```

**Test Case 4:**
```
choices:     1242177805Cat1Ch2-|-1220360021Cat1Ch4-|-1707725160Cat1Ch2
ItemType = Item type values = Weight item|Quantity = Quantity values = Ignored quantity|Weight =
Weight values = Expected weight range
values:      Weight = '1/2', Quantity = '0', ItemType = 'WeightItem'
```

**Test Case 5:**
```
choices:     1242177805Cat1Ch1-|-1220360021Cat1Ch3-|-1707725160Cat1Ch4
ItemType = Item type values = Ordinary item|Quantity = Quantity values = Large|Weight = Weight
values = Ignored weight
values:      Weight = '0', Quantity = '6', ItemType = 'OrdinaryItem'
```

**Test Case 6:**
```
choices:     1242177805Cat1Ch2-|-1220360021Cat1Ch4-|-1707725160Cat1Ch3
ItemType = Item type values = Weight item|Quantity = Quantity values = Ignored quantity|Weight =
Weight values = suspicious weight value
values:      Weight = '11', Quantity = '0', ItemType = 'WeightItem'
```

**Test Case 7:**
```
choices:     1220360021Cat1Ch1
Quantity = Quantity values = Illegal quantity
values:      Quantity = '-1'
```

**Test Case 8:**
```
choices:     1707725160Cat1Ch1
Weight = Weight values = Illegal weight
values:      Weight = '-1'
```

**Test Case 9:**
```
choices:      1242177805Cat1Ch4
ItemType = Item type values = Coupon
values:       ItemType = 'Coupon'
```

## *I.6      Pairwise Test Frames and Corresponding Test Cases when using ACTS*

**Test Case 1:**
```
choices:      1242177805Cat1Ch1−|−1220360021Cat1Ch2−|−1707725160Cat1Ch4
ItemType = Item type values = Ordinary item|Quantity = Quantity values = Small quantity|Weight =
Weight values = Ignored weight
values:       Weight = '0', Quantity = '1', ItemType = 'OrdinaryItem'
```
**Test Case 2:**
```
choices:      1242177805Cat1Ch2−|−1220360021Cat1Ch4−|−1707725160Cat1Ch2
ItemType = Item type values = Weight item|Quantity = Quantity values = Ignored quantity|Weight =
Weight values = Expected weight range
values:       Weight = '1/2', Quantity = '0', ItemType = 'WeightItem'
```
**Test Case 3:**
```
choices:      1242177805Cat1Ch1−|−1220360021Cat1Ch3−|−1707725160Cat1Ch4
ItemType = Item type values = Ordinary item|Quantity = Quantity values = Large|Weight = Weight
values = Ignored weight
values:       Weight = '0', Quantity = '6', ItemType = 'OrdinaryItem'
```
**Test Case 4:**
```
choices:      1242177805Cat1Ch2−|−1220360021Cat1Ch4−|−1707725160Cat1Ch3
ItemType = Item type values = Weight item|Quantity = Quantity values = Ignored quantity|Weight =
Weight values = suspicious weight value
values:       Weight = '11', Quantity = '0', ItemType = 'WeightItem'
```
**Test Case 5:**
```
choices:      1220360021Cat1Ch1
Quantity = Quantity values = Illegal quantity
values:       Quantity = '-1'
```
**Test Case 6:**
```
choices:      1707725160Cat1Ch1
Weight = Weight values = Illegal weight
values:       Weight = '-1'
```
**Test Case 7:**
```
choices:      1242177805Cat1Ch4
ItemType = Item type values = Coupon
values:       ItemType = 'Coupon'
```
**Test Case 8:**
```
choices:      1242177805Cat1Ch3
ItemType = Item type values = Health item
values:       ItemType = 'HealthItem'
```

## *I.7      Three-way Test Frames and Corresponding Test Cases when using CASA*

**Test Case 1:**
```
choices:      1242177805Cat1Ch3−|−1220360021Cat1Ch2−|−1707725160Cat1Ch4
ItemType = Item type values = Health item|Quantity = Quantity values = Small quantity|Weight =
Weight values = Ignored weight
values:       Weight = '0', Quantity = '1', ItemType = 'HealthItem'
```
**Test Case 2:**
```
choices:      1242177805Cat1Ch2−|−1220360021Cat1Ch4−|−1707725160Cat1Ch2
ItemType = Item type values = Weight item|Quantity = Quantity values = Ignored quantity|Weight =
Weight values = Expected weight range
values:       Weight = '1/2', Quantity = '0', ItemType = 'WeightItem'
```
**Test Case 3:**
```
choices:      1242177805Cat1Ch2−|−1220360021Cat1Ch4−|−1707725160Cat1Ch4
ItemType = Item type values = Weight item|Quantity = Quantity values = Ignored quantity|Weight =
Weight values = Ignored weight
values:       Weight = '0', Quantity = '0', ItemType = 'WeightItem'
```
**Test Case 4:**
```
choices:      1242177805Cat1Ch1−|−1220360021Cat1Ch2−|−1707725160Cat1Ch4
ItemType = Item type values = Ordinary item|Quantity = Quantity values = Small quantity|Weight =
Weight values = Ignored weight
```

```
values:      Weight = '0', Quantity = '1', ItemType = 'OrdinaryItem'
```

**Test Case 5:**
```
choices:     1242177805Cat1Ch2-|-1220360021Cat1Ch4-|-1707725160Cat1Ch3
ItemType = Item type values = Weight item|Quantity = Quantity values = Ignored quantity|Weight =
Weight values = suspicious weight value
values:      Weight = '11', Quantity = '0', ItemType = 'WeightItem'
```

**Test Case 6:**
```
choices:     1242177805Cat1Ch1-|-1220360021Cat1Ch3-|-1707725160Cat1Ch4
ItemType = Item type values = Ordinary item|Quantity = Quantity values = Large|Weight = Weight
values = Ignored weight
values:      Weight = '0', Quantity = '6', ItemType = 'OrdinaryItem'
```

**Test Case 7:**
```
choices:     1220360021Cat1Ch1
Quantity = Quantity values = Illegal quantity
values:      Quantity = '-1'
```

**Test Case 8:**
```
choices:     1707725160Cat1Ch1
Weight = Weight values = Illegal weight
values:      Weight = '-1'
```

**Test Case 9:**
```
choices:     1242177805Cat1Ch4
ItemType = Item type values = Coupon
values:      ItemType = 'Coupon'
```

## I.8      Three-way Test Frames and Corresponding Test Cases when using ACTS

**Test Case 1:**
```
choices:     1242177805Cat1Ch1-|-1220360021Cat1Ch2-|-1707725160Cat1Ch4
ItemType = Item type values = Ordinary item|Quantity = Quantity values = Small quantity|Weight =
Weight values = Ignored weight
values:      Weight = '0', Quantity = '1', ItemType = 'OrdinaryItem'
```

**Test Case 2:**
```
choices:     1242177805Cat1Ch2-|-1220360021Cat1Ch4-|-1707725160Cat1Ch2
ItemType = Item type values = Weight item|Quantity = Quantity values = Ignored quantity|Weight =
Weight values = Expected weight range
values:      Weight = '1/2', Quantity = '0', ItemType = 'WeightItem'
```

**Test Case 3:**
```
choices:     1242177805Cat1Ch1-|-1220360021Cat1Ch3-|-1707725160Cat1Ch4
ItemType = Item type values = Ordinary item|Quantity = Quantity values = Large|Weight = Weight
values = Ignored weight
values:      Weight = '0', Quantity = '6', ItemType = 'OrdinaryItem'
```

**Test Case 4:**
```
choices:     1242177805Cat1Ch2-|-1220360021Cat1Ch4-|-1707725160Cat1Ch3
ItemType = Item type values = Weight item|Quantity = Quantity values = Ignored quantity|Weight =
Weight values = suspicious weight value
values:      Weight = '11', Quantity = '0', ItemType = 'WeightItem'
```

**Test Case 5:**
```
choices:     1220360021Cat1Ch1
Quantity = Quantity values = Illegal quantity
values:      Quantity = '-1'
```

**Test Case 6:**
```
choices:     1707725160Cat1Ch1
Weight = Weight values = Illegal weight
values:      Weight = '-1'
```

**Test Case 7:**
```
choices:     1242177805Cat1Ch4
ItemType = Item type values = Coupon
values:      ItemType = 'Coupon'
```

**Test Case 8:**
```
choices:     1242177805Cat1Ch3
ItemType = Item type values = Health item
values:      ItemType = 'HealthItem'
```

## I.9    Base Choice Test Frames and Corresponding Test Cases

**Test Case 1:**
```
choices:      1242177805Cat1Ch2—|—1220360021Cat1Ch4—|—1707725160Cat1Ch2
ItemType = Item type values = Weight item|Quantity = Quantity values = Ignored quantity|Weight =
Weight values = Expected weight range
values:       Weight = '1/2', Quantity = '0', ItemType = 'WeightItem'
```
**Test Case 2:**
```
choices:      1242177805Cat1Ch1—|—1220360021Cat1Ch2—|—1707725160Cat1Ch4
ItemType = Item type values = Ordinary item|Quantity = Quantity values = Small quantity|Weight =
Weight values = Ignored weight
values:       Weight = '0', Quantity = '1', ItemType = 'OrdinaryItem'
```
**Test Case 3:**
```
choices:      1242177805Cat1Ch3
ItemType = Item type values = Health item
values:       ItemType = 'HealthItem'
```
**Test Case 4:**
```
choices:      1242177805Cat1Ch4
ItemType = Item type values = Coupon
values:       ItemType = 'Coupon'
```
**Test Case 5:**
```
choices:      1220360021Cat1Ch1
Quantity = Quantity values = Illegal quantity
values:       Quantity = '-1'
```
**Test Case 6:**
```
choices:      1242177805Cat1Ch1—|—1220360021Cat1Ch2—|—1707725160Cat1Ch4
ItemType = Item type values = Ordinary item|Quantity = Quantity values = Small quantity|Weight =
Weight values = Ignored weight
values:       Weight = '0', Quantity = '1', ItemType = 'OrdinaryItem'
```
**Test Case 7:**
```
choices:      1242177805Cat1Ch1—|—1220360021Cat1Ch3—|—1707725160Cat1Ch4
ItemType = Item type values = Ordinary item|Quantity = Quantity values = Large|Weight = Weight
values = Ignored weight
values:       Weight = '0', Quantity = '6', ItemType = 'OrdinaryItem'
```
**Test Case 8:**
```
choices:      1707725160Cat1Ch1
Weight = Weight values = Illegal weight
values:       Weight = '-1'
```
**Test Case 9:**
```
choices:      1242177805Cat1Ch2—|—1220360021Cat1Ch4—|—1707725160Cat1Ch3
ItemType = Item type values = Weight item|Quantity = Quantity values = Ignored quantity|Weight =
Weight values = suspicious weight value
values:       Weight = '11', Quantity = '0', ItemType = 'WeightItem'
```
**Test Case 10:**
```
choices:      1242177805Cat1Ch1—|—1220360021Cat1Ch2—|—1707725160Cat1Ch4
ItemType = Item type values = Ordinary item|Quantity = Quantity values = Small quantity|Weight =
Weight values = Ignored weight

values:       Weight = '0', Quantity = '1', ItemType = 'OrdinaryItem'
```

# Appendix J. TCAS

In the CP specification below, an asterisk ('*') beside a choice indicates the base choice for that category. Each parameter is uniquely identified by an integer identifier, which is then used in the composition of test frames; As a result, a choice for a parameter is uniquely identified by the parameter identifier, the category identifier and the choice identifier. Below we summarize the specification of the unit under test (section J.1), and provide the CP specification we used (section J.2). We then provide the test frames and corresponding test cases (inputs) we obtain when using CASA (Each-Choice in section J.3, Pairwise in section J.5, Three-way in section J.7), when using ACTS (Each-Choice in section J.4, Pairwise in section J.6, Three-way in section J.8), and when using the Base Choice criterion (section J.9). The CP specification below is adapted from a graduate course work completed Curtis Winstanley.

## J.1    Specification

The Traffic Alert and Collision Avoidance System (TCAS) is a core safety component to modern aircraft. It helps prevent mid-air collisions between aircraft. The system monitors the surrounding airspace for intruding aircraft that could pose a threat. If a threat is detected, a Traffic Advisory (TA) is forwarded to the pilot. This will let the pilot know that there is an aircraft within the vicinity of their own aircraft which could pose a threat. If the intruding aircraft comes within a certain threshold range of the TCAS equipped aircraft, the TCAS system will notify the pilot and offer a Resolution Advisory (RA). The RA will tell the pilot the best course of action to minimize the chance of a collision. The RA can be in the form of either an Upwards RA, Downwards RA, or Unresolved. Upwards RA means the pilot should direct the aircraft in a climbing maneuver whereas Downwards RA means the pilot should begin a descending maneuver. Unresolved means that the TCAS could not find a suitable maneuver.

## J.2    CP Specification

```
Parameter(1871720133): Current_Vertical_Sep of type Integer
  Category 1 - Values of Current_Vertical_Sep
    *Ch 1.1: Greater (strictly) than MAXALTDIF (600)
            Current_Vertical_Sep > 600
     Ch 1.2: Smaller than MAXALTDIF (600), but positive (strictly)
            Current_Vertical_Sep>0 && Current_Vertical_Sep<=600
     Ch 1.3: Illegal value
            Current_Vertical_Sep <= 0
            [error]
Parameter(374476473): High_Confidence of type Boolean
  Category 1 - Values of High_Confidence
    *Ch 1.1: True
            High_Confidence == true
     Ch 1.2: False
            High_Confidence == false
Parameter(1325844154): Two_of_Three_Reports_Valid of type Boolean
  Category 1 - Values of Two_of_Three_Reports_Valid
    *Ch 1.1: True
            Two_of_Three_Reports_Valid == true
     Ch 1.2: False
            Two_of_Three_Reports_Valid == false
Parameter(603277445): Own_Tracked_Alt of type Integer
  Category 1 - Values of Own_Tracked_Alt
    *Ch 1.1: Higher (strictly) than other (intruder)
            Own_Tracked_Alt>0 && Own_Tracked_Alt>Other_Tracked_Alt
            [properties OwnHigherThanOther]
     Ch 1.2: Same as other (intruder)
            Own_Tracked_Alt>0 && Own_Tracked_Alt==Other_Tracked_Alt
            [properties OwnSameAsOther]
     Ch 1.3: Lower (strictly) than other (intruder)
            Own_Tracked_Alt>0 && Own_Tracked_Alt<Other_Tracked_Alt
            [properties OwnLowerThanOther]
```

```
        Ch 1.4: Illegal
                Own_Tracked_Alt <= 0
                [error]
Parameter(1787817979): Other_Tracked_Alt of type Integer
  Category 1 - Values Other_Tracked_Alt
      Ch 1.1: Higher (strictly) than TCAS (own)
                Other_Tracked_Alt>0 && Other_Tracked_Alt>Own_Tracked_Alt
                [if OwnLowerThanOther]
      Ch 1.2: Same as TCAS (own)
                Other_Tracked_Alt>0 && Other_Tracked_Alt==Own_Tracked_Alt
                [if OwnSameAsOther]
    *Ch 1.3: Lower (strictly) than TCAS (own)
                Other_Tracked_Alt>0 && Other_Tracked_Alt<Own_Tracked_Alt
                [if OwnHigherThanOther]
      Ch 1.4: Illegal
                Other_Tracked_Alt <= 0
                [error]
Parameter(927131798): Up_Separation of type Integer
  Category 1 - Values of Up_Separation
    *Ch 1.1: Greater (strictly) than down separation and threshold
                Up_Separation>0 && Up_Separation>Down_Separation &&
                Up_Separation>Positive_RA_Alt_Thresh
                [properties UpSepGreater]
      Ch 1.2: Smaler (strictly) than down separation and threshold
                Up_Separation>0 && Up_Separation<Down_Separation &&
                Up_Separation<Positive_RA_Alt_Thresh
                [properties UpSepSmaller]
      Ch 1.3: Between down separation and threshold
                Up_Separation>0 && ( (Up_Separation>=Down_Separation &&
                Up_Separation<=Positive_RA_Alt_Thresh) || (Up_Separation<=Down_Separation &&
                Up_Separation>=Positive_RA_Alt_Thresh) )
                [properties UpSepBetween]
      Ch 1.4: Illegal
                Up_Separation <= 0
                [error]
Parameter(875777795): Down_Separation of type Integer
  Category 1 - Values of Down_Separation
      Ch 1.1: Greater (strictly) than up separation and threshold
                Down_Separation>0 && Down_Separation>Up_Separation &&
                Down_Separation>Positive_RA_Alt_Thresh
                [if !UpSepGreater]
    *Ch 1.2: Smaller (strictly) than up separation and threshold
                Down_Separation>0 && Down_Separation<Up_Separation &&
                Down_Separation<Positive_RA_Alt_Thresh
                [if !UpSepSmaller]
      Ch 1.3: Between up separation and threshold
                Down_Separation>0 && ( (Down_Separation>=Up_Separation &&
                Down_Separation<=Positive_RA_Alt_Thresh) || (Down_Separation<=Up_Separation &&
                Down_Separation>=Positive_RA_Alt_Thresh) )
                [if !UpSepBetween]
      Ch 1.4: Illegal
                Down_Separation <= 0
                [error]
Parameter(302439181): Alt_Layer_Value of type Enumeration
  Category 1 - Values of Alt_Layer_Value
    *Ch 1.1: Value 0
                Alt_Layer_Value == Value0
                [if ThreshValue0]
      Ch 1.2: Value 1
                Alt_Layer_Value == Value1
                [if ThreshValue1]
      Ch 1.3: Value 2
                Alt_Layer_Value == Value2
                [if ThreshValue2]
```

```
              Ch 1.4:  Value 3
                       Alt_Layer_Value == Value3
                       [if ThreshValue3]
Parameter(58393125):  Other_RAC of type Enumeration
  Category 1 - Values of Other_RAC
     *Ch 1.1:  Zero
                       Other_RAC == Zero
      Ch 1.2:  One
                       Other_RAC == One
      Ch 1.3:  Two
                       Other_RAC == Two
Parameter(62606991):  Other_Capability of type Enumeration
  Category 1 - Values of Other_Capability
     *Ch 1.1:  One
                       Other_Capability == One
      Ch 1.2:  Two
                       Other_Capability == Two
Parameter(430451043):  Climb_Inhibit of type Enumeration
  Category 1 - Values of Climb_Inhibit
     *Ch 1.1:  Zero
                       Climb_Inhibit == Zero
      Ch 1.2:  One
                       Climb_Inhibit == One

Variable(976751556):  Positive_RA_Alt_Thresh of type Integer
  Category 1 - Values of Positive_RA_Alt_Thresh
     *Ch 1.1:  Value0
                       Positive_RA_Alt_Thresh == 400
                       [properties ThreshValue0]
      Ch 1.2:  Value1
                       Positive_RA_Alt_Thresh == 500
                       [properties ThreshValue1]
      Ch 1.3:  Value2
                       Positive_RA_Alt_Thresh == 640
                       [properties ThreshValue2]
      Ch 1.4:  Value3
                       Positive_RA_Alt_Thresh == 740
                       [properties ThreshValue3]
      Ch 1.5:  Illegal
                       Positive_RA_Alt_Thresh!=400 && Positive_RA_Alt_Thresh!=500 &&
                       Positive_RA_Alt_Thresh!=640 && Positive_RA_Alt_Thresh!=740
                       [error]
```

## *J.3      Each-Choice Test Frames and Corresponding Test Cases when using CASA*

**Test Case 1:**
```
choices:      1871720133Cat1Ch2-|-374476473Cat1Ch1-|-1325844154Cat1Ch1-|-603277445Cat1Ch3-|-
              1787817979Cat1Ch1-|-927131798Cat1Ch3-|-875777795Cat1Ch1-|-302439181Cat1Ch4-|-
              58393125Cat1Ch1-|-626069991Cat1Ch1-|-430451043Cat1Ch2-|-976751556Cat1Ch4
Current_Vertical_Sep = Values of Current_Vertical_Sep = Smaller than MAXALTDIF (600), but
positive (strictly)|High_Confidence = Values of High_Confidence =
True|Two_of_Three_Reports_Valid = Values of Two_of_Three_Reports_Valid = True|Own_Tracked_Alt =
Values of Own_Tracked_Alt = Lower (strictly) than other (intruder)|Other_Tracked_Alt = Values
Other_Tracked_Alt = Higher (strictly) than TCAS (own)|Up_Separation = Values of Up_Separation =
Between down separation and threshold|Down_Separation = Values of Down_Separation = Greater
(strictly) than up separation and threshold|Alt_Layer_Value = Values of Alt_Layer_Value = Value
3|Other_RAC = Values of Other_RAC = Zero|Other_Capability = Values of Other_Capability =
One|Climb_Inhibit = Values of Climb_Inhibit = One|Positive_RA_Alt_Thresh = Values of
Positive_RA_Alt_Thresh = Value3
values:       Current_Vertical_Sep = '1', Alt_Layer_Value = 'Value3', Down_Separation = '741',
              Up_Separation = '740', Other_RAC = 'Zero', High_Confidence = 'true',
              Other_Tracked_Alt = '2', Other_Capability = 'One', Own_Tracked_Alt = '1',
              Climb_Inhibit = 'One', Positive_RA_Alt_Thresh = '740', Two_of_Three_Reports_Valid =
              'true'
```

**Test Case 2:**

```
choices:     1871720133Cat1Ch1-|-374476473Cat1Ch2-|-1325844154Cat1Ch2-|-603277445Cat1Ch3-|-
             1787817979Cat1Ch1-|-927131798Cat1Ch1-|-875777795Cat1Ch2-|-302439181Cat1Ch2-|-
             58393125Cat1Ch2-|-626069991Cat1Ch2-|-430451043Cat1Ch2-|-976751556Cat1Ch2
Current_Vertical_Sep = Values of Current_Vertical_Sep = Greater (strictly) than MAXALTDIF
(600)|High_Confidence = Values of High_Confidence = False|Two_of_Three_Reports_Valid = Values of
Two_of_Three_Reports_Valid = False|Own_Tracked_Alt = Values of Own_Tracked_Alt = Lower
(strictly) than other (intruder)|Other_Tracked_Alt = Values Other_Tracked_Alt = Higher
(strictly) than TCAS (own)|Up_Separation = Values of Up_Separation = Greater (strictly) than
down separation and threshold|Down_Separation = Values of Down_Separation = Smaller (strictly)
than up separation and threshold|Alt_Layer_Value = Values of Alt_Layer_Value = Value 1|Other_RAC
= Values of Other_RAC = One|Other_Capability = Values of Other_Capability = Two|Climb_Inhibit =
Values of Climb_Inhibit = One|Positive_RA_Alt_Thresh = Values of Positive_RA_Alt_Thresh = Value1
values:      Current_Vertical_Sep = '601', Alt_Layer_Value = 'Value1', Down_Separation = '499',
             Up_Separation = '501', Other_RAC = 'One', High_Confidence = 'false',
             Other_Tracked_Alt = '2', Other_Capability = 'Two', Own_Tracked_Alt = '1',
             Positive_RA_Alt_Thresh = '500', Climb_Inhibit = 'One', Two_of_Three_Reports_Valid =
             'false'
```

**Test Case 3:**

```
choices:     1871720133Cat1Ch1-|-374476473Cat1Ch2-|-1325844154Cat1Ch1-|-603277445Cat1Ch2-|-
             1787817979Cat1Ch2-|-927131798Cat1Ch2-|-875777795Cat1Ch3-|-302439181Cat1Ch1-|-
             58393125Cat1Ch3-|-626069991Cat1Ch2-|-430451043Cat1Ch2-|-976751556Cat1Ch1
Current_Vertical_Sep = Values of Current_Vertical_Sep = Greater (strictly) than MAXALTDIF
(600)|High_Confidence = Values of High_Confidence = False|Two_of_Three_Reports_Valid = Values of
Two_of_Three_Reports_Valid = True|Own_Tracked_Alt = Values of Own_Tracked_Alt = Same as other
(intruder)|Other_Tracked_Alt = Values Other_Tracked_Alt = Same as TCAS (own)|Up_Separation =
Values of Up_Separation = Smaler (strictly) than down separation and threshold|Down_Separation =
Values of Down_Separation = Between up separation and threshold|Alt_Layer_Value = Values of
Alt_Layer_Value = Value 0|Other_RAC = Values of Other_RAC = Two|Other_Capability = Values of
Other_Capability = Two|Climb_Inhibit = Values of Climb_Inhibit = One|Positive_RA_Alt_Thresh =
Values of Positive_RA_Alt_Thresh = Value0
values:      Current_Vertical_Sep = '601', Alt_Layer_Value = 'Value0', Down_Separation = '2',
             Up_Separation = '1', Other_RAC = 'Two', High_Confidence = 'false',
             Other_Tracked_Alt = '1', Other_Capability = 'Two', Own_Tracked_Alt = '1',
             Positive_RA_Alt_Thresh = '400', Climb_Inhibit = 'One', Two_of_Three_Reports_Valid =
             'true'
```

**Test Case 4:**

```
choices:     1871720133Cat1Ch2-|-374476473Cat1Ch2-|-1325844154Cat1Ch2-|-603277445Cat1Ch1-|-
             1787817979Cat1Ch3-|-927131798Cat1Ch1-|-875777795Cat1Ch3-|-302439181Cat1Ch3-|-
             58393125Cat1Ch1-|-626069991Cat1Ch2-|-430451043Cat1Ch1-|-976751556Cat1Ch3
Current_Vertical_Sep = Values of Current_Vertical_Sep = Smaller than MAXALTDIF (600), but
positive (strictly)|High_Confidence = Values of High_Confidence =
False|Two_of_Three_Reports_Valid = Values of Two_of_Three_Reports_Valid = False|Own_Tracked_Alt
= Values of Own_Tracked_Alt = Higher (strictly) than other (intruder)|Other_Tracked_Alt = Values
Other_Tracked_Alt = Lower (strictly) than TCAS (own)|Up_Separation = Values of Up_Separation =
Greater (strictly) than down separation and threshold|Down_Separation = Values of
Down_Separation = Between up separation and threshold|Alt_Layer_Value = Values of
Alt_Layer_Value = Value 2|Other_RAC = Values of Other_RAC = Zero|Other_Capability = Values of
Other_Capability = Two|Climb_Inhibit = Values of Climb_Inhibit = Zero|Positive_RA_Alt_Thresh =
Values of Positive_RA_Alt_Thresh = Value2
values:      Current_Vertical_Sep = '1', Alt_Layer_Value = 'Value2', Down_Separation = '640',
             Up_Separation = '641', Other_RAC = 'Zero', High_Confidence = 'false',
             Other_Tracked_Alt = '1', Other_Capability = 'Two', Own_Tracked_Alt = '2',
             Positive_RA_Alt_Thresh = '640', Climb_Inhibit = 'Zero', Two_of_Three_Reports_Valid
             = 'false'
```

**Test Case 5:**

```
choices:     1871720133Cat1Ch3
Current_Vertical_Sep = Values of Current_Vertical_Sep = Illegal value
values:      Current_Vertical_Sep = '0'
```

**Test Case 6:**

```
choices:     603277445Cat1Ch4
Own_Tracked_Alt = Values of Own_Tracked_Alt = Illegal
values:      Own_Tracked_Alt = '0'
```

**Test Case 7:**

```
choices:     1787817979Cat1Ch4
```

```
Other_Tracked_Alt = Values Other_Tracked_Alt = Illegal
values:        Other_Tracked_Alt = '0'
```
**Test Case 8:**
```
choices:       927131798Cat1Ch4
Up_Separation = Values of Up_Separation = Illegal
values:        Up_Separation = '0'
```
**Test Case 9:**
```
choices:       875777795Cat1Ch4
Down_Separation = Values of Down_Separation = Illegal
values:        Down_Separation = '0'
```
**Test Case 10:**
```
choices:       976751556Cat1Ch5
Positive_RA_Alt_Thresh = Values of Positive_RA_Alt_Thresh = Illegal
values:        Positive_RA_Alt_Thresh = '2'
```

## *J.4 Each-Choice Test Frames and Corresponding Test Cases when using ACTS*

**Test Case 1:**
```
choices:       302439181Cat1Ch1-|-976751556Cat1Ch1-|-603277445Cat1Ch2-|-1787817979Cat1Ch2-|-
               927131798Cat1Ch2-|-875777795Cat1Ch3-|-58393125Cat1Ch2-|-1871720133Cat1Ch2-|-
               374476473Cat1Ch2-|-1325844154Cat1Ch2-|-626069991Cat1Ch2-|-430451043Cat1Ch2
Alt_Layer_Value = Values of Alt_Layer_Value = Value 0|Positive_RA_Alt_Thresh = Values of
Positive_RA_Alt_Thresh = Value0|Own_Tracked_Alt = Values of Own_Tracked_Alt = Same as other
(intruder)|Other_Tracked_Alt = Values Other_Tracked_Alt = Same as TCAS (own)|Up_Separation =
Values of Up_Separation = Smaler (strictly) than down separation and threshold|Down_Separation =
Values of Down_Separation = Between up separation and threshold|Other_RAC = Values of Other_RAC
= One|Current_Vertical_Sep = Values of Current_Vertical_Sep = Smaller than MAXALTDIF (600), but
positive (strictly)|High_Confidence = Values of High_Confidence =
False|Two_of_Three_Reports_Valid = Values of Two_of_Three_Reports_Valid = False|Other_Capability
= Values of Other_Capability = Two|Climb_Inhibit = Values of Climb_Inhibit = One
values:        Current_Vertical_Sep = '1', Alt_Layer_Value = 'Value0', Down_Separation = '2',
               Up_Separation = '1', Other_RAC = 'One', High_Confidence = 'false',
               Other_Tracked_Alt = '1', Other_Capability = 'Two', Own_Tracked_Alt = '1',
               Positive_RA_Alt_Thresh = '400', Climb_Inhibit = 'One', Two_of_Three_Reports_Valid =
               'false'
```
**Test Case 2:**
```
choices:       302439181Cat1Ch2-|-976751556Cat1Ch2-|-603277445Cat1Ch3-|-1787817979Cat1Ch1-|-
               927131798Cat1Ch3-|-875777795Cat1Ch1-|-58393125Cat1Ch3-|-1871720133Cat1Ch1-|-
               374476473Cat1Ch1-|-1325844154Cat1Ch1-|-626069991Cat1Ch1-|-430451043Cat1Ch1
Alt_Layer_Value = Values of Alt_Layer_Value = Value 1|Positive_RA_Alt_Thresh = Values of
Positive_RA_Alt_Thresh = Value1|Own_Tracked_Alt = Values of Own_Tracked_Alt = Lower (strictly)
than other (intruder)|Other_Tracked_Alt = Values Other_Tracked_Alt = Higher (strictly) than TCAS
(own)|Up_Separation = Values of Up_Separation = Between down separation and
threshold|Down_Separation = Values of Down_Separation = Greater (strictly) than up separation
and threshold|Other_RAC = Values of Other_RAC = Two|Current_Vertical_Sep = Values of
Current_Vertical_Sep = Greater (strictly) than MAXALTDIF (600)|High_Confidence = Values of
High_Confidence = True|Two_of_Three_Reports_Valid = Values of Two_of_Three_Reports_Valid =
True|Other_Capability = Values of Other_Capability = One|Climb_Inhibit = Values of Climb_Inhibit
= Zero
values:        Current_Vertical_Sep = '601', Alt_Layer_Value = 'Value1', Down_Separation = '501',
               Up_Separation = '500', Other_RAC = 'Two', High_Confidence = 'true',
               Other_Tracked_Alt = '2', Other_Capability = 'One', Own_Tracked_Alt = '1',
               Climb_Inhibit = 'Zero', Positive_RA_Alt_Thresh = '500', Two_of_Three_Reports_Valid
               = 'true'
```
**Test Case 3:**
```
choices:       302439181Cat1Ch3-|-976751556Cat1Ch3-|-603277445Cat1Ch1-|-1787817979Cat1Ch3-|-
               927131798Cat1Ch1-|-875777795Cat1Ch2-|-58393125Cat1Ch1-|-1871720133Cat1Ch2-|-
               374476473Cat1Ch2-|-1325844154Cat1Ch1-|-626069991Cat1Ch2-|-430451043Cat1Ch2
Alt_Layer_Value = Values of Alt_Layer_Value = Value 2|Positive_RA_Alt_Thresh = Values of
Positive_RA_Alt_Thresh = Value2|Own_Tracked_Alt = Values of Own_Tracked_Alt = Higher (strictly)
than other (intruder)|Other_Tracked_Alt = Values Other_Tracked_Alt = Lower (strictly) than TCAS
(own)|Up_Separation = Values of Up_Separation = Greater (strictly) than down separation and
threshold|Down_Separation = Values of Down_Separation = Smaller (strictly) than up separation
and threshold|Other_RAC = Values of Other_RAC = Zero|Current_Vertical_Sep = Values of
Current_Vertical_Sep = Smaller than MAXALTDIF (600), but positive (strictly)|High_Confidence =
```

```
Values of High_Confidence = False|Two_of_Three_Reports_Valid = Values of
Two_of_Three_Reports_Valid = True|Other_Capability = Values of Other_Capability =
Two|Climb_Inhibit = Values of Climb_Inhibit = One
values:      Current_Vertical_Sep = '1', Alt_Layer_Value = 'Value2', Down_Separation = '1',
             Up_Separation = '641', Other_RAC = 'Zero', High_Confidence = 'false',
             Other_Tracked_Alt = '1', Other_Capability = 'Two', Own_Tracked_Alt = '2',
             Climb_Inhibit = 'One', Positive_RA_Alt_Thresh = '640', Two_of_Three_Reports_Valid =
             'true'
```

**Test Case 4:**
```
choices:     302439181Cat1Ch4-|—976751556Cat1Ch4-|—603277445Cat1Ch3-|—1787817979Cat1Ch1-|—
             927131798Cat1Ch1-|—875777795Cat1Ch3-|—58393125Cat1Ch3-|—1871720133Cat1Ch1-|—
             374476473Cat1Ch1-|—1325844154Cat1Ch2-|—626069991Cat1Ch2-|—430451043Cat1Ch2
Alt_Layer_Value = Values of Alt_Layer_Value = Value 3|Positive_RA_Alt_Thresh = Values of
Positive_RA_Alt_Thresh = Value3|Own_Tracked_Alt = Values of Own_Tracked_Alt = Lower (strictly)
than other (intruder)|Other_Tracked_Alt = Values Other_Tracked_Alt = Higher (strictly) than TCAS
(own)|Up_Separation = Values of Up_Separation = Greater (strictly) than down separation and
threshold|Down_Separation = Values of Down_Separation = Between up separation and
threshold|Other_RAC = Values of Other_RAC = Two|Current_Vertical_Sep = Values of
Current_Vertical_Sep = Greater (strictly) than MAXALTDIF (600)|High_Confidence = Values of
High_Confidence = True|Two_of_Three_Reports_Valid = Values of Two_of_Three_Reports_Valid =
False|Other_Capability = Values of Other_Capability = Two|Climb_Inhibit = Values of
Climb_Inhibit = One
values:      Current_Vertical_Sep = '601', Alt_Layer_Value = 'Value3', Down_Separation = '740',
             Up_Separation = '741', Other_RAC = 'Two', High_Confidence = 'true',
             Other_Tracked_Alt = '2', Other_Capability = 'Two', Own_Tracked_Alt = '1',
             Positive_RA_Alt_Thresh = '740', Climb_Inhibit = 'One', Two_of_Three_Reports_Valid =
             'false'
```

**Test Case 5:**
```
choices:     1871720133Cat1Ch3
Current_Vertical_Sep = Values of Current_Vertical_Sep = Illegal value
values:      Current_Vertical_Sep = '0'
```

**Test Case 6:**
```
choices:     603277445Cat1Ch4
Own_Tracked_Alt = Values of Own_Tracked_Alt = Illegal
values:      Own_Tracked_Alt = '0'
```

**Test Case 7:**
```
choices:     1787817979Cat1Ch4
Other_Tracked_Alt = Values Other_Tracked_Alt = Illegal
values:      Other_Tracked_Alt = '0'
```

**Test Case 8:**
```
choices:     927131798Cat1Ch4
Up_Separation = Values of Up_Separation = Illegal
values:      Up_Separation = '0'
```

**Test Case 9:**
```
choices:     875777795Cat1Ch4
Down_Separation = Values of Down_Separation = Illegal
values:      Down_Separation = '0'
```

**Test Case 10:**
```
choices:     976751556Cat1Ch5
Positive_RA_Alt_Thresh = Values of Positive_RA_Alt_Thresh = Illegal
values:      Positive_RA_Alt_Thresh = '2'
```

## J.5    Pairwise Test Frames and Corresponding Test Cases when using CASA

**Test Case 1:**
```
choices:     1871720133Cat1Ch1-|—374476473Cat1Ch1-|—1325844154Cat1Ch1-|—603277445Cat1Ch3-|—
             1787817979Cat1Ch1-|—927131798Cat1Ch1-|—875777795Cat1Ch3-|—302439181Cat1Ch3-|—
             58393125Cat1Ch2-|—626069991Cat1Ch2-|—430451043Cat1Ch2-|—976751556Cat1Ch3
Current_Vertical_Sep = Values of Current_Vertical_Sep = Greater (strictly) than MAXALTDIF
(600)|High_Confidence = Values of High_Confidence = True|Two_of_Three_Reports_Valid = Values of
Two_of_Three_Reports_Valid = True|Own_Tracked_Alt = Values of Own_Tracked_Alt = Lower (strictly)
than other (intruder)|Other_Tracked_Alt = Values Other_Tracked_Alt = Higher (strictly) than TCAS
(own)|Up_Separation = Values of Up_Separation = Greater (strictly) than down separation and
threshold|Down_Separation = Values of Down_Separation = Between up separation and
threshold|Alt_Layer_Value = Values of Alt_Layer_Value = Value 2|Other_RAC = Values of Other_RAC
```

= One|Other_Capability = Values of Other_Capability = Two|Climb_Inhibit = Values of
Climb_Inhibit = One|Positive_RA_Alt_Thresh = Values of Positive_RA_Alt_Thresh = Value2
values:      Current_Vertical_Sep = '601', Alt_Layer_Value = 'Value2', Down_Separation = '640',
             Up_Separation = '641', Other_RAC = 'One', High_Confidence = 'true',
             Other_Tracked_Alt = '2', Other_Capability = 'Two', Own_Tracked_Alt = '1',
             Climb_Inhibit = 'One', Positive_RA_Alt_Thresh = '640', Two_of_Three_Reports_Valid =
             'true'

**Test Case 2:**
choices:     1871720133Cat1Ch1-|-374476473Cat1Ch2-|-1325844154Cat1Ch2-|-603277445Cat1Ch1-|-
             1787817979Cat1Ch3-|-927131798Cat1Ch3-|-875777795Cat1Ch2-|-302439181Cat1Ch3-|-
             58393125Cat1Ch3-|-626069991Cat1Ch1-|-430451043Cat1Ch1-|-976751556Cat1Ch3
Current_Vertical_Sep = Values of Current_Vertical_Sep = Greater (strictly) than MAXALTDIF
(600)|High_Confidence = Values of High_Confidence = False|Two_of_Three_Reports_Valid = Values of
Two_of_Three_Reports_Valid = False|Own_Tracked_Alt = Values of Own_Tracked_Alt = Higher
(strictly) than other (intruder)|Other_Tracked_Alt = Values Other_Tracked_Alt = Lower (strictly)
than TCAS (own)|Up_Separation = Values of Up_Separation = Between down separation and
threshold|Down_Separation = Values of Down_Separation = Smaller (strictly) than up separation
and threshold|Alt_Layer_Value = Values of Alt_Layer_Value = Value 2|Other_RAC = Values of
Other_RAC = Two|Other_Capability = Values of Other_Capability = One|Climb_Inhibit = Values of
Climb_Inhibit = Zero|Positive_RA_Alt_Thresh = Values of Positive_RA_Alt_Thresh = Value2
values:      Current_Vertical_Sep = '601', Alt_Layer_Value = 'Value2', Down_Separation = '639',
             Up_Separation = '640', Other_RAC = 'Two', High_Confidence = 'false',
             Other_Tracked_Alt = '1', Other_Capability = 'One', Own_Tracked_Alt = '2',
             Positive_RA_Alt_Thresh = '640', Climb_Inhibit = 'Zero', Two_of_Three_Reports_Valid
             = 'false'

**Test Case 3:**
choices:     1871720133Cat1Ch1-|-374476473Cat1Ch2-|-1325844154Cat1Ch1-|-603277445Cat1Ch3-|-
             1787817979Cat1Ch1-|-927131798Cat1Ch2-|-875777795Cat1Ch1-|-302439181Cat1Ch2-|-
             58393125Cat1Ch3-|-626069991Cat1Ch2-|-430451043Cat1Ch1-|-976751556Cat1Ch2
Current_Vertical_Sep = Values of Current_Vertical_Sep = Greater (strictly) than MAXALTDIF
(600)|High_Confidence = Values of High_Confidence = False|Two_of_Three_Reports_Valid = Values of
Two_of_Three_Reports_Valid = True|Own_Tracked_Alt = Values of Own_Tracked_Alt = Lower (strictly)
than other (intruder)|Other_Tracked_Alt = Values Other_Tracked_Alt = Higher (strictly) than TCAS
(own)|Up_Separation = Values of Up_Separation = Smaler (strictly) than down separation and
threshold|Down_Separation = Values of Down_Separation = Greater (strictly) than up separation
and threshold|Alt_Layer_Value = Values of Alt_Layer_Value = Value 1|Other_RAC = Values of
Other_RAC = Two|Other_Capability = Values of Other_Capability = Two|Climb_Inhibit = Values of
Climb_Inhibit = Zero|Positive_RA_Alt_Thresh = Values of Positive_RA_Alt_Thresh = Value1
values:      Current_Vertical_Sep = '601', Alt_Layer_Value = 'Value1', Down_Separation = '501',
             Up_Separation = '1', Other_RAC = 'Two', High_Confidence = 'false',
             Other_Tracked_Alt = '2', Other_Capability = 'Two', Own_Tracked_Alt = '1',
             Positive_RA_Alt_Thresh = '500', Climb_Inhibit = 'Zero', Two_of_Three_Reports_Valid
             = 'true'

**Test Case 4:**
choices:     1871720133Cat1Ch1-|-374476473Cat1Ch2-|-1325844154Cat1Ch1-|-603277445Cat1Ch2-|-
             1787817979Cat1Ch2-|-927131798Cat1Ch1-|-875777795Cat1Ch2-|-302439181Cat1Ch4-|-
             58393125Cat1Ch1-|-626069991Cat1Ch2-|-430451043Cat1Ch1-|-976751556Cat1Ch4
Current_Vertical_Sep = Values of Current_Vertical_Sep = Greater (strictly) than MAXALTDIF
(600)|High_Confidence = Values of High_Confidence = False|Two_of_Three_Reports_Valid = Values of
Two_of_Three_Reports_Valid = True|Own_Tracked_Alt = Values of Own_Tracked_Alt = Same as other
(intruder)|Other_Tracked_Alt = Values Other_Tracked_Alt = Same as TCAS (own)|Up_Separation =
Values of Up_Separation = Greater (strictly) than down separation and threshold|Down_Separation
= Values of Down_Separation = Smaller (strictly) than up separation and
threshold|Alt_Layer_Value = Values of Alt_Layer_Value = Value 3|Other_RAC = Values of Other_RAC
= Zero|Other_Capability = Values of Other_Capability = Two|Climb_Inhibit = Values of
Climb_Inhibit = Zero|Positive_RA_Alt_Thresh = Values of Positive_RA_Alt_Thresh = Value3
values:      Current_Vertical_Sep = '601', Alt_Layer_Value = 'Value3', Down_Separation = '1',
             Up_Separation = '741', Other_RAC = 'Zero', High_Confidence = 'false',
             Other_Tracked_Alt = '1', Other_Capability = 'Two', Own_Tracked_Alt = '1',
             Positive_RA_Alt_Thresh = '740', Climb_Inhibit = 'Zero', Two_of_Three_Reports_Valid
             = 'true'

**Test Case 5:**
choices:     1871720133Cat1Ch2-|-374476473Cat1Ch1-|-1325844154Cat1Ch2-|-603277445Cat1Ch3-|-
             1787817979Cat1Ch1-|-927131798Cat1Ch3-|-875777795Cat1Ch2-|-302439181Cat1Ch1-|-
             58393125Cat1Ch1-|-626069991Cat1Ch1-|-430451043Cat1Ch2-|-976751556Cat1Ch1

```
Current_Vertical_Sep = Values of Current_Vertical_Sep = Smaller than MAXALTDIF (600), but
positive (strictly)|High_Confidence = Values of High_Confidence =
True|Two_of_Three_Reports_Valid = Values of Two_of_Three_Reports_Valid = False|Own_Tracked_Alt =
Values of Own_Tracked_Alt = Lower (strictly) than other (intruder)|Other_Tracked_Alt = Values
Other_Tracked_Alt = Higher (strictly) than TCAS (own)|Up_Separation = Values of Up_Separation =
Between down separation and threshold|Down_Separation = Values of Down_Separation = Smaller
(strictly) than up separation and threshold|Alt_Layer_Value = Values of Alt_Layer_Value = Value
0|Other_RAC = Values of Other_RAC = Zero|Other_Capability = Values of Other_Capability =
One|Climb_Inhibit = Values of Climb_Inhibit = One|Positive_RA_Alt_Thresh = Values of
Positive_RA_Alt_Thresh = Value0
values:        Current_Vertical_Sep = '1', Alt_Layer_Value = 'Value0', Down_Separation = '1',
               Up_Separation = '2', Other_RAC = 'Zero', High_Confidence = 'true',
               Other_Tracked_Alt = '2', Other_Capability = 'One', Own_Tracked_Alt = '1',
               Climb_Inhibit = 'One', Positive_RA_Alt_Thresh = '400', Two_of_Three_Reports_Valid =
               'false'
```

**Test Case 6:**
```
choices:       1871720133Cat1Ch2-|-374476473Cat1Ch1-|-1325844154Cat1Ch1-|-603277445Cat1Ch2-|-
               1787817979Cat1Ch2-|-927131798Cat1Ch3-|-875777795Cat1Ch2-|-302439181Cat1Ch2-|-
               58393125Cat1Ch2-|-626069991Cat1Ch1-|-430451043Cat1Ch2-|-976751556Cat1Ch2

Current_Vertical_Sep = Values of Current_Vertical_Sep = Smaller than MAXALTDIF (600), but
positive (strictly)|High_Confidence = Values of High_Confidence =
True|Two_of_Three_Reports_Valid = Values of Two_of_Three_Reports_Valid = True|Own_Tracked_Alt =
Values of Own_Tracked_Alt = Same as other (intruder)|Other_Tracked_Alt = Values
Other_Tracked_Alt = Same as TCAS (own)|Up_Separation = Values of Up_Separation = Between down
separation and threshold|Down_Separation = Values of Down_Separation = Smaller (strictly) than
up separation and threshold|Alt_Layer_Value = Values of Alt_Layer_Value = Value 1|Other_RAC =
Values of Other_RAC = One|Other_Capability = Values of Other_Capability = One|Climb_Inhibit =
Values of Climb_Inhibit = One|Positive_RA_Alt_Thresh = Values of Positive_RA_Alt_Thresh = Value1
values:        Current_Vertical_Sep = '1', Alt_Layer_Value = 'Value1', Down_Separation = '499',
               Up_Separation = '500', Other_RAC = 'One', High_Confidence = 'true',
               Other_Tracked_Alt = '1', Other_Capability = 'One', Own_Tracked_Alt = '1',
               Positive_RA_Alt_Thresh = '500', Climb_Inhibit = 'One', Two_of_Three_Reports_Valid =
               'true'
```

**Test Case 7:**
```
choices:       1871720133Cat1Ch2-|-374476473Cat1Ch1-|-1325844154Cat1Ch2-|-603277445Cat1Ch3-|-
               1787817979Cat1Ch1-|-927131798Cat1Ch2-|-875777795Cat1Ch3-|-302439181Cat1Ch4-|-
               58393125Cat1Ch3-|-626069991Cat1Ch1-|-430451043Cat1Ch2-|-976751556Cat1Ch4

Current_Vertical_Sep = Values of Current_Vertical_Sep = Smaller than MAXALTDIF (600), but
positive (strictly)|High_Confidence = Values of High_Confidence =
True|Two_of_Three_Reports_Valid = Values of Two_of_Three_Reports_Valid = False|Own_Tracked_Alt =
Values of Own_Tracked_Alt = Lower (strictly) than other (intruder)|Other_Tracked_Alt = Values
Other_Tracked_Alt = Higher (strictly) than TCAS (own)|Up_Separation = Values of Up_Separation =
Smaler (strictly) than down separation and threshold|Down_Separation = Values of Down_Separation
= Between up separation and threshold|Alt_Layer_Value = Values of Alt_Layer_Value = Value
3|Other_RAC = Values of Other_RAC = Two|Other_Capability = Values of Other_Capability =
One|Climb_Inhibit = Values of Climb_Inhibit = One|Positive_RA_Alt_Thresh = Values of
Positive_RA_Alt_Thresh = Value3
values:        Current_Vertical_Sep = '1', Alt_Layer_Value = 'Value3', Down_Separation = '740',
               Up_Separation = '739', Other_RAC = 'Two', High_Confidence = 'true',
               Other_Tracked_Alt = '2', Other_Capability = 'One', Own_Tracked_Alt = '1',
               Positive_RA_Alt_Thresh = '740', Climb_Inhibit = 'One', Two_of_Three_Reports_Valid =
               'false'
```

**Test Case 8:**
```
choices:       1871720133Cat1Ch2-|-374476473Cat1Ch1-|-1325844154Cat1Ch2-|-603277445Cat1Ch1-|-
               1787817979Cat1Ch3-|-927131798Cat1Ch3-|-875777795Cat1Ch1-|-302439181Cat1Ch4-|-
               58393125Cat1Ch2-|-626069991Cat1Ch2-|-430451043Cat1Ch1-|-976751556Cat1Ch4

Current_Vertical_Sep = Values of Current_Vertical_Sep = Smaller than MAXALTDIF (600), but
positive (strictly)|High_Confidence = Values of High_Confidence =
True|Two_of_Three_Reports_Valid = Values of Two_of_Three_Reports_Valid = False|Own_Tracked_Alt =
Values of Own_Tracked_Alt = Higher (strictly) than other (intruder)|Other_Tracked_Alt = Values
Other_Tracked_Alt = Lower (strictly) than TCAS (own)|Up_Separation = Values of Up_Separation =
Between down separation and threshold|Down_Separation = Values of Down_Separation = Greater
(strictly) than up separation and threshold|Alt_Layer_Value = Values of Alt_Layer_Value = Value
3|Other_RAC = Values of Other_RAC = One|Other_Capability = Values of Other_Capability =
```

```
Two|Climb_Inhibit = Values of Climb_Inhibit = Zero|Positive_RA_Alt_Thresh = Values of
Positive_RA_Alt_Thresh = Value3
values:        Current_Vertical_Sep = '1', Alt_Layer_Value = 'Value3', Down_Separation = '741',
               Up_Separation = '740', Other_RAC = 'One', High_Confidence = 'true',
               Other_Tracked_Alt = '1', Other_Capability = 'Two', Own_Tracked_Alt = '2',
               Positive_RA_Alt_Thresh = '740', Climb_Inhibit = 'Zero', Two_of_Three_Reports_Valid
               = 'false'
```

**Test Case 9:**

```
choices:       1871720133Cat1Ch2—|—374476473Cat1Ch2—|—1325844154Cat1Ch1—|—603277445Cat1Ch2—|—
               1787817979Cat1Ch2—|—927131798Cat1Ch1—|—875777795Cat1Ch1—|—302439181Cat1Ch3—|—
               58393125Cat1Ch1—|—626069991Cat1Ch2—|—430451043Cat1Ch1—|—976751556Cat1Ch3
Current_Vertical_Sep = Values of Current_Vertical_Sep = Smaller than MAXALTDIF (600), but
positive (strictly)|High_Confidence = Values of High_Confidence =
False|Two_of_Three_Reports_Valid = Values of Two_of_Three_Reports_Valid = True|Own_Tracked_Alt =
Values of Own_Tracked_Alt = Same as other (intruder)|Other_Tracked_Alt = Values
Other_Tracked_Alt = Same as TCAS (own)|Up_Separation = Values of Up_Separation = Smaler
(strictly) than down separation and threshold|Down_Separation = Values of Down_Separation =
Greater (strictly) than up separation and threshold|Alt_Layer_Value = Values of Alt_Layer_Value
= Value 2|Other_RAC = Values of Other_RAC = Zero|Other_Capability = Values of Other_Capability =
Two|Climb_Inhibit = Values of Climb_Inhibit = Zero|Positive_RA_Alt_Thresh = Values of
Positive_RA_Alt_Thresh = Value2
values:        Current_Vertical_Sep = '1', Alt_Layer_Value = 'Value2', Down_Separation = '641',
               Up_Separation = '1', Other_RAC = 'Zero', High_Confidence = 'false',
               Other_Tracked_Alt = '1', Other_Capability = 'Two', Own_Tracked_Alt = '1',
               Positive_RA_Alt_Thresh = '640', Climb_Inhibit = 'Zero', Two_of_Three_Reports_Valid
               = 'true'
```

**Test Case 10:**

```
choices:       1871720133Cat1Ch1—|—374476473Cat1Ch2—|—1325844154Cat1Ch2—|—603277445Cat1Ch1—|—
               1787817979Cat1Ch3—|—927131798Cat1Ch1—|—875777795Cat1Ch3—|—302439181Cat1Ch1—|—
               58393125Cat1Ch1—|—626069991Cat1Ch1—|—430451043Cat1Ch1—|—976751556Cat1Ch2
Current_Vertical_Sep = Values of Current_Vertical_Sep = Greater (strictly) than MAXALTDIF
(600)|High_Confidence = Values of High_Confidence = False|Two_of_Three_Reports_Valid = Values of
Two_of_Three_Reports_Valid = False|Own_Tracked_Alt = Values of Own_Tracked_Alt = Higher
(strictly) than other (intruder)|Other_Tracked_Alt = Values Other_Tracked_Alt = Lower (strictly)
than TCAS (own)|Up_Separation = Values of Up_Separation = Greater (strictly) than down
separation and threshold|Down_Separation = Values of Down_Separation = Between up separation and
threshold|Alt_Layer_Value = Values of Alt_Layer_Value = Value 1|Other_RAC = Values of Other_RAC
= Zero|Other_Capability = Values of Other_Capability = One|Climb_Inhibit = Values of
Climb_Inhibit = Zero|Positive_RA_Alt_Thresh = Values of Positive_RA_Alt_Thresh = Value1
values:        Current_Vertical_Sep = '601', Alt_Layer_Value = 'Value1', Down_Separation = '500',
               Up_Separation = '501', Other_RAC = 'Zero', High_Confidence = 'false',
               Other_Tracked_Alt = '1', Other_Capability = 'One', Own_Tracked_Alt = '2',
               Positive_RA_Alt_Thresh = '500', Climb_Inhibit = 'Zero', Two_of_Three_Reports_Valid
               = 'false'
```

**Test Case 11:**

```
choices:       1871720133Cat1Ch2—|—374476473Cat1Ch1—|—1325844154Cat1Ch2—|—603277445Cat1Ch2—|—
               1787817979Cat1Ch2—|—927131798Cat1Ch1—|—875777795Cat1Ch3—|—302439181Cat1Ch1—|—
               58393125Cat1Ch3—|—626069991Cat1Ch2—|—430451043Cat1Ch1—|—976751556Cat1Ch1
Current_Vertical_Sep = Values of Current_Vertical_Sep = Smaller than MAXALTDIF (600), but
positive (strictly)|High_Confidence = Values of High_Confidence =
True|Two_of_Three_Reports_Valid = Values of Two_of_Three_Reports_Valid = False|Own_Tracked_Alt =
Values of Own_Tracked_Alt = Same as other (intruder)|Other_Tracked_Alt = Values
Other_Tracked_Alt = Same as TCAS (own)|Up_Separation = Values of Up_Separation = Greater
(strictly) than down separation and threshold|Down_Separation = Values of Down_Separation =
Between up separation and threshold|Alt_Layer_Value = Values of Alt_Layer_Value = Value
0|Other_RAC = Values of Other_RAC = Two|Other_Capability = Values of Other_Capability =
Two|Climb_Inhibit = Values of Climb_Inhibit = Zero|Positive_RA_Alt_Thresh = Values of
Positive_RA_Alt_Thresh = Value0
values:        Current_Vertical_Sep = '1', Alt_Layer_Value = 'Value0', Down_Separation = '400',
               Up_Separation = '401', Other_RAC = 'Two', High_Confidence = 'true',
               Other_Tracked_Alt = '1', Other_Capability = 'Two', Own_Tracked_Alt = '1',
               Climb_Inhibit = 'Zero', Positive_RA_Alt_Thresh = '400', Two_of_Three_Reports_Valid
               = 'false'
```

**Test Case 12:**
```
choices:      1871720133Cat1Ch1-|-374476473Cat1Ch2-|-1325844154Cat1Ch1-|-603277445Cat1Ch1-|-
              1787817979Cat1Ch3-|-927131798Cat1Ch2-|-875777795Cat1Ch1-|-302439181Cat1Ch1-|-
              58393125Cat1Ch2-|-626069991Cat1Ch1-|-430451043Cat1Ch2-|-976751556Cat1Ch1
Current_Vertical_Sep = Values of Current_Vertical_Sep = Greater (strictly) than MAXALTDIF
(600)|High_Confidence = Values of High_Confidence = False|Two_of_Three_Reports_Valid = Values of
Two_of_Three_Reports_Valid = True|Own_Tracked_Alt = Values of Own_Tracked_Alt = Higher
(strictly) than other (intruder)|Other_Tracked_Alt = Values Other_Tracked_Alt = Lower (strictly)
than TCAS (own)|Up_Separation = Values of Up_Separation = Smaler (strictly) than down separation
and threshold|Down_Separation = Values of Down_Separation = Greater (strictly) than up
separation and threshold|Alt_Layer_Value = Values of Alt_Layer_Value = Value 0|Other_RAC =
Values of Other_RAC = One|Other_Capability = Values of Other_Capability = One|Climb_Inhibit =
Values of Climb_Inhibit = One|Positive_RA_Alt_Thresh = Values of Positive_RA_Alt_Thresh = Value0
values:       Current_Vertical_Sep = '601', Alt_Layer_Value = 'Value0', Down_Separation = '401',
              Up_Separation = '1', Other_RAC = 'One', High_Confidence = 'false',
              Other_Tracked_Alt = '1', Other_Capability = 'One', Own_Tracked_Alt = '2',
              Positive_RA_Alt_Thresh = '400', Climb_Inhibit = 'One', Two_of_Three_Reports_Valid =
              'true'
```
**Test Case 13:**
```
choices:      1871720133Cat1Ch3
Current_Vertical_Sep = Values of Current_Vertical_Sep = Illegal value
values:       Current_Vertical_Sep = '0'
```
**Test Case 14:**
```
choices:      603277445Cat1Ch4
Own_Tracked_Alt = Values of Own_Tracked_Alt = Illegal
values:       Own_Tracked_Alt = '0'
```
**Test Case 15:**
```
choices:      1787817979Cat1Ch4
Other_Tracked_Alt = Values of Other_Tracked_Alt = Illegal
values:       Other_Tracked_Alt = '0'
```
**Test Case 16:**
```
choices:      927131798Cat1Ch4
Up_Separation = Values of Up_Separation = Illegal
values:       Up_Separation = '0'
```
**Test Case 17:**
```
choices:      875777795Cat1Ch4
Down_Separation = Values of Down_Separation = Illegal
values:       Down_Separation = '0'
```
**Test Case 18:**
```
choices:      976751556Cat1Ch5
Positive_RA_Alt_Thresh = Values of Positive_RA_Alt_Thresh = Illegal
values:       Positive_RA_Alt_Thresh = '2'
```

## *J.6    Pairwise Test Frames and Corresponding Test Cases when using ACTS*

**Test Case 1:**
```
choices:      302439181Cat1Ch1-|-976751556Cat1Ch1-|-603277445Cat1Ch2-|-1787817979Cat1Ch2-|-
              927131798Cat1Ch2-|-875777795Cat1Ch3-|-58393125Cat1Ch2-|-1871720133Cat1Ch2-|-
              374476473Cat1Ch2-|-1325844154Cat1Ch2-|-626069991Cat1Ch2-|-430451043Cat1Ch2
Alt_Layer_Value = Values of Alt_Layer_Value = Value 0|Positive_RA_Alt_Thresh = Values of
Positive_RA_Alt_Thresh = Value0|Own_Tracked_Alt = Values of Own_Tracked_Alt = Same as other
(intruder)|Other_Tracked_Alt = Values Other_Tracked_Alt = Same as TCAS (own)|Up_Separation =
Values of Up_Separation = Smaler (strictly) than down separation and threshold|Down_Separation =
Values of Down_Separation = Between up separation and threshold|Other_RAC = Values of Other_RAC
= One|Current_Vertical_Sep = Values of Current_Vertical_Sep = Smaller than MAXALTDIF (600), but
positive (strictly)|High_Confidence = Values of High_Confidence =
False|Two_of_Three_Reports_Valid = Values of Two_of_Three_Reports_Valid = False|Other_Capability
= Values of Other_Capability = Two|Climb_Inhibit = Values of Climb_Inhibit = One
values:       Current_Vertical_Sep = '1', Alt_Layer_Value = 'Value0', Down_Separation = '2',
              Up_Separation = '1', Other_RAC = 'One', High_Confidence = 'false',
              Other_Tracked_Alt = '1', Other_Capability = 'Two', Own_Tracked_Alt = '1',
              Positive_RA_Alt_Thresh = '400', Climb_Inhibit = 'One', Two_of_Three_Reports_Valid =
              'false'
```

**Test Case 2:**

choices:     302439181Cat1Ch2—|—976751556Cat1Ch2—|—603277445Cat1Ch3—|—1787817979Cat1Ch1—|—
            927131798Cat1Ch3—|—875777795Cat1Ch3—|—58393125Cat1Ch3—|—1871720133Cat1Ch1—|—
            374476473Cat1Ch1—|—1325844154Cat1Ch1—|—626069991Cat1Ch1—|—430451043Cat1Ch1

Alt_Layer_Value = Values of Alt_Layer_Value = Value 1|Positive_RA_Alt_Thresh = Values of
Positive_RA_Alt_Thresh = Value1|Own_Tracked_Alt = Values of Own_Tracked_Alt = Lower (strictly)
than other (intruder)|Other_Tracked_Alt = Values Other_Tracked_Alt = Higher (strictly) than TCAS
(own)|Up_Separation = Values of Up_Separation = Between down separation and
threshold|Down_Separation = Values of Down_Separation = Greater (strictly) than up separation
and threshold|Other_RAC = Values of Other_RAC = Two|Current_Vertical_Sep = Values of
Current_Vertical_Sep = Greater (strictly) than MAXALTDIF (600)|High_Confidence = Values of
High_Confidence = True|Two_of_Three_Reports_Valid = Values of Two_of_Three_Reports_Valid =
True|Other_Capability = Values of Other_Capability = One|Climb_Inhibit = Values of Climb_Inhibit
= Zero

values:      Current_Vertical_Sep = '601', Alt_Layer_Value = 'Value1', Down_Separation = '501',
            Up_Separation = '500', Other_RAC = 'Two', High_Confidence = 'true',
            Other_Tracked_Alt = '2', Other_Capability = 'One', Own_Tracked_Alt = '1',
            Climb_Inhibit = 'Zero', Positive_RA_Alt_Thresh = '500', Two_of_Three_Reports_Valid
            = 'true'

**Test Case 3:**

choices:     302439181Cat1Ch3—|—976751556Cat1Ch3—|—603277445Cat1Ch1—|—1787817979Cat1Ch3—|—
            927131798Cat1Ch1—|—875777795Cat1Ch2—|—58393125Cat1Ch3—|—1871720133Cat1Ch1—|—
            374476473Cat1Ch1—|—1325844154Cat1Ch2—|—626069991Cat1Ch1—|—430451043Cat1Ch2

Alt_Layer_Value = Values of Alt_Layer_Value = Value 2|Positive_RA_Alt_Thresh = Values of
Positive_RA_Alt_Thresh = Value2|Own_Tracked_Alt = Values of Own_Tracked_Alt = Higher (strictly)
than other (intruder)|Other_Tracked_Alt = Values Other_Tracked_Alt = Lower (strictly) than TCAS
(own)|Up_Separation = Values of Up_Separation = Greater (strictly) than down separation and
threshold|Down_Separation = Values of Down_Separation = Smaller (strictly) than up separation
and threshold|Other_RAC = Values of Other_RAC = Zero|Current_Vertical_Sep = Values of
Current_Vertical_Sep = Smaller than MAXALTDIF (600), but positive (strictly)|High_Confidence =
Values of High_Confidence = True|Two_of_Three_Reports_Valid = Values of
Two_of_Three_Reports_Valid = False|Other_Capability = Values of Other_Capability =
One|Climb_Inhibit = Values of Climb_Inhibit = One

values:      Current_Vertical_Sep = '1', Alt_Layer_Value = 'Value2', Down_Separation = '1',
            Up_Separation = '641', Other_RAC = 'Zero', High_Confidence = 'true',
            Other_Tracked_Alt = '1', Other_Capability = 'One', Own_Tracked_Alt = '2',
            Climb_Inhibit = 'One', Positive_RA_Alt_Thresh = '640', Two_of_Three_Reports_Valid =
            'false'

**Test Case 4:**

choices:     302439181Cat1Ch4—|—976751556Cat1Ch4—|—603277445Cat1Ch2—|—1787817979Cat1Ch2—|—
            927131798Cat1Ch3—|—875777795Cat1Ch2—|—58393125Cat1Ch3—|—1871720133Cat1Ch1—|—
            374476473Cat1Ch2—|—1325844154Cat1Ch1—|—626069991Cat1Ch2—|—430451043Cat1Ch1

Alt_Layer_Value = Values of Alt_Layer_Value = Value 3|Positive_RA_Alt_Thresh = Values of
Positive_RA_Alt_Thresh = Value3|Own_Tracked_Alt = Values of Own_Tracked_Alt = Same as other
(intruder)|Other_Tracked_Alt = Values Other_Tracked_Alt = Same as TCAS (own)|Up_Separation =
Values of Up_Separation = Between down separation and threshold|Down_Separation = Values of
Down_Separation = Smaller (strictly) than up separation and threshold|Other_RAC = Values of
Other_RAC = Two|Current_Vertical_Sep = Values of Current_Vertical_Sep = Greater (strictly) than
MAXALTDIF (600)|High_Confidence = Values of High_Confidence = False|Two_of_Three_Reports_Valid =
Values of Two_of_Three_Reports_Valid = True|Other_Capability = Values of Other_Capability =
Two|Climb_Inhibit = Values of Climb_Inhibit = Zero

values:      Current_Vertical_Sep = '601', Alt_Layer_Value = 'Value3', Down_Separation = '1',
            Up_Separation = '2', Other_RAC = 'Two', High_Confidence = 'false',
            Other_Tracked_Alt = '1', Other_Capability = 'Two', Own_Tracked_Alt = '1',
            Positive_RA_Alt_Thresh = '740', Climb_Inhibit = 'Zero', Two_of_Three_Reports_Valid
            = 'true'

**Test Case 5:**

choices:     302439181Cat1Ch1—|—976751556Cat1Ch1—|—603277445Cat1Ch1—|—1787817979Cat1Ch3—|—
            927131798Cat1Ch3—|—875777795Cat1Ch3—|—58393125Cat1Ch1—|—1871720133Cat1Ch1—|—
            374476473Cat1Ch2—|—1325844154Cat1Ch1—|—626069991Cat1Ch2—|—430451043Cat1Ch1

Alt_Layer_Value = Values of Alt_Layer_Value = Value 0|Positive_RA_Alt_Thresh = Values of
Positive_RA_Alt_Thresh = Value0|Own_Tracked_Alt = Values of Own_Tracked_Alt = Higher (strictly)
than other (intruder)|Other_Tracked_Alt = Values Other_Tracked_Alt = Lower (strictly) than TCAS
(own)|Up_Separation = Values of Up_Separation = Between down separation and
threshold|Down_Separation = Values of Down_Separation = Greater (strictly) than up separation

and threshold|Other_RAC = Values of Other_RAC = Zero|Current_Vertical_Sep = Values of
Current_Vertical_Sep = Greater (strictly) than MAXALTDIF (600)|High_Confidence = Values of
High_Confidence = False|Two_of_Three_Reports_Valid = Values of Two_of_Three_Reports_Valid =
True|Other_Capability = Values of Other_Capability = Two|Climb_Inhibit = Values of Climb_Inhibit
= Zero

```
values:     Current_Vertical_Sep = '601', Alt_Layer_Value = 'Value0', Down_Separation = '401',
            Up_Separation = '400', Other_RAC = 'Zero', High_Confidence = 'false',
            Other_Tracked_Alt = '1', Other_Capability = 'Two', Own_Tracked_Alt = '2',
            Positive_RA_Alt_Thresh = '400', Climb_Inhibit = 'Zero', Two_of_Three_Reports_Valid
            = 'true'
```

**Test Case 6:**

choices:    302439181Cat1Ch2-|-976751556Cat1Ch2-|-603277445Cat1Ch1-|-1787817979Cat1Ch3-|-
            927131798Cat1Ch2-|-875777795Cat1Ch2-|-58393125Cat1Ch2-|-1871720133Cat1Ch1-|-
            374467473Cat1Ch1-|-1325844154Cat1Ch2-|-626069991Cat1Ch1-|-430451043Cat1Ch1

Alt_Layer_Value = Values of Alt_Layer_Value = Value 1|Positive_RA_Alt_Thresh = Values of
Positive_RA_Alt_Thresh = Value1|Own_Tracked_Alt = Values of Own_Tracked_Alt = Higher (strictly)
than other (intruder)|Other_Tracked_Alt = Values Other_Tracked_Alt = Lower (strictly) than TCAS
(own)|Up_Separation = Values of Up_Separation = Smaler (strictly) than down separation and
threshold|Down_Separation = Values of Down_Separation = Between up separation and
threshold|Other_RAC = Values of Other_RAC = One|Current_Vertical_Sep = Values of
Current_Vertical_Sep = Greater (strictly) than MAXALTDIF (600)|High_Confidence = Values of
High_Confidence = True|Two_of_Three_Reports_Valid = Values of Two_of_Three_Reports_Valid =
False|Other_Capability = Values of Other_Capability = One|Climb_Inhibit = Values of
Climb_Inhibit = Zero

```
values:     Current_Vertical_Sep = '601', Alt_Layer_Value = 'Value1', Down_Separation = '500',
            Up_Separation = '499', Other_RAC = 'One', High_Confidence = 'true',
            Other_Tracked_Alt = '1', Other_Capability = 'One', Own_Tracked_Alt = '2',
            Climb_Inhibit = 'Zero', Positive_RA_Alt_Thresh = '500', Two_of_Three_Reports_Valid
            = 'false'
```

**Test Case 7:**

choices:    302439181Cat1Ch4-|-976751556Cat1Ch4-|-603277445Cat1Ch1-|-1787817979Cat1Ch3-|-
            927131798Cat1Ch1-|-875777795Cat1Ch3-|-58393125Cat1Ch3-|-1871720133Cat1Ch2-|-
            374467473Cat1Ch1-|-1325844154Cat1Ch1-|-626069991Cat1Ch2-|-430451043Cat1Ch2

Alt_Layer_Value = Values of Alt_Layer_Value = Value 3|Positive_RA_Alt_Thresh = Values of
Positive_RA_Alt_Thresh = Value3|Own_Tracked_Alt = Values of Own_Tracked_Alt = Higher (strictly)
than other (intruder)|Other_Tracked_Alt = Values Other_Tracked_Alt = Lower (strictly) than TCAS
(own)|Up_Separation = Values of Up_Separation = Greater (strictly) than down separation and
threshold|Down_Separation = Values of Down_Separation = Between up separation and
threshold|Other_RAC = Values of Other_RAC = Two|Current_Vertical_Sep = Values of
Current_Vertical_Sep = Smaller than MAXALTDIF (600), but positive (strictly)|High_Confidence =
Values of High_Confidence = True|Two_of_Three_Reports_Valid = Values of
Two_of_Three_Reports_Valid = True|Other_Capability = Values of Other_Capability =
Two|Climb_Inhibit = Values of Climb_Inhibit = One

```
values:     Current_Vertical_Sep = '1', Alt_Layer_Value = 'Value3', Down_Separation = '740',
            Up_Separation = '741', Other_RAC = 'Two', High_Confidence = 'true',
            Other_Tracked_Alt = '1', Other_Capability = 'Two', Own_Tracked_Alt = '2',
            Climb_Inhibit = 'One', Positive_RA_Alt_Thresh = '740', Two_of_Three_Reports_Valid =
            'true'
```

**Test Case 8:**

choices:    302439181Cat1Ch2-|-976751556Cat1Ch2-|-603277445Cat1Ch2-|-1787817979Cat1Ch2-|-
            927131798Cat1Ch1-|-875777795Cat1Ch2-|-58393125Cat1Ch1-|-1871720133Cat1Ch2-|-
            374467473Cat1Ch2-|-1325844154Cat1Ch2-|-626069991Cat1Ch1-|-430451043Cat1Ch2

Alt_Layer_Value = Values of Alt_Layer_Value = Value 1|Positive_RA_Alt_Thresh = Values of
Positive_RA_Alt_Thresh = Value1|Own_Tracked_Alt = Values of Own_Tracked_Alt = Same as other
(intruder)|Other_Tracked_Alt = Values Other_Tracked_Alt = Same as TCAS (own)|Up_Separation =
Values of Up_Separation = Greater (strictly) than down separation and threshold|Down_Separation
= Values of Down_Separation = Smaller (strictly) than up separation and threshold|Other_RAC =
Values of Other_RAC = Zero|Current_Vertical_Sep = Values of Current_Vertical_Sep = Smaller than
MAXALTDIF (600), but positive (strictly)|High_Confidence = Values of High_Confidence =
False|Two_of_Three_Reports_Valid = Values of Two_of_Three_Reports_Valid = False|Other_Capability
= Values of Other_Capability = One|Climb_Inhibit = Values of Climb_Inhibit = One

```
values:     Current_Vertical_Sep = '1', Alt_Layer_Value = 'Value1', Down_Separation = '499',
            Up_Separation = '501', Other_RAC = 'Zero', High_Confidence = 'false',
            Other_Tracked_Alt = '1', Other_Capability = 'One', Own_Tracked_Alt = '1',
```

Positive_RA_Alt_Thresh = '500', Climb_Inhibit = 'One', Two_of_Three_Reports_Valid = 'false'

**Test Case 9:**
choices:      302439181Cat1Ch3−|−976751556Cat1Ch3−|−603277445Cat1Ch2−|−1787817979Cat1Ch2−|−927131798Cat1Ch2−|−875777795Cat1Ch1−|−58393125Cat1Ch2−|−1871720133Cat1Ch1−|−374476473Cat1Ch1−|−1325844154Cat1Ch1−|−626069991Cat1Ch2−|−430451043Cat1Ch1
Alt_Layer_Value = Values of Alt_Layer_Value = Value 2|Positive_RA_Alt_Thresh = Values of Positive_RA_Alt_Thresh = Value2|Own_Tracked_Alt = Values of Own_Tracked_Alt = Same as other (intruder)|Other_Tracked_Alt = Values Other_Tracked_Alt = Same as TCAS (own)|Up_Separation = Values of Up_Separation = Smaler (strictly) than own separation and threshold|Down_Separation = Values of Down_Separation = Greater (strictly) than up separation and threshold|Other_RAC = Values of Other_RAC = One|Current_Vertical_Sep = Values of Current_Vertical_Sep = Greater (strictly) than MAXALTDIF (600)|High_Confidence = Values of High_Confidence = True|Two_of_Three_Reports_Valid = Values of Two_of_Three_Reports_Valid = True|Other_Capability = Values of Other_Capability = Two|Climb_Inhibit = Values of Climb_Inhibit = Zero
values:      Current_Vertical_Sep = '601', Alt_Layer_Value = 'Value2', Down_Separation = '641', Up_Separation = '1', Other_RAC = 'One', High_Confidence = 'true', Other_Tracked_Alt = '1', Other_Capability = 'Two', Own_Tracked_Alt = '1', Climb_Inhibit = 'Zero', Positive_RA_Alt_Thresh = '640', Two_of_Three_Reports_Valid = 'true'

**Test Case 10:**
choices:      302439181Cat1Ch1−|−976751556Cat1Ch1−|−603277445Cat1Ch3−|−1787817979Cat1Ch1−|−927131798Cat1Ch1−|−875777795Cat1Ch1−|−58393125Cat1Ch2−|−1871720133Cat1Ch2−|−374476473Cat1Ch2−|−1325844154Cat1Ch2−|−626069991Cat1Ch1−|−430451043Cat1Ch2
Alt_Layer_Value = Values of Alt_Layer_Value = Value 0|Positive_RA_Alt_Thresh = Values of Positive_RA_Alt_Thresh = Value0|Own_Tracked_Alt = Values of Own_Tracked_Alt = Lower (strictly) than other (intruder)|Other_Tracked_Alt = Values Other_Tracked_Alt = Higher (strictly) than TCAS (own)|Up_Separation = Values of Up_Separation = Greater (strictly) than down separation and threshold|Down_Separation = Values of Down_Separation = Smaller (strictly) than up separation and threshold|Other_RAC = Values of Other_RAC = One|Current_Vertical_Sep = Values of Current_Vertical_Sep = Smaller than MAXALTDIF (600), but positive (strictly)|High_Confidence = Values of High_Confidence = False|Two_of_Three_Reports_Valid = Values of Two_of_Three_Reports_Valid = False|Other_Capability = Values of Other_Capability = One|Climb_Inhibit = Values of Climb_Inhibit = One
values:      Current_Vertical_Sep = '1', Alt_Layer_Value = 'Value0', Down_Separation = '399', Up_Separation = '401', Other_RAC = 'One', High_Confidence = 'false', Other_Tracked_Alt = '2', Other_Capability = 'One', Own_Tracked_Alt = '1', Climb_Inhibit = 'One', Positive_RA_Alt_Thresh = '400', Two_of_Three_Reports_Valid = 'false'

**Test Case 11:**
choices:      302439181Cat1Ch3−|−976751556Cat1Ch3−|−603277445Cat1Ch3−|−1787817979Cat1Ch1−|−927131798Cat1Ch2−|−875777795Cat1Ch3−|−58393125Cat1Ch1−|−1871720133Cat1Ch1−|−374476473Cat1Ch2−|−1325844154Cat1Ch2−|−626069991Cat1Ch2−|−430451043Cat1Ch2
Alt_Layer_Value = Values of Alt_Layer_Value = Value 2|Positive_RA_Alt_Thresh = Values of Positive_RA_Alt_Thresh = Value2|Own_Tracked_Alt = Values of Own_Tracked_Alt = Lower (strictly) than other (intruder)|Other_Tracked_Alt = Values Other_Tracked_Alt = Higher (strictly) than TCAS (own)|Up_Separation = Values of Up_Separation = Smaler (strictly) than down separation and threshold|Down_Separation = Values of Down_Separation = Between up separation and threshold|Other_RAC = Values of Other_RAC = Zero|Current_Vertical_Sep = Values of Current_Vertical_Sep = Greater (strictly) than MAXALTDIF (600)|High_Confidence = Values of High_Confidence = False|Two_of_Three_Reports_Valid = Values of Two_of_Three_Reports_Valid = False|Other_Capability = Values of Other_Capability = Two|Climb_Inhibit = Values of Climb_Inhibit = One
values:      Current_Vertical_Sep = '601', Alt_Layer_Value = 'Value2', Down_Separation = '2', Up_Separation = '1', Other_RAC = 'Zero', High_Confidence = 'false', Other_Tracked_Alt = '2', Other_Capability = 'Two', Own_Tracked_Alt = '1', Positive_RA_Alt_Thresh = '640', Climb_Inhibit = 'One', Two_of_Three_Reports_Valid = 'false'

**Test Case 12:**
choices:      302439181Cat1Ch4−|−976751556Cat1Ch4−|−603277445Cat1Ch3−|−1787817979Cat1Ch1−|−927131798Cat1Ch2−|−875777795Cat1Ch1−|−58393125Cat1Ch2−|−1871720133Cat1Ch2−|−374476473Cat1Ch1−|−1325844154Cat1Ch2−|−626069991Cat1Ch1−|−430451043Cat1Ch1
Alt_Layer_Value = Values of Alt_Layer_Value = Value 3|Positive_RA_Alt_Thresh = Values of Positive_RA_Alt_Thresh = Value3|Own_Tracked_Alt = Values of Own_Tracked_Alt = Lower (strictly) than other (intruder)|Other_Tracked_Alt = Values Other_Tracked_Alt = Higher (strictly) than TCAS (own)|Up_Separation = Values of Up_Separation = Smaler (strictly) than down separation and

threshold|Down_Separation = Values of Down_Separation = Greater (strictly) than up separation
and threshold|Other_RAC = Values of Other_RAC = One|Current_Vertical_Sep = Values of
Current_Vertical_Sep = Smaller than MAXALTDIF (600), but positive (strictly)|High_Confidence =
Values of High_Confidence = True|Two_of_Three_Reports_Valid = Values of
Two_of_Three_Reports_Valid = False|Other_Capability = Values of Other_Capability =
One|Climb_Inhibit = Values of Climb_Inhibit = Zero
values:          Current_Vertical_Sep = '1', Alt_Layer_Value = 'Value3', Down_Separation = '741',
                 Up_Separation = '1', Other_RAC = 'One', High_Confidence = 'true', Other_Tracked_Alt
                 = '2', Other_Capability = 'One', Own_Tracked_Alt = '1', Climb_Inhibit = 'Zero',
                 Positive_RA_Alt_Thresh = '740', Two_of_Three_Reports_Valid = 'false'

**Test Case 13:**
choices:         302439181Cat1Ch3−|−976751556Cat1Ch3−|−603277445Cat1Ch3−|−1787817979Cat1Ch1−|−
                 927131798Cat1Ch3−|−875777795Cat1Ch3−|−58393125Cat1Ch3−|−1871720133Cat1Ch2−|−
                 374476473Cat1Ch1−|−1325844154Cat1Ch2−|−626069991Cat1Ch2−|−430451043Cat1Ch2
Alt_Layer_Value = Values of Alt_Layer_Value = Value 2|Positive_RA_Alt_Thresh = Values of
Positive_RA_Alt_Thresh = Value2|Own_Tracked_Alt = Values of Own_Tracked_Alt = Lower (strictly)
than other (intruder)|Other_Tracked_Alt = Values Other_Tracked_Alt = Higher (strictly) than TCAS
(own)|Up_Separation = Values of Up_Separation = Between down separation and
threshold|Down_Separation = Values of Down_Separation = Greater (strictly) than up separation
and threshold|Other_RAC = Values of Other_RAC = Two|Current_Vertical_Sep = Values of
Current_Vertical_Sep = Smaller than MAXALTDIF (600), but positive (strictly)|High_Confidence =
Values of High_Confidence = True|Two_of_Three_Reports_Valid = Values of
Two_of_Three_Reports_Valid = False|Other_Capability = Values of Other_Capability =
Two|Climb_Inhibit = Values of Climb_Inhibit = One
values:          Current_Vertical_Sep = '1', Alt_Layer_Value = 'Value2', Down_Separation = '641',
                 Up_Separation = '640', Other_RAC = 'Two', High_Confidence = 'true',
                 Other_Tracked_Alt = '2', Other_Capability = 'Two', Own_Tracked_Alt = '1',
                 Positive_RA_Alt_Thresh = '640', Climb_Inhibit = 'One', Two_of_Three_Reports_Valid =
                 'false'

**Test Case 14:**
choices:         302439181Cat1Ch2−|−976751556Cat1Ch2−|−603277445Cat1Ch3−|−1787817979Cat1Ch1−|−
                 927131798Cat1Ch3−|−875777795Cat1Ch2−|−58393125Cat1Ch2−|−1871720133Cat1Ch2−|−
                 374476473Cat1Ch2−|−1325844154Cat1Ch2−|−626069991Cat1Ch2−|−430451043Cat1Ch1
Alt_Layer_Value = Values of Alt_Layer_Value = Value 1|Positive_RA_Alt_Thresh = Values of
Positive_RA_Alt_Thresh = Value1|Own_Tracked_Alt = Values of Own_Tracked_Alt = Lower (strictly)
than other (intruder)|Other_Tracked_Alt = Values Other_Tracked_Alt = Higher (strictly) than TCAS
(own)|Up_Separation = Values of Up_Separation = Between down separation and
threshold|Down_Separation = Values of Down_Separation = Smaller (strictly) than up separation
and threshold|Other_RAC = Values of Other_RAC = One|Current_Vertical_Sep = Values of
Current_Vertical_Sep = Smaller than MAXALTDIF (600), but positive (strictly)|High_Confidence =
Values of High_Confidence = False|Two_of_Three_Reports_Valid = Values of
Two_of_Three_Reports_Valid = False|Other_Capability = Values of Other_Capability =
Two|Climb_Inhibit = Values of Climb_Inhibit = Zero
values:          Current_Vertical_Sep = '1', Alt_Layer_Value = 'Value1', Down_Separation = '1',
                 Up_Separation = '2', Other_RAC = 'One', High_Confidence = 'false',
                 Other_Tracked_Alt = '2', Other_Capability = 'Two', Own_Tracked_Alt = '1',
                 Climb_Inhibit = 'Zero', Positive_RA_Alt_Thresh = '500', Two_of_Three_Reports_Valid
                 = 'false'

**Test Case 15:**
choices:         302439181Cat1Ch1−|−976751556Cat1Ch1−|−603277445Cat1Ch3−|−1787817979Cat1Ch1−|−
                 927131798Cat1Ch2−|−875777795Cat1Ch3−|−58393125Cat1Ch3−|−1871720133Cat1Ch1−|−
                 374476473Cat1Ch1−|−1325844154Cat1Ch1−|−626069991Cat1Ch2−|−430451043Cat1Ch1
Alt_Layer_Value = Values of Alt_Layer_Value = Value 0|Positive_RA_Alt_Thresh = Values of
Positive_RA_Alt_Thresh = Value0|Own_Tracked_Alt = Values of Own_Tracked_Alt = Lower (strictly)
than other (intruder)|Other_Tracked_Alt = Values Other_Tracked_Alt = Higher (strictly) than TCAS
(own)|Up_Separation = Values of Up_Separation = Smaler (strictly) than down separation and
threshold|Down_Separation = Values of Down_Separation = Between up separation and
threshold|Other_RAC = Values of Other_RAC = Two|Current_Vertical_Sep = Values of
Current_Vertical_Sep = Greater (strictly) than MAXALTDIF (600)|High_Confidence = Values of
High_Confidence = True|Two_of_Three_Reports_Valid = Values of Two_of_Three_Reports_Valid =
True|Other_Capability = Values of Other_Capability = Two|Climb_Inhibit = Values of Climb_Inhibit
= Zero
values:          Current_Vertical_Sep = '601', Alt_Layer_Value = 'Value0', Down_Separation = '2',
                 Up_Separation = '1', Other_RAC = 'Two', High_Confidence = 'true', Other_Tracked_Alt

```
                    = '2', Other_Capability = 'Two', Own_Tracked_Alt = '1', Climb_Inhibit = 'Zero',
                    Positive_RA_Alt_Thresh = '400', Two_of_Three_Reports_Valid = 'true'
```

**Test Case 16:**
```
choices:        302439181Cat1Ch4-|-976751556Cat1Ch4-|-603277445Cat1Ch2-|-1787817979Cat1Ch2-|-
                927131798Cat1Ch1-|-875777795Cat1Ch3-|-58393125Cat1Ch1-|-1871720133Cat1Ch1-|-
                374476473Cat1Ch1-|-1325844154Cat1Ch1-|-626069991Cat1Ch2-|-430451043Cat1Ch1
Alt_Layer_Value = Values of Alt_Layer_Value = Value 3|Positive_RA_Alt_Thresh = Values of
Positive_RA_Alt_Thresh = Value3|Own_Tracked_Alt = Values of Own_Tracked_Alt = Same as other
(intruder)|Other_Tracked_Alt = Values Other_Tracked_Alt = Same as TCAS (own)|Up_Separation =
Values of Up_Separation = Greater (strictly) than down separation and threshold|Down_Separation
= Values of Down_Separation = Between up separation and threshold|Other_RAC = Values of
Other_RAC = Zero|Current_Vertical_Sep = Values of Current_Vertical_Sep = Greater (strictly) than
MAXALTDIF (600)|High_Confidence = Values of High_Confidence = True|Two_of_Three_Reports_Valid =
Values of Two_of_Three_Reports_Valid = True|Other_Capability = Values of Other_Capability =
Two|Climb_Inhibit = Values of Climb_Inhibit = Zero
values:         Current_Vertical_Sep = '601', Alt_Layer_Value = 'Value3', Down_Separation = '740',
                Up_Separation = '741', Other_RAC = 'Zero', High_Confidence = 'true',
                Other_Tracked_Alt = '1', Other_Capability = 'Two', Own_Tracked_Alt = '1',
                Positive_RA_Alt_Thresh = '740', Climb_Inhibit = 'Zero', Two_of_Three_Reports_Valid
                = 'true'
```

**Test Case 17:**
```
choices:        1871720133Cat1Ch3
Current_Vertical_Sep = Values of Current_Vertical_Sep = Illegal value
values:         Current_Vertical_Sep = '0'
```

**Test Case 18:**
```
choices:        603277445Cat1Ch4
Own_Tracked_Alt = Values of Own_Tracked_Alt = Illegal
values:         Own_Tracked_Alt = '0'
```

**Test Case 19:**
```
choices:        1787817979Cat1Ch4
Other_Tracked_Alt = Values Other_Tracked_Alt = Illegal
values:         Other_Tracked_Alt = '0'
```

**Test Case 20:**
```
choices:        927131798Cat1Ch4
Up_Separation = Values of Up_Separation = Illegal
values:         Up_Separation = '0'
```

**Test Case 21:**
```
choices:        875777795Cat1Ch4
Down_Separation = Values of Down_Separation = Illegal
values:         Down_Separation = '0'
```

**Test Case 22:**
```
choices:        976751556Cat1Ch5
Positive_RA_Alt_Thresh = Values of Positive_RA_Alt_Thresh = Illegal
values:         Positive_RA_Alt_Thresh = '2'
```

## J.7    Three-way Test Frames and Corresponding Test Cases when using CASA

**Test Case 1:**
```
choices:        1871720133Cat1Ch2-|-374476473Cat1Ch2-|-1325844154Cat1Ch1-|-603277445Cat1Ch3-|-
                1787817979Cat1Ch1-|-927131798Cat1Ch3-|-875777795Cat1Ch2-|-302439181Cat1Ch4-|-
                58393125Cat1Ch2-|-626069991Cat1Ch2-|-430451043Cat1Ch1-|-976751556Cat1Ch4
Current_Vertical_Sep = Values of Current_Vertical_Sep = Smaller than MAXALTDIF (600), but
positive (strictly)|High_Confidence = Values of High_Confidence =
False|Two_of_Three_Reports_Valid = Values of Two_of_Three_Reports_Valid = True|Own_Tracked_Alt =
Values of Own_Tracked_Alt = Lower (strictly) than other (intruder)|Other_Tracked_Alt = Values
Other_Tracked_Alt = Higher (strictly) than TCAS (own)|Up_Separation = Values of Up_Separation =
Between down separation and threshold|Down_Separation = Values of Down_Separation = Smaller
(strictly) than up separation and threshold|Alt_Layer_Value = Values of Alt_Layer_Value = Value
3|Other_RAC = Values of Other_RAC = One|Other_Capability = Values of Other_Capability =
Two|Climb_Inhibit = Values of Climb_Inhibit = Zero|Positive_RA_Alt_Thresh = Values of
Positive_RA_Alt_Thresh = Value3
values:         Current_Vertical_Sep = '1', Alt_Layer_Value = 'Value3', Down_Separation = '1',
                Up_Separation = '2', Other_RAC = 'One', High_Confidence = 'false',
                Other_Tracked_Alt = '2', Other_Capability = 'Two', Own_Tracked_Alt = '1',
```

Climb_Inhibit = 'Zero', Positive_RA_Alt_Thresh = '740', Two_of_Three_Reports_Valid = 'true'

**Test Case 2:**
choices:     1871720133Cat1Ch1-|—374476473Cat1Ch2-|—1325844154Cat1Ch1-|—603277445Cat1Ch1-|—1787817979Cat1Ch3-|—927131798Cat1Ch2-|—875777795Cat1Ch1-|—302439181Cat1Ch4-|—58393125Cat1Ch1-|—626069991Cat1Ch2-|—430451043Cat1Ch1-|—976751556Cat1Ch4

Current_Vertical_Sep = Values of Current_Vertical_Sep = Greater (strictly) than MAXALTDIF (600)|High_Confidence = Values of High_Confidence = False|Two_of_Three_Reports_Valid = Values of Two_of_Three_Reports_Valid = True|Own_Tracked_Alt = Values of Own_Tracked_Alt = Higher (strictly) than other (intruder)|Other_Tracked_Alt = Values Other_Tracked_Alt = Lower (strictly) than TCAS (own)|Up_Separation = Values of Up_Separation = Smaler (strictly) than down separation and threshold|Down_Separation = Values of Down_Separation = Greater (strictly) than up separation and threshold|Alt_Layer_Value = Values of Alt_Layer_Value = Value 3|Other_RAC = Values of Other_RAC = Zero|Other_Capability = Values of Other_Capability = Two|Climb_Inhibit = Values of Climb_Inhibit = Zero|Positive_RA_Alt_Thresh = Values of Positive_RA_Alt_Thresh = Value3
values:      Current_Vertical_Sep = '601', Alt_Layer_Value = 'Value3', Down_Separation = '741', Up_Separation = '739', Other_RAC = 'Zero', High_Confidence = 'false', Other_Tracked_Alt = '1', Other_Capability = 'Two', Own_Tracked_Alt = '2', Positive_RA_Alt_Thresh = '740', Climb_Inhibit = 'Zero', Two_of_Three_Reports_Valid = 'true'

**Test Case 3:**
choices:     1871720133Cat1Ch2-|—374476473Cat1Ch1-|—1325844154Cat1Ch2-|—603277445Cat1Ch1-|—1787817979Cat1Ch3-|—927131798Cat1Ch3-|—875777795Cat1Ch3-|—302439181Cat1Ch4-|—58393125Cat1Ch3-|—626069991Cat1Ch2-|—430451043Cat1Ch2-|—976751556Cat1Ch4

Current_Vertical_Sep = Values of Current_Vertical_Sep = Smaller than MAXALTDIF (600), but positive (strictly)|High_Confidence = Values of High_Confidence = True|Two_of_Three_Reports_Valid = Values of Two_of_Three_Reports_Valid = False|Own_Tracked_Alt = Values of Own_Tracked_Alt = Higher (strictly) than other (intruder)|Other_Tracked_Alt = Values Other_Tracked_Alt = Lower (strictly) than TCAS (own)|Up_Separation = Values of Up_Separation = Between down separation and threshold|Down_Separation = Values of Down_Separation = Smaller (strictly) than up separation and threshold|Alt_Layer_Value = Values of Alt_Layer_Value = Value 3|Other_RAC = Values of Other_RAC = Two|Other_Capability = Values of Other_Capability = Two|Climb_Inhibit = Values of Climb_Inhibit = One|Positive_RA_Alt_Thresh = Values of Positive_RA_Alt_Thresh = Value3
values:      Current_Vertical_Sep = '1', Alt_Layer_Value = 'Value3', Down_Separation = '1', Up_Separation = '2', Other_RAC = 'Two', High_Confidence = 'true', Other_Tracked_Alt = '1', Other_Capability = 'Two', Own_Tracked_Alt = '2', Positive_RA_Alt_Thresh = '740', Climb_Inhibit = 'One', Two_of_Three_Reports_Valid = 'false'

**Test Case 4:**
choices:     1871720133Cat1Ch1-|—374476473Cat1Ch1-|—1325844154Cat1Ch2-|—603277445Cat1Ch3-|—1787817979Cat1Ch1-|—927131798Cat1Ch2-|—875777795Cat1Ch1-|—302439181Cat1Ch4-|—58393125Cat1Ch3-|—626069991Cat1Ch1-|—430451043Cat1Ch2-|—976751556Cat1Ch4

Current_Vertical_Sep = Values of Current_Vertical_Sep = Greater (strictly) than MAXALTDIF (600)|High_Confidence = Values of High_Confidence = True|Two_of_Three_Reports_Valid = Values of Two_of_Three_Reports_Valid = False|Own_Tracked_Alt = Values of Own_Tracked_Alt = Lower (strictly) than other (intruder)|Other_Tracked_Alt = Values Other_Tracked_Alt = Higher (strictly) than TCAS (own)|Up_Separation = Values of Up_Separation = Smaler (strictly) than down separation and threshold|Down_Separation = Values of Down_Separation = Greater (strictly) than up separation and threshold|Alt_Layer_Value = Values of Alt_Layer_Value = Value 3|Other_RAC = Values of Other_RAC = Two|Other_Capability = Values of Other_Capability = One|Climb_Inhibit = Values of Climb_Inhibit = One|Positive_RA_Alt_Thresh = Values of Positive_RA_Alt_Thresh = Value3
values:      Current_Vertical_Sep = '601', Alt_Layer_Value = 'Value3', Down_Separation = '741', Up_Separation = '1', Other_RAC = 'Two', High_Confidence = 'true', Other_Tracked_Alt = '2', Other_Capability = 'One', Own_Tracked_Alt = '1', Positive_RA_Alt_Thresh = '740', Climb_Inhibit = 'One', Two_of_Three_Reports_Valid = 'false'

**Test Case 5:**
choices:     1871720133Cat1Ch2-|—374476473Cat1Ch1-|—1325844154Cat1Ch2-|—603277445Cat1Ch3-|—1787817979Cat1Ch1-|—927131798Cat1Ch1-|—875777795Cat1Ch3-|—302439181Cat1Ch4-|—58393125Cat1Ch1-|—626069991Cat1Ch1-|—430451043Cat1Ch1-|—976751556Cat1Ch4

Current_Vertical_Sep = Values of Current_Vertical_Sep = Smaller than MAXALTDIF (600), but positive (strictly)|High_Confidence = Values of High_Confidence = True|Two_of_Three_Reports_Valid = Values of Two_of_Three_Reports_Valid = False|Own_Tracked_Alt = Values of Own_Tracked_Alt = Lower (strictly) than other (intruder)|Other_Tracked_Alt = Values Other_Tracked_Alt = Higher (strictly) than TCAS (own)|Up_Separation = Values of Up_Separation =

Greater (strictly) than down separation and threshold|Down_Separation = Values of
Down_Separation = Between up separation and threshold|Alt_Layer_Value = Values of
Alt_Layer_Value = Value 3|Other_RAC = Values of Other_RAC = Zero|Other_Capability = Values of
Other_Capability = One|Climb_Inhibit = Values of Climb_Inhibit = Zero|Positive_RA_Alt_Thresh =
Values of Positive_RA_Alt_Thresh = Value3
values:          Current_Vertical_Sep = '1', Alt_Layer_Value = 'Value3', Down_Separation = '740',
                 Up_Separation = '741', Other_RAC = 'Zero', High_Confidence = 'true',
                 Other_Tracked_Alt = '2', Other_Capability = 'One', Own_Tracked_Alt = '1',
                 Climb_Inhibit = 'Zero', Positive_RA_Alt_Thresh = '740', Two_of_Three_Reports_Valid
                 = 'false'

**Test Case 6:**
choices:         1871720133Cat1Ch1-|-374476473Cat1Ch1-|-1325844154Cat1Ch1-|-603277445Cat1Ch1-|-
                 1787817979Cat1Ch3-|-927131798Cat1Ch2-|-875777795Cat1Ch1-|-302439181Cat1Ch3-|-
                 58393125Cat1Ch2-|-626069991Cat1Ch1-|-430451043Cat1Ch2-|-976751556Cat1Ch3
Current_Vertical_Sep = Values of Current_Vertical_Sep = Greater (strictly) than MAXALTDIF
(600)|High_Confidence = Values of High_Confidence = True|Two_of_Three_Reports_Valid = Values of
Two_of_Three_Reports_Valid = True|Own_Tracked_Alt = Values of Own_Tracked_Alt = Higher
(strictly) than other (intruder)|Other_Tracked_Alt = Values Other_Tracked_Alt = Lower (strictly)
than TCAS (own)|Up_Separation = Values of Up_Separation = Smaler (strictly) than down separation
and threshold|Down_Separation = Values of Down_Separation = Greater (strictly) than up
separation and threshold|Alt_Layer_Value = Values of Alt_Layer_Value = Value 2|Other_RAC =
Values of Other_RAC = One|Other_Capability = Values of Other_Capability = One|Climb_Inhibit =
Values of Climb_Inhibit = One|Positive_RA_Alt_Thresh = Values of Positive_RA_Alt_Thresh = Value2
values:          Current_Vertical_Sep = '601', Alt_Layer_Value = 'Value2', Down_Separation = '641',
                 Up_Separation = '639', Other_RAC = 'One', High_Confidence = 'true',
                 Other_Tracked_Alt = '1', Other_Capability = 'One', Own_Tracked_Alt = '2',
                 Positive_RA_Alt_Thresh = '640', Climb_Inhibit = 'One', Two_of_Three_Reports_Valid =
                 'true'

**Test Case 7:**
choices:         1871720133Cat1Ch1-|-374476473Cat1Ch2-|-1325844154Cat1Ch1-|-603277445Cat1Ch2-|-
                 1787817979Cat1Ch2-|-927131798Cat1Ch1-|-875777795Cat1Ch3-|-302439181Cat1Ch4-|-
                 58393125Cat1Ch3-|-626069991Cat1Ch2-|-430451043Cat1Ch1-|-976751556Cat1Ch4
Current_Vertical_Sep = Values of Current_Vertical_Sep = Greater (strictly) than MAXALTDIF
(600)|High_Confidence = Values of High_Confidence = False|Two_of_Three_Reports_Valid = Values of
Two_of_Three_Reports_Valid = True|Own_Tracked_Alt = Values of Own_Tracked_Alt = Same as other
(intruder)|Other_Tracked_Alt = Values Other_Tracked_Alt = Same as TCAS (own)|Up_Separation =
Values of Up_Separation = Greater (strictly) than down separation and threshold|Down_Separation
= Values of Down_Separation = Between up separation and threshold|Alt_Layer_Value = Values of
Alt_Layer_Value = Value 3|Other_RAC = Values of Other_RAC = Two|Other_Capability = Values of
Other_Capability = Two|Climb_Inhibit = Values of Climb_Inhibit = Zero|Positive_RA_Alt_Thresh =
Values of Positive_RA_Alt_Thresh = Value3
values:          Current_Vertical_Sep = '601', Alt_Layer_Value = 'Value3', Down_Separation = '740',
                 Up_Separation = '741', Other_RAC = 'Two', High_Confidence = 'false',
                 Other_Tracked_Alt = '1', Other_Capability = 'Two', Own_Tracked_Alt = '1',
                 Positive_RA_Alt_Thresh = '740', Climb_Inhibit = 'Zero', Two_of_Three_Reports_Valid
                 = 'true'

**Test Case 8:**
choices:         1871720133Cat1Ch2-|-374476473Cat1Ch1-|-1325844154Cat1Ch2-|-603277445Cat1Ch2-|-
                 1787817979Cat1Ch2-|-927131798Cat1Ch2-|-875777795Cat1Ch3-|-302439181Cat1Ch2-|-
                 58393125Cat1Ch3-|-626069991Cat1Ch2-|-430451043Cat1Ch1-|-976751556Cat1Ch2
Current_Vertical_Sep = Values of Current_Vertical_Sep = Smaller than MAXALTDIF (600), but
positive (strictly)|High_Confidence = Values of High_Confidence =
True|Two_of_Three_Reports_Valid = Values of Two_of_Three_Reports_Valid = False|Own_Tracked_Alt =
Values of Own_Tracked_Alt = Same as other (intruder)|Other_Tracked_Alt = Values
Other_Tracked_Alt = Same as TCAS (own)|Up_Separation = Values of Up_Separation = Smaler
(strictly) than down separation and threshold|Down_Separation = Values of Down_Separation =
Between up separation and threshold|Alt_Layer_Value = Values of Alt_Layer_Value = Value
1|Other_RAC = Values of Other_RAC = Two|Other_Capability = Values of Other_Capability =
Two|Climb_Inhibit = Values of Climb_Inhibit = Zero|Positive_RA_Alt_Thresh = Values of
Positive_RA_Alt_Thresh = Value1
values:          Current_Vertical_Sep = '1', Alt_Layer_Value = 'Value1', Down_Separation = '2',
                 Up_Separation = '1', Other_RAC = 'Two', High_Confidence = 'true', Other_Tracked_Alt
                 = '1', Other_Capability = 'Two', Own_Tracked_Alt = '1', Positive_RA_Alt_Thresh =
                 '500', Climb_Inhibit = 'Zero', Two_of_Three_Reports_Valid = 'false'

**Test Case 9:**

choices:         1871720133Cat1Ch1−|−374476473Cat1Ch2−|−1325844154Cat1Ch2−|−603277445Cat1Ch3−|−
                 1787817979Cat1Ch1−|−927131798Cat1Ch1−|−875777795Cat1Ch2−|−302439181Cat1Ch2−|−
                 58393125Cat1Ch3−|−626069991Cat1Ch2−|−430451043Cat1Ch2−|−976751556Cat1Ch2

Current_Vertical_Sep = Values of Current_Vertical_Sep = Greater (strictly) than MAXALTDIF
(600)|High_Confidence = Values of High_Confidence = False|Two_of_Three_Reports_Valid = Values of
Two_of_Three_Reports_Valid = False|Own_Tracked_Alt = Values of Own_Tracked_Alt = Lower
(strictly) than other (intruder)|Other_Tracked_Alt = Values Other_Tracked_Alt = Higher
(strictly) than TCAS (own)|Up_Separation = Values of Up_Separation = Greater (strictly) than
down separation and threshold|Down_Separation = Values of Down_Separation = Smaller (strictly)
than up separation and threshold|Alt_Layer_Value = Values of Alt_Layer_Value = Value 1|Other_RAC
= Values of Other_RAC = Two|Other_Capability = Values of Other_Capability = Two|Climb_Inhibit =
Values of Climb_Inhibit = One|Positive_RA_Alt_Thresh = Values of Positive_RA_Alt_Thresh = Value1

values:          Current_Vertical_Sep = '601', Alt_Layer_Value = 'Value1', Down_Separation = '1',
                 Up_Separation = '501', Other_RAC = 'Two', High_Confidence = 'false',
                 Other_Tracked_Alt = '2', Other_Capability = 'Two', Own_Tracked_Alt = '1',
                 Positive_RA_Alt_Thresh = '500', Climb_Inhibit = 'One', Two_of_Three_Reports_Valid =
                 'false'

**Test Case 10:**

choices:         1871720133Cat1Ch1−|−374476473Cat1Ch2−|−1325844154Cat1Ch2−|−603277445Cat1Ch2−|−
                 1787817979Cat1Ch2−|−927131798Cat1Ch3−|−875777795Cat1Ch2−|−302439181Cat1Ch4−|−
                 58393125Cat1Ch1−|−626069991Cat1Ch1−|−430451043Cat1Ch2−|−976751556Cat1Ch4

Current_Vertical_Sep = Values of Current_Vertical_Sep = Greater (strictly) than MAXALTDIF
(600)|High_Confidence = Values of High_Confidence = False|Two_of_Three_Reports_Valid = Values of
Two_of_Three_Reports_Valid = False|Own_Tracked_Alt = Values of Own_Tracked_Alt = Same as other
(intruder)|Other_Tracked_Alt = Values Other_Tracked_Alt = Same as TCAS (own)|Up_Separation =
Values of Up_Separation = Between down separation and threshold|Down_Separation = Values of
Down_Separation = Smaller (strictly) than up separation and threshold|Alt_Layer_Value = Values
of Alt_Layer_Value = Value 3|Other_RAC = Values of Other_RAC = Zero|Other_Capability = Values of
Other_Capability = One|Climb_Inhibit = Values of Climb_Inhibit = One|Positive_RA_Alt_Thresh =
Values of Positive_RA_Alt_Thresh = Value3

values:          Current_Vertical_Sep = '601', Alt_Layer_Value = 'Value3', Down_Separation = '1',
                 Up_Separation = '2', Other_RAC = 'Zero', High_Confidence = 'false',
                 Other_Tracked_Alt = '1', Other_Capability = 'One', Own_Tracked_Alt = '1',
                 Positive_RA_Alt_Thresh = '740', Climb_Inhibit = 'One', Two_of_Three_Reports_Valid =
                 'false'

**Test Case 11:**

choices:         1871720133Cat1Ch1−|−374476473Cat1Ch2−|−1325844154Cat1Ch1−|−603277445Cat1Ch1−|−
                 1787817979Cat1Ch3−|−927131798Cat1Ch1−|−875777795Cat1Ch3−|−302439181Cat1Ch4−|−
                 58393125Cat1Ch2−|−626069991Cat1Ch1−|−430451043Cat1Ch2−|−976751556Cat1Ch4

Current_Vertical_Sep = Values of Current_Vertical_Sep = Greater (strictly) than MAXALTDIF
(600)|High_Confidence = Values of High_Confidence = False|Two_of_Three_Reports_Valid = Values of
Two_of_Three_Reports_Valid = True|Own_Tracked_Alt = Values of Own_Tracked_Alt = Higher
(strictly) than other (intruder)|Other_Tracked_Alt = Values Other_Tracked_Alt = Lower (strictly)
than TCAS (own)|Up_Separation = Values of Up_Separation = Greater (strictly) than down
separation and threshold|Down_Separation = Values of Down_Separation = Between up separation and
threshold|Alt_Layer_Value = Values of Alt_Layer_Value = Value 3|Other_RAC = Values of Other_RAC
= One|Other_Capability = Values of Other_Capability = One|Climb_Inhibit = Values of
Climb_Inhibit = One|Positive_RA_Alt_Thresh = Values of Positive_RA_Alt_Thresh = Value3

values:          Current_Vertical_Sep = '601', Alt_Layer_Value = 'Value3', Down_Separation = '740',
                 Up_Separation = '741', Other_RAC = 'One', High_Confidence = 'false',
                 Other_Tracked_Alt = '1', Other_Capability = 'One', Own_Tracked_Alt = '2',
                 Climb_Inhibit = 'One', Positive_RA_Alt_Thresh = '740', Two_of_Three_Reports_Valid =
                 'true'

**Test Case 12:**

choices:         1871720133Cat1Ch1−|−374476473Cat1Ch2−|−1325844154Cat1Ch2−|−603277445Cat1Ch3−|−
                 1787817979Cat1Ch1−|−927131798Cat1Ch1−|−875777795Cat1Ch3−|−302439181Cat1Ch3−|−
                 58393125Cat1Ch2−|−626069991Cat1Ch2−|−430451043Cat1Ch2−|−976751556Cat1Ch3

Current_Vertical_Sep = Values of Current_Vertical_Sep = Greater (strictly) than MAXALTDIF
(600)|High_Confidence = Values of High_Confidence = False|Two_of_Three_Reports_Valid = Values of
Two_of_Three_Reports_Valid = False|Own_Tracked_Alt = Values of Own_Tracked_Alt = Lower
(strictly) than other (intruder)|Other_Tracked_Alt = Values Other_Tracked_Alt = Higher
(strictly) than TCAS (own)|Up_Separation = Values of Up_Separation = Greater (strictly) than
down separation and threshold|Down_Separation = Values of Down_Separation = Between up
separation and threshold|Alt_Layer_Value = Values of Alt_Layer_Value = Value 2|Other_RAC =

```
Values of Other_RAC = One|Other_Capability = Values of Other_Capability = Two|Climb_Inhibit =
Values of Climb_Inhibit = One|Positive_RA_Alt_Thresh = Values of Positive_RA_Alt_Thresh = Value2
values:       Current_Vertical_Sep = '601', Alt_Layer_Value = 'Value2', Down_Separation = '640',
              Up_Separation = '641', Other_RAC = 'One', High_Confidence = 'false',
              Other_Tracked_Alt = '2', Other_Capability = 'Two', Own_Tracked_Alt = '1',
              Positive_RA_Alt_Thresh = '640', Climb_Inhibit = 'One', Two_of_Three_Reports_Valid =
              'false'
```

**Test Case 13:**

```
choices:      1871720133Cat1Ch1-|—374476473Cat1Ch1-|—1325844154Cat1Ch1-|—603277445Cat1Ch3-|—
              1787817979Cat1Ch1-|—927131798Cat1Ch3-|—875777795Cat1Ch1-|—302439181Cat1Ch3-|—
              58393125Cat1Ch1-|—626069991Cat1Ch1-|—430451043Cat1Ch2-|—976751556Cat1Ch3
Current_Vertical_Sep = Values of Current_Vertical_Sep = Greater (strictly) than MAXALTDIF
(600)|High_Confidence = Values of High_Confidence = True|Two_of_Three_Reports_Valid = Values of
Two_of_Three_Reports_Valid = True|Own_Tracked_Alt = Values of Own_Tracked_Alt = Lower (strictly)
than other (intruder)|Other_Tracked_Alt = Values Other_Tracked_Alt = Higher (strictly) than TCAS
(own)|Up_Separation = Values of Up_Separation = Between down separation and
threshold|Down_Separation = Values of Down_Separation = Smaller (strictly) than up separation
and threshold|Alt_Layer_Value = Values of Alt_Layer_Value = Value 2|Other_RAC = Values of
Other_RAC = Zero|Other_Capability = Values of Other_Capability = One|Climb_Inhibit = Values of
Climb_Inhibit = One|Positive_RA_Alt_Thresh = Values of Positive_RA_Alt_Thresh = Value2
values:       Current_Vertical_Sep = '601', Alt_Layer_Value = 'Value2', Down_Separation = '1',
              Up_Separation = '2', Other_RAC = 'Zero', High_Confidence = 'true',
              Other_Tracked_Alt = '2', Other_Capability = 'One', Own_Tracked_Alt = '1',
              Positive_RA_Alt_Thresh = '640', Climb_Inhibit = 'One', Two_of_Three_Reports_Valid =
              'true'
```

**Test Case 14:**

```
choices:      1871720133Cat1Ch2-|—374476473Cat1Ch2-|—1325844154Cat1Ch2-|—603277445Cat1Ch2-|—
              1787817979Cat1Ch2-|—927131798Cat1Ch3-|—875777795Cat1Ch1-|—302439181Cat1Ch2-|—
              58393125Cat1Ch1-|—626069991Cat1Ch1-|—430451043Cat1Ch2-|—976751556Cat1Ch2
Current_Vertical_Sep = Values of Current_Vertical_Sep = Smaller than MAXALTDIF (600), but
positive (strictly)|High_Confidence = Values of High_Confidence =
False|Two_of_Three_Reports_Valid = Values of Two_of_Three_Reports_Valid = False|Own_Tracked_Alt
= Values of Own_Tracked_Alt = Same as other (intruder)|Other_Tracked_Alt = Values
Other_Tracked_Alt = Same as TCAS (own)|Up_Separation = Values of Up_Separation = Between down
separation and threshold|Down_Separation = Values of Down_Separation = Greater (strictly) than
up separation and threshold|Alt_Layer_Value = Values of Alt_Layer_Value = Value 1|Other_RAC =
Values of Other_RAC = Zero|Other_Capability = Values of Other_Capability = One|Climb_Inhibit =
Values of Climb_Inhibit = One|Positive_RA_Alt_Thresh = Values of Positive_RA_Alt_Thresh = Value1
values:       Current_Vertical_Sep = '1', Alt_Layer_Value = 'Value1', Down_Separation = '501',
              Up_Separation = '500', Other_RAC = 'Zero', High_Confidence = 'false',
              Other_Tracked_Alt = '1', Other_Capability = 'One', Own_Tracked_Alt = '1',
              Climb_Inhibit = 'One', Positive_RA_Alt_Thresh = '500', Two_of_Three_Reports_Valid =
              'false'
```

**Test Case 15:**

```
choices:      1871720133Cat1Ch2-|—374476473Cat1Ch2-|—1325844154Cat1Ch1-|—603277445Cat1Ch3-|—
              1787817979Cat1Ch1-|—927131798Cat1Ch2-|—875777795Cat1Ch1-|—302439181Cat1Ch3-|—
              58393125Cat1Ch1-|—626069991Cat1Ch1-|—430451043Cat1Ch1-|—976751556Cat1Ch2
Current_Vertical_Sep = Values of Current_Vertical_Sep = Smaller than MAXALTDIF (600), but
positive (strictly)|High_Confidence = Values of High_Confidence =
False|Two_of_Three_Reports_Valid = Values of Two_of_Three_Reports_Valid = True|Own_Tracked_Alt =
Values of Own_Tracked_Alt = Lower (strictly) than other (intruder)|Other_Tracked_Alt = Values
Other_Tracked_Alt = Higher (strictly) than TCAS (own)|Up_Separation = Values of Up_Separation =
Smaler (strictly) than down separation and threshold|Down_Separation = Values of Down_Separation
= Greater (strictly) than up separation and threshold|Alt_Layer_Value = Values of
Alt_Layer_Value = Value 2|Other_RAC = Values of Other_RAC = Zero|Other_Capability = Values of
Other_Capability = One|Climb_Inhibit = Values of Climb_Inhibit = Zero|Positive_RA_Alt_Thresh =
Values of Positive_RA_Alt_Thresh = Value2
values:       Current_Vertical_Sep = '1', Alt_Layer_Value = 'Value2', Down_Separation = '641',
              Up_Separation = '639', Other_RAC = 'Zero', High_Confidence = 'false',
              Other_Tracked_Alt = '2', Other_Capability = 'One', Own_Tracked_Alt = '1',
              Positive_RA_Alt_Thresh = '640', Climb_Inhibit = 'Zero', Two_of_Three_Reports_Valid
              = 'true'
```

**Test Case 16:**

choices:    1871720133Cat1Ch2−|−374476473Cat1Ch1−|−1325844154Cat1Ch2−|−603277445Cat1Ch1−|−
            1787817979Cat1Ch3−|−927131798Cat1Ch3−|−875777795Cat1Ch1−|−302439181Cat1Ch1−|−
            58393125Cat1Ch2−|−626069991Cat1Ch1−|−430451043Cat1Ch2−|−976751556Cat1Ch1

Current_Vertical_Sep = Values of Current_Vertical_Sep = Smaller than MAXALTDIF (600), but
positive (strictly)|High_Confidence = Values of High_Confidence =
True|Two_of_Three_Reports_Valid = Values of Two_of_Three_Reports_Valid = False|Own_Tracked_Alt =
Values of Own_Tracked_Alt = Higher (strictly) than other (intruder)|Other_Tracked_Alt = Values
Other_Tracked_Alt = Lower (strictly) than TCAS (own)|Up_Separation = Values of Up_Separation =
Between down separation and threshold|Down_Separation = Values of Down_Separation = Greater
(strictly) than up separation and threshold|Alt_Layer_Value = Values of Alt_Layer_Value = Value
0|Other_RAC = Values of Other_RAC = One|Other_Capability = Values of Other_Capability =
One|Climb_Inhibit = Values of Climb_Inhibit = One|Positive_RA_Alt_Thresh = Values of
Positive_RA_Alt_Thresh = Value0

values:     Current_Vertical_Sep = '1', Alt_Layer_Value = 'Value0', Down_Separation = '401',
            Up_Separation = '400', Other_RAC = 'One', High_Confidence = 'true',
            Other_Tracked_Alt = '1', Other_Capability = 'One', Own_Tracked_Alt = '2',
            Climb_Inhibit = 'One', Positive_RA_Alt_Thresh = '400', Two_of_Three_Reports_Valid =
            'false'

**Test Case 17:**

choices:    1871720133Cat1Ch2−|−374476473Cat1Ch2−|−1325844154Cat1Ch1−|−603277445Cat1Ch3−|−
            1787817979Cat1Ch1−|−927131798Cat1Ch1−|−875777795Cat1Ch3−|−302439181Cat1Ch2−|−
            58393125Cat1Ch1−|−626069991Cat1Ch2−|−430451043Cat1Ch2−|−976751556Cat1Ch2

Current_Vertical_Sep = Values of Current_Vertical_Sep = Smaller than MAXALTDIF (600), but
positive (strictly)|High_Confidence = Values of High_Confidence =
False|Two_of_Three_Reports_Valid = Values of Two_of_Three_Reports_Valid = True|Own_Tracked_Alt =
Values of Own_Tracked_Alt = Lower (strictly) than other (intruder)|Other_Tracked_Alt = Values
Other_Tracked_Alt = Higher (strictly) than TCAS (own)|Up_Separation = Values of Up_Separation =
Greater (strictly) than down separation and threshold|Down_Separation = Values of
Down_Separation = Between up separation and threshold|Alt_Layer_Value = Values of
Alt_Layer_Value = Value 1|Other_RAC = Values of Other_RAC = Zero|Other_Capability = Values of
Other_Capability = Two|Climb_Inhibit = Values of Climb_Inhibit = One|Positive_RA_Alt_Thresh =
Values of Positive_RA_Alt_Thresh = Value1

values:     Current_Vertical_Sep = '1', Alt_Layer_Value = 'Value1', Down_Separation = '500',
            Up_Separation = '501', Other_RAC = 'Zero', High_Confidence = 'false',
            Other_Capability = 'Two', Other_Tracked_Alt = '2', Own_Tracked_Alt = '1',
            Positive_RA_Alt_Thresh = '500', Climb_Inhibit = 'One', Two_of_Three_Reports_Valid =
            'true'

**Test Case 18:**

choices:    1871720133Cat1Ch1−|−374476473Cat1Ch1−|−1325844154Cat1Ch1−|−603277445Cat1Ch2−|−
            1787817979Cat1Ch2−|−927131798Cat1Ch1−|−875777795Cat1Ch2−|−302439181Cat1Ch3−|−
            58393125Cat1Ch2−|−626069991Cat1Ch2−|−430451043Cat1Ch1−|−976751556Cat1Ch3

Current_Vertical_Sep = Values of Current_Vertical_Sep = Greater (strictly) than MAXALTDIF
(600)|High_Confidence = Values of High_Confidence = True|Two_of_Three_Reports_Valid = Values of
Two_of_Three_Reports_Valid = True|Own_Tracked_Alt = Values of Own_Tracked_Alt = Same as other
(intruder)|Other_Tracked_Alt = Values Other_Tracked_Alt = Same as TCAS (own)|Up_Separation =
Values of Up_Separation = Greater (strictly) than down separation and threshold|Down_Separation
= Values of Down_Separation = Smaller (strictly) than up separation and
threshold|Alt_Layer_Value = Values of Alt_Layer_Value = Value 2|Other_RAC = Values of Other_RAC
= One|Other_Capability = Values of Other_Capability = Two|Climb_Inhibit = Values of
Climb_Inhibit = Zero|Positive_RA_Alt_Thresh = Values of Positive_RA_Alt_Thresh = Value2

values:     Current_Vertical_Sep = '601', Alt_Layer_Value = 'Value2', Down_Separation = '1',
            Up_Separation = '641', Other_RAC = 'One', High_Confidence = 'true',
            Other_Tracked_Alt = '1', Other_Capability = 'Two', Own_Tracked_Alt = '1',
            Positive_RA_Alt_Thresh = '640', Climb_Inhibit = 'Zero', Two_of_Three_Reports_Valid
            = 'true'

**Test Case 19:**

choices:    1871720133Cat1Ch1−|−374476473Cat1Ch2−|−1325844154Cat1Ch2−|−603277445Cat1Ch3−|−
            1787817979Cat1Ch1−|−927131798Cat1Ch1−|−875777795Cat1Ch1−|−302439181Cat1Ch1−|−
            58393125Cat1Ch3−|−626069991Cat1Ch2−|−430451043Cat1Ch1−|−976751556Cat1Ch1

Current_Vertical_Sep = Values of Current_Vertical_Sep = Greater (strictly) than MAXALTDIF
(600)|High_Confidence = Values of High_Confidence = False|Two_of_Three_Reports_Valid = Values of
Two_of_Three_Reports_Valid = False|Own_Tracked_Alt = Values of Own_Tracked_Alt = Lower
(strictly) than other (intruder)|Other_Tracked_Alt = Values Other_Tracked_Alt = Higher
(strictly) than TCAS (own)|Up_Separation = Values of Up_Separation = Smaler (strictly) than down

separation and threshold|Down_Separation = Values of Down_Separation = Greater (strictly) than up separation and threshold|Alt_Layer_Value = Values of Alt_Layer_Value = Value 0|Other_RAC = Values of Other_RAC = Two|Other_Capability = Values of Other_Capability = Two|Climb_Inhibit = Values of Climb_Inhibit = Zero|Positive_RA_Alt_Thresh = Values of Positive_RA_Alt_Thresh = Value0

values:     Current_Vertical_Sep = '601', Alt_Layer_Value = 'Value0', Down_Separation = '401',
            Up_Separation = '1', Other_RAC = 'Two', High_Confidence = 'false',
            Other_Tracked_Alt = '2', Other_Capability = 'Two', Own_Tracked_Alt = '1',
            Positive_RA_Alt_Thresh = '400', Climb_Inhibit = 'Zero', Two_of_Three_Reports_Valid
            = 'false'

**Test Case 20:**
choices:    1871720133Cat1Ch1—|—374476473Cat1Ch2—|—1325844154Cat1Ch2—|—603277445Cat1Ch1—|—
            1787817979Cat1Ch3—|—927131798Cat1Ch3—|—875777795Cat1Ch3—|—302439181Cat1Ch2—|—
            58393125Cat1Ch2—|—626069991Cat1Ch1—|—430451043Cat1Ch1—|—976751556Cat1Ch2

Current_Vertical_Sep = Values of Current_Vertical_Sep = Greater (strictly) than MAXALTDIF (600)|High_Confidence = Values of High_Confidence = False|Two_of_Three_Reports_Valid = Values of Two_of_Three_Reports_Valid = False|Own_Tracked_Alt = Values of Own_Tracked_Alt = Higher (strictly) than other (intruder)|Other_Tracked_Alt = Values Other_Tracked_Alt = Lower (strictly) than TCAS (own)|Up_Separation = Values of Up_Separation = Greater (strictly) than down separation and threshold|Down_Separation = Values of Down_Separation = Between up separation and threshold|Alt_Layer_Value = Values of Alt_Layer_Value = Value 1|Other_RAC = Values of Other_RAC = One|Other_Capability = Values of Other_Capability = One|Climb_Inhibit = Values of Climb_Inhibit = Zero|Positive_RA_Alt_Thresh = Values of Positive_RA_Alt_Thresh = Value1

values:     Current_Vertical_Sep = '601', Alt_Layer_Value = 'Value1', Down_Separation = '500',
            Up_Separation = '501', Other_RAC = 'One', High_Confidence = 'false',
            Other_Capability = 'One', Other_Tracked_Alt = '1', Own_Tracked_Alt = '2',
            Positive_RA_Alt_Thresh = '500', Two_of_Three_Reports_Valid = 'false', Climb_Inhibit
            = 'Zero'

**Test Case 21:**
choices:    1871720133Cat1Ch2—|—374476473Cat1Ch2—|—1325844154Cat1Ch1—|—603277445Cat1Ch1—|—
            1787817979Cat1Ch3—|—927131798Cat1Ch3—|—875777795Cat1Ch2—|—302439181Cat1Ch1—|—
            58393125Cat1Ch1—|—626069991Cat1Ch2—|—430451043Cat1Ch1—|—976751556Cat1Ch1

Current_Vertical_Sep = Values of Current_Vertical_Sep = Smaller than MAXALTDIF (600), but positive (strictly)|High_Confidence = Values of High_Confidence = False|Two_of_Three_Reports_Valid = Values of Two_of_Three_Reports_Valid = True|Own_Tracked_Alt = Values of Own_Tracked_Alt = Higher (strictly) than other (intruder)|Other_Tracked_Alt = Values Other_Tracked_Alt = Lower (strictly) than TCAS (own)|Up_Separation = Values of Up_Separation = Between down separation and threshold|Down_Separation = Values of Down_Separation = Smaller (strictly) than up separation and threshold|Alt_Layer_Value = Values of Alt_Layer_Value = Value 0|Other_RAC = Values of Other_RAC = Zero|Other_Capability = Values of Other_Capability = Two|Climb_Inhibit = Values of Climb_Inhibit = Zero|Positive_RA_Alt_Thresh = Values of Positive_RA_Alt_Thresh = Value0

values:     Current_Vertical_Sep = '1', Alt_Layer_Value = 'Value0', Down_Separation = '1',
            Up_Separation = '2', Other_RAC = 'Zero', High_Confidence = 'false',
            Other_Tracked_Alt = '1', Other_Capability = 'Two', Own_Tracked_Alt = '2',
            Positive_RA_Alt_Thresh = '400', Climb_Inhibit = 'Zero', Two_of_Three_Reports_Valid
            = 'true'

**Test Case 22:**
choices:    1871720133Cat1Ch1—|—374476473Cat1Ch2—|—1325844154Cat1Ch2—|—603277445Cat1Ch2—|—
            1787817979Cat1Ch2—|—927131798Cat1Ch1—|—875777795Cat1Ch2—|—302439181Cat1Ch1—|—
            58393125Cat1Ch2—|—626069991Cat1Ch1—|—430451043Cat1Ch2—|—976751556Cat1Ch1

Current_Vertical_Sep = Values of Current_Vertical_Sep = Greater (strictly) than MAXALTDIF (600)|High_Confidence = Values of High_Confidence = False|Two_of_Three_Reports_Valid = Values of Two_of_Three_Reports_Valid = False|Own_Tracked_Alt = Values of Own_Tracked_Alt = Same as other (intruder)|Other_Tracked_Alt = Values Other_Tracked_Alt = Same as TCAS (own)|Up_Separation = Values of Up_Separation = Greater (strictly) than down separation and threshold|Down_Separation = Values of Down_Separation = Smaller (strictly) than up separation and threshold|Alt_Layer_Value = Values of Alt_Layer_Value = Value 0|Other_RAC = Values of Other_RAC = One|Other_Capability = Values of Other_Capability = One|Climb_Inhibit = Values of Climb_Inhibit = One|Positive_RA_Alt_Thresh = Values of Positive_RA_Alt_Thresh = Value0

values:     Current_Vertical_Sep = '601', Alt_Layer_Value = 'Value0', Down_Separation = '1',
            Up_Separation = '401', Other_RAC = 'One', High_Confidence = 'false',
            Other_Tracked_Alt = '1', Other_Capability = 'One', Own_Tracked_Alt = '1',
            Climb_Inhibit = 'One', Positive_RA_Alt_Thresh = '400', Two_of_Three_Reports_Valid =
            'false'

**Test Case 23:**

choices:      1871720133Cat1Ch2−|−374476473Cat1Ch2−|−1325844154Cat1Ch2−|−603277445Cat1Ch1−|−
              1787817979Cat1Ch3−|−927131798Cat1Ch3−|−875777795Cat1Ch2−|−302439181Cat1Ch3−|−
              58393125Cat1Ch2−|−626069991Cat1Ch2−|−430451043Cat1Ch1−|−976751556Cat1Ch3

Current_Vertical_Sep = Values of Current_Vertical_Sep = Smaller than MAXALTDIF (600), but
positive (strictly)|High_Confidence = Values of High_Confidence =
False|Two_of_Three_Reports_Valid = Values of Two_of_Three_Reports_Valid = False|Own_Tracked_Alt
= Values of Own_Tracked_Alt = Higher (strictly) than other (intruder)|Other_Tracked_Alt = Values
Other_Tracked_Alt = Lower (strictly) than TCAS (own)|Up_Separation = Values of Up_Separation =
Between down separation and threshold|Down_Separation = Values of Down_Separation = Smaller
(strictly) than up separation and threshold|Alt_Layer_Value = Values of Alt_Layer_Value = Value
2|Other_RAC = Values of Other_RAC = One|Other_Capability = Values of Other_Capability =
Two|Climb_Inhibit = Values of Climb_Inhibit = Zero|Positive_RA_Alt_Thresh = Values of
Positive_RA_Alt_Thresh = Value2

values:       Current_Vertical_Sep = '1', Alt_Layer_Value = 'Value2', Down_Separation = '1',
              Up_Separation = '2', Other_RAC = 'One', High_Confidence = 'false',
              Other_Tracked_Alt = '1', Other_Capability = 'Two', Own_Tracked_Alt = '2',
              Climb_Inhibit = 'Zero', Positive_RA_Alt_Thresh = '640', Two_of_Three_Reports_Valid
              = 'false'

**Test Case 24:**

choices:      1871720133Cat1Ch1−|−374476473Cat1Ch1−|−1325844154Cat1Ch1−|−603277445Cat1Ch3−|−
              1787817979Cat1Ch1−|−927131798Cat1Ch3−|−875777795Cat1Ch1−|−302439181Cat1Ch2−|−
              58393125Cat1Ch2−|−626069991Cat1Ch2−|−430451043Cat1Ch1−|−976751556Cat1Ch2

Current_Vertical_Sep = Values of Current_Vertical_Sep = Greater (strictly) than MAXALTDIF
(600)|High_Confidence = Values of High_Confidence = True|Two_of_Three_Reports_Valid = Values of
Two_of_Three_Reports_Valid = True|Own_Tracked_Alt = Values of Own_Tracked_Alt = Lower (strictly)
than other (intruder)|Other_Tracked_Alt = Values Other_Tracked_Alt = Higher (strictly) than TCAS
(own)|Up_Separation = Values of Up_Separation = Between down separation and
threshold|Down_Separation = Values of Down_Separation = Greater (strictly) than up separation
and threshold|Alt_Layer_Value = Values of Alt_Layer_Value = Value 1|Other_RAC = Values of
Other_RAC = One|Other_Capability = Values of Other_Capability = Two|Climb_Inhibit = Values of
Climb_Inhibit = Zero|Positive_RA_Alt_Thresh = Values of Positive_RA_Alt_Thresh = Value1

values:       Current_Vertical_Sep = '601', Alt_Layer_Value = 'Value1', Down_Separation = '501',
              Up_Separation = '500', Other_RAC = 'One', High_Confidence = 'true',
              Other_Tracked_Alt = '2', Other_Capability = 'Two', Own_Tracked_Alt = '1',
              Positive_RA_Alt_Thresh = '500', Climb_Inhibit = 'Zero', Two_of_Three_Reports_Valid
              = 'true'

**Test Case 25:**

choices:      1871720133Cat1Ch2−|−374476473Cat1Ch2−|−1325844154Cat1Ch2−|−603277445Cat1Ch3−|−
              1787817979Cat1Ch1−|−927131798Cat1Ch2−|−875777795Cat1Ch1−|−302439181Cat1Ch2−|−
              58393125Cat1Ch2−|−626069991Cat1Ch1−|−430451043Cat1Ch2−|−976751556Cat1Ch2

Current_Vertical_Sep = Values of Current_Vertical_Sep = Smaller than MAXALTDIF (600), but
positive (strictly)|High_Confidence = Values of High_Confidence =
False|Two_of_Three_Reports_Valid = Values of Two_of_Three_Reports_Valid = False|Own_Tracked_Alt
= Values of Own_Tracked_Alt = Lower (strictly) than other (intruder)|Other_Tracked_Alt = Values
Other_Tracked_Alt = Higher (strictly) than TCAS (own)|Up_Separation = Values of Up_Separation =
Smaler (strictly) than down separation and threshold|Down_Separation = Values of Down_Separation
= Greater (strictly) than up separation and threshold|Alt_Layer_Value = Values of
Alt_Layer_Value = Value 1|Other_RAC = Values of Other_RAC = One|Other_Capability = Values of
Other_Capability = One|Climb_Inhibit = Values of Climb_Inhibit = One|Positive_RA_Alt_Thresh =
Values of Positive_RA_Alt_Thresh = Value1

values:       Current_Vertical_Sep = '1', Alt_Layer_Value = 'Value1', Down_Separation = '501',
              Up_Separation = '1', Other_RAC = 'One', High_Confidence = 'false',
              Other_Tracked_Alt = '2', Other_Capability = 'One', Own_Tracked_Alt = '1',
              Positive_RA_Alt_Thresh = '500', Climb_Inhibit = 'One', Two_of_Three_Reports_Valid =
              'false'

**Test Case 26:**

choices:      1871720133Cat1Ch1−|−374476473Cat1Ch1−|−1325844154Cat1Ch1−|−603277445Cat1Ch1−|−
              1787817979Cat1Ch3−|−927131798Cat1Ch3−|−875777795Cat1Ch3−|−302439181Cat1Ch1−|−
              58393125Cat1Ch3−|−626069991Cat1Ch2−|−430451043Cat1Ch1−|−976751556Cat1Ch1

Current_Vertical_Sep = Values of Current_Vertical_Sep = Greater (strictly) than MAXALTDIF
(600)|High_Confidence = Values of High_Confidence = True|Two_of_Three_Reports_Valid = Values of
Two_of_Three_Reports_Valid = True|Own_Tracked_Alt = Values of Own_Tracked_Alt = Higher
(strictly) than other (intruder)|Other_Tracked_Alt = Values Other_Tracked_Alt = Lower (strictly)
than TCAS (own)|Up_Separation = Values of Up_Separation = Greater (strictly) than down

separation and threshold|Down_Separation = Values of Down_Separation = Between up separation and threshold|Alt_Layer_Value = Values of Alt_Layer_Value = Value 0|Other_RAC = Values of Other_RAC = Two|Other_Capability = Values of Other_Capability = Two|Climb_Inhibit = Values of Climb_Inhibit = Zero|Positive_RA_Alt_Thresh = Values of Positive_RA_Alt_Thresh = Value0

values:      Current_Vertical_Sep = '601', Alt_Layer_Value = 'Value0', Down_Separation = '400',
             Up_Separation = '401', Other_RAC = 'Two', High_Confidence = 'true',
             Other_Capability = 'Two', Other_Tracked_Alt = '1', Own_Tracked_Alt = '2',
             Positive_RA_Alt_Thresh = '400', Climb_Inhibit = 'Zero', Two_of_Three_Reports_Valid
             = 'true'

**Test Case 27:**

choices:     1871720133Cat1Ch1—|—374476473Cat1Ch1—|—1325844154Cat1Ch1—|—603277445Cat1Ch1—|—
             1787817979Cat1Ch3—|—927131798Cat1Ch3—|—875777795Cat1Ch2—|—302439181Cat1Ch2—|—
             58393125Cat1Ch3—|—626069991Cat1Ch3—|—430451043Cat1Ch1—|—976751556Cat1Ch2

Current_Vertical_Sep = Values of Current_Vertical_Sep = Greater (strictly) than MAXALTDIF (600)|High_Confidence = Values of High_Confidence = True|Two_of_Three_Reports_Valid = Values of Two_of_Three_Reports_Valid = True|Own_Tracked_Alt = Values of Own_Tracked_Alt = Higher (strictly) than other (intruder)|Other_Tracked_Alt = Values Other_Tracked_Alt = Lower (strictly) than TCAS (own)|Up_Separation = Values of Up_Separation = Between down separation and threshold|Down_Separation = Values of Down_Separation = Smaller (strictly) than up separation and threshold|Alt_Layer_Value = Values of Alt_Layer_Value = Value 1|Other_RAC = Values of Other_RAC = Two|Other_Capability = Values of Other_Capability = One|Climb_Inhibit = Values of Climb_Inhibit = Zero|Positive_RA_Alt_Thresh = Values of Positive_RA_Alt_Thresh = Value1

values:      Current_Vertical_Sep = '601', Alt_Layer_Value = 'Value1', Down_Separation = '1',
             Up_Separation = '1', Other_RAC = 'Two', High_Confidence = 'true', Other_Tracked_Alt
             = '1', Other_Capability = 'One', Own_Tracked_Alt = '2', Climb_Inhibit = 'Zero',
             Positive_RA_Alt_Thresh = '500', Two_of_Three_Reports_Valid = 'true'

**Test Case 28:**

choices:     1871720133Cat1Ch1—|—374476473Cat1Ch1—|—1325844154Cat1Ch1—|—603277445Cat1Ch2—|—
             1787817979Cat1Ch2—|—927131798Cat1Ch3—|—875777795Cat1Ch1—|—302439181Cat1Ch1—|—
             58393125Cat1Ch1—|—626069991Cat1Ch2—|—430451043Cat1Ch1—|—976751556Cat1Ch1

Current_Vertical_Sep = Values of Current_Vertical_Sep = Greater (strictly) than MAXALTDIF (600)|High_Confidence = Values of High_Confidence = True|Two_of_Three_Reports_Valid = Values of Two_of_Three_Reports_Valid = True|Own_Tracked_Alt = Values of Own_Tracked_Alt = Same as other (intruder)|Other_Tracked_Alt = Values Other_Tracked_Alt = Same as TCAS (own)|Up_Separation = Values of Up_Separation = Smaler (strictly) than down separation and threshold|Down_Separation = Values of Down_Separation = Greater (strictly) than up separation and threshold|Alt_Layer_Value = Values of Alt_Layer_Value = Value 0|Other_RAC = Values of Other_RAC = Zero|Other_Capability = Values of Other_Capability = Two|Climb_Inhibit = Values of Climb_Inhibit = Zero|Positive_RA_Alt_Thresh = Values of Positive_RA_Alt_Thresh = Value0

values:      Current_Vertical_Sep = '601', Alt_Layer_Value = 'Value0', Down_Separation = '401',
             Up_Separation = '1', Other_RAC = 'Zero', High_Confidence = 'true',
             Other_Tracked_Alt = '1', Other_Capability = 'Two', Own_Tracked_Alt = '1',
             Climb_Inhibit = 'Zero', Positive_RA_Alt_Thresh = '400', Two_of_Three_Reports_Valid
             = 'true'

**Test Case 29:**

choices:     1871720133Cat1Ch2—|—374476473Cat1Ch2—|—1325844154Cat1Ch2—|—603277445Cat1Ch2—|—
             1787817979Cat1Ch2—|—927131798Cat1Ch3—|—875777795Cat1Ch3—|—302439181Cat1Ch1—|—
             58393125Cat1Ch3—|—626069991Cat1Ch2—|—430451043Cat1Ch2—|—976751556Cat1Ch1

Current_Vertical_Sep = Values of Current_Vertical_Sep = Smaller than MAXALTDIF (600), but positive (strictly)|High_Confidence = Values of High_Confidence = False|Two_of_Three_Reports_Valid = Values of Two_of_Three_Reports_Valid = False|Own_Tracked_Alt = Values of Own_Tracked_Alt = Same as other (intruder)|Other_Tracked_Alt = Values Other_Tracked_Alt = Same as TCAS (own)|Up_Separation = Values of Up_Separation = Between down separation and threshold|Down_Separation = Values of Down_Separation = Smaller (strictly) than up separation and threshold|Alt_Layer_Value = Values of Alt_Layer_Value = Value 0|Other_RAC = Values of Other_RAC = Two|Other_Capability = Values of Other_Capability = Two|Climb_Inhibit = Values of Climb_Inhibit = One|Positive_RA_Alt_Thresh = Values of Positive_RA_Alt_Thresh = Value0

values:      Current_Vertical_Sep = '1', Alt_Layer_Value = 'Value0', Down_Separation = '1',
             Up_Separation = '2', Other_RAC = 'Two', High_Confidence = 'false',
             Other_Tracked_Alt = '1', Other_Capability = 'Two', Own_Tracked_Alt = '1',
             Positive_RA_Alt_Thresh = '400', Climb_Inhibit = 'One', Two_of_Three_Reports_Valid =
             'false'

**Test Case 30:**

choices:     1871720133Cat1Ch2−|−374476473Cat1Ch1−|−1325844154Cat1Ch2−|−603277445Cat1Ch1−|−
             1787817979Cat1Ch3−|−927131798Cat1Ch2−|−875777795Cat1Ch1−|−302439181Cat1Ch2−|−
             58393125Cat1Ch1−|−626069991Cat1Ch2−|−430451043Cat1Ch2−|−976751556Cat1Ch2

Current_Vertical_Sep = Values of Current_Vertical_Sep = Smaller than MAXALTDIF (600), but
positive (strictly)|High_Confidence = Values of High_Confidence =
True|Two_of_Three_Reports_Valid = Values of Two_of_Three_Reports_Valid = False|Own_Tracked_Alt =
Values of Own_Tracked_Alt = Higher (strictly) than other (intruder)|Other_Tracked_Alt = Values
Other_Tracked_Alt = Lower (strictly) than TCAS (own)|Up_Separation = Values of Up_Separation =
Smaller (strictly) than down separation and threshold|Down_Separation = Values of Down_Separation
= Greater (strictly) than up separation and threshold|Alt_Layer_Value = Values of
Alt_Layer_Value = Value 1|Other_RAC = Values of Other_RAC = Zero|Other_Capability = Values of
Other_Capability = Two|Climb_Inhibit = Values of Climb_Inhibit = One|Positive_RA_Alt_Thresh =
Values of Positive_RA_Alt_Thresh = Value1

values:      Current_Vertical_Sep = '1', Alt_Layer_Value = 'Value1', Down_Separation = '501',
             Up_Separation = '499', Other_RAC = 'Zero', High_Confidence = 'true',
             Other_Tracked_Alt = '1', Other_Capability = 'Two', Own_Tracked_Alt = '2',
             Positive_RA_Alt_Thresh = '500', Climb_Inhibit = 'One', Two_of_Three_Reports_Valid =
             'false'

**Test Case 31:**

choices:     1871720133Cat1Ch2−|−374476473Cat1Ch1−|−1325844154Cat1Ch1−|−603277445Cat1Ch2−|−
             1787817979Cat1Ch2−|−927131798Cat1Ch3−|−875777795Cat1Ch1−|−302439181Cat1Ch4−|−
             58393125Cat1Ch2−|−626069991Cat1Ch1−|−430451043Cat1Ch1−|−976751556Cat1Ch4

Current_Vertical_Sep = Values of Current_Vertical_Sep = Smaller than MAXALTDIF (600), but
positive (strictly)|High_Confidence = Values of High_Confidence =
True|Two_of_Three_Reports_Valid = Values of Two_of_Three_Reports_Valid = True|Own_Tracked_Alt =
Values of Own_Tracked_Alt = Same as other (intruder)|Other_Tracked_Alt = Values
Other_Tracked_Alt = Same as TCAS (own)|Up_Separation = Values of Up_Separation = Between down
separation and threshold|Down_Separation = Values of Down_Separation = Greater (strictly) than
up separation and threshold|Alt_Layer_Value = Values of Alt_Layer_Value = Value 3|Other_RAC =
Values of Other_RAC = One|Other_Capability = Values of Other_Capability = One|Climb_Inhibit =
Values of Climb_Inhibit = Zero|Positive_RA_Alt_Thresh = Values of Positive_RA_Alt_Thresh =
Value3

values:      Current_Vertical_Sep = '1', Alt_Layer_Value = 'Value3', Down_Separation = '741',
             Up_Separation = '740', Other_RAC = 'One', High_Confidence = 'true',
             Other_Tracked_Alt = '1', Other_Capability = 'One', Own_Tracked_Alt = '1',
             Positive_RA_Alt_Thresh = '740', Climb_Inhibit = 'Zero', Two_of_Three_Reports_Valid
             = 'true'

**Test Case 32:**

choices:     1871720133Cat1Ch1−|−374476473Cat1Ch2−|−1325844154Cat1Ch1−|−603277445Cat1Ch2−|−
             1787817979Cat1Ch2−|−927131798Cat1Ch1−|−875777795Cat1Ch2−|−302439181Cat1Ch2−|−
             58393125Cat1Ch1−|−626069991Cat1Ch1−|−430451043Cat1Ch1−|−976751556Cat1Ch2

Current_Vertical_Sep = Values of Current_Vertical_Sep = Greater (strictly) than MAXALTDIF
(600)|High_Confidence = Values of High_Confidence = False|Two_of_Three_Reports_Valid = Values of
Two_of_Three_Reports_Valid = True|Own_Tracked_Alt = Values of Own_Tracked_Alt = Same as other
(intruder)|Other_Tracked_Alt = Values Other_Tracked_Alt = Same as TCAS (own)|Up_Separation =
Values of Up_Separation = Greater (strictly) than down separation and threshold|Down_Separation
= Values of Down_Separation = Smaller (strictly) than up separation and
threshold|Alt_Layer_Value = Values of Alt_Layer_Value = Value 1|Other_RAC = Values of Other_RAC
= Zero|Other_Capability = Values of Other_Capability = One|Climb_Inhibit = Values of
Climb_Inhibit = Zero|Positive_RA_Alt_Thresh = Values of Positive_RA_Alt_Thresh = Value1

values:      Current_Vertical_Sep = '601', Alt_Layer_Value = 'Value1', Down_Separation = '499',
             Up_Separation = '501', Other_RAC = 'Zero', High_Confidence = 'false',
             Other_Tracked_Alt = '1', Other_Capability = 'One', Own_Tracked_Alt = '1',
             Climb_Inhibit = 'Zero', Positive_RA_Alt_Thresh = '500', Two_of_Three_Reports_Valid
             = 'true'

**Test Case 33:**

choices:     1871720133Cat1Ch2−|−374476473Cat1Ch1−|−1325844154Cat1Ch1−|−603277445Cat1Ch2−|−
             1787817979Cat1Ch2−|−927131798Cat1Ch3−|−875777795Cat1Ch1−|−302439181Cat1Ch3−|−
             58393125Cat1Ch3−|−626069991Cat1Ch1−|−430451043Cat1Ch2−|−976751556Cat1Ch3

Current_Vertical_Sep = Values of Current_Vertical_Sep = Smaller than MAXALTDIF (600), but
positive (strictly)|High_Confidence = Values of High_Confidence =
True|Two_of_Three_Reports_Valid = Values of Two_of_Three_Reports_Valid = True|Own_Tracked_Alt =
Values of Own_Tracked_Alt = Same as other (intruder)|Other_Tracked_Alt = Values
Other_Tracked_Alt = Same as TCAS (own)|Up_Separation = Values of Up_Separation = Between down

separation and threshold|Down_Separation = Values of Down_Separation = Greater (strictly) than
up separation and threshold|Alt_Layer_Value = Values of Alt_Layer_Value = Value 2|Other_RAC =
Values of Other_RAC = Two|Other_Capability = Values of Other_Capability = One|Climb_Inhibit =
Values of Climb_Inhibit = One|Positive_RA_Alt_Thresh = Values of Positive_RA_Alt_Thresh = Value2
values:        Current_Vertical_Sep = '1', Alt_Layer_Value = 'Value2', Down_Separation = '641',
               Up_Separation = '640', Other_RAC = 'Two', High_Confidence = 'true',
               Other_Tracked_Alt = '1', Other_Capability = 'One', Own_Tracked_Alt = '1',
               Positive_RA_Alt_Thresh = '640', Climb_Inhibit = 'One', Two_of_Three_Reports_Valid =
               'true'

**Test Case 34:**
choices:       1871720133Cat1Ch2-|—374476473Cat1Ch1-|—1325844154Cat1Ch1-|—603277445Cat1Ch3-|—
               1787817979Cat1Ch1-|—927131798Cat1Ch2-|—875777795Cat1Ch3-|—302439181Cat1Ch1-|—
               58393125Cat1Ch3-|—626069991Cat1Ch1-|—430451043Cat1Ch2-|—976751556Cat1Ch1
Current_Vertical_Sep = Values of Current_Vertical_Sep = Smaller than MAXALTDIF (600), but
positive (strictly)|High_Confidence = Values of High_Confidence =
True|Two_of_Three_Reports_Valid = Values of Two_of_Three_Reports_Valid = True|Own_Tracked_Alt =
Values of Own_Tracked_Alt = Lower (strictly) than other (intruder)|Other_Tracked_Alt = Values
Other_Tracked_Alt = Higher (strictly) than TCAS (own)|Up_Separation = Values of Up_Separation =
Smaler (strictly) than down separation and threshold|Down_Separation = Values of Down_Separation
= Between up separation and threshold|Alt_Layer_Value = Values of Alt_Layer_Value = Value
0|Other_RAC = Values of Other_RAC = Two|Other_Capability = Values of Other_Capability =
One|Climb_Inhibit = Values of Climb_Inhibit = One|Positive_RA_Alt_Thresh = Values of
Positive_RA_Alt_Thresh = Value0
values:        Current_Vertical_Sep = '1', Alt_Layer_Value = 'Value0', Down_Separation = '2',
               Up_Separation = '1', Other_RAC = 'Two', High_Confidence = 'true', Other_Tracked_Alt
               = '2', Other_Capability = 'One', Own_Tracked_Alt = '1', Positive_RA_Alt_Thresh =
               '400', Climb_Inhibit = 'One', Two_of_Three_Reports_Valid = 'true'

**Test Case 35:**
choices:       1871720133Cat1Ch2-|—374476473Cat1Ch2-|—1325844154Cat1Ch1-|—603277445Cat1Ch2-|—
               1787817979Cat1Ch2-|—927131798Cat1Ch2-|—875777795Cat1Ch3-|—302439181Cat1Ch3-|—
               58393125Cat1Ch3-|—626069991Cat1Ch1-|—430451043Cat1Ch1-|—976751556Cat1Ch1
Current_Vertical_Sep = Values of Current_Vertical_Sep = Smaller than MAXALTDIF (600), but
positive (strictly)|High_Confidence = Values of High_Confidence =
False|Two_of_Three_Reports_Valid = Values of Two_of_Three_Reports_Valid = True|Own_Tracked_Alt =
Values of Own_Tracked_Alt = Same as other (intruder)|Other_Tracked_Alt = Values
Other_Tracked_Alt = Same as TCAS (own)|Up_Separation = Values of Up_Separation = Smaler
(strictly) than down separation and threshold|Down_Separation = Values of Down_Separation =
Between up separation and threshold|Alt_Layer_Value = Values of Alt_Layer_Value = Value
2|Other_RAC = Values of Other_RAC = Two|Other_Capability = Values of Other_Capability =
One|Climb_Inhibit = Values of Climb_Inhibit = Zero|Positive_RA_Alt_Thresh = Values of
Positive_RA_Alt_Thresh = Value2
values:        Current_Vertical_Sep = '1', Alt_Layer_Value = 'Value2', Down_Separation = '2',
               Up_Separation = '1', Other_RAC = 'Two', High_Confidence = 'false',
               Other_Tracked_Alt = '1', Other_Capability = 'One', Own_Tracked_Alt = '1',
               Climb_Inhibit = 'Zero', Positive_RA_Alt_Thresh = '640', Two_of_Three_Reports_Valid
               = 'true'

**Test Case 36:**
choices:       1871720133Cat1Ch1-|—374476473Cat1Ch1-|—1325844154Cat1Ch1-|—603277445Cat1Ch3-|—
               1787817979Cat1Ch1-|—927131798Cat1Ch3-|—875777795Cat1Ch2-|—302439181Cat1Ch1-|—
               58393125Cat1Ch2-|—626069991Cat1Ch2-|—430451043Cat1Ch1-|—976751556Cat1Ch1
Current_Vertical_Sep = Values of Current_Vertical_Sep = Greater (strictly) than MAXALTDIF
(600)|High_Confidence = Values of High_Confidence = True|Two_of_Three_Reports_Valid = Values of
Two_of_Three_Reports_Valid = True|Own_Tracked_Alt = Values of Own_Tracked_Alt = Lower (strictly)
than other (intruder)|Other_Tracked_Alt = Values Other_Tracked_Alt = Higher (strictly) than TCAS
(own)|Up_Separation = Values of Up_Separation = Between down separation and
threshold|Down_Separation = Values of Down_Separation = Smaller (strictly) than up separation
and threshold|Alt_Layer_Value = Values of Alt_Layer_Value = Value 0|Other_RAC = Values of
Other_RAC = One|Other_Capability = Values of Other_Capability = Two|Climb_Inhibit = Values of
Climb_Inhibit = Zero|Positive_RA_Alt_Thresh = Values of Positive_RA_Alt_Thresh = Value0
values:        Current_Vertical_Sep = '601', Alt_Layer_Value = 'Value0', Down_Separation = '1',
               Up_Separation = '2', Other_RAC = 'One', High_Confidence = 'true', Other_Tracked_Alt
               = '2', Other_Capability = 'Two', Own_Tracked_Alt = '1', Positive_RA_Alt_Thresh =
               '400', Climb_Inhibit = 'Zero', Two_of_Three_Reports_Valid = 'true'

**Test Case 37:**
```
choices:     1871720133Cat1Ch1−|−374476473Cat1Ch1−|−1325844154Cat1Ch2−|−603277445Cat1Ch3−|−
             1787817979Cat1Ch1−|−927131798Cat1Ch3−|−875777795Cat1Ch1−|−302439181Cat1Ch3−|−
             58393125Cat1Ch3−|−626069991Cat1Ch2−|−430451043Cat1Ch2−|−976751556Cat1Ch3
```
Current_Vertical_Sep = Values of Current_Vertical_Sep = Greater (strictly) than MAXALTDIF
(600)|High_Confidence = Values of High_Confidence = True|Two_of_Three_Reports_Valid = Values of
Two_of_Three_Reports_Valid = False|Own_Tracked_Alt = Values of Own_Tracked_Alt = Lower
(strictly) than other (intruder)|Other_Tracked_Alt = Values Other_Tracked_Alt = Higher
(strictly) than TCAS (own)|Up_Separation = Values of Up_Separation = Between down separation and
threshold|Down_Separation = Values of Down_Separation = Greater (strictly) than up separation
and threshold|Alt_Layer_Value = Values of Alt_Layer_Value = Value 2|Other_RAC = Values of
Other_RAC = Two|Other_Capability = Values of Other_Capability = Two|Climb_Inhibit = Values of
Climb_Inhibit = One|Positive_RA_Alt_Thresh = Values of Positive_RA_Alt_Thresh = Value2
```
values:      Current_Vertical_Sep = '601', Alt_Layer_Value = 'Value2', Down_Separation = '641',
             Up_Separation = '640', Other_RAC = 'Two', High_Confidence = 'true',
             Other_Tracked_Alt = '2', Other_Capability = 'Two', Own_Tracked_Alt = '1',
             Positive_RA_Alt_Thresh = '640', Climb_Inhibit = 'One', Two_of_Three_Reports_Valid =
             'false'
```

**Test Case 38:**
```
choices:     1871720133Cat1Ch2−|−374476473Cat1Ch1−|−1325844154Cat1Ch2−|−603277445Cat1Ch2−|−
             1787817979Cat1Ch2−|−927131798Cat1Ch2−|−875777795Cat1Ch3−|−302439181Cat1Ch4−|−
             58393125Cat1Ch2−|−626069991Cat1Ch2−|−430451043Cat1Ch2−|−976751556Cat1Ch4
```
Current_Vertical_Sep = Values of Current_Vertical_Sep = Smaller than MAXALTDIF (600), but
positive (strictly)|High_Confidence = Values of High_Confidence =
True|Two_of_Three_Reports_Valid = Values of Two_of_Three_Reports_Valid = False|Own_Tracked_Alt =
Values of Own_Tracked_Alt = Same as other (intruder)|Other_Tracked_Alt = Values
Other_Tracked_Alt = Same as TCAS (own)|Up_Separation = Values of Up_Separation = Smaler
(strictly) than down separation and threshold|Down_Separation = Values of Down_Separation =
Between up separation and threshold|Alt_Layer_Value = Values of Alt_Layer_Value = Value
3|Other_RAC = Values of Other_RAC = One|Other_Capability = Values of Other_Capability =
Two|Climb_Inhibit = Values of Climb_Inhibit = One|Positive_RA_Alt_Thresh = Values of
Positive_RA_Alt_Thresh = Value3
```
values:      Current_Vertical_Sep = '1', Alt_Layer_Value = 'Value3', Down_Separation = '2',
             Up_Separation = '1', Other_RAC = 'One', High_Confidence = 'true', Other_Tracked_Alt
             = '1', Other_Capability = 'Two', Own_Tracked_Alt = '1', Positive_RA_Alt_Thresh =
             '740', Climb_Inhibit = 'One', Two_of_Three_Reports_Valid = 'false'
```

**Test Case 39:**
```
choices:     1871720133Cat1Ch2−|−374476473Cat1Ch2−|−1325844154Cat1Ch2−|−603277445Cat1Ch2−|−
             1787817979Cat1Ch2−|−927131798Cat1Ch2−|−875777795Cat1Ch3−|−302439181Cat1Ch3−|−
             58393125Cat1Ch1−|−626069991Cat1Ch2−|−430451043Cat1Ch1−|−976751556Cat1Ch3
```
Current_Vertical_Sep = Values of Current_Vertical_Sep = Smaller than MAXALTDIF (600), but
positive (strictly)|High_Confidence = Values of High_Confidence =
True|Two_of_Three_Reports_Valid = Values of Two_of_Three_Reports_Valid = False|Own_Tracked_Alt =
Values of Own_Tracked_Alt = Same as other (intruder)|Other_Tracked_Alt = Values
Other_Tracked_Alt = Same as TCAS (own)|Up_Separation = Values of Up_Separation = Smaler
(strictly) than down separation and threshold|Down_Separation = Values of Down_Separation =
Between up separation and threshold|Alt_Layer_Value = Values of Alt_Layer_Value = Value
2|Other_RAC = Values of Other_RAC = Zero|Other_Capability = Values of Other_Capability =
Two|Climb_Inhibit = Values of Climb_Inhibit = Zero|Positive_RA_Alt_Thresh = Values of
Positive_RA_Alt_Thresh = Value2
```
values:      Current_Vertical_Sep = '1', Alt_Layer_Value = 'Value2', Down_Separation = '2',
             Up_Separation = '1', Other_RAC = 'Zero', High_Confidence = 'true',
             Other_Tracked_Alt = '1', Other_Capability = 'Two', Own_Tracked_Alt = '1',
             Positive_RA_Alt_Thresh = '640', Climb_Inhibit = 'Zero', Two_of_Three_Reports_Valid
             = 'false'
```

**Test Case 40:**
```
choices:     1871720133Cat1Ch2−|−374476473Cat1Ch2−|−1325844154Cat1Ch1−|−603277445Cat1Ch3−|−
             1787817979Cat1Ch1−|−927131798Cat1Ch1−|−875777795Cat1Ch3−|−302439181Cat1Ch1−|−
             58393125Cat1Ch1−|−626069991Cat1Ch1−|−430451043Cat1Ch1−|−976751556Cat1Ch1
```
Current_Vertical_Sep = Values of Current_Vertical_Sep = Smaller than MAXALTDIF (600), but
positive (strictly)|High_Confidence = Values of High_Confidence =
False|Two_of_Three_Reports_Valid = Values of Two_of_Three_Reports_Valid = True|Own_Tracked_Alt =
Values of Own_Tracked_Alt = Lower (strictly) than other (intruder)|Other_Tracked_Alt = Values
Other_Tracked_Alt = Higher (strictly) than TCAS (own)|Up_Separation = Values of Up_Separation =
Greater (strictly) than down separation and threshold|Down_Separation = Values of

Down_Separation = Between up separation and threshold|Alt_Layer_Value = Values of
Alt_Layer_Value = Value 0|Other_RAC = Values of Other_RAC = Zero|Other_Capability = Values of
Other_Capability = One|Climb_Inhibit = Values of Climb_Inhibit = Zero|Positive_RA_Alt_Thresh =
Values of Positive_RA_Alt_Thresh = Value0
values:      Current_Vertical_Sep = '1', Alt_Layer_Value = 'Value0', Down_Separation = '400',
             Up_Separation = '401', Other_RAC = 'Zero', High_Confidence = 'false',
             Other_Capability = 'One', Other_Tracked_Alt = '2', Own_Tracked_Alt = '1',
             Positive_RA_Alt_Thresh = '400', Climb_Inhibit = 'Zero', Two_of_Three_Reports_Valid
             = 'true'

**Test Case 41:**
choices:     1871720133Cat1Ch1-|-374476473Cat1Ch1-|-1325844154Cat1Ch2-|-603277445Cat1Ch1-|-
             1787817979Cat1Ch3-|-927131798Cat1Ch2-|-875777795Cat1Ch3-|-302439181Cat1Ch1-|-
             58393125Cat1Ch1-|-626069991Cat1Ch1-|-430451043Cat1Ch2-|-976751556Cat1Ch1
Current_Vertical_Sep = Values of Current_Vertical_Sep = Greater (strictly) than MAXALTDIF
(600)|High_Confidence = Values of High_Confidence = True|Two_of_Three_Reports_Valid = Values of
Two_of_Three_Reports_Valid = False|Own_Tracked_Alt = Values of Own_Tracked_Alt = Higher
(strictly) than other (intruder)|Other_Tracked_Alt = Values Other_Tracked_Alt = Lower (strictly)
than TCAS (own)|Up_Separation = Values of Up_Separation = Smaler (strictly) than down separation
and threshold|Down_Separation = Values of Down_Separation = Between up separation and
threshold|Alt_Layer_Value = Values of Alt_Layer_Value = Value 0|Other_RAC = Values of Other_RAC
= Zero|Other_Capability = Values of Other_Capability = One|Climb_Inhibit = Values of
Climb_Inhibit = One|Positive_RA_Alt_Thresh = Values of Positive_RA_Alt_Thresh = Value0
values:      Current_Vertical_Sep = '601', Alt_Layer_Value = 'Value0', Down_Separation = '2',
             Up_Separation = '1', Other_RAC = 'Zero', High_Confidence = 'true',
             Other_Tracked_Alt = '1', Other_Capability = 'One', Own_Tracked_Alt = '2',
             Climb_Inhibit = 'One', Positive_RA_Alt_Thresh = '400', Two_of_Three_Reports_Valid =
             'false'

**Test Case 42:**
choices:     1871720133Cat1Ch2-|-374476473Cat1Ch1-|-1325844154Cat1Ch2-|-603277445Cat1Ch1-|-
             1787817979Cat1Ch3-|-927131798Cat1Ch1-|-875777795Cat1Ch3-|-302439181Cat1Ch3-|-
             58393125Cat1Ch1-|-626069991Cat1Ch1-|-430451043Cat1Ch2-|-976751556Cat1Ch3
Current_Vertical_Sep = Values of Current_Vertical_Sep = Smaller than MAXALTDIF (600), but
positive (strictly)|High_Confidence = Values of High_Confidence =
True|Two_of_Three_Reports_Valid = Values of Two_of_Three_Reports_Valid = False|Own_Tracked_Alt =
Values of Own_Tracked_Alt = Higher (strictly) than other (intruder)|Other_Tracked_Alt = Values
Other_Tracked_Alt = Lower (strictly) than TCAS (own)|Up_Separation = Values of Up_Separation =
Greater (strictly) than down separation and threshold|Down_Separation = Values of
Down_Separation = Between up separation and threshold|Alt_Layer_Value = Values of
Alt_Layer_Value = Value 2|Other_RAC = Values of Other_RAC = Zero|Other_Capability = Values of
Other_Capability = One|Climb_Inhibit = Values of Climb_Inhibit = One|Positive_RA_Alt_Thresh =
Values of Positive_RA_Alt_Thresh = Value2
values:      Current_Vertical_Sep = '1', Alt_Layer_Value = 'Value2', Down_Separation = '640',
             Up_Separation = '641', Other_RAC = 'Zero', High_Confidence = 'true',
             Other_Tracked_Alt = '1', Other_Capability = 'One', Own_Tracked_Alt = '2',
             Positive_RA_Alt_Thresh = '640', Climb_Inhibit = 'One', Two_of_Three_Reports_Valid =
             'false'

**Test Case 43:**
choices:     1871720133Cat1Ch2-|-374476473Cat1Ch2-|-1325844154Cat1Ch2-|-603277445Cat1Ch1-|-
             1787817979Cat1Ch3-|-927131798Cat1Ch1-|-875777795Cat1Ch2-|-302439181Cat1Ch3-|-
             58393125Cat1Ch3-|-626069991Cat1Ch1-|-430451043Cat1Ch1-|-976751556Cat1Ch3
Current_Vertical_Sep = Values of Current_Vertical_Sep = Smaller than MAXALTDIF (600), but
positive (strictly)|High_Confidence = Values of High_Confidence =
False|Two_of_Three_Reports_Valid = Values of Two_of_Three_Reports_Valid = False|Own_Tracked_Alt
= Values of Own_Tracked_Alt = Higher (strictly) than other (intruder)|Other_Tracked_Alt =
Other_Tracked_Alt = Lower (strictly) than TCAS (own)|Up_Separation = Values of Up_Separation =
Greater (strictly) than down separation and threshold|Down_Separation = Values of
Down_Separation = Smaller (strictly) than up separation and threshold|Alt_Layer_Value = Values
of Alt_Layer_Value = Value 2|Other_RAC = Values of Other_RAC = Two|Other_Capability = Values of
Other_Capability = One|Climb_Inhibit = Values of Climb_Inhibit = Zero|Positive_RA_Alt_Thresh =
Values of Positive_RA_Alt_Thresh = Value2
values:      Current_Vertical_Sep = '1', Alt_Layer_Value = 'Value2', Down_Separation = '1',
             Up_Separation = '641', Other_RAC = 'Two', High_Confidence = 'false',
             Other_Tracked_Alt = '1', Other_Capability = 'One', Own_Tracked_Alt = '2',
             Positive_RA_Alt_Thresh = '640', Climb_Inhibit = 'Zero', Two_of_Three_Reports_Valid
             = 'false'

**Test Case 44:**

choices:    1871720133Cat1Ch2-|-374476473Cat1Ch1-|-1325844154Cat1Ch2-|-603277445Cat1Ch2-|-
            1787817979Cat1Ch2-|-927131798Cat1Ch1-|-875777795Cat1Ch2-|-302439181Cat1Ch2-|-
            58393125Cat1Ch2-|-626069991Cat1Ch1-|-430451043Cat1Ch2-|-976751556Cat1Ch2

Current_Vertical_Sep = Values of Current_Vertical_Sep = Smaller than MAXALTDIF (600), but
positive (strictly)|High_Confidence = Values of High_Confidence =
True|Two_of_Three_Reports_Valid = Values of Two_of_Three_Reports_Valid = False|Own_Tracked_Alt =
Values of Own_Tracked_Alt = Same as other (intruder)|Other_Tracked_Alt = Values
Other_Tracked_Alt = Same as TCAS (own)|Up_Separation = Values of Up_Separation = Greater
(strictly) than down separation and threshold|Down_Separation = Values of Down_Separation =
Smaller (strictly) than up separation and threshold|Alt_Layer_Value = Values of Alt_Layer_Value
= Value 1|Other_RAC = Values of Other_RAC = One|Other_Capability = Values of Other_Capability =
One|Climb_Inhibit = Values of Climb_Inhibit = One|Positive_RA_Alt_Thresh = Values of
Positive_RA_Alt_Thresh = Value1

values:     Current_Vertical_Sep = '1', Alt_Layer_Value = 'Value1', Down_Separation = '1',
            Up_Separation = '501', Other_RAC = 'One', High_Confidence = 'true',
            Other_Tracked_Alt = '1', Other_Capability = 'One', Own_Tracked_Alt = '1',
            Positive_RA_Alt_Thresh = '500', Climb_Inhibit = 'One', Two_of_Three_Reports_Valid =
            'false'

**Test Case 45:**

choices:    1871720133Cat1Ch1-|-374476473Cat1Ch1-|-1325844154Cat1Ch1-|-603277445Cat1Ch1-|-
            1787817979Cat1Ch3-|-927131798Cat1Ch2-|-875777795Cat1Ch1-|-302439181Cat1Ch1-|-
            58393125Cat1Ch3-|-626069991Cat1Ch2-|-430451043Cat1Ch1-|-976751556Cat1Ch2

Current_Vertical_Sep = Values of Current_Vertical_Sep = Greater (strictly) than MAXALTDIF
(600)|High_Confidence = Values of High_Confidence = True|Two_of_Three_Reports_Valid = Values of
Two_of_Three_Reports_Valid = True|Own_Tracked_Alt = Values of Own_Tracked_Alt = Higher
(strictly) than other (intruder)|Other_Tracked_Alt = Values Other_Tracked_Alt = Lower (strictly)
than TCAS (own)|Up_Separation = Values of Up_Separation = Smaler (strictly) than down separation
and threshold|Down_Separation = Values of Down_Separation = Greater (strictly) than up
separation and threshold|Alt_Layer_Value = Values of Alt_Layer_Value = Value 1|Other_RAC =
Values of Other_RAC = Two|Other_Capability = Values of Other_Capability = Two|Climb_Inhibit =
Values of Climb_Inhibit = Zero|Positive_RA_Alt_Thresh = Values of Positive_RA_Alt_Thresh =
Value1

values:     Current_Vertical_Sep = '601', Alt_Layer_Value = 'Value1', Down_Separation = '501',
            Up_Separation = '499', Other_RAC = 'Two', High_Confidence = 'true',
            Other_Tracked_Alt = '1', Other_Capability = 'Two', Own_Tracked_Alt = '2',
            Positive_RA_Alt_Thresh = '500', Climb_Inhibit = 'Zero', Two_of_Three_Reports_Valid
            = 'true'

**Test Case 46:**

choices:    1871720133Cat1Ch2-|-374476473Cat1Ch1-|-1325844154Cat1Ch1-|-603277445Cat1Ch2-|-
            1787817979Cat1Ch2-|-927131798Cat1Ch2-|-875777795Cat1Ch3-|-302439181Cat1Ch1-|-
            58393125Cat1Ch2-|-626069991Cat1Ch2-|-430451043Cat1Ch1-|-976751556Cat1Ch1

Current_Vertical_Sep = Values of Current_Vertical_Sep = Smaller than MAXALTDIF (600), but
positive (strictly)|High_Confidence = Values of High_Confidence =
True|Two_of_Three_Reports_Valid = Values of Two_of_Three_Reports_Valid = True|Own_Tracked_Alt =
Values of Own_Tracked_Alt = Same as other (intruder)|Other_Tracked_Alt = Values
Other_Tracked_Alt = Same as TCAS (own)|Up_Separation = Values of Up_Separation = Smaler
(strictly) than down separation and threshold|Down_Separation = Values of Down_Separation =
Between up separation and threshold|Alt_Layer_Value = Values of Alt_Layer_Value = Value
0|Other_RAC = Values of Other_RAC = One|Other_Capability = Values of Other_Capability =
Two|Climb_Inhibit = Values of Climb_Inhibit = Zero|Positive_RA_Alt_Thresh = Values of
Positive_RA_Alt_Thresh = Value0

values:     Current_Vertical_Sep = '1', Alt_Layer_Value = 'Value0', Down_Separation = '2',
            Up_Separation = '1', Other_RAC = 'One', High_Confidence = 'true', Other_Tracked_Alt
            = '1', Other_Capability = 'Two', Own_Tracked_Alt = '1', Positive_RA_Alt_Thresh =
            '400', Climb_Inhibit = 'Zero', Two_of_Three_Reports_Valid = 'true'

**Test Case 47:**

choices:    1871720133Cat1Ch2-|-374476473Cat1Ch1-|-1325844154Cat1Ch2-|-603277445Cat1Ch1-|-
            1787817979Cat1Ch3-|-927131798Cat1Ch2-|-875777795Cat1Ch2-|-302439181Cat1Ch4-|-
            58393125Cat1Ch3-|-626069991Cat1Ch2-|-430451043Cat1Ch1-|-976751556Cat1Ch4

Current_Vertical_Sep = Values of Current_Vertical_Sep = Smaller than MAXALTDIF (600), but
positive (strictly)|High_Confidence = Values of High_Confidence =
True|Two_of_Three_Reports_Valid = Values of Two_of_Three_Reports_Valid = False|Own_Tracked_Alt =
Values of Own_Tracked_Alt = Higher (strictly) than other (intruder)|Other_Tracked_Alt = Values
Other_Tracked_Alt = Lower (strictly) than TCAS (own)|Up_Separation = Values of Up_Separation =

Greater (strictly) than down separation and threshold|Down_Separation = Values of
Down_Separation = Smaller (strictly) than up separation and threshold|Alt_Layer_Value = Values
of Alt_Layer_Value = Value 3|Other_RAC = Values of Other_RAC = Two|Other_Capability = Values of
Other_Capability = Two|Climb_Inhibit = Values of Climb_Inhibit = Zero|Positive_RA_Alt_Thresh =
Values of Positive_RA_Alt_Thresh = Value3
values:      Current_Vertical_Sep = '1', Alt_Layer_Value = 'Value3', Down_Separation = '1',
             Up_Separation = '741', Other_RAC = 'Two', High_Confidence = 'true',
             Other_Tracked_Alt = '1', Other_Capability = 'Two', Own_Tracked_Alt = '2',
             Climb_Inhibit = 'Zero', Positive_RA_Alt_Thresh = '740', Two_of_Three_Reports_Valid
             = 'false'

**Test Case 48:**
choices:     1871720133Cat1Ch3
Current_Vertical_Sep = Values of Current_Vertical_Sep = Illegal value
values:      Current_Vertical_Sep = '0'

**Test Case 49:**
choices:     603277445Cat1Ch4
Own_Tracked_Alt = Values of Own_Tracked_Alt = Illegal
values:      Own_Tracked_Alt = '0'

**Test Case 50:**
choices:     1787817979Cat1Ch4
Other_Tracked_Alt = Values Other_Tracked_Alt = Illegal
values:      Other_Tracked_Alt = '0'

**Test Case 51:**
choices:     927131798Cat1Ch4
Up_Separation = Values of Up_Separation = Illegal
values:      Up_Separation = '0'

**Test Case 52:**
choices:     875777795Cat1Ch4
Down_Separation = Values of Down_Separation = Illegal
values:      Down_Separation = '0'

**Test Case 53:**
choices:     976751556Cat1Ch5
Positive_RA_Alt_Thresh = Values of Positive_RA_Alt_Thresh = Illegal
values:      Positive_RA_Alt_Thresh = '2'

*J.8      Three-way Test Frames and Corresponding Test Cases when using ACTS*

**Test Case 1:**
choices:     302439181Cat1Ch1−|−976751556Cat1Ch1−|−603277445Cat1Ch1−|−1787817979Cat1Ch3−|−
             927131798Cat1Ch1−|−875777795Cat1Ch2−|−58393125Cat1Ch1−|−1871720133Cat1Ch1−|−
             374476473Cat1Ch1−|−1325844154Cat1Ch1−|−626069991Cat1Ch1−|−430451043Cat1Ch1

Alt_Layer_Value = Values of Alt_Layer_Value = Value 0|Positive_RA_Alt_Thresh = Values of
Positive_RA_Alt_Thresh = Value0|Own_Tracked_Alt = Values of Own_Tracked_Alt = Higher (strictly)
than other (intruder)|Other_Tracked_Alt = Values Other_Tracked_Alt = Lower (strictly) than TCAS
(own)|Up_Separation = Values of Up_Separation = Greater (strictly) than down separation and
threshold|Down_Separation = Values of Down_Separation = Smaller (strictly) than up separation
and threshold|Other_RAC = Values of Other_RAC = Zero|Current_Vertical_Sep = Values of
Current_Vertical_Sep = Greater (strictly) than MAXALTDIF (600)|High_Confidence = Values of
High_Confidence = True|Two_of_Three_Reports_Valid = Values of Two_of_Three_Reports_Valid =
True|Other_Capability = Values of Other_Capability = One|Climb_Inhibit = Values of Climb_Inhibit
= Zero
values:      Current_Vertical_Sep = '601', Alt_Layer_Value = 'Value0', Down_Separation = '399',
             Up_Separation = '401', Other_RAC = 'Zero', High_Confidence = 'true',
             Other_Tracked_Alt = '1', Other_Capability = 'One', Own_Tracked_Alt = '2',
             Positive_RA_Alt_Thresh = '400', Climb_Inhibit = 'Zero', Two_of_Three_Reports_Valid
             = 'true'

**Test Case 2:**
choices:     302439181Cat1Ch1−|−976751556Cat1Ch1−|−603277445Cat1Ch2−|−1787817979Cat1Ch2−|−
             927131798Cat1Ch2−|−875777795Cat1Ch1−|−58393125Cat1Ch2−|−1871720133Cat1Ch2−|−
             374476473Cat1Ch2−|−1325844154Cat1Ch2−|−626069991Cat1Ch2−|−430451043Cat1Ch2

Alt_Layer_Value = Values of Alt_Layer_Value = Value 0|Positive_RA_Alt_Thresh = Values of
Positive_RA_Alt_Thresh = Value0|Own_Tracked_Alt = Values of Own_Tracked_Alt = Same as other
(intruder)|Other_Tracked_Alt = Values Other_Tracked_Alt = Same as TCAS (own)|Up_Separation =
Values of Up_Separation = Smaler (strictly) than down separation and threshold|Down_Separation =
Values of Down_Separation = Greater (strictly) than up separation and threshold|Other_RAC =

Values of Other_RAC = One|Current_Vertical_Sep = Values of Current_Vertical_Sep = Smaller than
MAXALTDIF (600), but positive (strictly)|High_Confidence = Values of High_Confidence =
False|Two_of_Three_Reports_Valid = Values of Two_of_Three_Reports_Valid = False|Other_Capability
= Values of Other_Capability = Two|Climb_Inhibit = Values of Climb_Inhibit = One
values:      Current_Vertical_Sep = '1', Alt_Layer_Value = 'Value0', Down_Separation = '401',
             Up_Separation = '1', Other_RAC = 'One', High_Confidence = 'false',
             Other_Tracked_Alt = '1', Other_Capability = 'Two', Own_Tracked_Alt = '1',
             Climb_Inhibit = 'One', Positive_RA_Alt_Thresh = '400', Two_of_Three_Reports_Valid =
             'false'

**Test Case 3:**
choices:     302439181Cat1Ch1—|—976751556Cat1Ch1—|—603277445Cat1Ch3—|—1787817979Cat1Ch1—|—
             927131798Cat1Ch3—|—875777795Cat1Ch1—|—58393125Cat1Ch3—|—1871720133Cat1Ch1—|—
             374476473Cat1Ch1—|—1325844154Cat1Ch1—|—626069991Cat1Ch1—|—430451043Cat1Ch2

Alt_Layer_Value = Values of Alt_Layer_Value = Value 0|Positive_RA_Alt_Thresh = Values of
Positive_RA_Alt_Thresh = Value0|Own_Tracked_Alt = Values of Own_Tracked_Alt = Lower (strictly)
than other (intruder)|Other_Tracked_Alt = Values Other_Tracked_Alt = Higher (strictly) than TCAS
(own)|Up_Separation = Values of Up_Separation = Between down separation and
threshold|Down_Separation = Values of Down_Separation = Greater (strictly) than up separation
and threshold|Other_RAC = Values of Other_RAC = Two|Current_Vertical_Sep = Values of
Current_Vertical_Sep = Greater (strictly) than MAXALTDIF (600)|High_Confidence = Values of
High_Confidence = True|Two_of_Three_Reports_Valid = Values of Two_of_Three_Reports_Valid =
False|Other_Capability = Values of Other_Capability = One|Climb_Inhibit = Values of
Climb_Inhibit = One
values:      Current_Vertical_Sep = '601', Alt_Layer_Value = 'Value0', Down_Separation = '401',
             Up_Separation = '400', Other_RAC = 'Two', High_Confidence = 'true',
             Other_Tracked_Alt = '2', Other_Capability = 'One', Own_Tracked_Alt = '1',
             Climb_Inhibit = 'One', Positive_RA_Alt_Thresh = '400', Two_of_Three_Reports_Valid =
             'false'

**Test Case 4:**
choices:     302439181Cat1Ch2—|—976751556Cat1Ch2—|—603277445Cat1Ch1—|—1787817979Cat1Ch3—|—
             927131798Cat1Ch2—|—875777795Cat1Ch3—|—58393125Cat1Ch2—|—1871720133Cat1Ch1—|—
             374476473Cat1Ch1—|—1325844154Cat1Ch1—|—626069991Cat1Ch2—|—430451043Cat1Ch1

Alt_Layer_Value = Values of Alt_Layer_Value = Value 1|Positive_RA_Alt_Thresh = Values of
Positive_RA_Alt_Thresh = Value1|Own_Tracked_Alt = Values of Own_Tracked_Alt = Higher (strictly)
than other (intruder)|Other_Tracked_Alt = Values Other_Tracked_Alt = Lower (strictly) than TCAS
(own)|Up_Separation = Values of Up_Separation = Smaller (strictly) than down separation and
threshold|Other_RAC = Values of Other_RAC = One|Current_Vertical_Sep = Values of
Current_Vertical_Sep = Smaller than MAXALTDIF (600), but positive (strictly)|High_Confidence =
Values of High_Confidence = True|Two_of_Three_Reports_Valid = Values of
Two_of_Three_Reports_Valid = True|Other_Capability = Values of Other_Capability =
Two|Climb_Inhibit = Values of Climb_Inhibit = Zero
values:      Current_Vertical_Sep = '1', Alt_Layer_Value = 'Value1', Down_Separation = '2',
             Up_Separation = '1', Other_RAC = 'One', High_Confidence = 'true', Other_Tracked_Alt
             = '1', Other_Capability = 'Two', Own_Tracked_Alt = '2', Positive_RA_Alt_Thresh =
             '500', Climb_Inhibit = 'Zero', Two_of_Three_Reports_Valid = 'true'

**Test Case 5:**
choices:     302439181Cat1Ch2—|—976751556Cat1Ch2—|—603277445Cat1Ch2—|—1787817979Cat1Ch2—|—
             927131798Cat1Ch1—|—875777795Cat1Ch2—|—58393125Cat1Ch3—|—1871720133Cat1Ch1—|—
             374476473Cat1Ch2—|—1325844154Cat1Ch1—|—626069991Cat1Ch2—|—430451043Cat1Ch2

Alt_Layer_Value = Values of Alt_Layer_Value = Value 1|Positive_RA_Alt_Thresh = Values of
Positive_RA_Alt_Thresh = Value1|Own_Tracked_Alt = Values of Own_Tracked_Alt = Same as other
(intruder)|Other_Tracked_Alt = Values Other_Tracked_Alt = Same as TCAS (own)|Up_Separation =
Values of Up_Separation = Greater (strictly) than down separation and threshold|Down_Separation
= Values of Down_Separation = Smaller (strictly) than up separation and threshold|Other_RAC =
Values of Other_RAC = Two|Current_Vertical_Sep = Values of Current_Vertical_Sep = Greater
(strictly) than MAXALTDIF (600)|High_Confidence = Values of High_Confidence =
False|Two_of_Three_Reports_Valid = Values of Two_of_Three_Reports_Valid = True|Other_Capability
= Values of Other_Capability = Two|Climb_Inhibit = Values of Climb_Inhibit = One
values:      Current_Vertical_Sep = '601', Alt_Layer_Value = 'Value1', Down_Separation = '499',
             Up_Separation = '501', Other_RAC = 'Two', High_Confidence = 'false',
             Other_Tracked_Alt = '1', Other_Capability = 'Two', Own_Tracked_Alt = '1',
             Positive_RA_Alt_Thresh = '500', Climb_Inhibit = 'One', Two_of_Three_Reports_Valid =
             'true'

**Test Case 6:**
choices:      302439181Cat1Ch2-|-976751556Cat1Ch2-|-603277445Cat1Ch3-|-1787817979Cat1Ch1-|-
              927131798Cat1Ch3-|-875777795Cat1Ch2-|-58393125Cat1Ch1-|-1871720133Cat1Ch2-|-
              374476473Cat1Ch2-|-1325844154Cat1Ch2-|-626069991Cat1Ch1-|-430451043Cat1Ch1
Alt_Layer_Value = Values of Alt_Layer_Value = Value 1|Positive_RA_Alt_Thresh = Values of
Positive_RA_Alt_Thresh = Value1|Own_Tracked_Alt = Values of Own_Tracked_Alt = Lower (strictly)
than other (intruder)|Other_Tracked_Alt = Values Other_Tracked_Alt = Higher (strictly) than TCAS
(own)|Up_Separation = Values of Up_Separation = Between down separation and
threshold|Down_Separation = Values of Down_Separation = Smaller (strictly) than up separation
and threshold|Other_RAC = Values of Other_RAC = Zero|Current_Vertical_Sep = Values of
Current_Vertical_Sep = Smaller than MAXALTDIF (600), but positive (strictly)|High_Confidence =
Values of High_Confidence = False|Two_of_Three_Reports_Valid = Values of
Two_of_Three_Reports_Valid = False|Other_Capability = Values of Other_Capability =
One|Climb_Inhibit = Values of Climb_Inhibit = Zero
values:       Current_Vertical_Sep = '1', Alt_Layer_Value = 'Value1', Down_Separation = '1',
              Up_Separation = '2', Other_RAC = 'Zero', High_Confidence = 'false',
              Other_Tracked_Alt = '2', Other_Capability = 'One', Own_Tracked_Alt = '1',
              Climb_Inhibit = 'Zero', Positive_RA_Alt_Thresh = '500', Two_of_Three_Reports_Valid
              = 'false'

**Test Case 7:**
choices:      302439181Cat1Ch3-|-976751556Cat1Ch3-|-603277445Cat1Ch1-|-1787817979Cat1Ch3-|-
              927131798Cat1Ch3-|-875777795Cat1Ch3-|-58393125Cat1Ch1-|-1871720133Cat1Ch3-|-
              374476473Cat1Ch2-|-1325844154Cat1Ch1-|-626069991Cat1Ch2-|-430451043Cat1Ch2
Alt_Layer_Value = Values of Alt_Layer_Value = Value 2|Positive_RA_Alt_Thresh = Values of
Positive_RA_Alt_Thresh = Value2|Own_Tracked_Alt = Values of Own_Tracked_Alt = Higher (strictly)
than other (intruder)|Other_Tracked_Alt = Values Other_Tracked_Alt = Lower (strictly) than TCAS
(own)|Up_Separation = Values of Up_Separation = Between down separation and
threshold|Down_Separation = Values of Down_Separation = Greater (strictly) than up separation
and threshold|Other_RAC = Values of Other_RAC = Zero|Current_Vertical_Sep = Values of
Current_Vertical_Sep = Smaller than MAXALTDIF (600), but positive (strictly)|High_Confidence =
Values of High_Confidence = False|Two_of_Three_Reports_Valid = Values of
Two_of_Three_Reports_Valid = True|Other_Capability = Values of Other_Capability =
Two|Climb_Inhibit = Values of Climb_Inhibit = One
values:       Current_Vertical_Sep = '1', Alt_Layer_Value = 'Value2', Down_Separation = '641',
              Up_Separation = '640', Other_RAC = 'Zero', High_Confidence = 'false',
              Other_Tracked_Alt = '1', Other_Capability = 'Two', Own_Tracked_Alt = '2',
              Climb_Inhibit = 'One', Positive_RA_Alt_Thresh = '640', Two_of_Three_Reports_Valid =
              'true'

**Test Case 8:**
choices:      302439181Cat1Ch3-|-976751556Cat1Ch3-|-603277445Cat1Ch2-|-1787817979Cat1Ch2-|-
              927131798Cat1Ch1-|-875777795Cat1Ch3-|-58393125Cat1Ch2-|-1871720133Cat1Ch1-|-
              374476473Cat1Ch1-|-1325844154Cat1Ch2-|-626069991Cat1Ch1-|-430451043Cat1Ch1
Alt_Layer_Value = Values of Alt_Layer_Value = Value 2|Positive_RA_Alt_Thresh = Values of
Positive_RA_Alt_Thresh = Value2|Own_Tracked_Alt = Values of Own_Tracked_Alt = Same as other
(intruder)|Other_Tracked_Alt = Values Other_Tracked_Alt = Same as TCAS (own)|Up_Separation =
Values of Up_Separation = Between up separation and threshold|Down_Separation = Values of
Down_Separation = Greater (strictly) than down separation and threshold|Down_Separation
= Values of Down_Separation = Between up separation and threshold|Other_RAC = Values of
Other_RAC = One|Current_Vertical_Sep = Values of Current_Vertical_Sep = Greater (strictly) than
MAXALTDIF (600)|High_Confidence = Values of High_Confidence = True|Two_of_Three_Reports_Valid =
Values of Two_of_Three_Reports_Valid = False|Other_Capability = Values of Other_Capability =
One|Climb_Inhibit = Values of Climb_Inhibit = Zero
values:       Current_Vertical_Sep = '601', Alt_Layer_Value = 'Value2', Down_Separation = '640',
              Up_Separation = '641', Other_RAC = 'One', High_Confidence = 'true',
              Other_Tracked_Alt = '1', Other_Capability = 'One', Own_Tracked_Alt = '1',
              Positive_RA_Alt_Thresh = '640', Climb_Inhibit = 'Zero', Two_of_Three_Reports_Valid
              = 'false'

**Test Case 9:**
choices:      302439181Cat1Ch3-|-976751556Cat1Ch3-|-603277445Cat1Ch3-|-1787817979Cat1Ch1-|-
              927131798Cat1Ch3-|-875777795Cat1Ch3-|-58393125Cat1Ch3-|-1871720133Cat1Ch2-|-
              374476473Cat1Ch2-|-1325844154Cat1Ch1-|-626069991Cat1Ch1-|-430451043Cat1Ch2
Alt_Layer_Value = Values of Alt_Layer_Value = Value 2|Positive_RA_Alt_Thresh = Values of
Positive_RA_Alt_Thresh = Value2|Own_Tracked_Alt = Values of Own_Tracked_Alt = Lower (strictly)
than other (intruder)|Other_Tracked_Alt = Values Other_Tracked_Alt = Higher (strictly) than TCAS
(own)|Up_Separation = Values of Up_Separation = Smaler (strictly) than down separation and
threshold|Down_Separation = Values of Down_Separation = Between up separation and

threshold|Other_RAC = Values of Other_RAC = Two|Current_Vertical_Sep = Values of
Current_Vertical_Sep = Smaller than MAXALTDIF (600), but positive (strictly)|High_Confidence =
Values of High_Confidence = False|Two_of_Three_Reports_Valid = Values of
Two_of_Three_Reports_Valid = True|Other_Capability = Values of Other_Capability =
One|Climb_Inhibit = Values of Climb_Inhibit = One

values:        Current_Vertical_Sep = '1', Alt_Layer_Value = 'Value2', Down_Separation = '2',
               Up_Separation = '1', Other_RAC = 'Two', High_Confidence = 'false',
               Other_Tracked_Alt = '2', Other_Capability = 'One', Own_Tracked_Alt = '1',
               Climb_Inhibit = 'One', Positive_RA_Alt_Thresh = '640', Two_of_Three_Reports_Valid =
               'true'

**Test Case 10:**
choices:        302439181Cat1Ch4-|-976751556Cat1Ch4-|-603277445Cat1Ch1-|-1787817979Cat1Ch3-|-
                927131798Cat1Ch1-|-875777795Cat1Ch3-|-58393125Cat1Ch3-|-1871720133Cat1Ch2-|-
                374476473Cat1Ch2-|-1325844154Cat1Ch2-|-626069991Cat1Ch2-|-430451043Cat1Ch1
Alt_Layer_Value = Values of Alt_Layer_Value = Value 3|Positive_RA_Alt_Thresh = Values of
Positive_RA_Alt_Thresh = Value3|Own_Tracked_Alt = Values of Own_Tracked_Alt = Higher (strictly)
than other (intruder)|Other_Tracked_Alt = Values Other_Tracked_Alt = Lower (strictly) than TCAS
(own)|Up_Separation = Values of Up_Separation = Greater (strictly) than down separation and
threshold|Down_Separation = Values of Down_Separation = Between up separation and
threshold|Other_RAC = Values of Other_RAC = Two|Current_Vertical_Sep = Values of
Current_Vertical_Sep = Smaller than MAXALTDIF (600), but positive (strictly)|High_Confidence =
Values of High_Confidence = False|Two_of_Three_Reports_Valid = Values of
Two_of_Three_Reports_Valid = False|Other_Capability = Values of Other_Capability =
Two|Climb_Inhibit = Values of Climb_Inhibit = Zero

values:        Current_Vertical_Sep = '1', Alt_Layer_Value = 'Value3', Down_Separation = '740',
               Up_Separation = '741', Other_RAC = 'Two', High_Confidence = 'false',
               Other_Tracked_Alt = '1', Other_Capability = 'Two', Own_Tracked_Alt = '2',
               Climb_Inhibit = 'Zero', Positive_RA_Alt_Thresh = '740', Two_of_Three_Reports_Valid
               = 'false'

**Test Case 11:**
choices:        302439181Cat1Ch4-|-976751556Cat1Ch4-|-603277445Cat1Ch2-|-1787817979Cat1Ch2-|-
                927131798Cat1Ch3-|-875777795Cat1Ch1-|-58393125Cat1Ch1-|-1871720133Cat1Ch1-|-
                374476473Cat1Ch1-|-1325844154Cat1Ch1-|-626069991Cat1Ch2-|-430451043Cat1Ch1
Alt_Layer_Value = Values of Alt_Layer_Value = Value 3|Positive_RA_Alt_Thresh = Values of
Positive_RA_Alt_Thresh = Value3|Own_Tracked_Alt = Values of Own_Tracked_Alt = Same as other
(intruder)|Other_Tracked_Alt = Values Other_Tracked_Alt = Same as TCAS (own)|Up_Separation =
Values of Up_Separation = Between down separation and threshold|Down_Separation = Values of
Down_Separation = Greater (strictly) than up separation and threshold|Other_RAC = Values of
Other_RAC = Zero|Current_Vertical_Sep = Values of Current_Vertical_Sep = Greater (strictly) than
MAXALTDIF (600)|High_Confidence = Values of High_Confidence = True|Two_of_Three_Reports_Valid =
Values of Two_of_Three_Reports_Valid = True|Other_Capability = Values of Other_Capability =
Two|Climb_Inhibit = Values of Climb_Inhibit = Zero

values:        Current_Vertical_Sep = '601', Alt_Layer_Value = 'Value3', Down_Separation = '741',
               Up_Separation = '740', Other_RAC = 'Zero', High_Confidence = 'true',
               Other_Tracked_Alt = '1', Other_Capability = 'Two', Own_Tracked_Alt = '1',
               Positive_RA_Alt_Thresh = '740', Climb_Inhibit = 'Zero', Two_of_Three_Reports_Valid
               = 'true'

**Test Case 12:**
choices:        302439181Cat1Ch4-|-976751556Cat1Ch4-|-603277445Cat1Ch3-|-1787817979Cat1Ch1-|-
                927131798Cat1Ch2-|-875777795Cat1Ch1-|-58393125Cat1Ch2-|-1871720133Cat1Ch1-|-
                374476473Cat1Ch2-|-1325844154Cat1Ch1-|-626069991Cat1Ch1-|-430451043Cat1Ch1
Alt_Layer_Value = Values of Alt_Layer_Value = Value 3|Positive_RA_Alt_Thresh = Values of
Positive_RA_Alt_Thresh = Value3|Own_Tracked_Alt = Values of Own_Tracked_Alt = Lower (strictly)
than other (intruder)|Other_Tracked_Alt = Values Other_Tracked_Alt = Higher (strictly) than TCAS
(own)|Up_Separation = Values of Up_Separation = Smaler (strictly) than down separation and
threshold|Down_Separation = Values of Down_Separation = Greater (strictly) than up separation
and threshold|Other_RAC = Values of Other_RAC = One|Current_Vertical_Sep = Values of
Current_Vertical_Sep = Greater (strictly) than MAXALTDIF (600)|High_Confidence = Values of
High_Confidence = False|Two_of_Three_Reports_Valid = Values of Two_of_Three_Reports_Valid =
True|Other_Capability = Values of Other_Capability = One|Climb_Inhibit = Values of Climb_Inhibit
= Zero

values:        Current_Vertical_Sep = '601', Alt_Layer_Value = 'Value3', Down_Separation = '741',
               Up_Separation = '1', Other_RAC = 'One', High_Confidence = 'false',
               Other_Tracked_Alt = '2', Other_Capability = 'One', Own_Tracked_Alt = '1',

```
                 Climb_Inhibit = 'Zero', Positive_RA_Alt_Thresh = '740', Two_of_Three_Reports_Valid
                 = 'true'
```

**Test Case 13:**
```
choices:         302439181Cat1Ch1-|-976751556Cat1Ch1-|-603277445Cat1Ch3-|-1787817979Cat1Ch1-|-
                 927131798Cat1Ch1-|-875777795Cat1Ch3-|-58393125Cat1Ch2-|-1871720133Cat1Ch2-|-
                 374476473Cat1Ch1-|-1325844154Cat1Ch2-|-626069991Cat1Ch2-|-430451043Cat1Ch2
Alt_Layer_Value = Values of Alt_Layer_Value = Value 0|Positive_RA_Alt_Thresh = Values of
Positive_RA_Alt_Thresh = Value0|Own_Tracked_Alt = Values of Own_Tracked_Alt = Lower (strictly)
than other (intruder)|Other_Tracked_Alt = Values Other_Tracked_Alt = Higher (strictly) than TCAS
(own)|Up_Separation = Values of Up_Separation = Greater (strictly) than down separation and
threshold|Down_Separation = Values of Down_Separation = Between up separation and
threshold|Other_RAC = Values of Other_RAC = One|Current_Vertical_Sep = Values of
Current_Vertical_Sep = Smaller than MAXALTDIF (600), but positive (strictly)|High_Confidence =
Values of High_Confidence = True|Two_of_Three_Reports_Valid = Values of
Two_of_Three_Reports_Valid = True|Other_Capability = Values of Other_Capability =
Two|Climb_Inhibit = Values of Climb_Inhibit = One
values:          Current_Vertical_Sep = '1', Alt_Layer_Value = 'Value0', Down_Separation = '400',
                 Up_Separation = '401', Other_RAC = 'One', High_Confidence = 'true',
                 Other_Tracked_Alt = '2', Other_Capability = 'Two', Own_Tracked_Alt = '1',
                 Positive_RA_Alt_Thresh = '400', Climb_Inhibit = 'One', Two_of_Three_Reports_Valid =
                 'true'
```

**Test Case 14:**
```
choices:         302439181Cat1Ch2-|-976751556Cat1Ch2-|-603277445Cat1Ch3-|-1787817979Cat1Ch1-|-
                 927131798Cat1Ch1-|-875777795Cat1Ch1-|-58393125Cat1Ch1-|-1871720133Cat1Ch1-|-
                 374476473Cat1Ch1-|-1325844154Cat1Ch2-|-626069991Cat1Ch2-|-430451043Cat1Ch2
Alt_Layer_Value = Values of Alt_Layer_Value = Value 1|Positive_RA_Alt_Thresh = Values of
Positive_RA_Alt_Thresh = Value1|Own_Tracked_Alt = Values of Own_Tracked_Alt = Lower (strictly)
than other (intruder)|Other_Tracked_Alt = Values Other_Tracked_Alt = Higher (strictly) than TCAS
(own)|Up_Separation = Values of Up_Separation = Greater (strictly) than down separation and
threshold|Down_Separation = Values of Down_Separation = Between up separation and
threshold|Other_RAC = Values of Other_RAC = Zero|Current_Vertical_Sep = Values of
Current_Vertical_Sep = Greater (strictly) than MAXALTDIF (600)|High_Confidence = Values of
High_Confidence = True|Two_of_Three_Reports_Valid = Values of Two_of_Three_Reports_Valid =
False|Other_Capability = Values of Other_Capability = Two|Climb_Inhibit = Values of
Climb_Inhibit = One
values:          Current_Vertical_Sep = '601', Alt_Layer_Value = 'Value1', Down_Separation = '500',
                 Up_Separation = '501', Other_RAC = 'Zero', High_Confidence = 'true',
                 Other_Tracked_Alt = '2', Other_Capability = 'Two', Own_Tracked_Alt = '1',
                 Climb_Inhibit = 'One', Positive_RA_Alt_Thresh = '500', Two_of_Three_Reports_Valid =
                 'false'
```

**Test Case 15:**
```
choices:         302439181Cat1Ch3-|-976751556Cat1Ch3-|-603277445Cat1Ch3-|-1787817979Cat1Ch1-|-
                 927131798Cat1Ch1-|-875777795Cat1Ch1-|-58393125Cat1Ch2-|-1871720133Cat1Ch2-|-
                 374476473Cat1Ch2-|-1325844154Cat1Ch2-|-626069991Cat1Ch2-|-430451043Cat1Ch1
Alt_Layer_Value = Values of Alt_Layer_Value = Value 2|Positive_RA_Alt_Thresh = Values of
Positive_RA_Alt_Thresh = Value2|Own_Tracked_Alt = Values of Own_Tracked_Alt = Lower (strictly)
than other (intruder)|Other_Tracked_Alt = Values Other_Tracked_Alt = Higher (strictly) than TCAS
(own)|Up_Separation = Values of Up_Separation = Greater (strictly) than down separation and
threshold|Down_Separation = Values of Down_Separation = Smaller (strictly) than up separation
and threshold|Other_RAC = Values of Other_RAC = One|Current_Vertical_Sep = Values of
Current_Vertical_Sep = Greater (strictly) than MAXALTDIF (600)|High_Confidence = Values of
High_Confidence = False|Two_of_Three_Reports_Valid = Values of Two_of_Three_Reports_Valid =
False|Other_Capability = Values of Other_Capability = Two|Climb_Inhibit = Values of
Climb_Inhibit = Zero
values:          Current_Vertical_Sep = '601', Alt_Layer_Value = 'Value2', Down_Separation = '1',
                 Up_Separation = '641', Other_RAC = 'One', High_Confidence = 'false',
                 Other_Tracked_Alt = '2', Other_Capability = 'Two', Own_Tracked_Alt = '1',
                 Climb_Inhibit = 'Zero', Positive_RA_Alt_Thresh = '640', Two_of_Three_Reports_Valid
                 = 'false'
```

**Test Case 16:**
```
choices:         302439181Cat1Ch4-|-976751556Cat1Ch4-|-603277445Cat1Ch3-|-1787817979Cat1Ch1-|-
                 927131798Cat1Ch1-|-875777795Cat1Ch2-|-58393125Cat1Ch3-|-1871720133Cat1Ch2-|-
                 374476473Cat1Ch1-|-1325844154Cat1Ch1-|-626069991Cat1Ch1-|-430451043Cat1Ch2
Alt_Layer_Value = Values of Alt_Layer_Value = Value 3|Positive_RA_Alt_Thresh = Values of
Positive_RA_Alt_Thresh = Value3|Own_Tracked_Alt = Values of Own_Tracked_Alt = Lower (strictly)
```

than other (intruder)|Other_Tracked_Alt = Values Other_Tracked_Alt = Higher (strictly) than TCAS (own)|Up_Separation = Values of Up_Separation = Greater (strictly) than down separation and threshold|Down_Separation = Values of Down_Separation = Smaller (strictly) than up separation and threshold|Other_RAC = Values of Other_RAC = Two|Current_Vertical_Sep = Values of Current_Vertical_Sep = Smaller than MAXALTDIF (600), but positive (strictly)|High_Confidence = Values of High_Confidence = True|Two_of_Three_Reports_Valid = Values of Two_of_Three_Reports_Valid = True|Other_Capability = Values of Other_Capability = One|Climb_Inhibit = Values of Climb_Inhibit = One

values:      Current_Vertical_Sep = '1', Alt_Layer_Value = 'Value3', Down_Separation = '1',
             Up_Separation = '741', Other_RAC = 'Two', High_Confidence = 'true',
             Other_Tracked_Alt = '2', Other_Capability = 'One', Own_Tracked_Alt = '1',
             Positive_RA_Alt_Thresh = '740', Climb_Inhibit = 'One', Two_of_Three_Reports_Valid =
             'true'

**Test Case 17:**
choices:     302439181Cat1Ch1—|—976751556Cat1Ch1—|—603277445Cat1Ch2—|—1787817979Cat1Ch2—|—
             927131798Cat1Ch1—|—875777795Cat1Ch2—|—58393125Cat1Ch1—|—1871720133Cat1Ch2—|—
             374476473Cat1Ch1—|—1325844154Cat1Ch2—|—626069991Cat1Ch1—|—430451043Cat1Ch2
Alt_Layer_Value = Values of Alt_Layer_Value = Value 0|Positive_RA_Alt_Thresh = Values of Positive_RA_Alt_Thresh = Value0|Own_Tracked_Alt = Values of Own_Tracked_Alt = Same as other (intruder)|Other_Tracked_Alt = Values Other_Tracked_Alt = Same as TCAS (own)|Up_Separation = Values of Up_Separation = Greater (strictly) than down separation and threshold|Down_Separation = Values of Down_Separation = Smaller (strictly) than up separation and threshold|Other_RAC = Values of Other_RAC = Zero|Current_Vertical_Sep = Values of Current_Vertical_Sep = Smaller than MAXALTDIF (600), but positive (strictly)|High_Confidence = Values of High_Confidence = True|Two_of_Three_Reports_Valid = Values of Two_of_Three_Reports_Valid = False|Other_Capability = Values of Other_Capability = One|Climb_Inhibit = Values of Climb_Inhibit = One

values:      Current_Vertical_Sep = '1', Alt_Layer_Value = 'Value0', Down_Separation = '399',
             Up_Separation = '401', Other_RAC = 'Zero', High_Confidence = 'true',
             Other_Capability = 'One', Other_Tracked_Alt = '1', Own_Tracked_Alt = '1',
             Climb_Inhibit = 'One', Positive_RA_Alt_Thresh = '400', Two_of_Three_Reports_Valid =
             'false'

**Test Case 18:**
choices:     302439181Cat1Ch4—|—976751556Cat1Ch4—|—603277445Cat1Ch2—|—1787817979Cat1Ch2—|—
             927131798Cat1Ch1—|—875777795Cat1Ch2—|—58393125Cat1Ch2—|—1871720133Cat1Ch2—|—
             374476473Cat1Ch2—|—1325844154Cat1Ch2—|—626069991Cat1Ch1—|—430451043Cat1Ch2
Alt_Layer_Value = Values of Alt_Layer_Value = Value 3|Positive_RA_Alt_Thresh = Values of Positive_RA_Alt_Thresh = Value3|Own_Tracked_Alt = Values of Own_Tracked_Alt = Same as other (intruder)|Other_Tracked_Alt = Values Other_Tracked_Alt = Same as TCAS (own)|Up_Separation = Values of Up_Separation = Greater (strictly) than down separation and threshold|Down_Separation = Values of Down_Separation = Smaller (strictly) than up separation and threshold|Other_RAC = Values of Other_RAC = One|Current_Vertical_Sep = Values of Current_Vertical_Sep = Smaller than MAXALTDIF (600), but positive (strictly)|High_Confidence = Values of High_Confidence = False|Two_of_Three_Reports_Valid = Values of Two_of_Three_Reports_Valid = False|Other_Capability = Values of Other_Capability = One|Climb_Inhibit = Values of Climb_Inhibit = One

values:      Current_Vertical_Sep = '1', Alt_Layer_Value = 'Value3', Down_Separation = '739',
             Up_Separation = '741', Other_RAC = 'One', High_Confidence = 'false',
             Other_Tracked_Alt = '1', Other_Capability = 'One', Own_Tracked_Alt = '1',
             Positive_RA_Alt_Thresh = '740', Climb_Inhibit = 'One', Two_of_Three_Reports_Valid =
             'false'

**Test Case 19:**
choices:     302439181Cat1Ch2—|—976751556Cat1Ch2—|—603277445Cat1Ch1—|—1787817979Cat1Ch3—|—
             927131798Cat1Ch1—|—875777795Cat1Ch2—|—58393125Cat1Ch2—|—1871720133Cat1Ch1—|—
             374476473Cat1Ch2—|—1325844154Cat1Ch2—|—626069991Cat1Ch1—|—430451043Cat1Ch2
Alt_Layer_Value = Values of Alt_Layer_Value = Value 1|Positive_RA_Alt_Thresh = Values of Positive_RA_Alt_Thresh = Value1|Own_Tracked_Alt = Values of Own_Tracked_Alt = Higher (strictly) than other (intruder)|Other_Tracked_Alt = Values Other_Tracked_Alt = Lower (strictly) than TCAS (own)|Up_Separation = Values of Up_Separation = Greater (strictly) than down separation and threshold|Down_Separation = Values of Down_Separation = Smaller (strictly) than up separation and threshold|Other_RAC = Values of Other_RAC = One|Current_Vertical_Sep = Values of Current_Vertical_Sep = Greater (strictly) than MAXALTDIF (600)|High_Confidence = Values of High_Confidence = False|Two_of_Three_Reports_Valid = Values of Two_of_Three_Reports_Valid = False|Other_Capability = Values of Other_Capability = One|Climb_Inhibit = Values of Climb_Inhibit = One

values:      Current_Vertical_Sep = '601', Alt_Layer_Value = 'Value1', Down_Separation = '1',
             Up_Separation = '501', Other_RAC = 'One', High_Confidence = 'false',

```
                    Other_Tracked_Alt = '1', Other_Capability = 'One', Own_Tracked_Alt = '2',
                    Positive_RA_Alt_Thresh = '500', Climb_Inhibit = 'One', Two_of_Three_Reports_Valid =
                    'false'
```

**Test Case 20:**
```
choices:        302439181Cat1Ch3-|-976751556Cat1Ch3-|-603277445Cat1Ch1-|-1787817979Cat1Ch3-|-
                927131798Cat1Ch1-|-875777795Cat1Ch3-|-58393125Cat1Ch3-|-1871720133Cat1Ch1-|-
                374476473Cat1Ch1-|-1325844154Cat1Ch1-|-626069991Cat1Ch2-|-430451043Cat1Ch2
Alt_Layer_Value = Values of Alt_Layer_Value = Value 2|Positive_RA_Alt_Thresh = Values of
Positive_RA_Alt_Thresh = Value2|Own_Tracked_Alt = Values of Own_Tracked_Alt = Higher (strictly)
than other (intruder)|Other_Tracked_Alt = Values Other_Tracked_Alt = Lower (strictly) than TCAS
(own)|Up_Separation = Values of Up_Separation = Greater (strictly) than down separation and
threshold|Down_Separation = Values of Down_Separation = Between up separation and
threshold|Other_RAC = Values of Other_RAC = Two|Current_Vertical_Sep = Values of
Current_Vertical_Sep = Greater (strictly) than MAXALTDIF (600)|High_Confidence = Values of
High_Confidence = True|Two_of_Three_Reports_Valid = Values of Two_of_Three_Reports_Valid =
True|Other_Capability = Values of Other_Capability = Two|Climb_Inhibit = Values of Climb_Inhibit
= One
values:         Current_Vertical_Sep = '601', Alt_Layer_Value = 'Value2', Down_Separation = '640',
                Up_Separation = '641', Other_RAC = 'Two', High_Confidence = 'true',
                Other_Capability = 'Two', Other_Tracked_Alt = '1', Own_Tracked_Alt = '2',
                Positive_RA_Alt_Thresh = '640', Climb_Inhibit = 'One', Two_of_Three_Reports_Valid =
                'true'
```

**Test Case 21:**
```
choices:        302439181Cat1Ch1-|-976751556Cat1Ch1-|-603277445Cat1Ch3-|-1787817979Cat1Ch1-|-
                927131798Cat1Ch2-|-875777795Cat1Ch3-|-58393125Cat1Ch1-|-1871720133Cat1Ch1-|-
                374476473Cat1Ch2-|-1325844154Cat1Ch1-|-626069991Cat1Ch1-|-430451043Cat1Ch1
Alt_Layer_Value = Values of Alt_Layer_Value = Value 0|Positive_RA_Alt_Thresh = Values of
Positive_RA_Alt_Thresh = Value0|Own_Tracked_Alt = Values of Own_Tracked_Alt = Lower (strictly)
than other (intruder)|Other_Tracked_Alt = Values Other_Tracked_Alt = Higher (strictly) than TCAS
(own)|Up_Separation = Values of Up_Separation = Smaler (strictly) than down separation and
threshold|Down_Separation = Values of Down_Separation = Between up separation and
threshold|Other_RAC = Values of Other_RAC = Zero|Current_Vertical_Sep = Values of
Current_Vertical_Sep = Greater (strictly) than MAXALTDIF (600)|High_Confidence = Values of
High_Confidence = False|Two_of_Three_Reports_Valid = Values of Two_of_Three_Reports_Valid =
True|Other_Capability = Values of Other_Capability = One|Climb_Inhibit = Values of Climb_Inhibit
= Zero
values:         Current_Vertical_Sep = '601', Alt_Layer_Value = 'Value0', Down_Separation = '2',
                Up_Separation = '1', Other_RAC = 'Zero', High_Confidence = 'false',
                Other_Tracked_Alt = '2', Other_Capability = 'One', Own_Tracked_Alt = '1',
                Positive_RA_Alt_Thresh = '400', Climb_Inhibit = 'Zero', Two_of_Three_Reports_Valid
                = 'true'
```

**Test Case 22:**
```
choices:        302439181Cat1Ch2-|-976751556Cat1Ch2-|-603277445Cat1Ch3-|-1787817979Cat1Ch1-|-
                927131798Cat1Ch2-|-875777795Cat1Ch1-|-58393125Cat1Ch3-|-1871720133Cat1Ch2-|-
                374476473Cat1Ch1-|-1325844154Cat1Ch2-|-626069991Cat1Ch1-|-430451043Cat1Ch1
Alt_Layer_Value = Values of Alt_Layer_Value = Value 1|Positive_RA_Alt_Thresh = Values of
Positive_RA_Alt_Thresh = Value1|Own_Tracked_Alt = Values of Own_Tracked_Alt = Lower (strictly)
than other (intruder)|Other_Tracked_Alt = Values Other_Tracked_Alt = Higher (strictly) than TCAS
(own)|Up_Separation = Values of Up_Separation = Smaler (strictly) than down separation and
threshold|Down_Separation = Values of Down_Separation = Greater (strictly) than up separation
and threshold|Other_RAC = Values of Other_RAC = Two|Current_Vertical_Sep = Values of
Current_Vertical_Sep = Smaller than MAXALTDIF (600), but positive (strictly)|High_Confidence =
Values of High_Confidence = True|Two_of_Three_Reports_Valid = Values of
Two_of_Three_Reports_Valid = False|Other_Capability = Values of Other_Capability =
One|Climb_Inhibit = Values of Climb_Inhibit = Zero
values:         Current_Vertical_Sep = '1', Alt_Layer_Value = 'Value1', Down_Separation = '501',
                Up_Separation = '499', Other_RAC = 'Two', High_Confidence = 'true',
                Other_Tracked_Alt = '2', Other_Capability = 'One', Own_Tracked_Alt = '1',
                Climb_Inhibit = 'Zero', Positive_RA_Alt_Thresh = '500', Two_of_Three_Reports_Valid
                = 'false'
```

**Test Case 23:**
```
choices:        302439181Cat1Ch2-|-976751556Cat1Ch2-|-603277445Cat1Ch2-|-1787817979Cat1Ch2-|-
                927131798Cat1Ch2-|-875777795Cat1Ch3-|-58393125Cat1Ch1-|-1871720133Cat1Ch2-|-
                374476473Cat1Ch2-|-1325844154Cat1Ch1-|-626069991Cat1Ch1-|-430451043Cat1Ch1
```

```
Alt_Layer_Value = Values of Alt_Layer_Value = Value 1|Positive_RA_Alt_Thresh = Values of
Positive_RA_Alt_Thresh = Value1|Own_Tracked_Alt = Values of Own_Tracked_Alt = Same as other
(intruder)|Other_Tracked_Alt = Values Other_Tracked_Alt = Same as TCAS (own)|Up_Separation =
Values of Up_Separation = Smaler (strictly) than down separation and threshold|Down_Separation =
Values of Down_Separation = Between up separation and threshold|Other_RAC = Values of Other_RAC
= Zero|Current_Vertical_Sep = Values of Current_Vertical_Sep = Smaller than MAXALTDIF (600), but
positive (strictly)|High_Confidence = Values of High_Confidence =
False|Two_of_Three_Reports_Valid = Values of Two_of_Three_Reports_Valid = True|Other_Capability
= Values of Other_Capability = One|Climb_Inhibit = Values of Climb_Inhibit = Zero
values:        Current_Vertical_Sep = '1', Alt_Layer_Value = 'Value1', Down_Separation = '500',
               Up_Separation = '499', Other_RAC = 'Zero', High_Confidence = 'false',
               Other_Tracked_Alt = '1', Other_Capability = 'One', Own_Tracked_Alt = '1',
               Positive_RA_Alt_Thresh = '500', Climb_Inhibit = 'Zero', Two_of_Three_Reports_Valid
               = 'true'
```

**Test Case 24:**

```
choices:       302439181Cat1Ch3-|-976751556Cat1Ch3-|-603277445Cat1Ch2-|-1787817979Cat1Ch2-|-
               927131798Cat1Ch2-|-875777795Cat1Ch3-|-58393125Cat1Ch3-|-1871720133Cat1Ch1-|-
               374476473Cat1Ch1-|-1325844154Cat1Ch2-|-626069991Cat1Ch2-|-430451043Cat1Ch1

Alt_Layer_Value = Values of Alt_Layer_Value = Value 2|Positive_RA_Alt_Thresh = Values of
Positive_RA_Alt_Thresh = Value2|Own_Tracked_Alt = Values of Own_Tracked_Alt = Same as other
(intruder)|Other_Tracked_Alt = Values Other_Tracked_Alt = Same as TCAS (own)|Up_Separation =
Values of Up_Separation = Smaler (strictly) than down separation and threshold|Down_Separation =
Values of Down_Separation = Greater (strictly) than up separation and threshold|Other_RAC =
Values of Other_RAC = Two|Current_Vertical_Sep = Values of Current_Vertical_Sep = Greater
(strictly) than MAXALTDIF (600)|High_Confidence = Values of High_Confidence =
True|Two_of_Three_Reports_Valid = Values of Two_of_Three_Reports_Valid = False|Other_Capability
= Values of Other_Capability = Two|Climb_Inhibit = Values of Climb_Inhibit = Zero
values:        Current_Vertical_Sep = '601', Alt_Layer_Value = 'Value2', Down_Separation = '641',
               Up_Separation = '1', Other_RAC = 'Two', High_Confidence = 'true', Other_Tracked_Alt
               = '1', Other_Capability = 'Two', Own_Tracked_Alt = '1', Positive_RA_Alt_Thresh =
               '640', Climb_Inhibit = 'Zero', Two_of_Three_Reports_Valid = 'false'
```

**Test Case 25:**

```
choices:       302439181Cat1Ch4-|-976751556Cat1Ch4-|-603277445Cat1Ch2-|-1787817979Cat1Ch2-|-
               927131798Cat1Ch2-|-875777795Cat1Ch3-|-58393125Cat1Ch3-|-1871720133Cat1Ch1-|-
               374476473Cat1Ch1-|-1325844154Cat1Ch2-|-626069991Cat1Ch2-|-430451043Cat1Ch2

Alt_Layer_Value = Values of Alt_Layer_Value = Value 3|Positive_RA_Alt_Thresh = Values of
Positive_RA_Alt_Thresh = Value3|Own_Tracked_Alt = Values of Own_Tracked_Alt = Same as other
(intruder)|Other_Tracked_Alt = Values Other_Tracked_Alt = Same as TCAS (own)|Up_Separation =
Values of Up_Separation = Smaler (strictly) than down separation and threshold|Down_Separation =
Values of Down_Separation = Between up separation and threshold|Other_RAC = Values of Other_RAC
= Two|Current_Vertical_Sep = Values of Current_Vertical_Sep = Greater (strictly) than MAXALTDIF
(600)|High_Confidence = Values of High_Confidence = True|Two_of_Three_Reports_Valid = Values of
Two_of_Three_Reports_Valid = False|Other_Capability = Values of Other_Capability =
Two|Climb_Inhibit = Values of Climb_Inhibit = One
values:        Current_Vertical_Sep = '601', Alt_Layer_Value = 'Value3', Down_Separation = '2',
               Up_Separation = '1', Other_RAC = 'Two', High_Confidence = 'true', Other_Tracked_Alt
               = '1', Other_Capability = 'Two', Own_Tracked_Alt = '1', Positive_RA_Alt_Thresh =
               '740', Climb_Inhibit = 'One', Two_of_Three_Reports_Valid = 'false'
```

**Test Case 26:**

```
choices:       302439181Cat1Ch1-|-976751556Cat1Ch1-|-603277445Cat1Ch1-|-1787817979Cat1Ch3-|-
               927131798Cat1Ch2-|-875777795Cat1Ch1-|-58393125Cat1Ch3-|-1871720133Cat1Ch2-|-
               374476473Cat1Ch2-|-1325844154Cat1Ch2-|-626069991Cat1Ch1-|-430451043Cat1Ch1

Alt_Layer_Value = Values of Alt_Layer_Value = Value 0|Positive_RA_Alt_Thresh = Values of
Positive_RA_Alt_Thresh = Value0|Own_Tracked_Alt = Values of Own_Tracked_Alt = Higher (strictly)
than other (intruder)|Other_Tracked_Alt = Values Other_Tracked_Alt = Lower (strictly) than TCAS
(own)|Up_Separation = Values of Up_Separation = Smaler (strictly) than down separation and
threshold|Down_Separation = Values of Down_Separation = Greater (strictly) than up separation
and threshold|Other_RAC = Values of Other_RAC = Two|Current_Vertical_Sep = Values of
Current_Vertical_Sep = Smaller than MAXALTDIF (600), but positive (strictly)|High_Confidence =
Values of High_Confidence = False|Two_of_Three_Reports_Valid = Values of
Two_of_Three_Reports_Valid = False|Other_Capability = Values of Other_Capability =
One|Climb_Inhibit = Values of Climb_Inhibit = Zero
values:        Current_Vertical_Sep = '1', Alt_Layer_Value = 'Value0', Down_Separation = '401',
               Up_Separation = '399', Other_RAC = 'Two', High_Confidence = 'false',
               Other_Tracked_Alt = '1', Other_Capability = 'One', Own_Tracked_Alt = '2',
```

```
                 Climb_Inhibit = 'Zero', Positive_RA_Alt_Thresh = '400', Two_of_Three_Reports_Valid
                 = 'false'
```

**Test Case 27:**
```
choices:         302439181Cat1Ch3-|-976751556Cat1Ch3-|-603277445Cat1Ch1-|-1787817979Cat1Ch3-|-
                 927131798Cat1Ch2-|-875777795Cat1Ch1-|-58393125Cat1Ch2-|-1871720133Cat1Ch1-|-
                 374476473Cat1Ch1-|-1325844154Cat1Ch2-|-626069991Cat1Ch1-|-430451043Cat1Ch2
Alt_Layer_Value = Values of Alt_Layer_Value = Value 2|Positive_RA_Alt_Thresh = Values of
Positive_RA_Alt_Thresh = Value2|Own_Tracked_Alt = Values of Own_Tracked_Alt = Higher (strictly)
than other (intruder)|Other_Tracked_Alt = Values Other_Tracked_Alt = Lower (strictly) than TCAS
(own)|Up_Separation = Values of Up_Separation = Smaller (strictly) than down separation and
threshold|Down_Separation = Values of Down_Separation = Greater (strictly) than up separation
and threshold|Other_RAC = Values of Other_RAC = One|Current_Vertical_Sep = Values of
Current_Vertical_Sep = Greater (strictly) than MAXALTDIF (600)|High_Confidence = Values of
High_Confidence = True|Two_of_Three_Reports_Valid = Values of Two_of_Three_Reports_Valid =
False|Other_Capability = Values of Other_Capability = One|Climb_Inhibit = Values of
Climb_Inhibit = One
values:          Current_Vertical_Sep = '601', Alt_Layer_Value = 'Value2', Down_Separation = '641',
                 Up_Separation = '639', Other_RAC = 'One', High_Confidence = 'true',
                 Other_Tracked_Alt = '1', Other_Capability = 'One', Own_Tracked_Alt = '2',
                 Positive_RA_Alt_Thresh = '640', Climb_Inhibit = 'One', Two_of_Three_Reports_Valid =
                 'false'
```

**Test Case 28:**
```
choices:         302439181Cat1Ch4-|-976751556Cat1Ch4-|-603277445Cat1Ch1-|-1787817979Cat1Ch3-|-
                 927131798Cat1Ch2-|-875777795Cat1Ch1-|-58393125Cat1Ch2-|-1871720133Cat1Ch2-|-
                 374476473Cat1Ch1-|-1325844154Cat1Ch2-|-626069991Cat1Ch1-|-430451043Cat1Ch2
Alt_Layer_Value = Values of Alt_Layer_Value = Value 3|Positive_RA_Alt_Thresh = Values of
Positive_RA_Alt_Thresh = Value3|Own_Tracked_Alt = Values of Own_Tracked_Alt = Higher (strictly)
than other (intruder)|Other_Tracked_Alt = Values Other_Tracked_Alt = Lower (strictly) than TCAS
(own)|Up_Separation = Values of Up_Separation = Smaller (strictly) than down separation and
threshold|Down_Separation = Values of Down_Separation = Greater (strictly) than up separation
and threshold|Other_RAC = Values of Other_RAC = Zero|Current_Vertical_Sep = Values of
Current_Vertical_Sep = Smaller than MAXALTDIF (600), but positive (strictly)|High_Confidence =
Values of High_Confidence = True|Two_of_Three_Reports_Valid = Values of
Two_of_Three_Reports_Valid = False|Other_Capability = Values of Other_Capability =
One|Climb_Inhibit = Values of Climb_Inhibit = One
values:          Current_Vertical_Sep = '1', Alt_Layer_Value = 'Value3', Down_Separation = '741',
                 Up_Separation = '1', Other_RAC = 'Zero', High_Confidence = 'true',
                 Other_Tracked_Alt = '1', Other_Capability = 'One', Own_Tracked_Alt = '2',
                 Positive_RA_Alt_Thresh = '740', Climb_Inhibit = 'One', Two_of_Three_Reports_Valid =
                 'false'
```

**Test Case 29:**
```
choices:         302439181Cat1Ch3-|-976751556Cat1Ch3-|-603277445Cat1Ch3-|-1787817979Cat1Ch1-|-
                 927131798Cat1Ch2-|-875777795Cat1Ch1-|-58393125Cat1Ch2-|-1871720133Cat1Ch1-|-
                 374476473Cat1Ch1-|-1325844154Cat1Ch1-|-626069991Cat1Ch2-|-430451043Cat1Ch1
Alt_Layer_Value = Values of Alt_Layer_Value = Value 2|Positive_RA_Alt_Thresh = Values of
Positive_RA_Alt_Thresh = Value2|Own_Tracked_Alt = Values of Own_Tracked_Alt = Lower (strictly)
than other (intruder)|Other_Tracked_Alt = Values Other_Tracked_Alt = Higher (strictly) than TCAS
(own)|Up_Separation = Values of Up_Separation = Between down separation and
threshold|Down_Separation = Values of Down_Separation = Greater (strictly) than up separation
and threshold|Other_RAC = Values of Other_RAC = One|Current_Vertical_Sep = Values of
Current_Vertical_Sep = Greater (strictly) than MAXALTDIF (600)|High_Confidence = Values of
High_Confidence = True|Two_of_Three_Reports_Valid = Values of Two_of_Three_Reports_Valid =
True|Other_Capability = Values of Other_Capability = Two|Climb_Inhibit = Values of Climb_Inhibit
= Zero
values:          Current_Vertical_Sep = '601', Alt_Layer_Value = 'Value2', Down_Separation = '641',
                 Up_Separation = '640', Other_RAC = 'One', High_Confidence = 'true',
                 Other_Tracked_Alt = '2', Other_Capability = 'Two', Own_Tracked_Alt = '1',
                 Positive_RA_Alt_Thresh = '640', Climb_Inhibit = 'Zero', Two_of_Three_Reports_Valid
                 = 'true'
```

**Test Case 30:**
```
choices:         302439181Cat1Ch4-|-976751556Cat1Ch4-|-603277445Cat1Ch3-|-1787817979Cat1Ch1-|-
                 927131798Cat1Ch3-|-875777795Cat1Ch2-|-58393125Cat1Ch1-|-1871720133Cat1Ch1-|-
                 374476473Cat1Ch2-|-1325844154Cat1Ch2-|-626069991Cat1Ch2-|-430451043Cat1Ch2
Alt_Layer_Value = Values of Alt_Layer_Value = Value 3|Positive_RA_Alt_Thresh = Values of
Positive_RA_Alt_Thresh = Value3|Own_Tracked_Alt = Values of Own_Tracked_Alt = Lower (strictly)
```

than other (intruder)|Other_Tracked_Alt = Values Other_Tracked_Alt = Higher (strictly) than TCAS
(own)|Up_Separation = Values of Up_Separation = Between down separation and
threshold|Down_Separation = Values of Down_Separation = Smaller (strictly) than up separation
and threshold|Other_RAC = Values of Other_RAC = Zero|Current_Vertical_Sep = Values of
Current_Vertical_Sep = Greater (strictly) than MAXALTDIF (600)|High_Confidence = Values of
High_Confidence = False|Two_of_Three_Reports_Valid = Values of Two_of_Three_Reports_Valid =
False|Other_Capability = Values of Other_Capability = Two|Climb_Inhibit = Values of
Climb_Inhibit = One
values:        Current_Vertical_Sep = '601', Alt_Layer_Value = 'Value3', Down_Separation = '739',
               Up_Separation = '740', Other_RAC = 'Zero', High_Confidence = 'false',
               Other_Tracked_Alt = '2', Other_Capability = 'Two', Own_Tracked_Alt = '1',
               Positive_RA_Alt_Thresh = '740', Climb_Inhibit = 'One', Two_of_Three_Reports_Valid =
               'false'

**Test Case 31:**
choices:        302439181Cat1Ch1−|−976751556Cat1Ch1−|−603277445Cat1Ch2−|−1787817979Cat1Ch2−|−
                927131798Cat1Ch3−|−875777795Cat1Ch2−|−58393125Cat1Ch3−|−1871720133Cat1Ch2−|−
                374476473Cat1Ch2−|−1325844154Cat1Ch1−|−626069991Cat1Ch2−|−430451043Cat1Ch1
Alt_Layer_Value = Values of Alt_Layer_Value = Value 0|Positive_RA_Alt_Thresh = Values of
Positive_RA_Alt_Thresh = Value0|Own_Tracked_Alt = Values of Own_Tracked_Alt = Same as other
(intruder)|Other_Tracked_Alt = Values Other_Tracked_Alt = Same as TCAS (own)|Up_Separation =
Values of Up_Separation = Between down separation and threshold|Down_Separation = Values of
Down_Separation = Smaller (strictly) than up separation and threshold|Other_RAC = Values of
Other_RAC = Two|Current_Vertical_Sep = Values of Current_Vertical_Sep = Smaller than MAXALTDIF
(600), but positive (strictly)|High_Confidence = Values of High_Confidence =
False|Two_of_Three_Reports_Valid = Values of Two_of_Three_Reports_Valid = True|Other_Capability
= Values of Other_Capability = Two|Climb_Inhibit = Values of Climb_Inhibit = Zero
values:        Current_Vertical_Sep = '1', Alt_Layer_Value = 'Value0', Down_Separation = '1',
               Up_Separation = '2', Other_RAC = 'Two', High_Confidence = 'false',
               Other_Tracked_Alt = '1', Other_Capability = 'Two', Own_Tracked_Alt = '1',
               Climb_Inhibit = 'Zero', Positive_RA_Alt_Thresh = '400', Two_of_Three_Reports_Valid
               = 'true'

**Test Case 32:**
choices:        302439181Cat1Ch2−|−976751556Cat1Ch2−|−603277445Cat1Ch2−|−1787817979Cat1Ch2−|−
                927131798Cat1Ch3−|−875777795Cat1Ch1−|−58393125Cat1Ch2−|−1871720133Cat1Ch1−|−
                374476473Cat1Ch1−|−1325844154Cat1Ch2−|−626069991Cat1Ch1−|−430451043Cat1Ch2
Alt_Layer_Value = Values of Alt_Layer_Value = Value 1|Positive_RA_Alt_Thresh = Values of
Positive_RA_Alt_Thresh = Value1|Own_Tracked_Alt = Values of Own_Tracked_Alt = Same as other
(intruder)|Other_Tracked_Alt = Values Other_Tracked_Alt = Same as TCAS (own)|Up_Separation =
Values of Up_Separation = Between down separation and threshold|Down_Separation = Values of
Down_Separation = Greater (strictly) than up separation and threshold|Other_RAC = Values of
Other_RAC = One|Current_Vertical_Sep = Values of Current_Vertical_Sep = Greater (strictly) than
MAXALTDIF (600)|High_Confidence = Values of High_Confidence = True|Two_of_Three_Reports_Valid =
Values of Two_of_Three_Reports_Valid = False|Other_Capability = Values of Other_Capability =
One|Climb_Inhibit = Values of Climb_Inhibit = One
values:        Current_Vertical_Sep = '601', Alt_Layer_Value = 'Value1', Down_Separation = '501',
               Up_Separation = '500', Other_RAC = 'One', High_Confidence = 'true',
               Other_Capability = 'One', Other_Tracked_Alt = '1', Own_Tracked_Alt = '1',
               Positive_RA_Alt_Thresh = '500', Climb_Inhibit = 'One', Two_of_Three_Reports_Valid =
               'false'

**Test Case 33:**
choices:        302439181Cat1Ch3−|−976751556Cat1Ch3−|−603277445Cat1Ch2−|−1787817979Cat1Ch2−|−
                927131798Cat1Ch3−|−875777795Cat1Ch2−|−58393125Cat1Ch1−|−1871720133Cat1Ch2−|−
                374476473Cat1Ch1−|−1325844154Cat1Ch1−|−626069991Cat1Ch1−|−430451043Cat1Ch2
Alt_Layer_Value = Values of Alt_Layer_Value = Value 2|Positive_RA_Alt_Thresh = Values of
Positive_RA_Alt_Thresh = Value2|Own_Tracked_Alt = Values of Own_Tracked_Alt = Same as other
(intruder)|Other_Tracked_Alt = Values Other_Tracked_Alt = Same as TCAS (own)|Up_Separation =
Values of Up_Separation = Between down separation and threshold|Down_Separation = Values of
Down_Separation = Smaller (strictly) than up separation and threshold|Other_RAC = Values of
Other_RAC = Zero|Current_Vertical_Sep = Values of Current_Vertical_Sep = Smaller than MAXALTDIF
(600), but positive (strictly)|High_Confidence = Values of High_Confidence =
True|Two_of_Three_Reports_Valid = Values of Two_of_Three_Reports_Valid = True|Other_Capability =
Values of Other_Capability = One|Climb_Inhibit = Values of Climb_Inhibit = One
values:        Current_Vertical_Sep = '1', Alt_Layer_Value = 'Value2', Down_Separation = '639',
               Up_Separation = '640', Other_RAC = 'Zero', High_Confidence = 'true',
               Other_Tracked_Alt = '1', Other_Capability = 'One', Own_Tracked_Alt = '1',

Climb_Inhibit = 'One', Positive_RA_Alt_Thresh = '640', Two_of_Three_Reports_Valid = 'true'

**Test Case 34:**
choices:       302439181Cat1Ch1—|—976751556Cat1Ch1—|—603277445Cat1Ch1—|—1787817979Cat1Ch3—|—927131798Cat1Ch3—|—875777795Cat1Ch2—|—58393125Cat1Ch2—|—1871720133Cat1Ch1—|—374476473Cat1Ch1—|—1325844154Cat1Ch1—|—626069991Cat1Ch2—|—430451043Cat1Ch1

Alt_Layer_Value = Values of Alt_Layer_Value = Value 0|Positive_RA_Alt_Thresh = Values of Positive_RA_Alt_Thresh = Value0|Own_Tracked_Alt = Values of Own_Tracked_Alt = Higher (strictly) than other (intruder)|Other_Tracked_Alt = Values Other_Tracked_Alt = Lower (strictly) than TCAS (own)|Up_Separation = Values of Up_Separation = Between down separation and threshold|Down_Separation = Values of Down_Separation = Smaller (strictly) than up separation and threshold|Other_RAC = Values of Other_RAC = One|Current_Vertical_Sep = Values of Current_Vertical_Sep = Greater (strictly) than MAXALTDIF (600)|High_Confidence = Values of High_Confidence = True|Two_of_Three_Reports_Valid = Values of Two_of_Three_Reports_Valid = False|Other_Capability = Values of Other_Capability = Two|Climb_Inhibit = Values of Climb_Inhibit = Zero

values:        Current_Vertical_Sep = '601', Alt_Layer_Value = 'Value0', Down_Separation = '1', Up_Separation = '2', Other_RAC = 'One', High_Confidence = 'true', Other_Capability = 'Two', Other_Tracked_Alt = '1', Own_Tracked_Alt = '2', Positive_RA_Alt_Thresh = '400', Climb_Inhibit = 'Zero', Two_of_Three_Reports_Valid = 'false'

**Test Case 35:**
choices:       302439181Cat1Ch2—|—976751556Cat1Ch2—|—603277445Cat1Ch1—|—1787817979Cat1Ch3—|—927131798Cat1Ch3—|—875777795Cat1Ch1—|—58393125Cat1Ch3—|—1871720133Cat1Ch2—|—374476473Cat1Ch2—|—1325844154Cat1Ch1—|—626069991Cat1Ch2—|—430451043Cat1Ch2

Alt_Layer_Value = Values of Alt_Layer_Value = Value 1|Positive_RA_Alt_Thresh = Values of Positive_RA_Alt_Thresh = Value1|Own_Tracked_Alt = Values of Own_Tracked_Alt = Higher (strictly) than other (intruder)|Other_Tracked_Alt = Values Other_Tracked_Alt = Lower (strictly) than TCAS (own)|Up_Separation = Values of Up_Separation = Between down separation and threshold|Down_Separation = Values of Down_Separation = Greater (strictly) than up separation and threshold|Other_RAC = Values of Other_RAC = Two|Current_Vertical_Sep = Values of Current_Vertical_Sep = Smaller than MAXALTDIF (600), but positive (strictly)|High_Confidence = Values of High_Confidence = False|Two_of_Three_Reports_Valid = Values of Two_of_Three_Reports_Valid = True|Other_Capability = Values of Other_Capability = Two|Climb_Inhibit = Values of Climb_Inhibit = One

values:        Current_Vertical_Sep = '1', Alt_Layer_Value = 'Value1', Down_Separation = '501', Up_Separation = '500', Other_RAC = 'Two', High_Confidence = 'false', Other_Tracked_Alt = '1', Other_Capability = 'Two', Own_Tracked_Alt = '2', Positive_RA_Alt_Thresh = '500', Climb_Inhibit = 'One', Two_of_Three_Reports_Valid = 'true'

**Test Case 36:**
choices:       302439181Cat1Ch4—|—976751556Cat1Ch4—|—603277445Cat1Ch1—|—1787817979Cat1Ch3—|—927131798Cat1Ch3—|—875777795Cat1Ch2—|—58393125Cat1Ch2—|—1871720133Cat1Ch2—|—374476473Cat1Ch1—|—1325844154Cat1Ch1—|—626069991Cat1Ch1—|—430451043Cat1Ch1

Alt_Layer_Value = Values of Alt_Layer_Value = Value 3|Positive_RA_Alt_Thresh = Values of Positive_RA_Alt_Thresh = Value3|Own_Tracked_Alt = Values of Own_Tracked_Alt = Higher (strictly) than other (intruder)|Other_Tracked_Alt = Values Other_Tracked_Alt = Lower (strictly) than TCAS (own)|Up_Separation = Values of Up_Separation = Between down separation and threshold|Down_Separation = Values of Down_Separation = Smaller (strictly) than up separation and threshold|Other_RAC = Values of Other_RAC = One|Current_Vertical_Sep = Values of Current_Vertical_Sep = Smaller than MAXALTDIF (600), but positive (strictly)|High_Confidence = Values of High_Confidence = True|Two_of_Three_Reports_Valid = Values of Two_of_Three_Reports_Valid = True|Other_Capability = Values of Other_Capability = One|Climb_Inhibit = Values of Climb_Inhibit = Zero

values:        Current_Vertical_Sep = '1', Alt_Layer_Value = 'Value3', Down_Separation = '739', Up_Separation = '740', Other_RAC = 'One', High_Confidence = 'true', Other_Tracked_Alt = '1', Other_Capability = 'One', Own_Tracked_Alt = '2', Positive_RA_Alt_Thresh = '740', Climb_Inhibit = 'Zero', Two_of_Three_Reports_Valid = 'true'

**Test Case 37:**
choices:       302439181Cat1Ch1—|—976751556Cat1Ch1—|—603277445Cat1Ch3—|—1787817979Cat1Ch1—|—927131798Cat1Ch1—|—875777795Cat1Ch2—|—58393125Cat1Ch1—|—1871720133Cat1Ch1—|—374476473Cat1Ch2—|—1325844154Cat1Ch2—|—626069991Cat1Ch2—|—430451043Cat1Ch1

Alt_Layer_Value = Values of Alt_Layer_Value = Value 0|Positive_RA_Alt_Thresh = Values of Positive_RA_Alt_Thresh = Value0|Own_Tracked_Alt = Values of Own_Tracked_Alt = Lower (strictly) than other (intruder)|Other_Tracked_Alt = Values Other_Tracked_Alt = Higher (strictly) than TCAS

(own)|Up_Separation = Values of Up_Separation = Between down separation and
threshold|Down_Separation = Values of Down_Separation = Smaller (strictly) than up separation
and threshold|Other_RAC = Values of Other_RAC = Zero|Current_Vertical_Sep = Values of
Current_Vertical_Sep = Greater (strictly) than MAXALTDIF (600)|High_Confidence = Values of
High_Confidence = False|Two_of_Three_Reports_Valid = Values of Two_of_Three_Reports_Valid =
False|Other_Capability = Values of Other_Capability = Two|Climb_Inhibit = Values of
Climb_Inhibit = Zero

```
values:        Current_Vertical_Sep = '601', Alt_Layer_Value = 'Value0', Down_Separation = '1',
               Up_Separation = '2', Other_RAC = 'Zero', High_Confidence = 'false',
               Other_Tracked_Alt = '2', Other_Capability = 'Two', Own_Tracked_Alt = '1',
               Positive_RA_Alt_Thresh = '400', Climb_Inhibit = 'Zero', Two_of_Three_Reports_Valid
               = 'false'
```

**Test Case 38:**

choices:       302439181Cat1Ch3-|-976751556Cat1Ch3-|-603277445Cat1Ch1-|-1787817979Cat1Ch3-|-
               927131798Cat1Ch3-|-875777795Cat1Ch2-|-58393125Cat1Ch3-|-1871720133Cat1Ch2-|-
               374476473Cat1Ch1-|-1325844154Cat1Ch2-|-626069991Cat1Ch2-|-430451043Cat1Ch1

Alt_Layer_Value = Values of Alt_Layer_Value = Value 2|Positive_RA_Alt_Thresh = Values of
Positive_RA_Alt_Thresh = Value2|Own_Tracked_Alt = Values of Own_Tracked_Alt = Higher (strictly)
than other (intruder)|Other_Tracked_Alt = Values Other_Tracked_Alt = Lower (strictly) than TCAS
(own)|Up_Separation = Values of Up_Separation = Between down separation and
threshold|Down_Separation = Values of Down_Separation = Smaller (strictly) than up separation
and threshold|Other_RAC = Values of Other_RAC = Two|Current_Vertical_Sep = Values of
Current_Vertical_Sep = Smaller than MAXALTDIF (600), but positive (strictly)|High_Confidence =
Values of High_Confidence = True|Two_of_Three_Reports_Valid = Values of
Two_of_Three_Reports_Valid = False|Other_Capability = Values of Other_Capability =
Two|Climb_Inhibit = Values of Climb_Inhibit = Zero

```
values:        Current_Vertical_Sep = '1', Alt_Layer_Value = 'Value2', Down_Separation = '639',
               Up_Separation = '640', Other_RAC = 'Two', High_Confidence = 'true',
               Other_Tracked_Alt = '1', Other_Capability = 'Two', Own_Tracked_Alt = '2',
               Positive_RA_Alt_Thresh = '640', Climb_Inhibit = 'Zero', Two_of_Three_Reports_Valid
               = 'false'
```

**Test Case 39:**

choices:       302439181Cat1Ch4-|-976751556Cat1Ch4-|-603277445Cat1Ch3-|-1787817979Cat1Ch1-|-
               927131798Cat1Ch1-|-875777795Cat1Ch3-|-58393125Cat1Ch1-|-1871720133Cat1Ch1-|-
               374476473Cat1Ch2-|-1325844154Cat1Ch1-|-626069991Cat1Ch1-|-430451043Cat1Ch2

Alt_Layer_Value = Values of Alt_Layer_Value = Value 3|Positive_RA_Alt_Thresh = Values of
Positive_RA_Alt_Thresh = Value3|Own_Tracked_Alt = Values of Own_Tracked_Alt = Lower (strictly)
than other (intruder)|Other_Tracked_Alt = Values Other_Tracked_Alt = Higher (strictly) than TCAS
(own)|Up_Separation = Values of Up_Separation = Greater (strictly) than down separation and
threshold|Down_Separation = Values of Down_Separation = Between up separation and
threshold|Other_RAC = Values of Other_RAC = Zero|Current_Vertical_Sep = Values of
Current_Vertical_Sep = Greater (strictly) than MAXALTDIF (600)|High_Confidence = Values of
High_Confidence = False|Two_of_Three_Reports_Valid = Values of Two_of_Three_Reports_Valid =
True|Other_Capability = Values of Other_Capability = One|Climb_Inhibit = Values of Climb_Inhibit
= One

```
values:        Current_Vertical_Sep = '601', Alt_Layer_Value = 'Value3', Down_Separation = '740',
               Up_Separation = '741', Other_RAC = 'Zero', High_Confidence = 'false',
               Other_Tracked_Alt = '2', Other_Capability = 'One', Own_Tracked_Alt = '1',
               Positive_RA_Alt_Thresh = '740', Climb_Inhibit = 'One', Two_of_Three_Reports_Valid =
               'true'
```

**Test Case 40:**

choices:       302439181Cat1Ch1-|-976751556Cat1Ch1-|-603277445Cat1Ch2-|-1787817979Cat1Ch2-|-
               927131798Cat1Ch1-|-875777795Cat1Ch3-|-58393125Cat1Ch3-|-1871720133Cat1Ch1-|-
               374476473Cat1Ch2-|-1325844154Cat1Ch2-|-626069991Cat1Ch1-|-430451043Cat1Ch1

Alt_Layer_Value = Values of Alt_Layer_Value = Value 0|Positive_RA_Alt_Thresh = Values of
Positive_RA_Alt_Thresh = Value0|Own_Tracked_Alt = Values of Own_Tracked_Alt = Same as other
(intruder)|Other_Tracked_Alt = Values Other_Tracked_Alt = Same as TCAS (own)|Up_Separation =
Values of Up_Separation = Greater (strictly) than down separation and threshold|Down_Separation
= Values of Down_Separation = Between up separation and threshold|Other_RAC = Values of
Other_RAC = Two|Current_Vertical_Sep = Values of Current_Vertical_Sep = Greater (strictly) than
MAXALTDIF (600)|High_Confidence = Values of High_Confidence = False|Two_of_Three_Reports_Valid =
Values of Two_of_Three_Reports_Valid = False|Other_Capability = Values of Other_Capability =
One|Climb_Inhibit = Values of Climb_Inhibit = Zero

```
values:        Current_Vertical_Sep = '601', Alt_Layer_Value = 'Value0', Down_Separation = '400',
               Up_Separation = '401', Other_RAC = 'Two', High_Confidence = 'false',
```

```
                    Other_Capability = 'One', Other_Tracked_Alt = '1', Own_Tracked_Alt = '1',
                    Climb_Inhibit = 'Zero', Positive_RA_Alt_Thresh = '400', Two_of_Three_Reports_Valid
                    = 'false'
```

**Test Case 41:**
```
choices:        302439181Cat1Ch1-|-976751556Cat1Ch1-|-603277445Cat1Ch1-|-1787817979Cat1Ch3-|-
                927131798Cat1Ch2-|-875777795Cat1Ch3-|-58393125Cat1Ch1-|-1871720133Cat1Ch1-|-
                374476473Cat1Ch1-|-1325844154Cat1Ch1-|-626069991Cat1Ch1-|-430451043Cat1Ch2
Alt_Layer_Value = Values of Alt_Layer_Value = Value 0|Positive_RA_Alt_Thresh = Values of
Positive_RA_Alt_Thresh = Value0|Own_Tracked_Alt = Values of Own_Tracked_Alt = Higher (strictly)
than other (intruder)|Other_Tracked_Alt = Values Other_Tracked_Alt = Lower (strictly) than TCAS
(own)|Up_Separation = Values of Up_Separation = Smaler (strictly) than down separation and
threshold|Down_Separation = Values of Down_Separation = Between up separation and
threshold|Other_RAC = Values of Other_RAC = Zero|Current_Vertical_Sep = Values of
Current_Vertical_Sep = Greater (strictly) than MAXALTDIF (600)|High_Confidence = Values of
High_Confidence = True|Two_of_Three_Reports_Valid = Values of Two_of_Three_Reports_Valid =
True|Other_Capability = Values of Other_Capability = One|Climb_Inhibit = Values of Climb_Inhibit
= One
values:         Current_Vertical_Sep = '601', Alt_Layer_Value = 'Value0', Down_Separation = '2',
                Up_Separation = '1', Other_RAC = 'Zero', High_Confidence = 'true',
                Other_Tracked_Alt = '1', Other_Capability = 'One', Own_Tracked_Alt = '2',
                Positive_RA_Alt_Thresh = '400', Climb_Inhibit = 'One', Two_of_Three_Reports_Valid =
                'true'
```

**Test Case 42:**
```
choices:        302439181Cat1Ch1-|-976751556Cat1Ch1-|-603277445Cat1Ch3-|-1787817979Cat1Ch1-|-
                927131798Cat1Ch2-|-875777795Cat1Ch1-|-58393125Cat1Ch1-|-1871720133Cat1Ch2-|-
                374476473Cat1Ch2-|-1325844154Cat1Ch1-|-626069991Cat1Ch2-|-430451043Cat1Ch2
Alt_Layer_Value = Values of Alt_Layer_Value = Value 0|Positive_RA_Alt_Thresh = Values of
Positive_RA_Alt_Thresh = Value0|Own_Tracked_Alt = Values of Own_Tracked_Alt = Lower (strictly)
than other (intruder)|Other_Tracked_Alt = Values Other_Tracked_Alt = Higher (strictly) than TCAS
(own)|Up_Separation = Values of Up_Separation = Smaler (strictly) than down separation and
threshold|Down_Separation = Values of Down_Separation = Greater (strictly) than up separation
and threshold|Other_RAC = Values of Other_RAC = Zero|Current_Vertical_Sep = Values of
Current_Vertical_Sep = Smaller than MAXALTDIF (600), but positive (strictly)|High_Confidence =
Values of High_Confidence = False|Two_of_Three_Reports_Valid = Values of
Two_of_Three_Reports_Valid = True|Other_Capability = Values of Other_Capability =
Two|Climb_Inhibit = Values of Climb_Inhibit = One
values:         Current_Vertical_Sep = '1', Alt_Layer_Value = 'Value0', Down_Separation = '401',
                Up_Separation = '399', Other_RAC = 'Zero', High_Confidence = 'false',
                Other_Tracked_Alt = '2', Other_Capability = 'Two', Own_Tracked_Alt = '1',
                Climb_Inhibit = 'One', Positive_RA_Alt_Thresh = '400', Two_of_Three_Reports_Valid =
                'true'
```

**Test Case 43:**
```
choices:        302439181Cat1Ch2-|-976751556Cat1Ch2-|-603277445Cat1Ch1-|-1787817979Cat1Ch3-|-
                927131798Cat1Ch2-|-875777795Cat1Ch1-|-58393125Cat1Ch1-|-1871720133Cat1Ch1-|-
                374476473Cat1Ch2-|-1325844154Cat1Ch1-|-626069991Cat1Ch2-|-430451043Cat1Ch1
Alt_Layer_Value = Values of Alt_Layer_Value = Value 1|Positive_RA_Alt_Thresh = Values of
Positive_RA_Alt_Thresh = Value1|Own_Tracked_Alt = Values of Own_Tracked_Alt = Higher (strictly)
than other (intruder)|Other_Tracked_Alt = Values Other_Tracked_Alt = Lower (strictly) than TCAS
(own)|Up_Separation = Values of Up_Separation = Smaler (strictly) than down separation and
threshold|Down_Separation = Values of Down_Separation = Greater (strictly) than up separation
and threshold|Other_RAC = Values of Other_RAC = Zero|Current_Vertical_Sep = Values of
Current_Vertical_Sep = Greater (strictly) than MAXALTDIF (600)|High_Confidence = Values of
High_Confidence = False|Two_of_Three_Reports_Valid = Values of Two_of_Three_Reports_Valid =
True|Other_Capability = Values of Other_Capability = Two|Climb_Inhibit = Values of Climb_Inhibit
= Zero
values:         Current_Vertical_Sep = '601', Alt_Layer_Value = 'Value1', Down_Separation = '501',
                Up_Separation = '1', Other_RAC = 'Zero', High_Confidence = 'false',
                Other_Tracked_Alt = '1', Other_Capability = 'Two', Own_Tracked_Alt = '2',
                Positive_RA_Alt_Thresh = '500', Climb_Inhibit = 'Zero', Two_of_Three_Reports_Valid
                = 'true'
```

**Test Case 44:**
```
choices:        302439181Cat1Ch3-|-976751556Cat1Ch3-|-603277445Cat1Ch3-|-1787817979Cat1Ch1-|-
                927131798Cat1Ch1-|-875777795Cat1Ch3-|-58393125Cat1Ch1-|-1871720133Cat1Ch2-|-
                374476473Cat1Ch2-|-1325844154Cat1Ch2-|-626069991Cat1Ch1-|-430451043Cat1Ch1
```

```
Alt_Layer_Value = Values of Alt_Layer_Value = Value 2|Positive_RA_Alt_Thresh = Values of
Positive_RA_Alt_Thresh = Value2|Own_Tracked_Alt = Values of Own_Tracked_Alt = Lower (strictly)
than other (intruder)|Other_Tracked_Alt = Values Other_Tracked_Alt = Higher (strictly) than TCAS
(own)|Up_Separation = Values of Up_Separation = Greater (strictly) than down separation and
threshold|Down_Separation = Values of Down_Separation = Between up separation and
threshold|Other_RAC = Values of Other_RAC = Zero|Current_Vertical_Sep = Values of
Current_Vertical_Sep = Smaller than MAXALTDIF (600), but positive (strictly)|High_Confidence =
Values of High_Confidence = False|Two_of_Three_Reports_Valid = Values of
Two_of_Three_Reports_Valid = False|Other_Capability = Values of Other_Capability =
One|Climb_Inhibit = Values of Climb_Inhibit = Zero
values:        Current_Vertical_Sep = '1', Alt_Layer_Value = 'Value2', Down_Separation = '640',
               Up_Separation = '641', Other_RAC = 'Zero', High_Confidence = 'false',
               Other_Tracked_Alt = '2', Other_Capability = 'One', Own_Tracked_Alt = '1',
               Positive_RA_Alt_Thresh = '640', Climb_Inhibit = 'Zero', Two_of_Three_Reports_Valid
               = 'false'
```

**Test Case 45:**

```
choices:        302439181Cat1Ch4-|-976751556Cat1Ch4-|-603277445Cat1Ch1-|-1787817979Cat1Ch3-|-
                927131798Cat1Ch2-|-875777795Cat1Ch3-|-58393125Cat1Ch2-|-1871720133Cat1Ch1-|-
                374476473Cat1Ch2-|-1325844154Cat1Ch1-|-626069991Cat1Ch2-|-430451043Cat1Ch1
Alt_Layer_Value = Values of Alt_Layer_Value = Value 3|Positive_RA_Alt_Thresh = Values of
Positive_RA_Alt_Thresh = Value3|Own_Tracked_Alt = Values of Own_Tracked_Alt = Higher (strictly)
than other (intruder)|Other_Tracked_Alt = Values Other_Tracked_Alt = Lower (strictly) than TCAS
(own)|Up_Separation = Values of Up_Separation = Smaler (strictly) than down separation and
threshold|Down_Separation = Values of Down_Separation = Between up separation and
threshold|Other_RAC = Values of Other_RAC = One|Current_Vertical_Sep = Values of
Current_Vertical_Sep = Greater (strictly) than MAXALTDIF (600)|High_Confidence = Values of
High_Confidence = False|Two_of_Three_Reports_Valid = Values of Two_of_Three_Reports_Valid =
True|Other_Capability = Values of Other_Capability = Two|Climb_Inhibit = Values of Climb_Inhibit
= Zero
values:        Current_Vertical_Sep = '601', Alt_Layer_Value = 'Value3', Down_Separation = '2',
               Up_Separation = '1', Other_RAC = 'One', High_Confidence = 'false',
               Other_Tracked_Alt = '1', Other_Capability = 'Two', Own_Tracked_Alt = '2',
               Climb_Inhibit = 'Zero', Positive_RA_Alt_Thresh = '740', Two_of_Three_Reports_Valid
               = 'true'
```

**Test Case 46:**

```
choices:        302439181Cat1Ch4-|-976751556Cat1Ch4-|-603277445Cat1Ch3-|-1787817979Cat1Ch1-|-
                927131798Cat1Ch3-|-875777795Cat1Ch1-|-58393125Cat1Ch3-|-1871720133Cat1Ch1-|-
                374476473Cat1Ch1-|-1325844154Cat1Ch2-|-626069991Cat1Ch3-|-430451043Cat1Ch1
Alt_Layer_Value = Values of Alt_Layer_Value = Value 3|Positive_RA_Alt_Thresh = Values of
Positive_RA_Alt_Thresh = Value3|Own_Tracked_Alt = Values of Own_Tracked_Alt = Lower (strictly)
than other (intruder)|Other_Tracked_Alt = Values Other_Tracked_Alt = Higher (strictly) than TCAS
(own)|Up_Separation = Values of Up_Separation = Between down separation and
threshold|Down_Separation = Values of Down_Separation = Greater (strictly) than up separation
and threshold|Other_RAC = Values of Other_RAC = Two|Current_Vertical_Sep = Values of
Current_Vertical_Sep = Greater (strictly) than MAXALTDIF (600)|High_Confidence = Values of
High_Confidence = True|Two_of_Three_Reports_Valid = Values of Two_of_Three_Reports_Valid =
False|Other_Capability = Values of Other_Capability = Two|Climb_Inhibit = Values of
Climb_Inhibit = Zero
values:        Current_Vertical_Sep = '601', Alt_Layer_Value = 'Value3', Down_Separation = '741',
               Up_Separation = '740', Other_RAC = 'Two', High_Confidence = 'true',
               Other_Tracked_Alt = '2', Other_Capability = 'Two', Own_Tracked_Alt = '1',
               Positive_RA_Alt_Thresh = '740', Climb_Inhibit = 'Zero', Two_of_Three_Reports_Valid
               = 'false'
```

**Test Case 47:**

```
choices:        302439181Cat1Ch2-|-976751556Cat1Ch2-|-603277445Cat1Ch2-|-1787817979Cat1Ch2-|-
                927131798Cat1Ch2-|-875777795Cat1Ch1-|-58393125Cat1Ch3-|-1871720133Cat1Ch2-|-
                374476473Cat1Ch2-|-1325844154Cat1Ch1-|-626069991Cat1Ch1-|-430451043Cat1Ch2
Alt_Layer_Value = Values of Alt_Layer_Value = Value 1|Positive_RA_Alt_Thresh = Values of
Positive_RA_Alt_Thresh = Value1|Own_Tracked_Alt = Values of Own_Tracked_Alt = Same as other
(intruder)|Other_Tracked_Alt = Values Other_Tracked_Alt = Same as TCAS (own)|Up_Separation =
Values of Up_Separation = Greater (strictly) than down separation and threshold|Down_Separation
= Values of Down_Separation = Between up separation and threshold|Other_RAC = Values of
Other_RAC = Two|Current_Vertical_Sep = Values of Current_Vertical_Sep = Smaller than MAXALTDIF
(600), but positive (strictly)|High_Confidence = Values of High_Confidence =
```

False|Two_of_Three_Reports_Valid = Values of Two_of_Three_Reports_Valid = True|Other_Capability
= Values of Other_Capability = One|Climb_Inhibit = Values of Climb_Inhibit = One
values:        Current_Vertical_Sep = '1', Alt_Layer_Value = 'Value1', Down_Separation = '500',
               Up_Separation = '501', Other_RAC = 'Two', High_Confidence = 'false',
               Other_Tracked_Alt = '1', Other_Capability = 'One', Own_Tracked_Alt = '1',
               Positive_RA_Alt_Thresh = '500', Climb_Inhibit = 'One', Two_of_Three_Reports_Valid =
               'true'

**Test Case 48:**

choices:       302439181Cat1Ch3-|-976751556Cat1Ch3-|-603277445Cat1Ch2-|-1787817979Cat1Ch2-|-
               927131798Cat1Ch2-|-875777795Cat1Ch3-|-58393125Cat1Ch1-|-1871720133Cat1Ch1-|-
               374476473Cat1Ch2-|-1325844154Cat1Ch1-|-626069991Cat1Ch1-|-430451043Cat1Ch1

Alt_Layer_Value = Values of Alt_Layer_Value = Value 2|Positive_RA_Alt_Thresh = Values of
Positive_RA_Alt_Thresh = Value2|Own_Tracked_Alt = Values of Own_Tracked_Alt = Same as other
(intruder)|Other_Tracked_Alt = Values Other_Tracked_Alt = Same as TCAS (own)|Up_Separation =
Values of Up_Separation = Smaler (strictly) than down separation and threshold|Down_Separation =
Values of Down_Separation = Between up separation and threshold|Other_RAC = Values of Other_RAC
= Zero|Current_Vertical_Sep = Values of Current_Vertical_Sep = Greater (strictly) than MAXALTDIF
(600)|High_Confidence = Values of High_Confidence = False|Two_of_Three_Reports_Valid = Values of
Two_of_Three_Reports_Valid = True|Other_Capability = Values of Other_Capability =
One|Climb_Inhibit = Values of Climb_Inhibit = Zero
values:        Two_of_Three_Reports_Valid = 'true', Current_Vertical_Sep = '601', Alt_Layer_Value
               = 'Value2', Down_Separation = '2', Up_Separation = '1', Other_RAC = 'Zero',
               High_Confidence = 'false', Other_Tracked_Alt = '1', Other_Capability = 'One',
               Own_Tracked_Alt = '1', Positive_RA_Alt_Thresh = '640', Climb_Inhibit = 'Zero'

**Test Case 49:**

choices:       302439181Cat1Ch2-|-976751556Cat1Ch2-|-603277445Cat1Ch2-|-1787817979Cat1Ch2-|-
               927131798Cat1Ch3-|-875777795Cat1Ch2-|-58393125Cat1Ch2-|-1871720133Cat1Ch2-|-
               374476473Cat1Ch2-|-1325844154Cat1Ch1-|-626069991Cat1Ch2-|-430451043Cat1Ch2

Alt_Layer_Value = Values of Alt_Layer_Value = Value 1|Positive_RA_Alt_Thresh = Values of
Positive_RA_Alt_Thresh = Value1|Own_Tracked_Alt = Values of Own_Tracked_Alt = Lower (strictly)
than other (intruder)|Other_Tracked_Alt = Values Other_Tracked_Alt = Higher (strictly) than TCAS
(own)|Up_Separation = Values of Up_Separation = Between down separation and
threshold|Down_Separation = Values of Down_Separation = Smaller (strictly) than up separation
and threshold|Other_RAC = Values of Other_RAC = One|Current_Vertical_Sep = Values of
Current_Vertical_Sep = Smaller than MAXALTDIF (600), but positive (strictly)|High_Confidence =
Values of High_Confidence = False|Two_of_Three_Reports_Valid = Values of
Two_of_Three_Reports_Valid = True|Other_Capability = Values of Other_Capability =
Two|Climb_Inhibit = Values of Climb_Inhibit = One
values:        Current_Vertical_Sep = '1', Alt_Layer_Value = 'Value1', Down_Separation = '1',
               Up_Separation = '2', Other_RAC = 'One', High_Confidence = 'false',
               Other_Tracked_Alt = '2', Other_Capability = 'Two', Own_Tracked_Alt = '1',
               Climb_Inhibit = 'One', Positive_RA_Alt_Thresh = '500', Two_of_Three_Reports_Valid =
               'true'

**Test Case 50:**

choices:       302439181Cat1Ch3-|-976751556Cat1Ch3-|-603277445Cat1Ch2-|-1787817979Cat1Ch2-|-
               927131798Cat1Ch3-|-875777795Cat1Ch2-|-58393125Cat1Ch2-|-1871720133Cat1Ch2-|-
               374476473Cat1Ch2-|-1325844154Cat1Ch1-|-626069991Cat1Ch3-|-430451043Cat1Ch2

Alt_Layer_Value = Values of Alt_Layer_Value = Value 2|Positive_RA_Alt_Thresh = Values of
Positive_RA_Alt_Thresh = Value2|Own_Tracked_Alt = Values of Own_Tracked_Alt = Same as other
(intruder)|Other_Tracked_Alt = Values Other_Tracked_Alt = Same as TCAS (own)|Up_Separation =
Values of Up_Separation = Between down separation and threshold|Down_Separation = Values of
Down_Separation = Smaller (strictly) than up separation and threshold|Other_RAC = Values of
Other_RAC = One|Current_Vertical_Sep = Values of Current_Vertical_Sep = Smaller than MAXALTDIF
(600), but positive (strictly)|High_Confidence = Values of High_Confidence =
False|Two_of_Three_Reports_Valid = Values of Two_of_Three_Reports_Valid = True|Other_Capability
= Values of Other_Capability = Two|Climb_Inhibit = Values of Climb_Inhibit = One
values:        Current_Vertical_Sep = '1', Alt_Layer_Value = 'Value2', Down_Separation = '1',
               Up_Separation = '2', Other_RAC = 'One', High_Confidence = 'false',
               Other_Tracked_Alt = '1', Other_Capability = 'Two', Own_Tracked_Alt = '1',
               Positive_RA_Alt_Thresh = '640', Climb_Inhibit = 'One', Two_of_Three_Reports_Valid =
               'true'

**Test Case 51:**

choices:       302439181Cat1Ch2-|-976751556Cat1Ch2-|-603277445Cat1Ch2-|-1787817979Cat1Ch2-|-
               927131798Cat1Ch1-|-875777795Cat1Ch2-|-58393125Cat1Ch1-|-1871720133Cat1Ch2-|-
               374476473Cat1Ch1-|-1325844154Cat1Ch2-|-626069991Cat1Ch2-|-430451043Cat1Ch1

```
Alt_Layer_Value = Values of Alt_Layer_Value = Value 1|Positive_RA_Alt_Thresh = Values of
Positive_RA_Alt_Thresh = Value1|Own_Tracked_Alt = Values of Own_Tracked_Alt = Same as other
(intruder)|Other_Tracked_Alt = Values Other_Tracked_Alt = Same as TCAS (own)|Up_Separation =
Values of Up_Separation = Greater (strictly) than down separation and threshold|Down_Separation
= Values of Down_Separation = Smaller (strictly) than up separation and threshold|Other_RAC =
Values of Other_RAC = Zero|Current_Vertical_Sep = Values of Current_Vertical_Sep = Smaller than
MAXALTDIF (600), but positive (strictly)|High_Confidence = Values of High_Confidence =
True|Two_of_Three_Reports_Valid = Values of Two_of_Three_Reports_Valid = False|Other_Capability
= Values of Other_Capability = Two|Climb_Inhibit = Values of Climb_Inhibit = Zero
values:        Current_Vertical_Sep = '1', Alt_Layer_Value = 'Value1', Down_Separation = '1',
               Up_Separation = '501', Other_RAC = 'Zero', High_Confidence = 'true',
               Other_Tracked_Alt = '1', Other_Capability = 'Two', Own_Tracked_Alt = '1',
               Positive_RA_Alt_Thresh = '500', Climb_Inhibit = 'Zero', Two_of_Three_Reports_Valid
               = 'false'
```

**Test Case 52:**
```
choices:       302439181Cat1Ch1—|—976751556Cat1Ch1—|—603277445Cat1Ch2—|—1787817979Cat1Ch2—|—
               927131798Cat1Ch3—|—875777795Cat1Ch3—|—58393125Cat1Ch2—|—1871720133Cat1Ch1—|—
               374476473Cat1Ch2—|—1325844154Cat1Ch1—|—626069991Cat1Ch1—|—430451043Cat1Ch2
Alt_Layer_Value = Values of Alt_Layer_Value = Value 0|Positive_RA_Alt_Thresh = Values of
Positive_RA_Alt_Thresh = Value0|Own_Tracked_Alt = Values of Own_Tracked_Alt = Same as other
(intruder)|Other_Tracked_Alt = Values Other_Tracked_Alt = Same as TCAS (own)|Up_Separation =
Values of Up_Separation = Between down separation and threshold|Other_RAC = Values of
Other_RAC = One|Current_Vertical_Sep = Values of Current_Vertical_Sep = Greater (strictly) than
MAXALTDIF (600)|High_Confidence = Values of High_Confidence = False|Two_of_Three_Reports_Valid =
Values of Two_of_Three_Reports_Valid = True|Other_Capability = Values of Other_Capability =
One|Climb_Inhibit = Values of Climb_Inhibit = One
values:        Current_Vertical_Sep = '601', Alt_Layer_Value = 'Value0', Down_Separation = '401',
               Up_Separation = '400', Other_RAC = 'One', High_Confidence = 'false',
               Other_Tracked_Alt = '1', Other_Capability = 'One', Own_Tracked_Alt = '1',
               Climb_Inhibit = 'One', Positive_RA_Alt_Thresh = '400', Two_of_Three_Reports_Valid =
               'true'
```

**Test Case 53:**
```
choices:       302439181Cat1Ch2—|—976751556Cat1Ch2—|—603277445Cat1Ch1—|—1787817979Cat1Ch3—|—
               927131798Cat1Ch2—|—875777795Cat1Ch3—|—58393125Cat1Ch2—|—1871720133Cat1Ch1—|—
               374476473Cat1Ch1—|—1325844154Cat1Ch2—|—626069991Cat1Ch1—|—430451043Cat1Ch2
Alt_Layer_Value = Values of Alt_Layer_Value = Value 1|Positive_RA_Alt_Thresh = Values of
Positive_RA_Alt_Thresh = Value1|Own_Tracked_Alt = Values of Own_Tracked_Alt = Higher (strictly)
than other (intruder)|Other_Tracked_Alt = Values Other_Tracked_Alt = Lower (strictly) than TCAS
(own)|Up_Separation = Values of Up_Separation = Smaler (strictly) than down separation and
threshold|Down_Separation = Values of Down_Separation = Between up separation and
threshold|Other_RAC = Values of Other_RAC = One|Current_Vertical_Sep = Values of
Current_Vertical_Sep = Greater (strictly) than MAXALTDIF (600)|High_Confidence = Values of
High_Confidence = True|Two_of_Three_Reports_Valid = Values of Two_of_Three_Reports_Valid =
False|Other_Capability = Values of Other_Capability = One|Climb_Inhibit = Values of
Climb_Inhibit = One
values:        Current_Vertical_Sep = '601', Alt_Layer_Value = 'Value1', Down_Separation = '2',
               Up_Separation = '1', Other_RAC = 'One', High_Confidence = 'true', Other_Tracked_Alt
               = '1', Other_Capability = 'One', Own_Tracked_Alt = '2', Positive_RA_Alt_Thresh =
               '500', Climb_Inhibit = 'One', Two_of_Three_Reports_Valid = 'false'
```

**Test Case 54:**
```
choices:       1871720133Cat1Ch3
Current_Vertical_Sep = Values of Current_Vertical_Sep = Illegal value
values:        Current_Vertical_Sep = '0'
```

**Test Case 55:**
```
choices:       603277445Cat1Ch4
Own_Tracked_Alt = Values of Own_Tracked_Alt = Illegal
values:        Own_Tracked_Alt = '0'
```

**Test Case 56:**
```
choices:       1787817979Cat1Ch4
Other_Tracked_Alt = Values Other_Tracked_Alt = Illegal
values:        Other_Tracked_Alt = '0'
```

**Test Case 57:**
```
choices:       927131798Cat1Ch4
Up_Separation = Values of Up_Separation = Illegal
```

```
values:         Up_Separation = '0'
```

```
choices:        875777795Cat1Ch4
Down_Separation = Values of Down_Separation = Illegal
values:         Down_Separation = '0'
```

**Test Case 59:**
```
choices:        976751556Cat1Ch5
Positive_RA_Alt_Thresh = Values of Positive_RA_Alt_Thresh = Illegal
values:         Positive_RA_Alt_Thresh = '2'
```

## *J.9    Base Choice Test Frames and Corresponding Test Cases*

**Test Case 1:**
```
choices:        1871720133Cat1Ch1-|-374476473Cat1Ch1-|-1325844154Cat1Ch1-|-603277445Cat1Ch1-|-
                1787817979Cat1Ch3-|-927131798Cat1Ch1-|-875777795Cat1Ch2-|-302439181Cat1Ch1-|-
                58393125Cat1Ch1-|-626069991Cat1Ch1-|-430451043Cat1Ch1-|-976751556Cat1Ch1
Current_Vertical_Sep = Values of Current_Vertical_Sep = Greater (strictly) than MAXALTDIF
(600)|High_Confidence = Values of High_Confidence = True|Two_of_Three_Reports_Valid = Values of
Two_of_Three_Reports_Valid = True|Own_Tracked_Alt = Values of Own_Tracked_Alt = Higher
(strictly) than other (intruder)|Other_Tracked_Alt = Values Other_Tracked_Alt = Lower (strictly)
than TCAS (own)|Up_Separation = Values of Up_Separation = Greater (strictly) than down
separation and threshold|Down_Separation = Values of Down_Separation = Smaller (strictly) than
up separation and threshold|Alt_Layer_Value = Values of Alt_Layer_Value = Value 0|Other_RAC =
Values of Other_RAC = Zero|Other_Capability = Values of Other_Capability = One|Climb_Inhibit =
Values of Climb_Inhibit = Zero|Positive_RA_Alt_Thresh = Values of Positive_RA_Alt_Thresh =
Value0
values:         Current_Vertical_Sep = '601', Alt_Layer_Value = 'Value0', Down_Separation = '1',
                Up_Separation = '401', Other_RAC = 'Zero', High_Confidence = 'true',
                Other_Tracked_Alt = '1', Other_Capability = 'One', Own_Tracked_Alt = '2',
                Climb_Inhibit = 'Zero', Positive_RA_Alt_Thresh = '400', Two_of_Three_Reports_Valid
                = 'true'
```

**Test Case 2:**
```
choices:        374476473Cat1Ch1-|-1325844154Cat1Ch1-|-603277445Cat1Ch1-|-1787817979Cat1Ch3-|-
                927131798Cat1Ch1-|-875777795Cat1Ch2-|-302439181Cat1Ch1-|-58393125Cat1Ch1-|-
                626069991Cat1Ch1-|-430451043Cat1Ch1-|-976751556Cat1Ch1-|-1871720133Cat1Ch2
High_Confidence = Values of High_Confidence = True|Two_of_Three_Reports_Valid = Values of
Two_of_Three_Reports_Valid = True|Own_Tracked_Alt = Values of Own_Tracked_Alt = Higher
(strictly) than other (intruder)|Other_Tracked_Alt = Values of Other_Tracked_Alt = Lower (strictly)
than TCAS (own)|Up_Separation = Values of Up_Separation = Greater (strictly) than down
separation and threshold|Down_Separation = Values of Down_Separation = Smaller (strictly) than
up separation and threshold|Alt_Layer_Value = Values of Alt_Layer_Value = Value 0|Other_RAC =
Values of Other_RAC = Zero|Other_Capability = Values of Other_Capability = One|Climb_Inhibit =
Values of Climb_Inhibit = Zero|Positive_RA_Alt_Thresh = Values of Positive_RA_Alt_Thresh =
Value0|Current_Vertical_Sep = Values of Current_Vertical_Sep = Smaller than MAXALTDIF (600), but
positive (strictly)
values:         Current_Vertical_Sep = '1', Alt_Layer_Value = 'Value0', Down_Separation = '399',
                Up_Separation = '401', Other_RAC = 'Zero', High_Confidence = 'true',
                Other_Tracked_Alt = '1', Other_Capability = 'One', Own_Tracked_Alt = '2',
                Positive_RA_Alt_Thresh = '400', Climb_Inhibit = 'Zero', Two_of_Three_Reports_Valid
                = 'true'
```

**Test Case 3:**
```
choices:        1871720133Cat1Ch3
Current_Vertical_Sep = Values of Current_Vertical_Sep = Illegal value
values:         Current_Vertical_Sep = '0'
```

**Test Case 4:**
```
choices:        1871720133Cat1Ch1-|-1325844154Cat1Ch1-|-603277445Cat1Ch1-|-1787817979Cat1Ch3-|-
                927131798Cat1Ch1-|-875777795Cat1Ch2-|-302439181Cat1Ch1-|-58393125Cat1Ch1-|-
                626069991Cat1Ch1-|-430451043Cat1Ch1-|-976751556Cat1Ch1-|-374476473Cat1Ch2
Current_Vertical_Sep = Values of Current_Vertical_Sep = Greater (strictly) than MAXALTDIF
(600)|Two_of_Three_Reports_Valid = Values of Two_of_Three_Reports_Valid = True|Own_Tracked_Alt =
Values of Own_Tracked_Alt = Higher (strictly) than other (intruder)|Other_Tracked_Alt = Values
Other_Tracked_Alt = Lower (strictly) than TCAS (own)|Up_Separation = Values of Up_Separation =
Greater (strictly) than down separation and threshold|Down_Separation = Values of
Down_Separation = Smaller (strictly) than up separation and threshold|Alt_Layer_Value = Values
of Alt_Layer_Value = Value 0|Other_RAC = Values of Other_RAC = Zero|Other_Capability = Values of
```

Other_Capability = One|Climb_Inhibit = Values of Climb_Inhibit = Zero|Positive_RA_Alt_Thresh =
Values of Positive_RA_Alt_Thresh = Value0|High_Confidence = Values of High_Confidence = False

values:     Current_Vertical_Sep = '601', Alt_Layer_Value = 'Value0', Down_Separation = '399',
            Up_Separation = '401', Other_RAC = 'Zero', High_Confidence = 'false',
            Other_Capability = 'One', Other_Tracked_Alt = '1', Own_Tracked_Alt = '2',
            Positive_RA_Alt_Thresh = '400', Climb_Inhibit = 'Zero', Two_of_Three_Reports_Valid
            = 'true'

**Test Case 5:**

choices:    1871720133Cat1Ch1-|-374476473Cat1Ch1-|-603277445Cat1Ch1-|-1787817979Cat1Ch3-|-
            927131798Cat1Ch1-|-875777795Cat1Ch2-|-302439181Cat1Ch1-|-58393125Cat1Ch1-|-
            626069991Cat1Ch1-|-430451043Cat1Ch1-|-976751556Cat1Ch1-|-1325844154Cat1Ch2

Current_Vertical_Sep = Values of Current_Vertical_Sep = Greater (strictly) than MAXALTDIF
(600)|High_Confidence = Values of High_Confidence = True|Own_Tracked_Alt = Values of
Own_Tracked_Alt = Higher (strictly) than other (intruder)|Other_Tracked_Alt = Values
Other_Tracked_Alt = Lower (strictly) than TCAS (own)|Up_Separation = Values of Up_Separation =
Greater (strictly) than down separation and threshold|Down_Separation = Values of
Down_Separation = Smaller (strictly) than up separation and threshold|Alt_Layer_Value = Values
of Alt_Layer_Value = Value 0|Other_RAC = Values of Other_RAC = Zero|Other_Capability = Values of
Other_Capability = One|Climb_Inhibit = Values of Climb_Inhibit = Zero|Positive_RA_Alt_Thresh =
Values of Positive_RA_Alt_Thresh = Value0|Two_of_Three_Reports_Valid = Values of
Two_of_Three_Reports_Valid = False

values:     Current_Vertical_Sep = '601', Alt_Layer_Value = 'Value0', Down_Separation = '1',
            Up_Separation = '401', Other_RAC = 'Zero', High_Confidence = 'true',
            Other_Tracked_Alt = '1', Other_Capability = 'One', Own_Tracked_Alt = '2',
            Positive_RA_Alt_Thresh = '400', Climb_Inhibit = 'Zero', Two_of_Three_Reports_Valid
            = 'false'

**Test Case 6:**

choices:    302439181Cat1Ch1-|-976751556Cat1Ch1-|-603277445Cat1Ch2-|-1787817979Cat1Ch2-|-
            927131798Cat1Ch2-|-875777795Cat1Ch2-|-58393125Cat1Ch2-|-1871720133Cat1Ch1-|-
            374476473Cat1Ch2-|-1325844154Cat1Ch2-|-626069991Cat1Ch2-|-430451043Cat1Ch2

Alt_Layer_Value = Values of Alt_Layer_Value = Value 0|Positive_RA_Alt_Thresh = Values of
Positive_RA_Alt_Thresh = Value0|Own_Tracked_Alt = Values of Own_Tracked_Alt = Same as other
(intruder)|Other_Tracked_Alt = Values Other_Tracked_Alt = Same as TCAS (own)|Up_Separation =
Values of Up_Separation = Smaler (strictly) than down separation and threshold|Down_Separation =
Values of Down_Separation = Between up separation and threshold|Other_RAC = Values of Other_RAC
= One|Current_Vertical_Sep = Values of Current_Vertical_Sep = Smaller than MAXALTDIF (600), but
positive (strictly)|High_Confidence = Values of High_Confidence =
False|Two_of_Three_Reports_Valid = Values of Two_of_Three_Reports_Valid = False|Other_Capability
= Values of Other_Capability = Two|Climb_Inhibit = Values of Climb_Inhibit = One

values:     Current_Vertical_Sep = '1', Alt_Layer_Value = 'Value0', Down_Separation = '2',
            Up_Separation = '1', Other_RAC = 'One', High_Confidence = 'false',
            Other_Tracked_Alt = '1', Other_Capability = 'Two', Own_Tracked_Alt = '1',
            Positive_RA_Alt_Thresh = '400', Climb_Inhibit = 'One', Two_of_Three_Reports_Valid =
            'false'

**Test Case 7:**

choices:    302439181Cat1Ch2-|-976751556Cat1Ch2-|-603277445Cat1Ch3-|-1787817979Cat1Ch1-|-
            927131798Cat1Ch3-|-875777795Cat1Ch3-|-58393125Cat1Ch3-|-1871720133Cat1Ch3-|-
            374476473Cat1Ch1-|-1325844154Cat1Ch1-|-626069991Cat1Ch1-|-430451043Cat1Ch1

Alt_Layer_Value = Values of Alt_Layer_Value = Value 1|Positive_RA_Alt_Thresh = Values of
Positive_RA_Alt_Thresh = Value1|Own_Tracked_Alt = Values of Own_Tracked_Alt = Lower (strictly)
than other (intruder)|Other_Tracked_Alt = Values Other_Tracked_Alt = Higher (strictly) than TCAS
(own)|Up_Separation = Values of Up_Separation = Between down separation and
threshold|Down_Separation = Values of Down_Separation = Greater (strictly) than up separation
and threshold|Other_RAC = Values of Other_RAC = Two|Current_Vertical_Sep = Values of
Current_Vertical_Sep = Greater (strictly) than MAXALTDIF (600)|High_Confidence = Values of
High_Confidence = True|Two_of_Three_Reports_Valid = Values of Two_of_Three_Reports_Valid =
True|Other_Capability = Values of Other_Capability = One|Climb_Inhibit = Values of Climb_Inhibit
= Zero

values:     Current_Vertical_Sep = '601', Alt_Layer_Value = 'Value1', Down_Separation = '501',
            Up_Separation = '500', Other_RAC = 'Two', High_Confidence = 'true',
            Other_Tracked_Alt = '2', Other_Capability = 'One', Own_Tracked_Alt = '1',
            Positive_RA_Alt_Thresh = '500', Climb_Inhibit = 'Zero', Two_of_Three_Reports_Valid
            = 'true'

**Test Case 8:**

choices:    603277445Cat1Ch4

```
Own_Tracked_Alt = Values of Own_Tracked_Alt = Illegal
values:        Own_Tracked_Alt = '0'
```

**Test Case 9:**
```
choices:       302439181Cat1Ch2-|-976751556Cat1Ch2-|-603277445Cat1Ch3-|-1787817979Cat1Ch1-|-
               927131798Cat1Ch3-|-875777795Cat1Ch1-|-58393125Cat1Ch3-|-1871720133Cat1Ch1-|-
               374476473Cat1Ch1-|-1325844154Cat1Ch1-|-626069991Cat1Ch1-|-430451043Cat1Ch1
Alt_Layer_Value = Values of Alt_Layer_Value = Value 1|Positive_RA_Alt_Thresh = Values of
Positive_RA_Alt_Thresh = Value1|Own_Tracked_Alt = Values of Own_Tracked_Alt = Lower (strictly)
than other (intruder)|Other_Tracked_Alt = Values Other_Tracked_Alt = Higher (strictly) than TCAS
(own)|Up_Separation = Values of Up_Separation = Between down separation and
threshold|Down_Separation = Values of Down_Separation = Greater (strictly) than up separation
and threshold|Other_RAC = Values of Other_RAC = Two|Current_Vertical_Sep = Values of
Current_Vertical_Sep = Greater (strictly) than MAXALTDIF (600)|High_Confidence = Values of
High_Confidence = True|Two_of_Three_Reports_Valid = Values of Two_of_Three_Reports_Valid =
True|Other_Capability = Values of Other_Capability = One|Climb_Inhibit = Values of Climb_Inhibit
= Zero
values:        Current_Vertical_Sep = '601', Alt_Layer_Value = 'Value1', Down_Separation = '501',
               Up_Separation = '500', Other_RAC = 'Two', High_Confidence = 'true',
               Other_Tracked_Alt = '2', Other_Capability = 'One', Own_Tracked_Alt = '1',
               Positive_RA_Alt_Thresh = '500', Climb_Inhibit = 'Zero', Two_of_Three_Reports_Valid
               = 'true'
```
**Test Case 10:**
```
choices:       302439181Cat1Ch1-|-976751556Cat1Ch1-|-603277445Cat1Ch2-|-1787817979Cat1Ch2-|-
               927131798Cat1Ch2-|-875777795Cat1Ch3-|-58393125Cat1Ch2-|-1871720133Cat1Ch2-|-
               374476473Cat1Ch2-|-1325844154Cat1Ch2-|-626069991Cat1Ch2-|-430451043Cat1Ch2
Alt_Layer_Value = Values of Alt_Layer_Value = Value 0|Positive_RA_Alt_Thresh = Values of
Positive_RA_Alt_Thresh = Value0|Own_Tracked_Alt = Values of Own_Tracked_Alt = Same as other
(intruder)|Other_Tracked_Alt = Values Other_Tracked_Alt = Same as TCAS (own)|Up_Separation =
Values of Up_Separation = Smaler (strictly) than down separation and threshold|Down_Separation =
Values of Down_Separation = Between up separation and threshold|Other_RAC = Values of Other_RAC
= One|Current_Vertical_Sep = Values of Current_Vertical_Sep = Smaller than MAXALTDIF (600), but
positive (strictly)|High_Confidence = Values of High_Confidence =
False|Two_of_Three_Reports_Valid = Values of Two_of_Three_Reports_Valid = False|Other_Capability
= Values of Other_Capability = Two|Climb_Inhibit = Values of Climb_Inhibit = One
values:        Current_Vertical_Sep = '1', Alt_Layer_Value = 'Value0', Down_Separation = '2',
               Up_Separation = '1', Other_RAC = 'One', High_Confidence = 'false',
               Other_Tracked_Alt = '1', Other_Capability = 'Two', Own_Tracked_Alt = '1',
               Positive_RA_Alt_Thresh = '400', Climb_Inhibit = 'One', Two_of_Three_Reports_Valid =
               'false'
```
**Test Case 11:**
```
choices:       1787817979Cat1Ch4
Other_Tracked_Alt = Values Other_Tracked_Alt = Illegal
values:        Other_Tracked_Alt = '0'
```
**Test Case 12:**
```
choices:       302439181Cat1Ch1-|-976751556Cat1Ch1-|-603277445Cat1Ch2-|-1787817979Cat1Ch2-|-
               927131798Cat1Ch2-|-875777795Cat1Ch3-|-58393125Cat1Ch2-|-1871720133Cat1Ch2-|-
               374476473Cat1Ch2-|-1325844154Cat1Ch2-|-626069991Cat1Ch2-|-430451043Cat1Ch2
Alt_Layer_Value = Values of Alt_Layer_Value = Value 0|Positive_RA_Alt_Thresh = Values of
Positive_RA_Alt_Thresh = Value0|Own_Tracked_Alt = Values of Own_Tracked_Alt = Same as other
(intruder)|Other_Tracked_Alt = Values Other_Tracked_Alt = Same as TCAS (own)|Up_Separation =
Values of Up_Separation = Smaler (strictly) than down separation and threshold|Down_Separation =
Values of Down_Separation = Between up separation and threshold|Other_RAC = Values of Other_RAC
= One|Current_Vertical_Sep = Values of Current_Vertical_Sep = Smaller than MAXALTDIF (600), but
positive (strictly)|High_Confidence = Values of High_Confidence =
False|Two_of_Three_Reports_Valid = Values of Two_of_Three_Reports_Valid = False|Other_Capability
= Values of Other_Capability = Two|Climb_Inhibit = Values of Climb_Inhibit = One
values:        Current_Vertical_Sep = '1', Alt_Layer_Value = 'Value0', Down_Separation = '2',
               Up_Separation = '1', Other_RAC = 'One', High_Confidence = 'false',
               Other_Tracked_Alt = '1', Other_Capability = 'Two', Own_Tracked_Alt = '1',
               Positive_RA_Alt_Thresh = '400', Climb_Inhibit = 'One', Two_of_Three_Reports_Valid =
               'false'
```
**Test Case 13:**
```
choices:       1871720133Cat1Ch1-|-374476473Cat1Ch1-|-1325844154Cat1Ch1-|-603277445Cat1Ch1-|-
               1787817979Cat1Ch3-|-875777795Cat1Ch2-|-302439181Cat1Ch1-|-58393125Cat1Ch1-|-
               626069991Cat1Ch1-|-430451043Cat1Ch1-|-976751556Cat1Ch1-|-927131798Cat1Ch3
```

Current_Vertical_Sep = Values of Current_Vertical_Sep = Greater (strictly) than MAXALTDIF
(600)|High_Confidence = Values of High_Confidence = True|Two_of_Three_Reports_Valid = Values of
Two_of_Three_Reports_Valid = True|Own_Tracked_Alt = Values of Own_Tracked_Alt = Higher
(strictly) than other (intruder)|Other_Tracked_Alt = Values Other_Tracked_Alt = Lower (strictly)
than TCAS (own)|Down_Separation = Values of Down_Separation = Smaller (strictly) than up
separation and threshold|Alt_Layer_Value = Values of Alt_Layer_Value = Value 0|Other_RAC =
Values of Other_RAC = Zero|Other_Capability = Values of Other_Capability = One|Climb_Inhibit =
Values of Climb_Inhibit = Zero|Positive_RA_Alt_Thresh = Values of Positive_RA_Alt_Thresh =
Value0|Up_Separation = Values of Up_Separation = Between down separation and threshold
values:        Two_of_Three_Reports_Valid = 'true', Current_Vertical_Sep = '601', Alt_Layer_Value
               = 'Value0', Down_Separation = '1', Up_Separation = '2', Other_RAC = 'Zero',
               High_Confidence = 'true', Other_Tracked_Alt = '1', Other_Capability = 'One',
               Own_Tracked_Alt = '2', Positive_RA_Alt_Thresh = '400', Climb_Inhibit = 'Zero'

**Test Case 14:**
choices:       927131798Cat1Ch4
Up_Separation = Values of Up_Separation = Illegal
values:        Up_Separation = '0'

**Test Case 15:**
choices:       302439181Cat1Ch2—|—976751556Cat1Ch2—|—603277445Cat1Ch3—|—1787817979Cat1Ch1—|—
               927131798Cat1Ch3—|—875777795Cat1Ch1—|—58393125Cat1Ch3—|—1871720133Cat1Ch1—|—
               374476473Cat1Ch1—|—1325844154Cat1Ch1—|—626069991Cat1Ch1—|—430451043Cat1Ch1
Alt_Layer_Value = Values of Alt_Layer_Value = Value 1|Positive_RA_Alt_Thresh = Values of
Positive_RA_Alt_Thresh = Value1|Own_Tracked_Alt = Values of Own_Tracked_Alt = Lower (strictly)
than other (intruder)|Other_Tracked_Alt = Values Other_Tracked_Alt = Higher (strictly) than TCAS
(own)|Up_Separation = Values of Up_Separation = Between down separation and
threshold|Down_Separation = Values of Down_Separation = Greater (strictly) than up separation
and threshold|Other_RAC = Values of Other_RAC = Two|Current_Vertical_Sep = Values of
Current_Vertical_Sep = Greater (strictly) than MAXALTDIF (600)|High_Confidence = Values of
High_Confidence = True|Two_of_Three_Reports_Valid = Values of Two_of_Three_Reports_Valid =
True|Other_Capability = Values of Other_Capability = One|Climb_Inhibit = Values of Climb_Inhibit
= Zero
values:        Current_Vertical_Sep = '601', Alt_Layer_Value = 'Value1', Down_Separation = '501',
               Up_Separation = '500', Other_RAC = 'Two', High_Confidence = 'true',
               Other_Tracked_Alt = '2', Other_Capability = 'One', Own_Tracked_Alt = '1',
               Positive_RA_Alt_Thresh = '500', Climb_Inhibit = 'Zero', Two_of_Three_Reports_Valid
               = 'true'

**Test Case 16:**
choices:       1871720133Cat1Ch1—|—374476473Cat1Ch1—|—1325844154Cat1Ch1—|—603277445Cat1Ch1—|—
               1787817979Cat1Ch3—|—927131798Cat1Ch1—|—302439181Cat1Ch1—|—58393125Cat1Ch1—|—
               626069991Cat1Ch1—|—430451043Cat1Ch1—|—875777795Cat1Ch3
Current_Vertical_Sep = Values of Current_Vertical_Sep = Greater (strictly) than MAXALTDIF
(600)|High_Confidence = Values of High_Confidence = True|Two_of_Three_Reports_Valid = Values of
Two_of_Three_Reports_Valid = True|Own_Tracked_Alt = Values of Own_Tracked_Alt = Higher
(strictly) than other (intruder)|Other_Tracked_Alt = Values Other_Tracked_Alt = Lower (strictly)
than TCAS (own)|Up_Separation = Values of Up_Separation = Greater (strictly) than down
separation and threshold|Alt_Layer_Value = Values of Alt_Layer_Value = Value 0|Other_RAC =
Values of Other_RAC = Zero|Other_Capability = Values of Other_Capability = One|Climb_Inhibit =
Values of Climb_Inhibit = Zero|Positive_RA_Alt_Thresh = Values of Positive_RA_Alt_Thresh =
Value0|Down_Separation = Values of Down_Separation = Between up separation and threshold
values:        Current_Vertical_Sep = '601', Alt_Layer_Value = 'Value0', Down_Separation = '400',
               Up_Separation = '401', Other_RAC = 'Zero', High_Confidence = 'true',
               Other_Tracked_Alt = '1', Other_Capability = 'One', Own_Tracked_Alt = '2',
               Climb_Inhibit = 'Zero', Positive_RA_Alt_Thresh = '400', Two_of_Three_Reports_Valid
               = 'true'

**Test Case 17:**
choices:       875777795Cat1Ch4
Down_Separation = Values of Down_Separation = Illegal
values:        Down_Separation = '0'

**Test Case 18:**
choices:       302439181Cat1Ch2—|—976751556Cat1Ch2—|—603277445Cat1Ch3—|—1787817979Cat1Ch1—|—
               927131798Cat1Ch3—|—875777795Cat1Ch1—|—58393125Cat1Ch3—|—1871720133Cat1Ch1—|—
               374476473Cat1Ch1—|—1325844154Cat1Ch1—|—626069991Cat1Ch1—|—430451043Cat1Ch1
Alt_Layer_Value = Values of Alt_Layer_Value = Value 1|Positive_RA_Alt_Thresh = Values of
Positive_RA_Alt_Thresh = Value1|Own_Tracked_Alt = Values of Own_Tracked_Alt = Lower (strictly)
than other (intruder)|Other_Tracked_Alt = Values Other_Tracked_Alt = Higher (strictly) than TCAS

(own)|Up_Separation = Values of Up_Separation = Between down separation and threshold|Down_Separation = Values of Down_Separation = Greater (strictly) than up separation and threshold|Other_RAC = Values of Other_RAC = Two|Current_Vertical_Sep = Values of Current_Vertical_Sep = Greater (strictly) than MAXALTDIF (600)|High_Confidence = Values of High_Confidence = True|Two_of_Three_Reports_Valid = Values of Two_of_Three_Reports_Valid = True|Other_Capability = Values of Other_Capability = One|Climb_Inhibit = Values of Climb_Inhibit = Zero

values:    Current_Vertical_Sep = '601', Alt_Layer_Value = 'Value1', Down_Separation = '501', Up_Separation = '500', Other_RAC = 'Two', High_Confidence = 'true', Other_Tracked_Alt = '2', Other_Capability = 'One', Own_Tracked_Alt = '1', Positive_RA_Alt_Thresh = '500', Climb_Inhibit = 'Zero', Two_of_Three_Reports_Valid = 'true'

**Test Case 19:**

choices:    302439181Cat1Ch3−|−976751556Cat1Ch3−|−603277445Cat1Ch1−|−1787817979Cat1Ch3−|−927131798Cat1Ch1−|−875777795Cat1Ch2−|−58393125Cat1Ch1−|−1871720133Cat1Ch2−|−374476473Cat1Ch2−|−1325844154Cat1Ch1−|−626069991Cat1Ch2−|−430451043Cat1Ch2

Alt_Layer_Value = Values of Alt_Layer_Value = Value 2|Positive_RA_Alt_Thresh = Values of Positive_RA_Alt_Thresh = Value2|Own_Tracked_Alt = Values of Own_Tracked_Alt = Higher (strictly) than other (intruder)|Other_Tracked_Alt = Values Other_Tracked_Alt = Lower (strictly) than TCAS (own)|Up_Separation = Values of Up_Separation = Greater (strictly) than down separation and threshold|Down_Separation = Values of Down_Separation = Smaller (strictly) than up separation and threshold|Other_RAC = Values of Other_RAC = Zero|Current_Vertical_Sep = Values of Current_Vertical_Sep = Smaller than MAXALTDIF (600), but positive (strictly)|High_Confidence = Values of High_Confidence = False|Two_of_Three_Reports_Valid = Values of Two_of_Three_Reports_Valid = True|Other_Capability = Values of Other_Capability = Two|Climb_Inhibit = Values of Climb_Inhibit = One

values:    Current_Vertical_Sep = '1', Alt_Layer_Value = 'Value2', Down_Separation = '1', Up_Separation = '641', Other_RAC = 'Zero', High_Confidence = 'false', Other_Tracked_Alt = '1', Other_Capability = 'Two', Own_Tracked_Alt = '2', Climb_Inhibit = 'One', Positive_RA_Alt_Thresh = '640', Two_of_Three_Reports_Valid = 'true'

**Test Case 20:**

choices:    302439181Cat1Ch4−|−976751556Cat1Ch4−|−603277445Cat1Ch3−|−1787817979Cat1Ch1−|−927131798Cat1Ch1−|−875777795Cat1Ch3−|−58393125Cat1Ch3−|−1871720133Cat1Ch1−|−374476473Cat1Ch1−|−1325844154Cat1Ch2−|−626069991Cat1Ch2−|−430451043Cat1Ch2

Alt_Layer_Value = Values of Alt_Layer_Value = Value 3|Positive_RA_Alt_Thresh = Values of Positive_RA_Alt_Thresh = Value3|Own_Tracked_Alt = Values of Own_Tracked_Alt = Lower (strictly) than other (intruder)|Other_Tracked_Alt = Values Other_Tracked_Alt = Higher (strictly) than TCAS (own)|Up_Separation = Values of Up_Separation = Greater (strictly) than down separation and threshold|Down_Separation = Values of Down_Separation = Between up separation and threshold|Other_RAC = Values of Other_RAC = Two|Current_Vertical_Sep = Values of Current_Vertical_Sep = Greater (strictly) than MAXALTDIF (600)|High_Confidence = Values of High_Confidence = True|Two_of_Three_Reports_Valid = Values of Two_of_Three_Reports_Valid = False|Other_Capability = Values of Other_Capability = Two|Climb_Inhibit = Values of Climb_Inhibit = One

values:    Current_Vertical_Sep = '601', Alt_Layer_Value = 'Value3', Down_Separation = '740', Up_Separation = '741', Other_RAC = 'Two', High_Confidence = 'true', Other_Tracked_Alt = '2', Other_Capability = 'Two', Own_Tracked_Alt = '1', Positive_RA_Alt_Thresh = '740', Climb_Inhibit = 'One', Two_of_Three_Reports_Valid = 'false'

**Test Case 21:**

choices:    1871720133Cat1Ch1−|−374476473Cat1Ch1−|−1325844154Cat1Ch1−|−603277445Cat1Ch1−|−1787817979Cat1Ch3−|−927131798Cat1Ch1−|−875777795Cat1Ch2−|−302439181Cat1Ch1−|−626069991Cat1Ch1−|−430451043Cat1Ch1−|−976751556Cat1Ch1−|−58393125Cat1Ch2

Current_Vertical_Sep = Values of Current_Vertical_Sep = Greater (strictly) than MAXALTDIF (600)|High_Confidence = Values of High_Confidence = True|Two_of_Three_Reports_Valid = Values of Two_of_Three_Reports_Valid = True|Own_Tracked_Alt = Values of Own_Tracked_Alt = Higher (strictly) than other (intruder)|Other_Tracked_Alt = Values Other_Tracked_Alt = Lower (strictly) than TCAS (own)|Up_Separation = Values of Up_Separation = Greater (strictly) than down separation and threshold|Down_Separation = Values of Down_Separation = Smaller (strictly) than up separation and threshold|Alt_Layer_Value = Values of Alt_Layer_Value = Value 0|Other_Capability = Values of Other_Capability = One|Climb_Inhibit = Values of Climb_Inhibit = Zero|Positive_RA_Alt_Thresh = Values of Positive_RA_Alt_Thresh = Value0|Other_RAC = Values of Other_RAC = One

```
values:          Current_Vertical_Sep = '601', Alt_Layer_Value = 'Value0', Down_Separation = '399',
                 Up_Separation = '401', Other_RAC = 'One', High_Confidence = 'true',
                 Other_Tracked_Alt = '1', Other_Capability = 'One', Own_Tracked_Alt = '2',
                 Positive_RA_Alt_Thresh = '400', Climb_Inhibit = 'Zero', Two_of_Three_Reports_Valid
                 = 'true'
```

**Test Case 22:**

```
choices:         1871720133Cat1Ch1-|-374476473Cat1Ch1-|-1325844154Cat1Ch1-|-603277445Cat1Ch1-|-
                 1787817979Cat1Ch3-|-927131798Cat1Ch1-|-875777795Cat1Ch2-|-302439181Cat1Ch1-|-
                 626069991Cat1Ch1-|-430451043Cat1Ch1-|-976751556Cat1Ch1-|-58393125Cat1Ch3
```

Current_Vertical_Sep = Values of Current_Vertical_Sep = Greater (strictly) than MAXALTDIF
(600)|High_Confidence = Values of High_Confidence = True|Two_of_Three_Reports_Valid = Values of
Two_of_Three_Reports_Valid = True|Own_Tracked_Alt = Values of Own_Tracked_Alt = Higher
(strictly) than other (intruder)|Other_Tracked_Alt = Values Other_Tracked_Alt = Lower (strictly)
than TCAS (own)|Up_Separation = Values of Up_Separation = Greater (strictly) than down
separation and threshold|Down_Separation = Values of Down_Separation = Smaller (strictly) than
up separation and threshold|Alt_Layer_Value = Values of Alt_Layer_Value = Value
0|Other_Capability = Values of Other_Capability = One|Climb_Inhibit = Values of Climb_Inhibit =
Zero|Positive_RA_Alt_Thresh = Values of Positive_RA_Alt_Thresh = Value0|Other_RAC = Values of
Other_RAC = Two

```
values:          Current_Vertical_Sep = '601', Alt_Layer_Value = 'Value0', Down_Separation = '1',
                 Up_Separation = '401', Other_RAC = 'Two', High_Confidence = 'true',
                 Other_Tracked_Alt = '1', Other_Capability = 'One', Own_Tracked_Alt = '2',
                 Positive_RA_Alt_Thresh = '400', Climb_Inhibit = 'Zero', Two_of_Three_Reports_Valid
                 = 'true'
```

**Test Case 23:**

```
choices:         1871720133Cat1Ch1-|-374476473Cat1Ch1-|-1325844154Cat1Ch1-|-603277445Cat1Ch1-|-
                 1787817979Cat1Ch3-|-927131798Cat1Ch1-|-875777795Cat1Ch2-|-302439181Cat1Ch1-|-
                 58393125Cat1Ch1-|-430451043Cat1Ch1-|-976751556Cat1Ch1-|-626069991Cat1Ch2
```

Current_Vertical_Sep = Values of Current_Vertical_Sep = Greater (strictly) than MAXALTDIF
(600)|High_Confidence = Values of High_Confidence = True|Two_of_Three_Reports_Valid = Values of
Two_of_Three_Reports_Valid = True|Own_Tracked_Alt = Values of Own_Tracked_Alt = Higher
(strictly) than other (intruder)|Other_Tracked_Alt = Values Other_Tracked_Alt = Lower (strictly)
than TCAS (own)|Up_Separation = Values of Up_Separation = Greater (strictly) than down
separation and threshold|Down_Separation = Values of Down_Separation = Smaller (strictly) than
up separation and threshold|Alt_Layer_Value = Values of Alt_Layer_Value = Value 0|Other_RAC =
Values of Other_RAC = Zero|Climb_Inhibit = Values of Climb_Inhibit = Zero|Positive_RA_Alt_Thresh
= Values of Positive_RA_Alt_Thresh = Value0|Other_Capability = Values of Other_Capability = Two

```
values:          Current_Vertical_Sep = '601', Alt_Layer_Value = 'Value0', Down_Separation = '1',
                 Up_Separation = '401', Other_RAC = 'Zero', High_Confidence = 'true',
                 Other_Tracked_Alt = '1', Other_Capability = 'Two', Own_Tracked_Alt = '2',
                 Positive_RA_Alt_Thresh = '400', Climb_Inhibit = 'Zero', Two_of_Three_Reports_Valid
                 = 'true'
```

**Test Case 24:**

```
choices:         1871720133Cat1Ch1-|-374476473Cat1Ch1-|-1325844154Cat1Ch1-|-603277445Cat1Ch1-|-
                 1787817979Cat1Ch3-|-927131798Cat1Ch1-|-875777795Cat1Ch2-|-302439181Cat1Ch1-|-
                 58393125Cat1Ch1-|-626069991Cat1Ch1-|-976751556Cat1Ch1-|-430451043Cat1Ch2
```

Current_Vertical_Sep = Values of Current_Vertical_Sep = Greater (strictly) than MAXALTDIF
(600)|High_Confidence = Values of High_Confidence = True|Two_of_Three_Reports_Valid = Values of
Two_of_Three_Reports_Valid = True|Own_Tracked_Alt = Values of Own_Tracked_Alt = Higher
(strictly) than other (intruder)|Other_Tracked_Alt = Values Other_Tracked_Alt = Lower (strictly)
than TCAS (own)|Up_Separation = Values of Up_Separation = Greater (strictly) than down
separation and threshold|Down_Separation = Values of Down_Separation = Smaller (strictly) than
up separation and threshold|Alt_Layer_Value = Values of Alt_Layer_Value = Value 0|Other_RAC =
Values of Other_RAC = Zero|Other_Capability = Values of Other_Capability =
One|Positive_RA_Alt_Thresh = Values of Positive_RA_Alt_Thresh = Value0|Climb_Inhibit = Values of
Climb_Inhibit = One

```
values:          Current_Vertical_Sep = '601', Alt_Layer_Value = 'Value0', Down_Separation = '399',
                 Up_Separation = '401', Other_RAC = 'Zero', High_Confidence = 'true',
                 Other_Tracked_Alt = '1', Other_Capability = 'One', Own_Tracked_Alt = '2',
                 Positive_RA_Alt_Thresh = '400', Climb_Inhibit = 'One', Two_of_Three_Reports_Valid =
                 'true'
```

**Test Case 25:**

```
choices:         302439181Cat1Ch2-|-976751556Cat1Ch2-|-603277445Cat1Ch3-|-1787817979Cat1Ch1-|-
                 927131798Cat1Ch3-|-875777795Cat1Ch1-|-58393125Cat1Ch3-|-1871720133Cat1Ch1-|-
                 374476473Cat1Ch1-|-1325844154Cat1Ch1-|-626069991Cat1Ch1-|-430451043Cat1Ch1
```

```
Alt_Layer_Value = Values of Alt_Layer_Value = Value 1|Positive_RA_Alt_Thresh = Values of
Positive_RA_Alt_Thresh = Value1|Own_Tracked_Alt = Values of Own_Tracked_Alt = Lower (strictly)
than other (intruder)|Other_Tracked_Alt = Values of Other_Tracked_Alt = Higher (strictly) than TCAS
(own)|Up_Separation = Values of Up_Separation = Between down separation and
threshold|Down_Separation = Values of Down_Separation = Greater (strictly) than up separation
and threshold|Other_RAC = Values of Other_RAC = Two|Current_Vertical_Sep = Values of
Current_Vertical_Sep = Greater (strictly) than MAXALTDIF (600)|High_Confidence = Values of
High_Confidence = True|Two_of_Three_Reports_Valid = Values of Two_of_Three_Reports_Valid =
True|Other_Capability = Values of Other_Capability = One|Climb_Inhibit = Values of Climb_Inhibit
= Zero
values:      Current_Vertical_Sep = '601', Alt_Layer_Value = 'Value1', Down_Separation = '501',
             Up_Separation = '500', Other_RAC = 'Two', High_Confidence = 'true',
             Other_Tracked_Alt = '2', Other_Capability = 'One', Own_Tracked_Alt = '1',
             Positive_RA_Alt_Thresh = '500', Climb_Inhibit = 'Zero', Two_of_Three_Reports_Valid
             = 'true'
```

**Test Case 26:**
```
choices:     302439181Cat1Ch3-|-976751556Cat1Ch3-|-603277445Cat1Ch1-|-1787817979Cat1Ch3-|-
             927131798Cat1Ch1-|-875777795Cat1Ch2-|-58393125Cat1Ch1-|-1871720133Cat1Ch2-|-
             374476473Cat1Ch2-|-1325844154Cat1Ch1-|-626069991Cat1Ch2-|-430451043Cat1Ch2
Alt_Layer_Value = Values of Alt_Layer_Value = Value 2|Positive_RA_Alt_Thresh = Values of
Positive_RA_Alt_Thresh = Value2|Own_Tracked_Alt = Values of Own_Tracked_Alt = Higher (strictly)
than other (intruder)|Other_Tracked_Alt = Values of Other_Tracked_Alt = Lower (strictly) than TCAS
(own)|Up_Separation = Values of Up_Separation = Greater (strictly) than down separation and
threshold|Down_Separation = Values of Down_Separation = Smaller (strictly) than up separation
and threshold|Other_RAC = Values of Other_RAC = Zero|Current_Vertical_Sep = Values of
Current_Vertical_Sep = Smaller than MAXALTDIF (600), but positive (strictly)|High_Confidence =
Values of High_Confidence = False|Two_of_Three_Reports_Valid = Values of
Two_of_Three_Reports_Valid = True|Other_Capability = Values of Other_Capability =
Two|Climb_Inhibit = Values of Climb_Inhibit = One
values:      Current_Vertical_Sep = '1', Alt_Layer_Value = 'Value2', Down_Separation = '1',
             Up_Separation = '641', Other_RAC = 'Zero', High_Confidence = 'false',
             Other_Tracked_Alt = '1', Other_Capability = 'Two', Own_Tracked_Alt = '2',
             Climb_Inhibit = 'One', Positive_RA_Alt_Thresh = '640', Two_of_Three_Reports_Valid =
             'true'
```

**Test Case 27:**
```
choices:     302439181Cat1Ch4-|-976751556Cat1Ch4-|-603277445Cat1Ch3-|-1787817979Cat1Ch1-|-
             927131798Cat1Ch1-|-875777795Cat1Ch3-|-58393125Cat1Ch3-|-1871720133Cat1Ch1-|-
             374476473Cat1Ch1-|-1325844154Cat1Ch2-|-626069991Cat1Ch2-|-430451043Cat1Ch2
Alt_Layer_Value = Values of Alt_Layer_Value = Value 3|Positive_RA_Alt_Thresh = Values of
Positive_RA_Alt_Thresh = Value3|Own_Tracked_Alt = Values of Own_Tracked_Alt = Lower (strictly)
than other (intruder)|Other_Tracked_Alt = Values Other_Tracked_Alt = Higher (strictly) than TCAS
(own)|Up_Separation = Values of Up_Separation = Greater (strictly) than down separation and
threshold|Down_Separation = Values of Down_Separation = Between up separation and
threshold|Other_RAC = Values of Other_RAC = Two|Current_Vertical_Sep = Values of
Current_Vertical_Sep = Greater (strictly) than MAXALTDIF (600)|High_Confidence = Values of
High_Confidence = True|Two_of_Three_Reports_Valid = Values of Two_of_Three_Reports_Valid =
False|Other_Capability = Values of Other_Capability = Two|Climb_Inhibit = Values of
Climb_Inhibit = One
values:      Current_Vertical_Sep = '601', Alt_Layer_Value = 'Value3', Down_Separation = '740',
             Up_Separation = '741', Other_RAC = 'Two', High_Confidence = 'true',
             Other_Tracked_Alt = '2', Other_Capability = 'Two', Own_Tracked_Alt = '1',
             Positive_RA_Alt_Thresh = '740', Climb_Inhibit = 'One', Two_of_Three_Reports_Valid =
             'false'
```

**Test Case 28:**
```
choices:     976751556Cat1Ch5
Positive_RA_Alt_Thresh = Values of Positive_RA_Alt_Thresh = Illegal
values:      Positive_RA_Alt_Thresh = '2'
```